\documentclass[aps,prd,superscriptaddress,showpacs,nofootinbib,eqsecnum,amsfonts,amsmath,amssymb,ucs,youngtab]{revtex4-1}

\usepackage{epsfig}
\usepackage{longtable}
\usepackage{youngtab}
\usepackage{graphicx}
\usepackage[pagebackref,
colorlinks=true,linkcolor=blue,linktoc=page]{hyperref}
\usepackage{color}
\usepackage{multirow}

\def\be{\begin{equation}}
\def\ee{\end{equation}}
\def\bea{\begin{eqnarray}}
\def\eea{\end{eqnarray}}

\def\la{\langle}
\def\ra{\rangle}

\def\t{\tilde }

\newcommand{\lsim}{\raisebox{-0.13cm}{~\shortstack{$<$ \\[-0.07cm] $\sim$}}~}
\newcommand{\gsim}{\raisebox{-0.13cm}{~\shortstack{$>$ \\[-0.07cm] $\sim$}}~}

\begin{document}


\title{ Non-perturbative analysis of the spectrum of meson resonances
\vspace{.1cm}\\
in an ultraviolet-complete composite-Higgs model
\vspace{.5cm}}

\author{Nicolas Bizot}
\author{Michele Frigerio}
\affiliation{Laboratoire Charles Coulomb (L2C), UMR 5221 CNRS-Universit\'e de Montpellier, Montpellier, France}
\author{Marc Knecht}
\affiliation{Centre de Physique Th\'eorique (CPT),
 UMR 7332 CNRS/Aix-Marseille Univ./Univ. du Sud Toulon-Var, Marseille, France 
 \vspace{1cm}}
 \author{Jean-Lo\"{\i}c Kneur}
\affiliation{Laboratoire Charles Coulomb (L2C), UMR 5221 CNRS-Universit\'e de Montpellier, Montpellier, France}

\begin{abstract} 
We consider a vector-like gauge theory of fermions that confines at the multi-TeV scale, and that realizes the Higgs particle as
a composite Goldstone boson. The weak interactions are embedded in the unbroken subgroup $Sp(4)$ of a 
spontaneously broken $SU(4)$ flavour group. The meson resonances appear as poles in the two-point correlators of 
fermion bilinears, and include the Goldstone bosons plus a massive pseudoscalar $\eta'$, as well as scalars, vectors and axial vectors.
We compute the mass spectrum of these mesons, as well as their decay constants, in the chiral limit, 
in the approximation where the hypercolour $Sp(2N)$ dynamics is described by four-fermion  operators,
\`a la Nambu-Jona Lasinio. By resumming the leading diagrams in the $1/N$ expansion, we
find that the spin-one states lie beyond the LHC reach, while spin-zero electroweak-singlet states may be as light as 
the Goldstone-boson decay constant, $f\sim 1$ TeV. We also confront our results with a set of available spectral sum rules.
In order to supply composite top-quark partners, the theory contains additional fermions carrying both hypercolour and ordinary colour,
with an associated flavour symmetry-breaking pattern $SU(6)/SO(6)$. We identify and analyse several non-trivial features
of the complete two-sector gauge theory: 
the 't~Hooft anomaly 
matching conditions; the higher-dimension operator which incorporates the  effects of the hypercolour axial-singlet anomaly;
the coupled mass-gap equations; the mixing between the singlet mesons of the two sectors, resulting in an extra Goldstone boson $\eta_0$, and novel spectral sum rules. Assuming that the strength of 
the four-fermion interaction is the same in the two sectors,
we find that the coloured vector and scalar mesons have masses $\gtrsim 4 f$, while the masses of coloured
pseudo-Goldstone bosons, induced by gluon loops, are $\gtrsim 1.5f$. We discuss the scaling of the meson masses 
with the values of $N$, of the four-fermion couplings, and of a possible fermion mass.
\end{abstract}


\maketitle

\tableofcontents

\section{Introduction}

After the first LHC 13 TeV data have been analysed, we are left with a 125 GeV Higgs boson and no evidence for other new states.
Yet, it is too early to remove from consideration sufficiently weakly-coupled new particles in the sub-TeV range, or even new 
coloured particles in the multi-TeV range.
Even though the little hierarchy between the Higgs mass and the new states seem to require an adjustment of parameters,
the theories addressing the quantum stability of the electroweak scale may still solve larger hierarchy problems.
A classical possibility is a strongly-coupled sector that dynamically generates the electroweak scale. 
The observation of a scalar state, significantly lighter than the strong-coupling scale, suggests 
that the Higgs particle may be composite and,
in good approximation, a Nambu-Goldstone boson (NGB) associated to the global symmetries of the new sector 
\cite{Kaplan:1983fs,Kaplan:1983sm,Dugan:1984hq,Agashe:2004rs}.
While an effective description of the composite Higgs couplings is possible without specifying the strong dynamics, 
the spectrum of additional composite states essentially depends on the underlying ultraviolet theory.
Barring extra space-time dimensions, the simplest, well-understood, explicit realization is provided by a gauge theory of 
fermions that confines at the multi-TeV scale, 
with quantum chromodynamics (QCD) as a prototype. 
The historical incarnation being technicolor \cite{Weinberg:1975gm,Susskind:1978ms}, in recent years models of this sort featuring
the Higgs as a composite NGB have been built 
\cite{Galloway:2010bp,Barnard:2013zea,Cacciapaglia:2014uja,Ferretti:2014qta,
Vecchi:2015fma,Ma:2015gra} 
and classified in some generality \cite{Ferretti:2013kya}\cite{Vecchi:2015fma}. 
Alternative ultraviolet completions of composite Higgs models are discussed in Refs.~\cite{Caracciolo:2012je,vonGersdorff:2015fta,Fichet:2016xvs,Galloway:2016fuo}.

Our motivations to analyse in detail such a scenario are manifold. A characterisation of the spectrum of composite states 
is critical to confront with the LHC program:
does one foresee Standard Model (SM) singlet resonances close to one TeV? what are the expectations for the masses of the 
lightest charged and colour states? These
intrinsically non-perturbative questions are especially  
pressing, in order to compare with the well-defined predictions of weakly-coupled theories. 
In addition, a quantitative description of the composite masses and
couplings would allow for an explicit computation of the Higgs low energy properties, 
improving on the predictivity of the composite Higgs effective theory.
Furthermore, decades of QCD studies have provided us with a notable collection of non-perturbative, 
analytic techniques to study strongly-coupled gauge theories,
that have been hardly exploited in the context of models for the electroweak scale. 
A partial list includes anomaly matching \cite{'tHooft:1979bh}, spectral sum rules \cite{Weinberg:1967kj}, 
large-$N$ expansions \cite{'tHooft:1973jz,Witten:1979kh}, and the Nambu-Jona Lasinio (NJL) effective model 
\cite{Nambu:1961tp,Nambu:1961fr} (see also Refs.~\cite{Klevansky:1992qe,Hatsuda:1994pi}). 
With this approach one can reach several non-trivial  results, holding within well-defined approximations, 
with a relatively small computational effort, and thus one may broadly characterise several, different, possible models. 
This is complementary to lattice simulations,
which are suitable for potentially more precise computations, in specific and/or simplified scenarios.
Interestingly, we will also find that the peculiar structure of
composite Higgs models requires a gauge theory that is qualitatively different from QCD, in a handful of significant features.

We engage into this program by choosing, as a case study, an electroweak sector with global symmetry $SU(4)$ 
spontaneously broken to $Sp(4)$. This is the most economical
possibility to obtain a Nambu-Goldstone Higgs doublet with custodial symmetry, starting from a set of constituent fermions. 
This model, with a hypercolour gauge group $Sp(2N)$, emerges as the minimal benchmark for an 
ultraviolet-complete composite Higgs sector. The most significant challenge facing this class of theories
is to generate the large top quark Yukawa coupling, as it requires non-renormalisable operators to couple 
the top to the electroweak symmetry breaking (EWSB) order parameter.
A promising way to circumvent the potential suppression of the top Yukawa is partial compositeness \cite{Kaplan:1991dc}, 
which calls for composite fermion resonances with the quantum number of the top quark. A minimal realization of top partial 
compositeness is provided by an additional sector of hypercolour fermions, which are charged under QCD,
with global symmetry $SU(6)$ spontaneously broken to $SO(6)$.
While this particular choice for the colour sector appears less compelling than the one 
for the electroweak sector, we will show that it is instructive to study it explicitly in detail. 
Indeed, one needs to surmount a number of model-building difficulties, which
require quite technical complications: on the one hand this assesses the price to pay for top partners, on the 
other hand the interplay of the two sectors reveals a few novel physical phenomena, 
whose interest transcends the specific model under consideration.

Our analysis builds on an early, enlightening study \cite{Barnard:2013zea}, which employed four-fermion 
operators to understand the dynamics of this $SU(4)\times SU(6)$ model with hypercolour group $Sp(2N)$,
in close analogy with the NJL description of QCD
(NJL techniques have been applied to different ultraviolet-complete composite-Higgs models as well \cite{vonGersdorff:2015fta}). 
We will provide the first, thorough computation 
of the spectrum of the meson resonances in this scenario. To this end, we will perform a detailed
scrutiny of the symmetry structure of the model, which allows for several non-trivial consistency checks, 
as well as for an accurate determination of the allowed range of parameters.
In most of our analysis, we will stick to the chiral limit, where the constituent 
fermions have no bare masses, and the SM gauge and Yukawa couplings are neglected.
In this limit the Higgs and the other NGBs are massless. When relevant, we will discuss in some 
detail the effect of fermion masses and of switching on the SM gauge fields, however we will not study 
the generation of Yukawa couplings and of the NGB effective potential: the usual effective 
theory techniques to address these issues \cite{Contino:2010rs,Panico:2015jxa} hold in the present scenario as well,
but we leave for future work a more specific treatment of this subject.

The paper is organised as follows. In section \ref{general} we review exact results on vector-like gauge theories, 
especially concerning the spontaneous breaking of the flavour symmetries,
the associated spectral sum rules, the NGB couplings to external gauge fields. 
The reader more interested in the phenomenology of a specific model may just consult this part to inspect general formulas and conventions.
In section \ref{The electroweak sector} we study the electroweak sector with coset $SU(4)/Sp(4)$, in terms of four-fermion operators, \`a la NJL. 
The symmetry breaking is examined through the gap equation for the dynamical fermion mass, while the spin-zero and spin-one meson masses are extracted from the poles of resummed two-point correlators.
The spectrum of resonances is analysed in units of the NGB decay constant, and compared with available lattice results, as well  as with spectral sum rules.
This analysis of the electroweak sector in isolation is self-sufficient and it already illustrates the main potentialities of our approach.
The following sections require some extra model-building and rather technical computations, that however may be skipped to move directly to the phenomenological results.
In section \ref{coloured-sector} we introduce additional, coloured constituent fermions, in a different representation of $Sp(2N)$, to provide partners for the top quark. 
The consequences include non-trivial anomaly matching conditions, mixed sum rules across the two sectors, and mixed operators induced by the hypercolour gauge anomaly.
In section \ref{The spectrum of mesonic resonances in the coloured sector} we study the system of coupled mass-gap equations for the two sectors and 
derive the masses of coloured mesons.
In addition, the mixing between the two flavour singlet (pseudo)scalars leads to a peculiar mass spectrum and phenomenology. Finally, in section \ref{conclusion}
we summarise the main results of the analysis and delineate future directions. 
Technical material is collected in the appendices: the generators of the flavour symmetry group in appendix \ref{generators}, the relevant loop functions in appendix \ref{loop-functions},
some details on the computation of two-point correlators in appendix \ref{SDresum}, and the Fierz identities relating different four-fermion operators in appendix \ref{fierz}.

\section{General properties of flavour symmetries in vector-like gauge theories}\label{general}

The composite-Higgs model that we will study belongs to the class of vector-like gauge theories, namely
an asymptotically free and confining gauge theory, with
a set of $N_f$ Dirac fermions transforming under a (possibly reducible) self-contragredient
(i.e. unitarily equivalent to its complex conjugate) representation of the gauge group,
in such a way that it is possible to make all fermions massive in a gauge
invariant way\footnote{It is also possible to give all fermions gauge invariant masses in the
case of an odd number of Weyl fermions in the same {\it real} representation of the gauge group. Such theories do not have a conserved
fermion number, and are not vector-like \cite{Vafa:1983tf,Kosower:1984aw}. Although it can provide interesting composite-Higgs
models, as discussed, for instance, in Ref. \cite{Ferretti:2014qta}, this class of theories will not be addressed here.}.
Exact results concerning non-perturbative dynamical aspects in these theories are scarce,
and in this section we briefly review some of those that are actually available. They concern issues
related to the spontaneous breaking of the global flavour symmetries and the spectrum of low-lying
bound states.

\subsection{Restrictions on the pattern of spontaneous symmetry breaking}
\label{VW}

An important result for the spontaneous breaking of the global flavour symmetry group $G$ for
fermions with vector-like couplings to gauge fields has been obtained by Vafa and Witten
\cite{Vafa:1983tf}. The theorem they have proven makes the following statement:
{\it in any vector-like gauge theory with massless fermions and vanishing vacuum angles,
the subgroup $H_m$ of the flavour group $G$ that corresponds to the remaining global symmetry when all
fermion flavours are given identical gauge invariant masses, cannot be spontaneously broken}.
In other words, if $G$ undergoes spontaneous breaking towards some subgroup $H$,
then $H_m \subseteq H$ (in the absence of any vacuum angle). This theorem is particularly
powerful when $H_m$ corresponds to a maximal subgroup of $G$, since it then allows
only two alternatives: either $G$ is not spontaneously
broken at all, or $G$ is spontaneously broken towards $H_m$. This is actually what happens
in the three cases that we can encounter in vector-like theories \cite{Peskin:1980gc,Kogan:1984nb}: $G = SU(N_f)_L \times SU(N_f)_R $
and $H_m = SU(N_f)_V $\footnote{The issue of the $U(1)_V$ symmetry is somewhat subtle,
but we will not need to discuss it here.};
$G = SU(2 N_f)$ and $H_m = SO(2 N_f)$;  $G = SU(2 N_f)$ and $H_m = Sp(2 N_f)$.

Of particular interest for the discussion that follows are the Noether currents $\mathcal{J}^A_\mu$,
corresponding to the generators $T^A$ of the unbroken subgroup $H_m$, and ${\cal J}^{\hat A}_\mu$, corresponding to the generators
$T^{\hat A}$ in the coset $G/H_m$. Since the latter is a symmetric space for the three cases that
have just been listed, we will usually refer to the currents $\mathcal{J}^A_\mu$
(${\cal J}^{\hat A}_\mu$) as vector (axial) currents. When the fermions transform
under an irreducible but real ($\epsilon = +1$ below) or pseudo-real ($\epsilon = -1$)
representation of the gauge group, $G=SU(2N_f)$, and $H_m = SO (2 N_f)$ or $H_m = Sp (2N_f)$,
respectively.
In these two cases, it is convenient to write the fermion fields in terms of left-handed Weyl spinors $\psi_\alpha$.
The currents are then defined as follow [$\overline{\psi}_i \equiv \psi_j^\dagger \left(\Omega_\varepsilon \right)_{ji}$,
where $i$ and $j$ denote gauge indices, while spinor and flavour indices are omitted]:
\begin{equation}
\mathcal{J}^A_\mu= {\displaystyle\frac{1}{2}} \left(\Omega_\varepsilon \right)_{ij}
\left[  \varepsilon  \overline{\psi}_i \overline{\sigma}_\mu  T^A  \psi_j -\psi_i \sigma_\mu \big(T^A \big)^T \overline{\psi}_j \right] ~,
\qquad
\qquad
{\cal J}^{\hat A}_\mu= {\displaystyle\frac{1}{2}} \left(\Omega_\varepsilon \right)_{ij}
\left[ \varepsilon  \overline{\psi}_i \overline{\sigma}_\mu  T^{\hat A}  \psi_j -\psi_i \sigma_\mu \big(T^{\hat A} \big)^T \overline{\psi}_j   \right] ~.
\label{Jdef}
\end{equation}
The gauge contraction $\Omega_\varepsilon$ is an invariant tensor under the action of
the gauge group, which is symmetric for $\varepsilon=+1$ and  antisymmetric for $\varepsilon=-1$,
with $\big(\Omega_\varepsilon^2\big)_{ij} = \varepsilon \delta_{ij}$. The generators
$T^A$ and $T^{\hat A}$ are characterised by the properties
\begin{equation}
T^A\Sigma_\varepsilon +\Sigma_\varepsilon \big( T^A \big)^T=0~,
\qquad\qquad
T^{\hat A} \Sigma_\varepsilon -\Sigma_\varepsilon \big( T^{\hat A} \big)^T=0~,
\label{Tacom}
\end{equation}
and are normalised as
\be
{\rm Tr} (T^A T^B) = \frac{1}{2} \delta^{AB}~,
\qquad\qquad
{\rm Tr} (T^{\hat A} T^{\hat B}) = \frac{1}{2} \delta^{{\hat A}{\hat B}}~,
\qquad\qquad
{\rm Tr} (T^A T^{\hat B}) = 0~.
\label{norm_T}
\ee
The $2 N_f \times 2 N_f$ matrix $\Sigma_\varepsilon$ is an invariant tensor
of the subgroup $H_m$ of the flavour group. It plays for this subgroup a role analogous to the
role played by $\Omega_\varepsilon$ for the gauge group. In particular, it can be chosen real,
it is symmetric for $\varepsilon=+1$ and  antisymmetric for $\varepsilon=-1$,
and satisfies $\Sigma_\varepsilon^2 = \varepsilon 1\!\!1$, where $1\!\!1$ denotes the
$2 N_f \times 2 N_f$ unit matrix in flavour space.

\subsection{'t Hooft's anomaly matching condition}
\label{anomat}

Whereas the theorem of Vafa and Witten restricts the pattern of spontaneous breaking
of the global flavour symmetry group $G$, it does not by itself provide information
on which alternative will eventually be realized. Additional information is required
to that effect. The anomaly matching condition proposed by 't Hooft \cite{'tHooft:1979bh}
can prove helpful in this respect. This condition uses the fact that the Ward identities
satisfied by the three-point functions of the Noether currents corresponding to the
symmetry group $G$ receive anomalous contributions from the massless elementary fermions
\cite{BJanom,Aanom,AB2anom}
\be
i (q_1 + q_2)^\rho \int d^4 x_1 \int d^4 x_2 \, e^{i q_1 \cdot x_1 + i q_2 \cdot x_2}
\langle {\rm vac} \vert T \{ {\cal J}^A_\mu (x_1) {\cal J}^B_\nu (x_2) {\cal J}^{\hat C}_\rho (0)\}
\vert {\rm vac} \rangle
=
- \frac{d_{HC}}{8 \pi^2} \, \epsilon_{\mu\nu\alpha\beta} q_1^\alpha q_2^\beta d^{AB{\hat C}}~,
\label{Anom_WI}
\ee
with $d^{AB{\hat C}} = 2 {\rm tr}(\{T^A , T^B \} T^{\hat C})$, where the trace is over
the flavour group only, and $d_{HC}$ denotes the dimension
of the representation of the gauge group under which the fermions transform.
These anomalous contributions imply that
the corresponding three-point functions have very specific physical singularities at vanishing
momentum transfer \cite{'tHooft:1979bh,Frishman:1980dq,Coleman:1982yg}. Moreover, this
type of singularities can only be produced by physical
intermediate states consisting either of a single massless spin zero particle, or of a pair of 
massless spin one-half particles. If the symmetries of $G$ are
not spontaneously broken, the first option is excluded. If the theory confines, this
then implies that it has to produce massless spin one-half bound states (that we will call baryons).
These fermionic bound states will occur in multiplets of $G$, and their multiplicities
must be chosen such as to exactly reproduce the coefficient of the singularities
in the current three-point functions. If it is not possible to saturate this anomaly
coefficient with the exchange of massless fermionic bound states only, then massless
spin-zero bound states coupled to the currents of $G$ are required, and hence $G$ is spontaneously
broken. If this anomaly matching condition can be satisfied with massless spin one-half
bound states only, the spontaneous breaking of $G$ towards $H_m$ is not a necessity, but it cannot
be excluded either.

In particular, the global symmetry is necessarily spontaneously broken if, after confinement, the
theory cannot produce fermionic bound states at all.
If we restrict ourselves to constituent fermions in the fundamental representation of the gauge group,
this happens when the gauge group is  $SU(2N)$, $SO(2N)$, or $Sp(2N)$. In these cases,
the flavour group $G$ therefore necessarily suffers spontaneous breaking towards $H_m$.
On the contrary, fermionic bound states can be formed in the case of $SU(N)$ or $SO(N)$ gauge groups with $N$ odd.
Novel fermionic bound states may be possible if one admits elementary fermions transforming
in other representations than the fundamental under the gauge group.
We will discuss one such scenario below in section IV.

\subsection{Mass inequalities}\label{inequalities}

Various inequalities \cite{Weingarten:1983uj,Witten:1983ut,Nussinov:1983vh,Espriu:1984mq,Nussinov:1984kr}
involving the masses of the gauge-singlet bound states in confining vector-like gauge theories provide additional
insight into the fate of flavour symmetries in these theories, complementary to
the constraints arising from the Vafa-Witten theorem and from 't Hooft's anomaly matching condition.
The most rigorous versions of these inequalities hold under the same positivity constraint
on the path-integral measure in euclidian space as required for the
proof of the Vafa-Witten theorem, namely the absence of any vacuum angle.
A review on these inequalities is provided by Ref. \cite{Nussinov:1999sx}. Of particular interest in the present context
is the inequality of the type \cite{Weingarten:1983uj,Nussinov:1983vh,Espriu:1984mq,Nussinov:1984kr}
\be
M_{1/2} \ge C(N,N_f)  M_0~,
\ee
involving, on the one hand, the mass $M_{1/2}$ of any baryon state and,
on the other hand, the mass $M_0$  of the lightest quark-antiquark spin-zero state
having the flavour quantum numbers of the  $G/H_m$ currents.
The precise value of the (positive) constant $C(N,N_f)$
and its dependence on the number of hypercolours $N$ and/or number of flavours $N_f$
is not so important here, the main point being that such an inequality
again provides a strong indication that the flavour symmetry $G$ is necessarily
spontaneously broken towards $G/H_m$.

\subsection{Super-convergent spectral sum rules}
\label{SR}

Assuming that $G$ is spontaneously broken towards $H_m$, correlation functions that are at the same time
order parameters become of particular interest, since they enjoy a smooth behaviour
at short distances. These improved high-energy properties allow in turn
to write super-convergent sum rules for the corresponding spectral densities.
The paradigmatic example is provided by the Weinberg sum rules \cite{Weinberg:1967kj},
once interpreted \cite{Bernard:1975cd} and justified in the framework of QCD and of the
operator-product expansion \cite{Wilson:1969zs}, including non-perturbative power
corrections \cite{Shifman:1978by}.

Here we will consider two-point functions of certain fermion-bilinear operators,
when the fermions transform under an irreducible but real or pseudo-real representation of the gauge group.
Specifically, these operators comprise the Noether currents defined in Eq. (\ref{Jdef}),
to which we add the scalar and pseudoscalar densities defined as
\bea
\label{S_and_P}
&
{\cal S}^{\hat A} = {\displaystyle\frac{1}{2}} \left(\Omega_\varepsilon \right)_{ij}
\left[
\overline{\psi}_i T^{\hat A} \Sigma_\varepsilon \overline{\psi}_j
+
\psi_i \Sigma_\varepsilon T^{\hat A} \psi_j
\right] ~,
\qquad\qquad
&
{\cal S}^{0} = {\displaystyle\frac{1}{2}} \left(\Omega_\varepsilon \right)_{ij}
\left[
\overline{\psi}_i T^{0} \Sigma_\varepsilon \overline{\psi}_j
+
\psi_i \Sigma_\varepsilon T^{0} \psi_j
\right] ~,
\nonumber\\
\\
&
{\cal P}^{\hat A} = {\displaystyle\frac{1}{2i}} \left(\Omega_\varepsilon \right)_{ij}
\left[
\overline{\psi}_i T^{\hat A} \Sigma_\varepsilon \overline{\psi}_j
-
\psi_i \Sigma_\varepsilon T^{\hat A} \psi_j
\right] ~,
\qquad\qquad
&
{\cal P}^{0} = {\displaystyle\frac{1}{2i}} \left(\Omega_\varepsilon \right)_{ij}
\left[
\overline{\psi}_i T^{0} \Sigma_\varepsilon \overline{\psi}_j
-
\psi_i \Sigma_\varepsilon T^{0} \psi_j
\right] ~.
\nonumber
\eea
The singlet densities are normalised consistently with the other densities
by taking $T^0= 1\!\!1 /(2\sqrt{ N_f})$.
The two-point correlation functions of interest are then defined as
\bea
\Pi_V (q^2) \delta^{AB} (q_\mu q_\nu - \eta_{\mu\nu} q^2)
& = &
i \int d^4 x \, e^{i q \cdot x} \langle {\rm vac} \vert T \{ {\cal J}^A_\mu (x) {\cal J}^B_\nu (0) \}  \vert {\rm vac} \rangle
~,
\nonumber\\
\Pi_A (q^2) \delta^{{\hat A}{\hat B}} (q_\mu q_\nu - \eta_{\mu\nu} q^2)
& = &
i \int d^4 x \, e^{i q \cdot x} \langle {\rm vac} \vert T \{ {\cal J}_\mu^{\hat A} (x) {\cal J}^{\hat B}_\nu (0) \}  \vert {\rm vac} \rangle
~,
\label{PiVAdef}
\eea
\bea
\Pi_S (q^2) \delta^{{\hat A}{\hat B}} & = &
i \int d^4 x \, e^{i q \cdot x} \langle {\rm vac} \vert T \{ {\cal S}^{\hat A} (x) {\cal S}^{\hat B} (0) \}  \vert {\rm vac} \rangle
~,
\nonumber\\
\Pi_P (q^2) \delta^{{\hat A}{\hat B}} & = &
i \int d^4 x \, e^{i q \cdot x} \langle {\rm vac} \vert T \{ {\cal P}^{\hat A} (x) {\cal P}^{\hat B} (0) \}  \vert {\rm vac} \rangle
~,
\label{PiPSdef}
\eea
where ${\hat A} \neq 0$, ${\hat B} \neq 0$, and
\bea
\Pi_{S^0} (q^2) & = &
i \int d^4 x \, e^{i q \cdot x} \langle {\rm vac} \vert T \{ {\cal S}^0 (x) {\cal S}^0 (0) \}  \vert {\rm vac} \rangle
~,
\nonumber\\
\Pi_{P^0} (q^2) & = &
i \int d^4 x \, e^{i q \cdot x} \langle {\rm vac} \vert T \{ {\cal P}^0 (x) {\cal P}^0 (0) \}  \vert {\rm vac} \rangle
~.
\label{PiP0S0def}
\eea
The combinations
\be
\Pi_{V{\mbox -}A} (q^2) \equiv \Pi_V (q^2) - \Pi_A (q^2)
~,
\ee
\be
\ \Pi_{S{\mbox -}P} (q^2) \equiv \Pi_S (q^2) - \Pi_P (q^2)
~,
\qquad
\ \Pi_{S{\mbox -}P^0} (q^2) \equiv \Pi_S (q^2) - \Pi_{P^0} (q^2)
~,
\qquad
\ \Pi_{S^0{\mbox -}P} (q^2) \equiv \Pi_{S^0} (q^2) - \Pi_P (q^2)
~,
\ee
are order parameters\footnote{Concerning $\Pi_{S{\mbox -}P} (q^2)$, this statement 
and the ensuing sum rule hold
only to the extent that the tensor $d^{{\hat A}{\hat B}{\hat C}} \equiv 2 {\rm tr}(\{T^{\hat A} , T^{\hat B} \} T^{\hat C})$
does not vanish identically, which is not the case, for instance, when $G=SU(2)_L \times SU(2)_R$ and $H_m=SU(2)_V$,
but also, more interestingly for our purposes, when $G=SU(4)$ and $H_m=Sp(4)$.\label{fnte_SP}}
for the spontaneous breaking of $SU(2 N_f)$ towards $H_m$ for all values of $q^2$.
As a consequence, these two-point functions behave smoothly at short distances $(Q^2 \equiv - q^2 > 0)$:
\be
\lim_{Q^2 \to + \infty} \left( Q^2 \right)^2 \times \Pi_{V{\mbox -}A} (-Q^2) = 0
~,
\qquad
\lim_{Q^2 \to + \infty} Q^2 \times \{ \Pi_{S{\mbox -}P}  (-Q^2) \, ; \, \Pi_{S^0{\mbox -}P} (-Q^2)
\, ; \,  \Pi_{S{\mbox -}P^0} (-Q^2) \} = \{ 0 \, ; \, 0 \, ; \, 0 \}~.
\ee
From these short-distance properties, one then derives the following super-convergent spectral sum rules
\be
\int_0^\infty \!\! dt \, {\rm Im} \Pi_{V{\mbox -}A} (t) = 0~,
\qquad\qquad
\int_0^\infty \!\! dt \, t \,{\rm Im} \Pi_{V{\mbox -}A} (t) = 0~,
\label{WSRVA}
\ee
\be
\int_0^\infty \!\! dt \, {\rm Im} \Pi_{S{\mbox -}P} (t) = 0~,
\qquad\qquad
\int_0^\infty \!\! dt \, {\rm Im} \Pi_{S^0{\mbox -}P} (t) = 0~,
\qquad\qquad
\int_0^\infty \!\! dt \, {\rm Im} \Pi_{S{\mbox -}P^0} (t) = 0~.
\label{scalSR}
\ee
We will examine in the following to which extent these Weinberg-type sum rules, whose validity
is quite general in view of the short-distance properties of asymptotically-free vector-like
gauge theories, are actually satisfied in the specific NJL four-fermion interaction approximation.
For the sake of completeness, let us mention that the two-point function
\be
\Pi_{AP} (q^2) \delta^{{\hat A}{\hat B}} q_\mu =
\int d^4 x \, e^{i q \cdot x} \langle {\rm vac} \vert T \{ {\cal J}_\mu^{\hat A} (x) {\cal P}^{\hat B} (0) \}  \vert {\rm vac} \rangle~,
\label{PiAPdef}
\ee
also defines an order-parameter. However, there is no associated sum rule, since,
as a consequence of the Ward identities,
this correlator is entirely saturated by the Goldstone-boson pole
($\langle {\mathcal S}^0 \rangle$ denotes the vacuum expectation
value of ${\mathcal S}^0$)
\be
\Pi_{AP} (q^2) = \frac{1}{q^2} \, \frac{\langle {\mathcal S}^0 \rangle}{\sqrt{N_f}}
~.
\label{PiAP}
\ee

It may be useful to stress, at this stage, that the sum rules displayed above are only valid
in the absence of any explicit symmetry breaking effects. Introducing, for instance, masses
for the fermions would modify the short-distance properties of these correlators,
and thus spoil the convergence of the integrals of the corresponding spectral functions.
Let us briefly illustrate the changes that occur by giving the fermions a common mass $m$,
so that the currents belonging to the subgroup $H_m$ remain conserved. For the remaining
currents, one now has
\be
\partial^\mu {\cal J}^{\hat A}_\mu = 2 m {\cal P}^{\hat A}
~.
\ee
As far as the current-current correlators are concerned, while the two-point function of the vector
currents remains transverse, the correlator of two axial currents receives a longitudinal
part,
\be
i \int d^4 x \, e^{i q \cdot x} \langle {\rm vac} \vert T \{ {\cal J}_\mu^{\hat A} (x) {\cal J}^{\hat B}_\nu (0) \}  \vert {\rm vac} \rangle
=
\delta^{{\hat A}{\hat B}} \left[
\Pi_A (q^2) (q_\mu q_\nu - \eta_{\mu\nu} q^2) + \Pi_A^L (q^2) q_\mu q_\nu \right]~.
\label{PATL}
\ee
If one considers only corrections that are at most linear in $m$, then one can still write
a convergent sum rule \cite{Floratos:1978jb},
\be
\int_0^\infty \!\! dt \left[
{\rm Im} \Pi_V (t) - {\rm Im} \Pi_A (t)  - {\rm Im} \Pi_A^L (t) \right] = {\cal O} (m^2)
~.
\label{WSRmexpl}
\ee
Notice that the Ward identities relate this longitudinal piece to the two-point function of the
pseudoscalar densities and to the scalar condensate,
\be
(q^2)^2 \Pi_A^L (q^2) = 4 m^2 \Pi_P (q^2) + 2 m \frac{\langle {\mathcal S}^0 \rangle}{\sqrt{N_f}}
~.
\ee
The presence of a fermion mass $m$ also shifts the masses of the Goldstone bosons away from zero, by an amount
$\Delta_m M_{G}^2$ whose expression, at first order in $m$, actually follows from this identity and reads
\be
F_G^2 \Delta_m M_{G}^2 = - 2 m \,\frac{\langle {\mathcal S}^0 \rangle}{\sqrt{N_f}} + {\mathcal O} (m^2 \ln m)
~.
\label{Delta_m}
\ee
This formula involves the Goldstone-boson decay constant $F_G$ in the limit where $m$ vanishes, defined as
\be
\langle \, {\rm vac} \, \vert \, {\cal J}_\mu^{\hat A} (0) \, \vert G^{\hat B} (p) \rangle =
i p_\mu F_G \delta^{{\hat A}{\hat B}}
~,
\qquad 
\qquad p^2 = 0
~.
\label{FGdef}
\ee
Defining the coupling of the Goldstone bosons to the pseudoscalar
densities,
\be
\langle \, {\rm vac} \, \vert \, {\cal P}^{\hat A} (0) \, \vert G^{\hat B} (p) \rangle =
G_G \delta^{{\hat A}{\hat B}}
~,\qquad \qquad
p^2 = 0~,
\label{GGdef}
\ee
the identity obtained in Eq. (\ref{PiAP}) implies
\be
F_G G_G = - \, \frac{\langle {\mathcal S}^0 \rangle}{\sqrt{N_f}}~,
\label{FGrel}
\ee
in the massless limit.

In contrast to the symmetry currents and to quantities derived from them, like $F_G$
or $\Pi_{V/A} (q^2)$ for instance, the (pseudo)scalar densities and their matrix elements,
whether $\Pi_{S/P} (q^2)$ or $G_G$, need to be multiplicatively renormalised,
and are therefore not invariant under the action of the renormalisation
group. This dependence on the short-distance renormalisation scale does not impinge
on the validity or usefulness of the sum rules in Eqs.~(\ref{scalSR}) or (\ref{WSRmexpl}),
which hold at every scale.
Likewise, this scale dependence is exactly balanced out between the right- and left-hand
sides of relations like (\ref{PiAP}) or (\ref{FGrel}).

\subsection{Coupling to external gauge fields} \label{gauging}

Eventually, some currents of the global symmetry group $G$ become weakly coupled
to the standard model gauge fields. If, in the absence of these weakly coupled gauge
fields, the global symmetry group $G$ is spontaneously broken towards $H_m$,
turning on the gauge interactions will produce two effects. First,
the Goldstone bosons will acquire radiatively generated masses. Second,
transitions of a single Goldstone boson into a pair of gauge bosons
are induced and, at lowest order in the couplings to the external gauge fields,
the amplitude describing the transition towards a pair of zero-virtuality gauge bosons
is fixed by the anomalous Ward identities in Eq. (\ref{Anom_WI}).
Let us briefly discuss these two aspects in general terms.

Let $\vert G^{\hat A} (p) \rangle$ denote the massless Goldstone-boson states
corresponding to the generators $T^{\hat A}$ spanning the (symmetric) coset space $G/H_m$.
In the presence of a perturbation that explicitly breaks the global symmetry, these Goldstone
bosons become pseudo-Goldstone bosons, and their masses are shifted away from zero. At lowest
order in the external perturbation, these mass shifts are given by
\be
\Delta M^2_{G_{\hat A}} = - \langle G^{\hat A}(p) \vert \Delta {\cal L} (0) \vert G^{\hat A} (p) \rangle~,
\qquad \qquad
p^2=0
~,
\label{M2expl}
\ee
with $\Delta {\cal L} (x)$ the symmetry-breaking interaction term in the Lagrangian.
We are interested in particular in an interaction due to the
presence of massless gauge fields that is considered weak (in particular
non confining) at the scale under consideration, so that its
effect can be considered as a perturbation. These external gauge fields couple to some linear combinations
of the currents of the global symmetry group $G$. For a single gauge field ${\mathcal W}^\mu$,
this interaction reads
\be
{\cal L}_{\rm int} = -ig_{\mathcal W} {\mathcal W}^\mu {\mathcal J}_\mu^{\mathcal W}
~,
\qquad\qquad
\mathcal{J}^{\mathcal W}_\mu= {\displaystyle\frac{1}{2}} \left(\Omega_\varepsilon \right)_{ij}
\left[  \varepsilon  \overline{\psi}_i \overline{\sigma}_\mu T^{\mathcal W}  \psi_j
-\psi_i \sigma_\mu \big(T^{\mathcal W} \big)^T \overline{\psi}_j \right]
~,
\ee
where $T^{\mathcal W}$ is an element of the algebra of $G$.
At first non trivial order in the corresponding coupling $g_{\mathcal W}$, one has
\be
\Delta {\cal L} (x) = \frac{g_{\mathcal W}^2}{2} \int \frac{d^4 q}{(2 \pi)^4} \frac{\eta^{\mu\nu}}{q^2}
\int d^4 y \, e^{i q \cdot y}
T \{ {\mathcal J}_\mu^{\mathcal W} ( x+y) {\mathcal J}_\nu^{\mathcal W} (x) \}
~.
\label{DeltaL}
\ee
Decomposing $T^{\mathcal W}$
as $T^{\mathcal W} = T^W + T^{\hat W}$, where $T^W$ ($T^{\hat W}$) is a linear combination
of the generators $T^A$ ($T^{\hat A}$) of $H_m$ (of $G/H_m$), and taking further the
soft-Goldstone-boson limit in Eq. ~(\ref{M2expl}), then results in the following expressions for the mass shifts
\cite{Peskin:1980gc,Preskill:1980mz}
\be
\Delta M_{G_{\hat A}}^2 = - \frac{3}{4 \pi} \times \frac{1}{F_G^2} \times \frac{g_{\mathcal W}^2}{4 \pi}
\times
\int_0^\infty d Q^2 \, Q^2 \, \Pi_{V{\mbox -}A} (- Q^2)
\times
\left[ \sum_{{\hat B}} \left( f^{{\hat A}W{\hat B}} \right)^2 - \sum_{B} \left( f^{{\hat A}{\hat W}{B}} \right)^2 \right]
~.
\label{rad_masses}
\ee
Again, $F_G$ refers to the Goldstone-boson decay constant in the limit where any explicit
symmetry-breaking effects vanish, see Eq. (\ref{FGdef}), and we have used the short-hand notation
\be
{\rm Tr} \left( T^W [ T^{\hat A} , T^{\hat B} ] \right)\equiv \frac{1}{2i} \, f^{{\hat A}W{\hat B}}
~,
\qquad\qquad
{\rm Tr} \left( T^{\hat W} [ T^{\hat A} , T^B ] \right) \equiv \frac{1}{2i} \, f^{{\hat A}{\hat W}B}
~,
\ee
with the generators normalised as in Eq.~(\ref{norm_T}).
Since, according to the Witten inequality \cite{Witten:1983ut}, $- Q^2 \, \Pi_{V{\mbox -}A} (-Q^2)$ is positive,
the sign of $\Delta M_{G_{\hat A}}^2$, and hence the misalignment of the vacuum, hinges on the
sign of the last factor on the right-hand side of Eq.~(\ref{rad_masses}).
If it is positive, $\Delta M_{G_{\hat A}}^2$ is positive, and the vacuum is
stable under this perturbation by a weak gauge field. 
If it is negative, then
$\Delta M_{G_{\hat A}}^2$ is negative, which signals the instability of the unperturbed vacuum
under this perturbation. In particular, if the gauge field couples only to
the currents ${\cal J}^A_\mu$ corresponding to the unbroken generators (i.e. $T^{\hat W} = 0$),
then $\Delta M_{G_{\hat A}}^2 \ge 0$. This is the case, for instance, of the electromagnetic field
in QCD, which gives the charged pions a positive mass \cite{Das:1967it}
(see also the discussion in Ref.~\cite{Knecht:1997ts}),
\be
\Delta M_{\pi^\pm}^2 =
- \frac{3}{4} \times \frac{1}{F_\pi^2} \, \frac{\alpha}{\pi}
\times
\int_0^\infty d Q^2 \, Q^2 \, \Pi^{QCD}_{V{\mbox -}A} (- Q^2)
~,
\ee
while the neutral pion remains massless. If several gauge fields are present, the total mass shift
is given by a sum of contributions of the type (\ref{rad_masses}), one for each gauge field,
and the stability of the vacuum may then also depend on the relative strengths of the various gauge couplings.
For instance, if a subgroup $H_W$ of $H_m$ is gauged, and if the Goldstone
bosons transform as an irreducible representation $R_W$ under $H_W$, the (positive) induced mass shift
can be expressed \cite{Preskill:1980mz} in terms of the quadratic Casimir invariant of $H_W$ for the
representation $R_W$,
\be
\Delta M_{G_{\hat A}}^2 = - \frac{3}{4 \pi} \times \frac{1}{F_G^2} \times \frac{g_{\mathcal W}^2}{4 \pi}
\times
\int_0^\infty d Q^2 \, Q^2 \, \Pi_{V{\mbox -}A} (- Q^2)
\times
C_2^{(H_W)} (R_W)
~.
\label{rad_masses_Casimir}
\ee
The expression (\ref{rad_masses}) can also be rewritten as a contribution to the effective
potential induced by a gauge-field loop. In terms of the Goldstone field
\be
U(x) = e^{i G (x) /F_G} \Sigma_\varepsilon
~,\qquad\qquad
G (x) = 2 \sum_{\hat A} G^{\hat A} (x) T^{\hat A}
~,
\label{Udef}\ee
the relevant terms of the effective low-energy Lagrangian read \cite{Georgi:1986dw}
\be
{\mathcal L}_{\rm eff} = \frac{F_G^2}{4} \langle \partial_\mu U^\dagger \partial^\mu U \rangle
- C_{\mathcal W} \langle T^{\mathcal W} U \big( T^{\mathcal W}\big)^T U^\dagger \rangle
+ \cdots
~,
\ee
with $\langle \cdots \rangle$ denoting the flavour trace, and
\be
C_{\mathcal W} =  - \frac{3}{8 \pi}  \times \frac{g_{\mathcal W}^2}{4 \pi}
\times
\int_0^\infty d Q^2 \, Q^2 \, \Pi_{V{\mbox -}A} (- Q^2)
~.
\ee
As a side remark, let us notice that the procedure used here in order to
determine the induced mass shifts of the Goldstone bosons can also be
applied in the case where $\Delta {\mathcal L}$ in Eq. (\ref{M2expl})
stands for a mass term for the fermions, e.g.
\be
\Delta_m {\mathcal L} = - 2 \sqrt{N_f} \, m\, {\mathcal S}^0
~.
\ee
Going successively through the same steps, one then reproduces the
expression given in Eq. (\ref{Delta_m}).

We now turn to the second issue, namely the matrix element for the transition of a Goldstone bosons into
a pair of external gauge bosons with zero virtualities. At lowest order in the gauge couplings,
and for $q^2 = (p-q)^2 = 0$, this matrix element reads
\be
g_{\mathcal W}^2 \times i \! \int d^4 x \, e^{i q \cdot x}
\langle {\rm vac} \vert T \{ {\cal J}_\mu^{\mathcal W} (x) {\cal J}^{\mathcal W}_\nu (0) \}  \vert G^{\hat A} (p) \rangle
=
-  \frac{g_{\mathcal W}^2 d_{HC}}{8 \pi^2 F_G} \epsilon_{\mu\nu\rho\sigma} q^\rho p^\sigma d^{WW{\hat A}}
\left[ 1 + {\cal O}(m) \right]
~,
\ee
with $d^{WW{\hat A}} \equiv 2 {\rm Tr}(\{ T^W , T^W \} T^{\hat A} )$, and
$d_{HC}$ denotes the dimension of the representation of the hypercolour gauge group
to which the fermions making up the current ${\cal J}_\mu^{\mathcal W} (x)$ belong.
Here we are assuming (this will be the case of interest in the context of the composite
Higgs models discussed below) that only generators of $H_m$ are weakly coupled to the
external gauge fields (i.e. $T^{\hat W} =0$). The expression on the right-hand side is then obtained by
saturating the Ward identity in Eq. (\ref{Anom_WI}) with the Goldstone poles.
Again, if the fermions are given masses, there are corrections, indicated as ${\cal O}(m)$.
At the level of the low-energy theory, this coupling is reproduced by the Wess-Zumino-Witten
effective action \cite{Wess:1971yu,Witten:1983tw,Chu:1996fr}. Writing only the relevant term, one has
\be
{\mathcal L}_{\rm eff}^{\rm WZW} = -  \frac{g_{\mathcal W}^2 d_{HC}}{64 \pi^2 F_G}
\epsilon_{\mu\nu\rho\sigma} {\mathcal W}^{\mu\nu} (x) {\mathcal W}^{\rho\sigma} (x) \sum_{\hat A} d^{WW{\hat A}}  G^{\hat A} (x)
+ \cdots ~.
\label{WZW}
\ee

\section{The electroweak sector}
\label{The electroweak sector}

In this section we analyse a composite model for the Higgs sector of the SM.
We consider a flavour symmetry group $G=SU(4)\simeq SO(6)$, spontaneously broken towards
a subgroup $Sp(4) \simeq SO(5)$. The five Goldstone bosons transform as $(1_L,1_R)+(2_L,2_R)$ under
the custodial symmetry $SU(2)_L\times SU(2)_R\subset Sp(4)$,
corresponding to a real scalar singlet plus the complex Higgs doublet.
Composite Higgs models based on this coset have been studied in Refs.~\cite{Katz:2005au,Gripaios:2009pe,Frigerio:2012uc},
as effective theories with a non-specified strongly-coupled dynamics.
A simple UV completion is provided by a gauge theory with four Weyl fermions $\psi^a$ in a pseudo-real representation of the gauge group,
and which form a condensate $\langle \psi^a\psi^b \rangle\ne 0$.  Such a theory was considered in
Refs.~\cite{Ryttov:2008xe,Galloway:2010bp,Ferretti:2013kya,Cacciapaglia:2014uja}, as a minimal hypercolour  model.
The first analysis of the low energy dynamics of this theory in terms of four-fermion interactions (\`a la NJL)
was provided in Ref. \cite{Barnard:2013zea}.
We extend this former study by deriving additional phenomenological predictions.
We will particularise the general results of section \ref{general} to this specific case,
and in addition we will compute the masses of the spin-zero and spin-one bound states, as well as
their decay constants, by using NJL techniques.

\subsection{Scalar interactions of fermion bilinears and the mass gap}\label{colourless-sector}

Let us consider a $Sp(2N)$ hypercolour gauge theory and introduce four Weyl spinors $\psi^a$,
in the fundamental representation of $Sp(2N)$, which is pseudo-real.
The transformation properties of these elementary fermions are summarised in Table \ref{tabsu4}.
The dynamics of the $SU(4)/Sp(4)$ spontaneous symmetry breaking can be studied in terms of four-fermion interactions,
constructed out of hypercolour-invariant, spin-zero
fermion bilinears, in a NJL-like manner~\cite{Nambu:1961tp,Nambu:1961fr,Klevansky:1992qe,Hatsuda:1994pi}.
The Lagrangian
reads~\cite{Barnard:2013zea}
\begin{equation}
\mathcal{L}_{scal}^{\psi}=\frac{\kappa_A}{2N}(\psi^a \psi^b)(\overline{\psi}_a~ \overline{\psi}_b)
- \frac{\kappa_B}{8N} \left[ \epsilon_{abcd}(\psi^a \psi^b)(\psi^c \psi^d)+h.c. \right]
,
\label{LSbasic}
\end{equation}
where $a, b, \cdots=1,2,3,4$
are $SU(4)$ indices,
$\epsilon_{abcd}$ is the Levi-Civita symbol and $\kappa_{A,B}$ are real, dimensionful couplings. The phase of $\kappa_B$ can be
absorbed by the phase of $\psi$, so that we may take $\kappa_B$ real and positive without loss of generality.\footnote
{In comparison to Ref.~\cite{Barnard:2013zea}, we choose an opposite sign for $\kappa_B$,
and a different but equivalent vacuum alignment defined by Eq.~(\ref{Sigma}). Combining these two different conventions,
the mass gap defined by Eq.~(\ref{gap2}) has the same expression as in Ref.~\cite{Barnard:2013zea}.
This is because the two vacua are related by a $U(4)$ transformation with determinant minus one, that changes the sign of
$\epsilon_{abcd}$.}
Each fermion bilinear between brackets is contracted into a Lorentz and $Sp(2N)$ invariant quantity.
The hypercolour-invariant contraction is defined as
\be
(\psi^a \psi^b)\equiv  \psi_i^a \Omega_{ij} \psi_j^b = -(\psi^b \psi^a)
,
\label{bi_inv}
\ee
where $\Omega$ is the antisymmetric $2N \times 2N$ matrix
\be
\Omega= \begin{pmatrix}
0 & 1\!\!1_{N} \\ - 1\!\!1_{N} &0
\end{pmatrix}
.
\label{Omega}
\ee
The antisymmetry of the hypercolour contraction implies antisymmetry in the flavour $SU(4)$ indices.
Other four-fermion interactions,
involving spin-one fermion bilinears, are irrelevant for the discussion of spontaneous symmetry breaking.
We will introduce them later, in section \ref{psivectors}, when we discuss spin-one resonances.

\begin{table}[b]
\renewcommand{\arraystretch}{1.6}
\begin{center}
\begin{tabular}{|c|c|c|c|c|}
\hline
& Lorentz & $Sp(2 N)$ & $SU(4)$ & $Sp(4)$  \\
\hline \hline
 $\psi^a_i$ & $(1/2,0)$ & ${\Yvcentermath1 \tiny \yng(1)}_{\, i}$ & $4^a$ & 4 \\
\hline
 $\overline{\psi}_{ai} \equiv \psi^{\dagger}_{aj} \Omega_{ji}$ & $(0,1/2)$ & ${\Yvcentermath1 \tiny \yng(1)}_{\, i}$ & $\overline{4}_a$ & $4^*$ \\
\hline\hline
 $M^{ab}\sim (\psi^a \psi^b)$ & $(0,0)$ & $1$ & $6^{ab}$ & $5+1$ \\
\hline
$\overline{M}_{ab} \sim (\overline{\psi}_a \overline{\psi}_b)$ & $(0,0)$ & $1$ & $\overline{6}_{ab}$ & $5 + 1$ \\
\hline \hline
$ { a^\mu} \sim (\overline{\psi}_a \overline{\sigma}^\mu \psi^a)$ & $(1/2,1/2)$ & $1$ & $1$ & $1$ \\
\hline
~~~${ (V^\mu,A^\mu)_a^b}
\sim (\overline{\psi}_a \overline{\sigma}^\mu \psi^b)$ ~~~& ~~$(1/2,1/2)$~~ & ~~$1$~~ & ~~$15^a_b$~~ & ~~$10+5$~~ \\
\hline
\end{tabular}\end{center}
\caption{The transformation properties of the elementary fermions, and of the spin-0 and spin-1 fermion bilinears, 
in the electroweak sector of the model.
Spinor indexes are understood, and brackets stand for a hypercolour-invariant contraction
of the $Sp(2N)$ indexes. }
\label{tabsu4}
\end{table}

Note that for $\kappa_B=0$ there is an additional global $U(1)_\psi$ symmetry, which reflects
a classical invariance of the $Sp(2N)$ gauge theory, the associated Noether current being
\begin{equation}
\mathcal{J}^0_{\psi\mu} = - {\displaystyle\frac{1}{2}} \Omega_{ij}
\left[ \overline{\psi}_i \overline{\sigma}_\mu   \psi_j + \psi_i \sigma_\mu  \overline{\psi}_j \right]
,
\label{U1_psi}\end{equation}
as follows from Eq. (\ref{Jdef}) upon taking $\varepsilon = -1$ and a singlet generator normalised to $1\!\!1_4$.
At the quantum level, this current has a hypercolour gauge anomaly,
\begin{equation}
\partial^\mu \mathcal{J}^0_{\psi\mu} = {\displaystyle\frac{N_f^\psi g_{HC}^2}{32\pi^2}}
\sum_{ I=1}^{N(2N+1)} \epsilon_{\mu\nu\rho\sigma} G_{HC}^{I,\mu\nu} G_{HC}^{I,\rho\sigma}
,
\label{U1_psi_div}
\end{equation}
and the corresponding symmetry is explicitly broken by instantons \cite{'tHooft:1976up,'tHooft:1976fv}. 
Here $N_f^\psi=2$ denotes the number of Dirac flavours. The effect of the instantons can
be represented by an effective vertex \cite{'tHooft:1976up,'tHooft:1976fv,'tHooft:1986nc}
that breaks the $U(1)_\psi$ invariance. The important observation here is that for $2N_f^\psi = 4$ Weyl fermions in
the fundamental representation of the $Sp(2N)$ gauge group,
this effective vertex is precisely given by the term proportional to $\kappa_B$.
It is both quartic in the fermion fields, which provides the amount
of $U(1)_\psi$ breaking required, for $N_f^\psi = 2$, by the index theorem and the instanton solution
with unit winding number, and invariant under the $SU(4)$ global
symmetry \cite{Diakonov:1997sj}. It plays the same role as the analogous six-fermion 't Hooft determinant
effective Lagrangian~\cite{'tHooft:1976up,'tHooft:1976fv,'tHooft:1986nc} for QCD with three flavours, which
parameterises the instanton-induced anomaly interactions, explaining an $\eta'$ mass much larger than
the masses of the other Goldstone boson states. 
Such a term was originally constructed in the quark model \cite{Kobayashi:1971qz}, 
and later also introduced in the NJL model \cite{Bernard:1987gw,Bernard:1987sg}, see also \cite{Klimt:1989pm}.
Similarly, in the present case, $\kappa_B\ne 0$ is therefore crucial 
in order to evade the additional $U(1)_\psi$ Goldstone boson.

While this picture is essentially correct when considering the electroweak
$SU(4)$ sector in isolation, we stress that it will be significantly modified
when a coloured sector is introduced, in order to provide
composite partners for the top quark, as we will discuss in section \ref{coloured-sector}.
This sector also has an anomalous
extra $U(1)_X$ symmetry, but one linear combination of the two $U(1)$ currents
remains anomaly free, which implies that the effective 't Hooft determinant term is no longer given by the $\kappa_B$ operator.
This will have some important consequences on the spectrum of resonances,
but at a first stage we prefer to neglect the mixing with the coloured sector,
as the results are much more transparent and it will be easy to generalise them.

We assume that the $SU(4)$ global symmetry is exact, that is, we work in the chiral limit where $\psi^a$ has no elementary mass term.
The $SU(4)$ Noether currents are given by Eq.~(\ref{Jdef}), with $\Omega_\epsilon= \Omega$ defined in Eq.~(\ref{Omega}).
The $SU(4)$ generators decompose into
five broken ones, $T^{\hat{A}}$, living in the $SU(4)/Sp(4)$ coset, and ten unbroken ones, $T^A$,
whose explicit expressions are given in appendix \ref{generators}. They satisfy
the conditions spelled out in Eq.~(\ref{Tacom}), where $\Sigma_\epsilon$ stands for
\be
\Sigma_0 \equiv  \begin{pmatrix}
0 & 0 & 1 & 0 \\ 0 &0 & 0 & 1 \\ -1 & 0 & 0 & 0 \\ 0 & -1 & 0 & 0
\end{pmatrix}
~.
\label{Sigma}
\ee

By introducing in a standard manner~\cite{Klevansky:1992qe,Hatsuda:1994pi,Barnard:2013zea}
an auxiliary (antisymmetric) scalar field $M$, transforming as a gauge singlet and a flavour $SU(4)$ sextet,
the Lagrangian (\ref{LSbasic}) can be rewritten equivalently as
\begin{eqnarray}
\mathcal{L}_{scal}^{\psi}
&=&
-\frac{1}{\kappa_A +\kappa_B} \left[\left(\kappa_A M^*_{ab} - \frac{\kappa_B}{2}\epsilon_{abcd} M^{cd}\right)(\psi^a \psi^b) +h.c.\right]
\nonumber \\
&&
-\frac{2 N \kappa_A}{(\kappa_A+\kappa_B)^2}M^{ab} M^*_{ab} + 
\frac{1}{2}\frac{ N \kappa_B}{(\kappa_A+\kappa_B)^2}(\epsilon_{abcd} M^{ab} M^{cd} +h.c.)
~,
\label{LSaux}
\end{eqnarray}
where the equation of motion for $M$ gives
\be
M^{ab}= -\frac{\kappa_A +\kappa_B}{2 N} \left(\psi^a \psi^b \right)\;.
\ee
The matrix field $M$, being complex and antisymmetric,
can always be rotated by an $SU(4)$ transformation into the form
\be
M = \begin{pmatrix}
0 & 0 & m_1 &0  \\ 0 & 0 & 0 & m_2 \\
- m_1 & 0 & 0 & 0 \\
0 & -m_2 & 0 & 0
\end{pmatrix}
~.
\ee
Once a $(\psi^a \psi^b)$ condensate forms, $M$ acquires a vacuum expectation value (vev) and the Yukawa couplings induce
dynamical fermion masses. One can derive from Eq.~(\ref{LSaux}) the one-loop Coleman-Weinberg effective
potential~\cite{Coleman:1973jx}, by integrating over fermions,
and study the occurrence of spontaneous symmetry
breaking by looking for a non-trivial minimum with $\la m_{1,2}\ra\ne 0$ \cite{Barnard:2013zea}.
One finds that spontaneous symmetry breaking is only possible for $2\la m_1\ra =2\la m_2\ra \equiv M_\psi$,
in agreement with the Vafa-Witten theorem.
Below we provide an alternative derivation of the same result, which will also be useful for studying the spectrum of scalar resonances.

It is convenient to introduce the combination
\begin{equation}
\overline{M}_{ab} = \frac{1}{\kappa_A +\kappa_B} \left( \kappa_A M^*_{ab} -\frac{\kappa_B}{2}\epsilon_{abcd} M^{cd} \right)
,
\end{equation}
which can be expanded around the vacuum as
\begin{equation}
\overline{M} =\frac{1}{2} M_\psi \Sigma_0 +
\left( \sigma+i \eta^\prime \right) \Sigma_0 T^0_\psi  + \left(S^{\hat{A}} +i G^{\hat{A}} \right) \Sigma_0 T^{\hat{A}}
~.
\label{Mexp}
\end{equation}
The matrix $\overline{M}$ decomposes, according to $6_{SU(4)} = (1+5)_{Sp(4)}$,
into a scalar singlet $\sigma$, a pseudoscalar singlet $\eta^\prime$, a scalar quintuplet $S^{\hat{A}}$,
and a pseudoscalar quintuplet $G^{\hat{A}}$, which will be identified with the physical meson resonances.
Using the identity $ \epsilon_{abcd}= -(\Sigma_0)_{ab}(\Sigma_0)_{cd}
+(\Sigma_0)_{ac} (\Sigma_0)_{bd} -(\Sigma_0)_{ad} (\Sigma_0)_{bc}$,
and since, as already noted, $\kappa_B$ can be taken real and positive without loss
of generality, the Lagrangian (\ref{LSaux}) can be rewritten as
\begin{eqnarray}
\mathcal{L}_{scal}^{\psi} =
-(\psi  \overline{M} \psi +h.c.)
-N \left[ P_- (\sigma^2 +G_{\hat{A}}^2)+ P_+ (\eta^{\prime 2} +S_{\hat{A}}^2) \right]
~,
\label{Lscal}
\end{eqnarray}
where
\begin{equation}
P_\pm =\frac{\kappa_A}{\kappa_A^2-\kappa_B^2} \pm \frac{\kappa_B}{\left|\kappa_A^2-\kappa_B^2 \right|}
=\frac{1}{\kappa_A\mp \kappa_B}
~.
\label{Sprop}
\end{equation}
The sign in the last equality corresponds to the case $\kappa_A^2 > \kappa_B^2 $,
which will turn out to be the relevant region of parameter space.
Eqs.~(\ref{Mexp}) and (\ref{Lscal}) define the Feynman rules for the fermion Yukawa couplings to
the mesons:
the four-fermion interactions mediated by $\sigma$ and $G^{\hat{A}}$ are proportional to $P_-^{-1}$,
while the interactions mediated by $\eta^\prime$ and $S^{\hat{A}}$ are proportional to $P_+^{-1}$.

Indeed, the Lagrangian in Eq.~(\ref{LSbasic}) can be directly written in terms of the fermion bilinears
coupled to the mesons,
upon using Fierz identities for $SU(4)$ and $Sp(4)$, derived in Appendix \ref{fierz}.
The replacements
$\delta_{a}^c \delta_{b}^d - \delta_a^d\delta_b^c = 4(\Sigma_0 T^0_\psi)_{ab} ( T^0_\psi \Sigma_0)^{cd} + 4(\Sigma_0 T^{\hat A})_{ab}
( T^{\hat A} \Sigma_0)^{cd}$
and
$\epsilon_{abcd}=-4 (\Sigma_0 T^0_\psi)_{ab} (\Sigma_0 T^0_\psi)_{cd}+4  (\Sigma_0 T^{\hat A})_{ab} (\Sigma_0 T^{\hat A})_{cd}$
in Eq.~(\ref{LSbasic}), lead to
\begin{eqnarray}
{\cal L}_{scal}^\psi &=&2 \frac{\kappa_A}{(2N)}
\left[ \left(\psi \Sigma_0 T^0_\psi \psi \right) \left( \overline{\psi} T^0_\psi \Sigma_0\overline{\psi} \right)
+\left(\psi \Sigma_0 T^{\hat A} \psi \right) \left( \overline{\psi} T^{\hat A}\Sigma_0  \overline{\psi} \right)\right]
\nonumber
\\
&+& \frac{\kappa_B}{(2N)}
\left[\left(\psi \Sigma_0 T^0_\psi \psi\right) \left(\psi \Sigma_0 T^0_\psi \psi \right)
- \left(\psi\Sigma_0 T^{\hat A} \psi \right)\left(\psi\Sigma_0 T^{\hat A} \psi \right)+ h.c.\right]
~.
\label{LSphys}
\end{eqnarray}
Most of the resonance spectrum calculations could be performed directly from the
four-fermion interactions in  Eq.~(\ref{LSphys}). Nonetheless, the introduction of auxiliary fields
is convenient, because Eq.~(\ref{Mexp}) identifies the relevant scalar degrees of freedom, which
will become dynamical resonances upon $1/N$ resummation of the interactions in their respective channels,
as we will examine below.

\begin{figure}[b]
\begin{center}\includegraphics[scale=0.8, trim= 20 90 0 70]{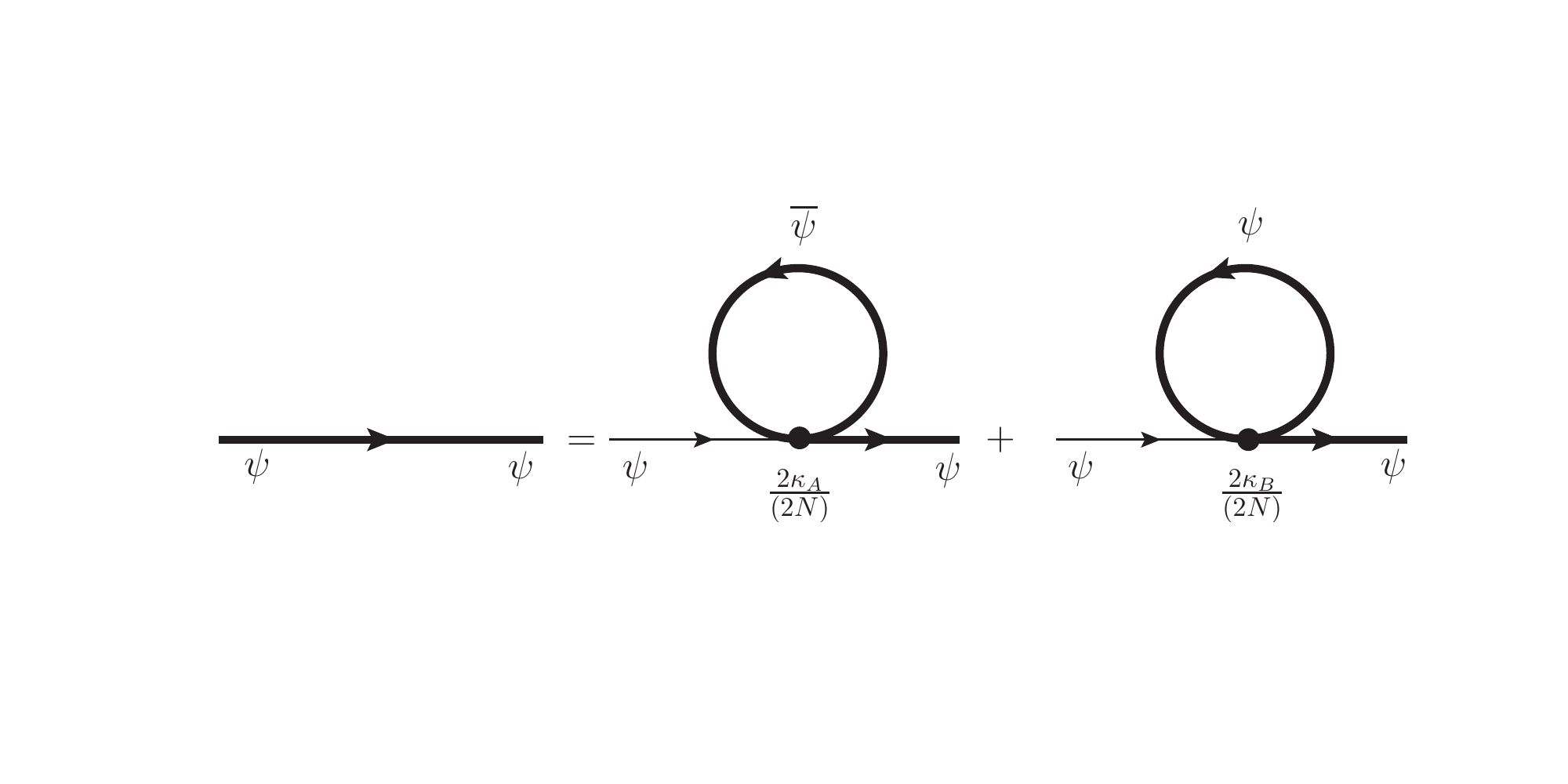}\end{center}
\caption{Graphical illustration of the mass gap equation, in the leading $1/N$-approximation. 
Thick and thin lines represent dressed and bare fermion propagators, respectively.}
\label{figgap}
\end{figure}

The first important step for the dynamical calculations of the resonance spectrum is
to determine the mass gap, namely whether a non-trivial dynamical fermion mass, signalling the spontaneous breaking
of $SU(4)$ to $Sp(4)$, develops  within the NJL approximation.
Let us consider the self-consistent mass gap equation \cite{Nambu:1961tp,Klevansky:1992qe,Hatsuda:1994pi},
obtained from the one-loop tadpole graph, as illustrated
in Fig.~\ref{figgap}. It is well-known that
this is equivalent to computing the minimum of the one-loop effective potential.
Note that, just like for the standard NJL model,
only the $\sigma$-exchange does contribute, namely only the spin-zero, parity-even, $Sp(4)$-singlet fermion bilinear
can take a vev. Therefore the mass-gap equation involves solely the inverse coupling $P_-$.
The computation of the diagrams in  Fig.  \ref{figgap}
leads to a self-consistent condition on the dynamical fermion mass $M_\psi$,
\be
-i M_\psi = 2\left( i\frac{2P_{-}^{-1}}{8(2N)}\right) (-2) Tr[\Omega^2]  Tr[\Sigma_0^2] \int^\Lambda
\frac{d^4 k}{(2\pi)^4}
\frac{i  M_\psi}{k^2-M^2_\psi+i\varepsilon}\;,
\label{gap1}
\ee
where  the first factor $2$ accounts for the normalisation $M_\psi\equiv 2\la m_{1,2} \ra$,
$(-2)$ is the trace over Weyl spinor indices in the loop, $Tr[\Omega^2]=-2 N$ is the trace over hypercolour,
and $Tr [\Sigma_0^2]=-4$ the one over flavour.
Note that the factors $2N$ cancel,
thanks to the appropriate large-$N$ normalisation of the original couplings $\kappa_{A,B}$
in Eq.~(\ref{LSphys}). Thus, one obtains
\begin{equation}
1-4 P_{-}^{-1} \t A_0( M_\psi^2) = 0
~,
\label{gap}
\end{equation}
where the basic one-loop scalar integral $\t A_0$ is defined in appendix \ref{loop-functions}.
In order to regularise the otherwise divergent integral, we introduce a (covariant 4-dimensional) cut-off
$\Lambda$,  which parameterises the scale at which the effective four-fermion interaction ceases to be valid and
all degrees of freedom of the underlying gauge theory become relevant.
Computing the integral, the gap equation takes the explicit form
\begin{equation}
1-\frac{ M_\psi^2}{\Lambda^2}
\ln \left(\frac{\Lambda^2 +M_\psi^2}{M_\psi^2}\right)
=\frac{4 \pi^2}{\Lambda^2} P_-
\equiv \frac{1}{\xi}~,
\label{gap2}
\end{equation}
in full agreement with the minimisation of the one-loop effective potential
in Ref.~\cite{Barnard:2013zea}.

Eq.~(\ref{gap2}) has a non-trivial solution, $M_\psi\ne 0$, as long as $\xi >1$, which implies
$\kappa_A^2>\kappa_B^2$ and
$P_-^{-1}=\kappa_A + \kappa_B > 4\pi^2/\Lambda^2$.
The existence of a minimal, critical coupling to realise spontaneous symmetry breaking
is a characteristic property of the NJL model. On the other hand, the consistency requirement
$M_\psi/\Lambda \lesssim 1$ implies an upper bound on the coupling, 
$\xi \equiv \Lambda^2 (\kappa_A + \kappa_B)/(4\pi^2)\lesssim (1-\ln 2)^{-1} \simeq 3.25$,
see also Fig.~\ref{FGLam} below.
Note that if the underlying $Sp(2N)$ gauge theory confines, it necessarily breaks $SU(4)$ into $Sp(4)$
as a consequence of the anomaly matching discussed in section \ref{anomat},
because the fermions $\psi$ cannot form baryons.
This means that the true strong dynamics has to correspond to a super-critical value of $\kappa_A +\kappa_B$.
This conclusion holds for the $\psi$-sector in isolation, but it may not be the case when a coloured $X$-sector
will be added in section \ref{coloured-sector}, and baryons become possible, see the discussion in section \ref{total-break}.
Note also that, in the NJL large-$N$ approximation, the mass gap $M_\psi$
and the fermion condensate, 
\be
\frac{1}{2} \langle (\psi^a \psi^b ) + ({\overline\psi}^a {\overline\psi}^b) \rangle
\equiv
\la \Psi \Psi \ra \Sigma_0^{ab}
,
\qquad\qquad
\la \Psi \Psi \ra = \frac{1}{\sqrt{N^\psi_f} } \la S^\psi_0 \ra  ,
\ee
corresponding to the tadpole in Fig. \ref{figgap}, are trivially related:
\be
\la \Psi \Psi \ra  \equiv -  2 (2N) M_\psi \t A_0(M^2_\psi)  = -\frac{N}{\kappa_A + \kappa_B} M_\psi 
.
\label{psicond}
\ee
We have also indicated the direct relation between the quark condensate and the vacuum
expectation value $\la S_0^\psi \ra$ of the singlet scalar density, at this level of NJL 
approximation, with $S_0^\psi$  defined in Eq.~(\ref{S_and_P}).

\begin{figure}[b]
\includegraphics[scale=0.95, trim= 60 90 0 70]{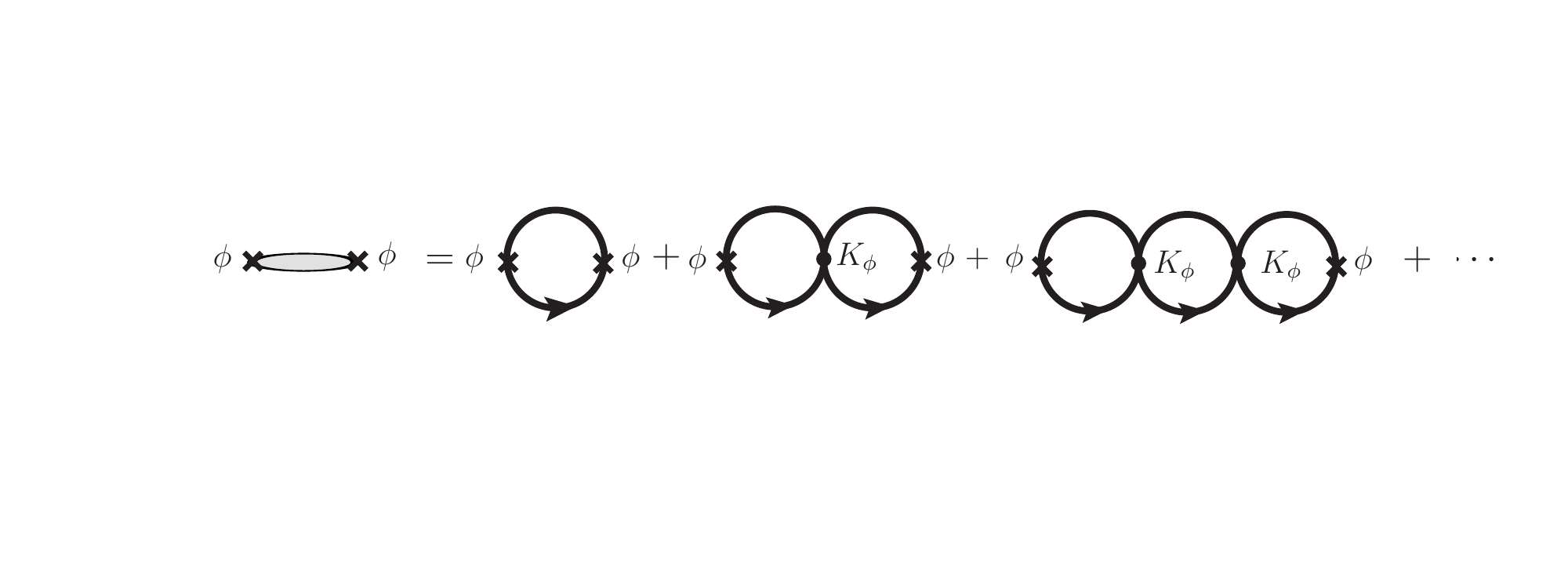}
\caption{Resummation of leading $1/N$ graphs for a mesonic two-point correlator, corresponding to a composite meson exchange.}
\label{BS}
\end{figure}

\subsection{Masses of scalar resonances}
\label{Masses and couplings of scalar resonances}

The masses and the couplings of the composite mesonic resonances can  be computed, at first  order in $1/N$,
by performing the resummation of the dominant large-$N$ graphs contributing to the two-point functions with the
appropriate quantum numbers, according to
a well-known procedure~\cite{Nambu:1961tp,Klevansky:1992qe,Hatsuda:1994pi,Klimt:1989pm,Bijnens:1993ap}.
The resummation takes the form of a geometric series,
as illustrated in Fig.~\ref{BS}. For the two-point functions  defined in Eqs.~(\ref{PiPSdef}) and (\ref{PiP0S0def}), 
the outcome of this procedure translates into the
generic formula
\be
\overline \Pi_{\phi}(q^2) \equiv \frac{\t \Pi_{\phi}(q^2)}{1- 2 K_{\phi} \t \Pi_{\phi}(q^2)}
~,
\label{PiSPsum}
\ee
where $K_{\phi}$ are combinations of the four-fermion couplings in Eq.~(\ref{LSphys}).
The expressions of $K_{\phi}$  and of the one-loop correlators $\t\Pi_\phi(q^2)$
have been collected in Table \ref{tab_phi}. They involve the one-loop two-point function
${\tilde B}_0 (q^2 , M_\psi^2)$ defined in appendix \ref{loop-functions}.
In this section, we will discuss the scalar and pseudoscalar channels, while the spin-one channels will be discussed in section \ref{Masses and couplings of vector resonances}.

Before starting this discussion, we would like to make a few remarks on the resummed correlators, some of which being also
relevant for the spin-one channels.
\begin{itemize}

 \item Expression (\ref{PiSPsum}) is not applicable in this simple form in the pseudoscalar
channel, $\phi=G^{\hat A},\,\eta^\prime$, due to the fact that, at one loop, the axial two-point function also receives
a longitudinal part, which will then mix with the pseudoscalar two-point function when the resummation in Fig.~\ref{BS}
is performed. For the time being, we can ignore these aspects, which will be treated in detail in Section 
\ref{Resummed correlators and the Goldstone decay constant}, and, in the meantime, we proceed with the general discussion
of masses and couplings on the basis of Eq. (\ref{PiSPsum}).

\item The corresponding resonance masses $M_\phi$ are determined by the poles of
the resummed propagators,
\be
1-2 K_\phi\, \t\Pi_{\phi}(q^2=M^2_\phi) = 0 ~.
\label{pole}
\ee
In order to discuss some general features of this type of equation, let us
point out that the functions $\t\Pi_{\phi}(q^2)$ can be defined in the cut complex $q^2$-plane, 
where the cut lies on the real positive axis
and starts at $q^2=4M_\psi^2$. The cut results from a logarithmic branch point,
so that the functions $\t\Pi_{\phi}(q^2)$ become multi-valued through analytic
continuation across the cut. These properties simply reflect those of the function
${\t B}_0(q^2, M_\psi^2)$ itself. In general, Eq. (\ref{pole}) has solutions
for complex values of $q^2$, lying on the second Riemann sheet, which are interpreted
as resonances, generated dynamically through the resummation procedure.

\item Other solutions to Eq. (\ref{pole}) than poles on the second sheet are possible. 
For instance, there can exist a critical value $K_\phi^{\rm crit}$, such
that if the coupling $K_\phi$ satisfies  $K_\phi\ge K_\phi^{\rm crit} >0$, then
Eq. (\ref{pole}) possesses (in addition) a real solution $0\le M_\phi \le 2 M_\psi$ \cite{Takizawa:1991mx}, corresponding 
to a two-fermion bound state. As we will see below, this situation
arises in the  singlet pseudoscalar channel (and also in the vector channel, but this time for $K_\phi\le K_\phi^{\rm crit} <0$). 
As $K_\phi$
moves towards $K_\phi^{\rm crit}$ from above, the bound-state mass moves from zero towards
the value $2 M_\psi$. When $K_\phi < K_\phi^{\rm crit}$, this solution of Eq. (\ref{pole})
moves back towards the origin, but now on the real axis of the second Riemann sheet,
and thus becomes a ``virtual-state'' solution \cite{Takizawa:1991mx}.

\item Another aspect concerning the solutions of Eq.~(\ref{pole}) is
intimately connected to the fact that, in order to make this equation
meaningful, it has been necessary to introduce a regularisation for the function
${\t B}_0(q^2, M_\psi^2)$. As a consequence, there are solutions corresponding
to real, but negative, values of $q^2$, $q^2 = - M_{{\rm gh}{\mbox -}\phi}^2 \gsim -3 \Lambda^2$. 
These ``ghost'' singularities\footnote{These pathologies are absent if the Pauli-Villars regularisation is
adopted \cite{Klevansky:1997dk}, but they reappear in another guise.} of the functions 
$\overline \Pi_{\phi}(q^2)$ occur quite far from the 
physical region, and have only a small influence on, for instance, the values of
the resonance masses. When determining the latter, we thus systematically discard them.
But they have to be taken into account when considering
more global properties of the functions $\overline \Pi_{\phi}(q^2)$, like the spectral sum rules of Section \ref{SR}. 
These will be discussed within the framework of the NJL approximation below, in Section \ref{secWSR}.

\item From a practical point of view, resonance solutions to Eq.~(\ref{pole}) will
not be determined by looking for poles on the second sheet, but rather  by solving
a real equation as follows.
We rewrite the denominator of Eq.~(\ref{PiSPsum}) as $1-2K_\phi \tilde{\Pi}_\phi(q^2)=c_0^\phi(q^2)+c_1^\phi(q^2) q^2$, 
where the $q^2$-dependence of the coefficients $c_{0,1}^\phi(q^2)$  comes from the loop function $\tilde{B}_0(q^2,M_\psi^2)$ only, see table 
\ref{tab_phi}.
The meson mass is then defined implicitly by 
\begin{equation}
M_\phi^2= {\rm Re}[g_\phi(M_\phi^2)]~,
\qquad\qquad
g_\phi(q^2)\equiv -\frac{c_0^\phi(q^2)}{c_1^\phi(q^2)}~.
\label{res_sol}
\end{equation}
The value $M_\phi$ obtained this way remains a good approximation to the mass given by 
the real part of the resonance pole, 
as long as the imaginary part of $g_\phi(M_\phi^2)$ remains small,
\begin{equation}
\left|\frac{ {\rm Im}[g_\phi(M_\phi^2)]}{ {\rm Re} [ g_\phi(M_\phi^2)]} \right|<1~.
\label{res_width}
\end{equation}
Indeed, the solution of Eq.~(\ref{res_sol}) may be larger than the threshold, $M_\phi^2 > 4 M_\psi^2$, so that the loop function $\t B_0(M^2_\phi, M_\psi^2)$ develops an imaginary part. 
This may happen in the case of the $Sp(4)$-singlet pseudoscalar state, see Eq.~(\ref{Meta}), and it always happens in 
the case of the non-singlet scalar state, see Eq.~(\ref{Ma2}).
This imaginary part corresponds to the unphysical decay of a meson into two constituent fermions, and reflects the
well known fact that the NJL model does not account for confinement. In what follows, it will be understood that 
resonance masses are defined as the solutions of Eq.~(\ref{res_sol}) and, in order to define a consistency condition for 
the NJL approximation to be reliable, we will require that  Eq.~(\ref{res_width}) holds.
Note also that, when extracting the expressions of the pole masses, 
it will be often convenient to take advantage of the gap equation (\ref{gap}), in order to obtain a simpler form of the solutions.

\end{itemize}

\begin{table}[b]
\renewcommand{\arraystretch}{1.8}
\begin{center}
\begin{tabular}{|c|c|c|}
\hline
$\phi$   &   $K_\phi$   &   $\t\Pi_{\phi}(q^2)$
\\
\hline
\hline
 $G^{\hat{A}}$ & $2(\kappa_A + \kappa_B)/(2N)$ &
\multirow{2}{*}{$ \t\Pi_{P}(q^2) =(2N) \big[\t A_0( M_\psi^2) -\frac{q^2}{2} \t B_0(q^2,M_\psi^2)\big]$}
\\
\cline{1-2}
$\eta^\prime$  & $2(\kappa_A - \kappa_B)/(2N)$   & \\
\hline
\hline
 $S^{\hat{A}}$  & $2(\kappa_A - \kappa_B)/(2N)$  &
\multirow{2}{*}{$ \t\Pi_{S}(q^2) =(2N) \big[\t A_0( M_\psi^2) - \frac{1}{2} (q^2 - 4 M_\psi^2) \t B_0(q^2,M_\psi^2)\big]$}
\\
\cline{1-2}
 $\sigma$  & $2(\kappa_A + \kappa_B)/(2N)$     & \\
\hline
\hline
 $V_\mu^A$     & $-2\kappa_D/(2N)$   &  $\t\Pi_{V}(q^2) =\frac{1}{3} (2N) \big[- 2 M_\psi^2 \t B_0(0,M_\psi^2) + (q^2 + 2 M_\psi^2) \t B_0(q^2,M_\psi^2)\big]$ \\
\hline
\hline
 $A_\mu^{\hat{A}}$     & $-2\kappa_D/(2N)$                 & $\t\Pi_{A}(q^2) = \frac{1}{3} (2N) \big[- 2 M_\psi^2 \t B_0(0,M_\psi^2) + (q^2 - 4 M_\psi^2) \t B_0(q^2,M_\psi^2)\big]$ \\
 \cline{1-2}
  $a_\mu$       & $-2\kappa_C/(2N)$                       & $\t\Pi_{A}^L(q^2) = - 2 (2N) M_\psi^2 \t B_0(q^2,M_\psi^2)$   \\
\hline
\hline
 $A_\mu^{\hat{A}}-G^{\hat{A}}$     &               & \multirow{2}{*}{$\t\Pi_{AP} (q^2) = -  (2N) M_\psi \t B_0(q^2 , M_\psi^2)$} \\
 \cline{1-2}
  $a_\mu-\eta^\prime$       &                       &    \\
\hline
\end{tabular}\end{center}
\caption{The couplings $K_\phi$ and the expressions of the one-loop spin-0 and spin-1 two-point functions. 
We also give the expression of the mixed (one-loop) pseudoscalar-longitudinal axial correlator, that enters
in the analysis of both the quintuplet and singlet sectors. 
The explicit calculation of the correlators $\t \Pi_\phi (q^2)$ is detailed in appendix \ref{SDresum}. }
\label{tab_phi}
\end{table}

After these general considerations, we now turn 
to the analysis of the scalar and pseudoscalar channels of the model. 
The functions $\t \Pi_{S/P}(q^2)$ correspond to the one-loop
estimates of the two-point functions $\Pi_{S/P} (q^2)$ defined in Eq. (\ref{PiPSdef}). 
Notice that one needs  $K_\phi \propto 1/N$, in order for the $1/N$-expansion to be well-defined.
Indeed, according to section \ref{colourless-sector} (see also Table \ref{tab_phi}), 
we have $K_{\sigma,G} = 2(\kappa_A+\kappa_B)/(2N)$ and
$K_{S,\eta^\prime} = 2(\kappa_A-\kappa_B)/(2N)$.

Let us consider first the pseudoscalar channels, ignoring, for the time being, the issue of mixing
with the longitudinal part of the axial correlator. 
After taking the traces and evaluating the momentum integral, the pseudoscalar two-point correlator 
in the $SU(4)$ sector takes the form
\be
\t\Pi_P(q^2) = 
 \left(2N \right) \,\left[\t A_0( M_\psi^2) -\frac{q^2}{2} \t B_0(q^2,M_\psi^2)\right]
~. 
\label{PiG}
\ee
In the case of the Goldstone states $G^{\hat A}$, Eq.~(\ref{pole}) becomes
\be
1-4\frac{(\kappa_A+\kappa_B)}{2N}\t\Pi_P (M_G^2) = 1-4(\kappa_A+\kappa_B)
\left[\t A_0( M_\psi^2) -\frac{M_G^2}{2} \t B_0(M_G^2,M_\psi^2)\right] = 2M_G^2\:(\kappa_A+\kappa_B)
\t B_0(M_G^2,M_\psi^2)=0
~,
\label{PiGexp}
\ee
and the term proportional to $\t A_0$ cancels out upon using the mass-gap equation, Eq.~(\ref{gap}),
a well-known feature of the standard NJL model~\cite{Nambu:1961tp,Klevansky:1992qe}.
As a consequence, one is left with an exactly massless inverse propagator, $M_G=0$,
as it should be for the Goldstone boson state.

A similar computation for the $Sp(4)$-singlet pseudoscalar $\eta^\prime$,
using the information provided by Table \ref{tab_phi}, leads to
\be
M_{\eta^\prime}^2 
 = g_{\eta^\prime}(M_{\eta^\prime}^2)
= \frac{2\t A_0( M_\psi^2)}{\t B_0
(M_{\eta^\prime}^2, M_\psi^2)} \left(1- \frac{P_+}{P_-} \right) =
-\frac{\kappa_B}{\kappa^2_A-\kappa^2_B}\:\frac{1}{\t B_0 (M_{\eta^\prime}^2,M_\psi^2)}
~,
\label{Meta}
\ee
where we have again used Eq.~(\ref{gap}).
In the above equation and in the following expressions of the resonance masses, it is implicitly assumed that only the  
real part of $g_\phi(M_\phi^2)$ is taken into account, according to Eq.~(\ref{res_sol}). 
Note that the constraint $\kappa_A^2>\kappa_B^2$, needed for the existence of a non-trivial solution of the gap equation, 
also ensures that $M_{\eta^\prime}^2$ is positive.
As it will be discussed in subsection \ref{mixing-singlets}, a similar but stronger constraint holds when the coloured sector is introduced.
To roughly estimate the expected range for $M_{\eta^\prime}$, one may
notice that $\t B_0 (q^2,M_\psi^2)$ is real and has a rather moderate $q^2$ dependence for $q^2\ll4M_\psi^2$, so that if
$M^2_{\eta^\prime}$ lies in this range, one can use the approximate expression
\be
M^2_{\eta^\prime} \simeq -\frac{\kappa_B}{\kappa^2_A-\kappa^2_B}\:\frac{1}{\t B_0 (0,M_\psi^2)} \simeq
\frac{4}{\xi}\ \frac{\kappa_B/\kappa_A}{1-\kappa_B/\kappa_A}\
\dfrac{\Lambda^2}{\ln (\Lambda^2/M^2_\psi)-1}
~,
\label{Metasimple}
\ee
where the expression for $\t B_0(0,M^2_\psi)$ is given in Eq.~(\ref{B00}).
Thus $M_{\eta^\prime}$ may become arbitrarily small for $\kappa_B/\kappa_A\rightarrow 0$,
as the extra $U(1)_\psi$ symmetry is restored when $\kappa_B=0$, and $\eta^\prime$ turns into the associated Goldstone boson.
However, $M_{\eta^\prime}$ rapidly increases with $\kappa_B/\kappa_A$ to become of order $\Lambda$.
Note that, in the large-$N$ limit, one expects $M_{\eta^\prime}^2\sim 1/N$, as for the $\eta^\prime$ mass in QCD \cite{Witten:1979vv}.
This indicates that the four-fermion couplings, normalised as in Eq.~(\ref{LSbasic}), should scale as $\kappa_B/\kappa_A\sim 1/N$.
Large-$N$ arguments indicate that $\kappa_A$ is $N$-independent,
as the associated four-fermion operator is generated from the hypercolour current-current 
interaction (for details see appendix \ref{$Sp(2N)$ current-current operators}).
Therefore, the correct scaling is reproduced for  $\kappa_B = {\overline\kappa}_B/(2N)$, with an  
$N$-independent $\overline\kappa_B$,
and the associated four-fermion operator, induced by the hypercolour anomaly, scales as $1/N^2$.

For the scalar channels, the two-point function is to be found in Table \ref{tab_phi},
and the corresponding scalar resonance masses are
\be
M_\sigma^2= 4 M_\psi^2~,
\qquad\qquad
M^2_S 
= 4 M_\psi^2 + M_{\eta^\prime}^2 \frac{\t B_0 (M_{\eta^\prime}^2,M_\psi^2)}{\t B_0 (M_S^2,M_\psi^2)} 
\simeq M_\sigma^2 +M^2_{\eta^\prime}
~,
\label{Ma2}\ee
where one recognises the same relation $M_\sigma= 2 M_\psi$, as in the standard NJL model for QCD with two flavours.
The relation $M_S^2 \simeq M_{\eta^\prime}^2 +M_\sigma^2$ holds again if one can neglect the difference
between the function $\t B_0 (p^2,M_\psi^2)$ evaluated at $p^2=M_{\eta^\prime}^2$ and at $p^2=M_S^2$.

We stress that all previous expressions for the spectrum of spin-zero resonances
hold in the pure chiral limit,
where the $SU(4)/Sp(4)$ Goldstone bosons $G^{\hat A}$, including the Higgs, are massless.
Eventually, they will receive a non-zero effective potential, radiatively induced by
the SM gauge and Yukawa couplings, which break explicitly the $SU(4)$ symmetry.
In particular, the top quark Yukawa coupling is generically expected to
destabilise the vacuum, and to trigger EWSB, see Refs.~\cite{Contino:2010rs,Panico:2015jxa} for reviews.
This implies that the masses of some resonances, obtained in the NJL large-$N$ approximation, may receive
corrections of order  ${\cal O}(m^2_{top}/\Lambda^2)$.
These represent typically mild  corrections for the non-Goldstone resonances, whose masses  $\sim \Lambda$ are significantly larger
than the electroweak scale. Thus, the qualitative features of the spectrum of meson resonances are
not expected to depart from those exhibited here, once the effect of the explicit symmetry-breaking
couplings is added to the picture. One should also remember that, in any case, the NJL large-$N$ approximation
already constitutes a limitation to the precision that can be achieved. 
The radiative contribution to the pseudo-Goldstone Higgs mass, induced
from the external electroweak gauge fields, is given in Eq.~(\ref{rad_ew}) (see also the general discussion in
section \ref{gauging}).
However, this contribution plays a secondary role in EWSB:
since it is positive, it cannot destabilise the $Sp(4)$-invariant vacuum,
and it should be overcome by the one from the top Yukawa coupling \cite{Contino:2010rs,Panico:2015jxa}.

In the traditional NJL literature~\cite{Nambu:1961tp,Klevansky:1992qe,Hatsuda:1994pi,Klimt:1989pm}, 
the resonance masses are determined from the resummed
scattering amplitudes for $\psi\psi\to\psi\psi$ in the various channels. These amplitudes
involve the same couplings $K_\phi$ and functions $\t\Pi_{\phi}(p^2)$ as in Eq. (\ref{PiSPsum}). Moreover, they
also allow to define couplings between the elementary fermions and the resonances. The
interested reader will find a brief discussion of these issues, not directly related to
our main purposes, in App. \ref{SDresum}.

\subsection{Vector interactions of fermion bilinears}\label{psivectors}

Let us now consider vector bilinears, in order to study spin-one resonances. There are
two independent four-fermion vector-vector operators, that can be written as
\begin{equation}
\mathcal{L}_{vect}^{\psi}=
\frac{\kappa^\prime_C}{2N}\left(\overline{\psi}_a \overline{\sigma}^\mu \psi^a  \right)
\left(\overline{\psi}_b \overline{\sigma}_\mu \psi^b  \right)
+\frac{\kappa^\prime_D}{2N} \left(\overline{\psi}_a \overline{\sigma}^\mu \psi^b \right) \left(\overline{\psi}_b \overline{\sigma}_\mu \psi^a \right) ~,
\label{L4Fvec}
\end{equation}
where the coupling constants $\kappa^\prime_C$ and $\kappa^\prime_D$ are real.
It turns out that consistent  (non-tachyonic) spin-one resonance masses are obtained
for  $\kappa^\prime_{C,D} >0$, in the same way as for the NJL vector interaction in QCD.
Applying the $SU(4)$ Fierz identity given by Eq.~(\ref{SUNfierz}),
the Lagrangian can be rewritten in the `physical' channels, corresponding to definite $Sp(4)$ representations,
\be
\mathcal{L}_{vect}^{\psi}=
\frac{\kappa_C}{2N} \left(\overline{\psi}\, T^0_\psi \, \overline{\sigma}^\mu \psi \right)^2
+\frac{\kappa_D}{2N}
\left(\overline{\psi} T^A \overline{\sigma}^\mu \psi  \right)^2
+\frac{\kappa_D}{2N}
\left(\overline{\psi} \, T^{\hat A} \, \overline{\sigma}^\mu \psi  \right)^2
~,
\label{L4Fv}
\ee
where $\kappa_D = 2\kappa^\prime_D$, $\kappa_C = 8\kappa^\prime_C+ 2\kappa^\prime_D$, and contracted flavour indexes are understood,
as well as summations over generator labels $A$ and ${\hat A}$.
Introducing auxiliary vector fields, the vector sector Lagrangian takes the form
\be
\mathcal{L}_{vect}^{\psi}
= - a_\mu (\overline{\psi} \,T^0_\psi\, \overline{\sigma}^\mu \psi)
- V_\mu^A (\overline{\psi} \,T^A\, \overline{\sigma}^\mu \psi)
- A_\mu^{\hat A} (\overline{\psi} \,T^{\hat A}\, \overline{\sigma}^\mu \psi)
-\frac{N}{2\kappa_C} a^{\mu} a_{\mu}
-\frac{N}{2\kappa_D} \left( V_\mu^A V^{A\mu} + A_\mu^{\hat A} A^{{\hat A}\mu}\right)
, 
\label{Lvaux}
\ee
with vectors $V_{\mu}^A \sim 10_{Sp(4)}$, and axial vectors $(a_\mu,A_{\mu}^{\hat A}) \sim (1+5)_{Sp(4)}$.
Their transformation properties are summarised in Table~\ref{tabsu4}. This Lagrangian
defines the strength of the four-fermion interactions in the three physical channels mediated by
$a_\mu$, $V_\mu^A$ and $A_\mu^{\hat A}$.

We remark that additional spin-one resonances
can be associated to the fermion bilinear $(\psi^a \sigma^{\mu\nu} \psi^b)\sim 10_{Sp(4)}$, or to its conjugate.
However, one can check that the corresponding four-fermion interactions vanish because of Lorentz and/or $SU(4)$ invariance.
Therefore, to describe these resonances one should consider higher-dimensional operators.
Although such an exercise is feasible with analogous NJL techniques, it goes beyond the scope of this paper.

In general, the couplings $\kappa_C$ and $\kappa_D$ are additional free parameters with respect to those in the spin-zero sector,
and in the following we will provide expressions for the vector masses and couplings as functions of these couplings.
However, $\kappa_C$ and $\kappa_D$
may be related to the scalar sector coupling $\kappa_A$, if one assumes that the low-energy effective interactions,
between two hypercolour-singlet
fermion bilinears, originate from a one-hypergluon exchange current-current interaction,
as determined by the underlying hypercolour gauge interaction.  This may be justified in the large-$N$ approximation
(or equivalently `ladder' approximation for the current-current interaction) and it proves to be a reasonably
good approximation in the NJL-QCD case~\cite{Klimt:1989pm,Bijnens:1992uz}.
Under such an assumption, one can apply Fierz identities for Weyl,
as well as for $SU(4)$ and $Sp(2N)$, indices, as detailed in appendix \ref{fierz},
in order to relate the coefficients of the various four-fermion operators.
We obtain that the vector couplings of Eq.~(\ref{L4Fv}) are simply related to the scalar coupling of
Eq.~(\ref{LSphys}) by
\be
\kappa_A = \kappa_C=\kappa_D
~.
\label{SVfierz4}
\ee
An analogous relation holds in the NJL-QCD case~\cite{Klimt:1989pm},
where the couplings of the scalar-scalar and vector-vector interactions are identical.
We will use Eq.~(\ref{SVfierz4}) as a
benchmark for numerical illustration, however one should keep in mind
that the true dynamics may appreciably depart from this naive relation.

\subsection{Masses of vector resonances}
\label{Masses and couplings of vector resonances}

The vector meson masses can be computed, at leading order
in the $1/N$ expansion, similarly to the scalar meson channels, from the resummed 
two-point functions, and the geometric series illustrated in Fig. \ref{BS} now leads, 
in this approximation, to the following expressions for the vector or axial two-point 
correlators $\Pi_{V,A}(p^2)$ defined in Eq.~(\ref{PiVAdef}),
\be
\overline \Pi_{V/A}(q^2) \equiv -  \frac{\t \Pi_{V/A}(q^2)}{q^2 [ 1-2 K_{V/A} \t \Pi_{V/A}(q^2) ]}
,
\label{PiVAsum}
\ee 
We have introduced one-loop correlators $\t \Pi_{V/A}(q^2)$
with a  normalisation that is more convenient for our purposes, so that
$\t \Pi_{V/A}(q^2)\equiv -q^2\Pi_{V/A}(q^2)|_{1-loop}$. 
Similarly, for the one-loop axial longitudinal part we have $\t \Pi_{A}^L(q^2)\equiv q^2\Pi_{A}^L(q^2)|_{1-loop}$, 
where $\Pi_{A}^L(q^2)$ is defined in Eq.~(\ref{PATL}). More precisely,
upon taking the traces over spinor indices, flavour and hypercolour,
the one-loop two-point vector and axial correlators take the form,
\be
\t \Pi_{V}^{\mu\nu,AB}(q) =\t \Pi_{V}(q^2) T^{\mu\nu} \delta^{AB}
,
\qquad\qquad
\t \Pi_{A}^{\mu\nu,\hat{A} \hat{B}}(q) = \left[\t \Pi_{A}(q^2) T^{\mu\nu} + \t \Pi_{A}^L(q^2) L^{\mu\nu} \right] \delta^{\hat{A} \hat{B}}
,
\label{PiV1loop}
\ee
where the transverse and longitudinal projectors are defined as
\begin{equation}
T^{\mu \nu}= \eta^{\mu \nu}- \frac{q^\mu q^\nu}{q^2}~, 
\qquad\qquad L^{\mu \nu}= \frac{q^\mu q^\nu}{q^2}
,
\label{tensors-spin1}
\end{equation}
and where the expressions of the functions $\t \Pi_{V/A}(q^2)$ and $\t \Pi_{A}^L(q^2)$
are given in Table \ref{tab_phi}.
One should be cautious to adopt a regularisation that preserves $SU(4)$ current conservation
for the one-loop correlators, which is not the case with the standard NJL cutoff regularisation.
There are various ways to deal with this well-known problem~\cite{Klevansky:1992qe},
the simplest being to use dimensional regularisation for the intermediate stages of the calculation.
In this way the one-loop vector correlator is  automatically transverse.
In the final expression for the correlators, the formally divergent loop function $\t B_0$
can be written as a function of the $D=4$ cutoff $\Lambda$, see Eq.~(\ref{B0exp}).
The latter is then interpreted as the physical cutoff of the NJL model.

As compared to the two-point axial correlator in the massless limit, defined by Eq.~(\ref{PiVAdef}),
and as already mentioned in Section \ref{Masses and couplings of scalar resonances},
the one-loop expression (\ref{PiV1loop}) also exhibits a longitudinal part. This is a specific trait
of the NJL model, where the dynamically generated mass $M_\psi$ acts here like an explicit
symmetry-breaking term. We will come back later on the manner this longitudinal piece is taken care of.
For the time being, one may notice that the transverse part of the two-point axial correlator
reproduces the expected physical features. Indeed, the resummed function $\overline\Pi_A (q^2)$  
exhibits the massless pole\footnote{As expected, such a massless pole does not occur in
$\overline\Pi_V (q^2)$, defined in Eq. (\ref{PiVAsum}), since, as can be inferred from
Table \ref{tab_phi}, $\t \Pi_{V}(q^2)$ vanishes for $q^2 = 0$.} due to the contribution
of the Goldstone bosons, but it also has a pole from the axial-vector state $A_\mu^{\hat A}$.
This second pole mass is extracted from Eq.~(\ref{pole}), by injecting the 
coupling\footnote{Note the relative minus sign between the four-fermion couplings in the Lagrangian of Eq.~(\ref{L4Fv})
$K_A= -2\kappa_D/(2N)$, and the couplings
$K_{V,A}$ that enter in the denominator of the resummed correlators in
 Eq.~(\ref{PiVAsum}). This follows from the proper definition of the argument of the associated geometric series.}
and the transverse part of the correlator, $\t \Pi_A(q^2)$. One obtains
\be
M_{A}^2=-\frac{3}{4 \kappa_D \t B_0(M_{A}^2,M_\psi^2)}
+ 2 M_\psi^2 \frac{\t B_0 (0,M_\psi^2)}{\t B_0(M_{A}^2,M_\psi^2)}
+ 4 M_\psi^2
~.
\label{MA}
\ee
The pole mass equation
for the axial vector singlet $a_\mu$ is obtained with the replacements $\kappa_D\to \kappa_C$ and  $M_A\to M_a$.

The $V_\mu^A$ pole mass can likewise be extracted from Eq.~(\ref{pole}), with the
replacements $K_\phi\to K_V= -2\kappa_D/(2N)$ and $\t\Pi_{\phi}(p^2)\to \t \Pi_V(p^2)$.
This leads to
\begin{equation}
M_{V}^2=-\frac{3}{4 \kappa_D \t B_0(M_V^2,M_\psi^2)}+
2 M_\psi^2 \frac{ \t B_0(0,M_\psi^2)}{\t B_0(M_V^2,M_\psi^2)} -2 M_\psi^2
~.
\label{MV}
\end{equation}

In estimating the sizes of the spin-one resonance masses, note that $ \t B_0(p^2,M_\psi^2)$ is real for
$0 \le p^2 \le 4M_\psi^2$, and negative in
the physically relevant range of $0 < M_\psi^2 < \Lambda^2$, with $|\t B_0(p^2,M_\psi^2)|\ge |\t B_0(0,M_\psi^2)|$. The
term proportional to  $1/\kappa_D$ on the right-hand side of Eqs.~(\ref{MV}) and (\ref{MA}) is
positive for  $\kappa_D>0$, and gives the dominant contribution to $M_{V,A}$
for, roughly, $\kappa_D M_\psi^2 \lesssim 4\pi^2$, that is $(M_\psi/\Lambda)^2\lesssim 1/\xi$ when one takes
$\kappa_D\simeq\kappa_A\gg\kappa_B$.
By neglecting the difference between $\t B_0(M^2_V,M_\psi^2)$ and $\t B_0(M^2_A,M_\psi^2)$,
we obtain the usual NJL  relation between the axial and vector masses,
\begin{equation}
M_A^2 \simeq M_V^2 +6 M_\psi^2 ~.
\label{MVMA}
\end{equation}
When one adopts the exact self-consistent pole mass definitions, 
$M_A$ is somewhat below the prediction of Eq.~(\ref{MVMA}), by typically $5-10\%$.
Also, the singlet mass $M_{a}$ is equal to $M_A$ when $\kappa_D =\kappa_C$ as in Eq.~(\ref{SVfierz4}).
As already mentioned in the general considerations at the beginning
of Section \ref{Masses and couplings of scalar resonances}, depending on the values
of the couplings, one may have resonance masses satisfying $M_\phi^2 > 4 M^2_\psi$, 
in which case $\tilde B_0 (M_\phi^2,M_\psi^2)$ develops an imaginary part.
Indeed, this is always the case for $M_A$, as one reads off Eq.~(\ref{MVMA}). 
In such cases, the resonance mass is obtained upon solving Eq. (\ref{res_sol}),
and we consider that the NJL predictions remain sensible as long as the width $\Gamma_\phi$
of the resonance, defined in Eq. (\ref{res_width}), does not exceed its mass.

\subsection{Goldstone decay constant and pseudoscalar-axial mixing}
\label{Resummed correlators and the Goldstone decay constant}

A key parameter of the composite sector is the Goldstone boson decay constant $F_G$, the analogous of
$F_\pi$ in QCD.
We recall that, when the Higgs is a composite pseudo-Goldstone boson,
the electroweak precision parameters, such as $S$, $T$  (see section \ref{S parameter}), and the Higgs couplings receive corrections
of order $(v/f)^2$ with respect to their SM value, 
where $v\simeq 246$ GeV and $f\equiv \sqrt{2} F_G$. Here $f$ is the Goldstone decay constant in the normalisation that is generally adopted in the composite Higgs 
literature.\footnote{\label{ftn7} The relation $f\equiv\sqrt{2} F_G$ follows from our definitions of $F_G$, see Eq.~(\ref{FGdef}), and of the Goldstone matrix $U$, see Eq.~(\ref{Udef}).
After the gauging of the SM group, the covariant derivative acting on the Goldstone bosons reads $D_\mu U = \partial_\mu U - i {\cal V}_\mu U - i U {\cal V}_\mu^T$, where the external source ${\cal V}_\mu$
is defined by Eq.~(\ref{Vsource}). This determines the non-linear corrections to the electroweak precision parameters in terms of $v/f$.}
Thus, $f$ is the physical scale most directly
constrained by precision measurements, $f\gtrsim (0.5 - 1)$ TeV,
the exact bound depending on the spontaneous symmetry breaking pattern, 
as well as on the flavour representations of the  spin-one and spin-one-half composite resonances
coupled to the SM fields.
Therefore, it will be convenient to express all the resonance masses in units of $f$,
and in the following we will adopt the more conservative bound  $f\gtrsim 1$ TeV.

\begin{figure}[b]
\includegraphics[scale=0.3]
{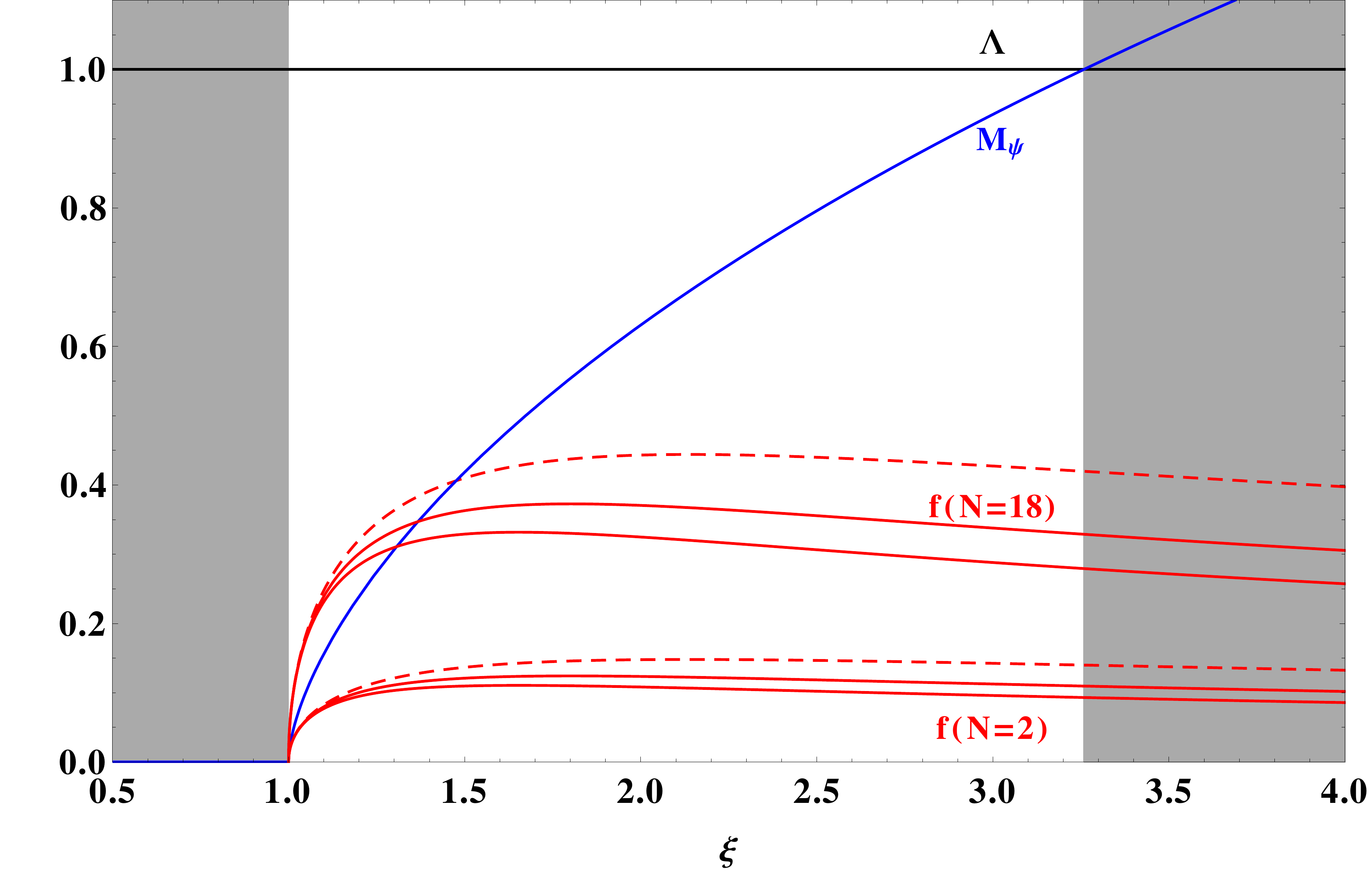}
\caption[long]{The mass gap $M_\psi$ and the Goldstone decay constant $f=\sqrt{2}F_G$, in units of the cutoff $\Lambda$,
as a function of the dimensionless coupling  $\xi \equiv (\kappa_A+\kappa_B)\Lambda^2/(4\pi^2)$.
For $\xi\le1$ there is no spontaneous symmetry breaking, $M_\psi=0$, while for $\xi\ge (1 - \ln 2)^{-1} \sim 3.25$ one
has $M_\psi\gtrsim \Lambda$ and the NJL description is no longer reliable. The decay constant $f$ is proportional
to $\sqrt{N}$, where $Sp(2N)$ is the hypercolour gauge group. In the complete model including a coloured sector (see section \ref{coloured-sector}),
one finds that $N\ge 2$ is required to allow for fermion-trilinear top partners,
and $N\leq18$ is needed
to preserve hypercolour asymptotic freedom \cite{Barnard:2013zea}.
 One further needs $N\leq 6$ to avoid Landau poles in the SM gauge couplings below $100$ TeV (see section \ref{The mass spectrum of the resonances}).
The red dashed line indicates the non-resummed decay constant $\t f =\sqrt{2}\t F_G$, while the upper (lower) red solid line corresponds to
the resummed $f$,  for $\kappa_D=\kappa_A$ and $\kappa_B=0$ ($\kappa_B=\kappa_A$).
}
\label{FGLam}
\end{figure}

The decay constant $F_G$, as defined by Eq.~(\ref{FGdef}), can most directly be extracted from the two-point axial transverse correlator,
introduced in Eq.~(\ref{PiVAdef}), through the residue of the Goldstone boson pole. Identifying this correlator
in the NJL approximation with the resummed correlator defined by Eq.~(\ref{PiVAsum}) and  using the explicit expression in Table~\ref{tab_phi}, one obtains
\be
F_G^2 = \lim_{q^2\to 0} \left[-q^2 \overline {\Pi}_A(q^2) \right]
=\frac{\t \Pi_A(0)}{1 -2 K_A \t \Pi_A(0)}
= \frac{\t F^2_G}{1- 2K_A \t F^2_G} = g_A(0) \t F^2_G~,
\label{FGsum}
\ee 
where we have defined the axial coupling form factor
\be
g_A(q^2)\equiv [1- 2K_A \t \Pi_A^L(q^2)]^{-1} = \left[1+\frac{4\kappa_D}{2N}\t \Pi_A^L(q^2)\right]^{-1} 
\label{gAdef}
\ee
and the one-loop decay constant
\be
\t F_G^2 \equiv \t \Pi_A(0) = -2 \left(2 N \right) M_\psi^2 \t B_0(0,M_\psi^2) = \t \Pi^L_A (0)
.
\label{FG4}
\ee
At this point, one should remark that $\t F_G$ would be the complete NJL result for the Goldstone decay constant
only if one would consider the scalar sector in isolation, i.e. by switching off the axial vector coupling $\kappa_{D}$.
However, since by definition the Goldstone boson couples to the axial current,
a non-zero $\kappa_D$ implies a non-trivial mixing of the pseudoscalar and axial vector channels, 
that affects the expression of the decay constant.
In order to take into account this effect and to define consistently $F_G$,
one needs to consider the resummed transverse axial-vector correlator $\overline \Pi_A(q^2)$ of Eq.~(\ref{PiVAsum}),
as shown in (\ref{FGsum}) above. This equation gives the complete NJL approximation for $F_G$,
which should be matched with its experimental value, once it becomes available,
as is the case of $F_\pi$ in the NJL approximation of QCD~\cite{Klevansky:1992qe,Klimt:1989pm}.

The behaviour of $F_G$  is illustrated in Fig.~\ref{FGLam}, as a function of the dimensionless coupling $\xi$.
Combining the definition of $\xi$ in Eq.~(\ref{gap2}) with the explicit form of $\t B_0(0, M_\psi^2 )$ given in Eq.~(\ref{B00}),
one obtains
\be
\t F_G^2 = \frac{N}{4\pi^2} \Lambda^2 \left(
\frac{\xi-1}{\xi} - \frac{M_\psi^2}{\Lambda^2 + M_\psi^2}\right)~.
\ee
Closely above the critical coupling, $\xi = 1$, the mass gap is much smaller than the cutoff, $M_\psi\ll \Lambda$,
and $\t F_G$ grows rapidly with $\xi$.
As $\xi-1$ becomes of order one, the mass gap approaches the cutoff, $M_\psi \lesssim \Lambda$,
while $\t F_G$ stops growing and remains below the cutoff by a factor of a few, $\tilde f \equiv \sqrt{2}\t F_G\simeq \sqrt{N}\Lambda/10$.
The resummed $F_G$,  see Eq.~(\ref{FGsum}), is smaller, as $K_A$ is negative. In Fig.~\ref{FGLam} we 
assumed Eq.~(\ref{SVfierz4}) to hold, so that
$K_A =  -4\pi^2\xi/[N\Lambda^2(1+\kappa_B/\kappa_A)]$, which leads to $f\simeq (0.6-0.8) \t f$.

As already mentioned at several places in this section, a non-vanishing axial-vector coupling $\kappa_D\ne 0$ 
implies a nontrivial mixing between the pseudoscalar and the axial longitudinal channel.  Therefore, 
the definition of the resummed pseudoscalar  correlator $\overline{\Pi}_P(q^2)$ in  Eq. (\ref{PiSPsum}) 
should be appropriately generalised in order to account for this mixing. In the process, we will also define
a resummed axial longitudinal correlator $\overline{\Pi}_A^L(q^2)$, we will recover consistency relations among the Goldstone
decay constants, and determine more precisely the properties of the non-Goldstone pseudoscalar $\eta^\prime$.
We discuss first the quintuplet $G-A^\mu$ mixing,
while the similar analysis of the singlet $\eta^\prime-a^\mu$ mixing is presented at the end of this section.

The mixing phenomenon is best described using a matrix formalism,
so that we are led to consider
\be
\mathbf{K}_G=
\begin{pmatrix}
K_G & 0
\\
 0 & K_A
\end{pmatrix}~,
\qquad\qquad
\mathbf{\Pi}  (q^2)
=\begin{pmatrix}
\tilde{\Pi}_P (q^2) & \sqrt{q^2}\,\tilde{\Pi}_{AP} (q^2) \\
\sqrt{q^2}\,\tilde{\Pi}_{AP} (q^2) &\tilde{\Pi}_A^L (q^2) 
\end{pmatrix}
.
\label{Pi22}
\ee
Explicit expressions for all the entries of these matrices can be found in Table \ref{tab_phi}. 
Notice the appearance of $\tilde{\Pi}_{AP} (q^2)$, the one-loop expression of the mixed
correlator $\Pi_{AP} (q^2)$ introduced in Eq. (\ref{PiAPdef}), and of 
the one-loop longitudinal axial correlator $\tilde{\Pi}_A^L(q^2)$ defined in Eq.~(\ref{PiV1loop}).
Note that, consistently with the normalisation of $\tilde{\Pi}_A^{L} (q^2)$ in Eq. (\ref{PiV1loop}), 
the matrix $\mathbf{\Pi}  (q^2)$ has been defined so that all its entries have the same dimensions, 
whence the factor of $\sqrt{q^2}$ in front of $\tilde{\Pi}_{AP} (q^2)$. 
The resummed large-$N$ two-point matrix correlator $\overline {\mathbf \Pi}_G$ in this basis
is then given by
\be
 \mathbf{\overline{\Pi}}_G \equiv\mathbf{\Pi} + \mathbf{\Pi}\, (2 \mathbf{K}_G) \, \mathbf{\Pi} +\cdots= (1\!\!1- 2\mathbf{\Pi} \mathbf{K}_G )^{-1} \mathbf{\Pi}~,
\label{Pisum22}
\ee 
which is the analog of Eqs.~(\ref{PiSPsum}) and (\ref{PiVAsum}).
From Eqs~(\ref{Pi22}), (\ref{Pisum22}) one then obtains
\bea
&&  \mathbf{\overline{\Pi}}_G (q^2)
\equiv  \begin{pmatrix}
\overline{\Pi}_G (q^2) & \sqrt{q^2}\,\overline{\Pi}_{AG} (q^2)\\
\sqrt{q^2}\,\overline{\Pi}_{AG} (q^2)& q^2\,\overline{\Pi}_A^{L }  (q^2)
\end{pmatrix}
\nonumber\\
&& =\frac{1}{D_G(q^2)} \begin{pmatrix}
\tilde{\Pi}_P (q^2)[1- 2 K_A \tilde {\Pi}_A^{L}(q^2)] +  2K_A q^2\tilde{\Pi}^2_{AP}(q^2) & \sqrt{q^2}\,\tilde{\Pi}_{AP}(q^2) \\
\sqrt{q^2}\,\tilde{\Pi}_{AP}(q^2) &\tilde{\Pi}_A^{L}(q^2) [1-2 K_G \tilde {\Pi}_P(q^2)] + 2K_G q^2\tilde{\Pi}^2_{AP}(q^2) 
\end{pmatrix}
~,
\eea
with
\be
D_G \equiv {\rm det}(1\!\!1 - 2 \mathbf{\Pi}\mathbf{K}_G )= (1- 2 K_G \t \Pi_P) (1-2 K_A \t \Pi_A^L) 
-4 K_G K_A q^2 \t \Pi_{AP}^2 = 2(\kappa_A+\kappa_B) q^2\, \t B_0(q^2,M^2_\psi)
.
\label{DG}
\ee
The last expression in this equation is obtained after using the gap-equation (\ref{gap}) and 
the relation $\t \Pi_{AP}^2(q^2) = -(1/2)(2N) \t B_0(q^2,M^2_\psi) \t \Pi_A^L(q^2)$. 
Using the relevant expressions in Table \ref{tab_phi}, gives explicitly
\be
\overline{\Pi}_G(q^2) = \frac{1}{2} (2N) \frac{2\t A_0(M^2_\psi) g^{-1}_A(q^2) - q^2 \t B_0(q^2,M^2_\psi) }{D_G (q^2)}, 
\qquad\qquad
 \overline{\Pi}_{AG}(q^2) =  \frac{\t \Pi_{AP} (q^2)}{D_G (q^2)},  
 \qquad\qquad
 \overline{\Pi}_A^{L}(q^2) = 0 .
\label{Pisum22fin}
\ee
Note in particular that 
the {\em resummed} longitudinal axial correlator $\overline{\Pi}_A^{L}(q^2)$
vanishes identically, thus consistently recovering the conservation of the axial current 
in the exact chiral limit, in spite of the nonzero mass gap, which induces a non-vanishing 
longitudinal axial correlator at the one-loop level,  $\t \Pi_A^L \propto M^2_\psi$ . 
Also the resummed mixed correlator $\overline{\Pi}_{AG}(q^2)$ satisfies the relation
(\ref{PiAP}), which shows that it is entirely saturated by the Goldstone-boson pole. 

Now one can extract the NJL prediction for the Goldstone constants $F_G$ and $G_G$, defined by Eqs.~(\ref{FGdef}) and (\ref{GGdef}) respectively.
The residue of 
$\overline{\Pi}_G (p^2)$ with respect to the Goldstone boson pole 
gives the pseudoscalar decay constant,
\be
G^2_G = -\lim_{q^2\to 0} q^2 \overline{\Pi}_G(q^2) = -\frac{(2N)}
{8(\kappa_A+\kappa_B)^2 \t B_0(0,M^2_\psi)} g^{-1}_A(0)~. 
\label{GGsum}
\ee
Next, the residue of 
$\overline{\Pi}_{AG} (q^2)$ determines $F_G G_G$, 
\be
F_G G_G = -\lim_{q^2\to 0} q^2 \overline{\Pi}_{AG}(q^2) = \frac{(2N)}{2} \frac{M_\psi}
{(\kappa_A+\kappa_B)} = 2 (2N) M_\psi\: \t A_0(M^2_\psi) ~,
\label{FGGGsum}
\ee
that satisfies Eq.~(\ref{FGrel}), by taking the expression for $\la S^\psi_0 \ra$ derived from Eq.~(\ref{psicond}).
Combining Eqs.~(\ref{GGsum}) and (\ref{FGGGsum}), and using the gap equation, one consistently recovers the very same  
expression of $F_G$  in Eq.~(\ref{FGsum}), as obtained from the resummed axial transverse correlator. 
Note that, if one had computed $G_G$ in the limit of vanishing axial-vector coupling, $\kappa_D=0$, by taking the residue of $\overline{\Pi}_P$ in Eq.~(\ref{PiSPsum}),
one would have missed the (inverse) axial form factor $g_A(0)$, see Eq.~(\ref{GGsum}).  
Such a correction is important e.g. when analysing the possible saturation of
the scalar spectral sum rules, which will be discussed in section \ref{secWSR}.

Obviously, a similar pseudoscalar-axial mixing mechanism also affects the singlet sector of the model,
as soon as the axial singlet coupling $\kappa_C$ is non-vanishing. The resummed correlator matrix 
for the singlet sector, $\mathbf{\overline{\Pi}}_{\eta^\prime}$,
is defined in complete analogy with Eq.~(\ref{Pisum22}), by taking the same one-loop correlator matrix  $\mathbf{\Pi}$, but
replacing the couplings, $K_G\to K_{\eta^\prime}$ and $K_A\to K_a$ (i.e. $\kappa_D\to \kappa_C$),
respectively for the pseudoscalar and axial-vector channels, according to Table \ref{tab_phi}.
One main consequence of the mixing is that the pseudoscalar singlet mass $M_{\eta^\prime}$ is modified with respect to 
Eq.~(\ref{Meta}), which holds for the pseudoscalar sector ``in isolation''. The $\eta^\prime$ mass rather 
corresponds to the pole of the determinant
\be
 D_{\eta^\prime} \equiv {\det}(1\!\!1 - 2\mathbf{\Pi} \mathbf{K}_{\eta^\prime} )= (1- 2K_{\eta^\prime} \t \Pi_P) g_a^{-1} 
-4 K_{\eta^\prime} K_a q^2 \t \Pi_{AP}^2  = 8 \kappa_B \t A_0(M^2_\psi)  g_a^{-1}
+ 2(\kappa_A-\kappa_B) q^2 \t B_0(q^2,M^2_\psi) ~,
\ee
where we defined an axial singlet form factor,
\be
g_a(q^2) = \left[1+\frac{4\kappa_C}{2N} \t \Pi_A^L(q^2) \right]^{-1}~,
\ee
in complete analogy with Eq.~(\ref{gAdef}) for the non-singlet sector. Therefore  
Eq.~(\ref{Meta}) gets  modified (``renormalised'')
by the (inverse) axial singlet form factor, 
\be
M_{\eta^\prime}^2 
= 
-\frac{\kappa_B}{\kappa^2_A-\kappa^2_B}\:\frac{1}{\t B_0 (M_{\eta^\prime}^2,M_\psi^2)}\:g_a^{-1}(M_{\eta^\prime}^2)
~,
\label{Metabis}
\ee
which is the final expression that we will use in numerical illustrations of the mass spectrum in the next subsection.

\subsection{The mass spectrum of the resonances}
\label{The mass spectrum of the resonances}

The resonance masses have to be proportional to the unique independent energy scale of the theory, 
which is conveniently choosen as $f\equiv \sqrt{2} F_G$, defined in Eq.~(\ref{FGsum}), 
as explained above. 
In order to fix the ideas, one can take $f$ just above the lower bound imposed by electroweak precision tests, 
which is conservatively given by $f = 1$ TeV.
Since the resonance masses are $N$-independent and $f \sim \sqrt{N}$, in principle 
the resonances become lighter and lighter in the large-$N$ limit. 
However, if the model is augmented with coloured fermions to provide top partners, as we will do in section \ref{coloured-sector}, 
the $Sp(2N)$ asymptotic freedom is lost (at one loop) for  $N\ge19$ \cite{Barnard:2013zea}.
Moreover, these coloured fermions are also charged under $U(1)_Y$, resulting in Landau poles in the SM gauge 
couplings ($\alpha_1$ and $\alpha_3$) possibly too close to the condensation scale of the strong sector.
A naive one-loop estimation of the running of the SM gauge couplings in presence of the hypercolour fermions leads to the appearance of 
Landau poles around $100$ ($500$) TeV for $N=6$ $(5)$ while for $N=4$, the Landau poles appear above $4\cdot 10^3$ TeV.
Then, a more reasonable  interval for the number of hypercolours is $2\leq N\leq 6$. For the numerical illustration, we take the 
conservative value $N=4$.

The resonance masses are a function of the couplings $\kappa_{A,B,C,D}$ of the four-fermion operators. 
For the numerical illustration, we will assume Eq.~(\ref{SVfierz4}) to hold, $\kappa_C=\kappa_D=\kappa_A$, and we will trade the two 
remaining, independent couplings
for the dimensionless parameters $\xi \equiv (\kappa_A+\kappa_B)\Lambda^2/(4\pi^2)$ and $\kappa_B/\kappa_A$.

Let us describe the main feature of the mass spectrum. Since we work in the chiral limit approximation, 
the resonances are complete multiplets of the unbroken $Sp(4)$ symmetry, and the Goldstone bosons $G_{\hat A}$ are massless.
In the spin-zero sector, there are three independent massive states: the singlet scalar $\sigma$ and the five-plet scalar $S_{\hat A}$,
see Eq.~(\ref{Ma2}), as well as the
singlet pseudoscalar $\eta^\prime$, see Eq.~(\ref{Meta}). 
The latter is the would-be Goldstone boson of the anomalous $U(1)_\psi$, therefore $M_{\eta^\prime}$
vanishes when this symmetry is restored, that is when $\kappa_B/\kappa_A\rightarrow 0$.
In the spin-one sector, there are two independent masses: 
the singlet axial vector $a^\mu$ and the five-plet axial vector $A^\mu_{\hat A}$ are mass-degenerate 
as we assume $\kappa_C=\kappa_D$,
with mass given by Eq.~(\ref{MA}), while the ten-plet vector $V^\mu_A$ has a different mass, see Eq.~(\ref{MV}).
Even though we neglect the mass splitting among the different electroweak components,
in view of collider searches it is important to keep in mind the electroweak charges of the resonances, that are
fixed by the decomposition of the $Sp(4)$ representations under the $SU(2)_w\times U(1)_Y$ gauged subgroup:
\be
1_{Sp(4)}=1_0~,
\qquad 5_{Sp(4)} = (2_{1/2} +h.c.)+1_0~,
\qquad 10_{Sp(4)} = 3_0 + (2_{1/2} +h.c.) + (1_1+h.c.) + 1_0 ~.
\ee

\begin{figure}[b]
\includegraphics[scale=0.25]{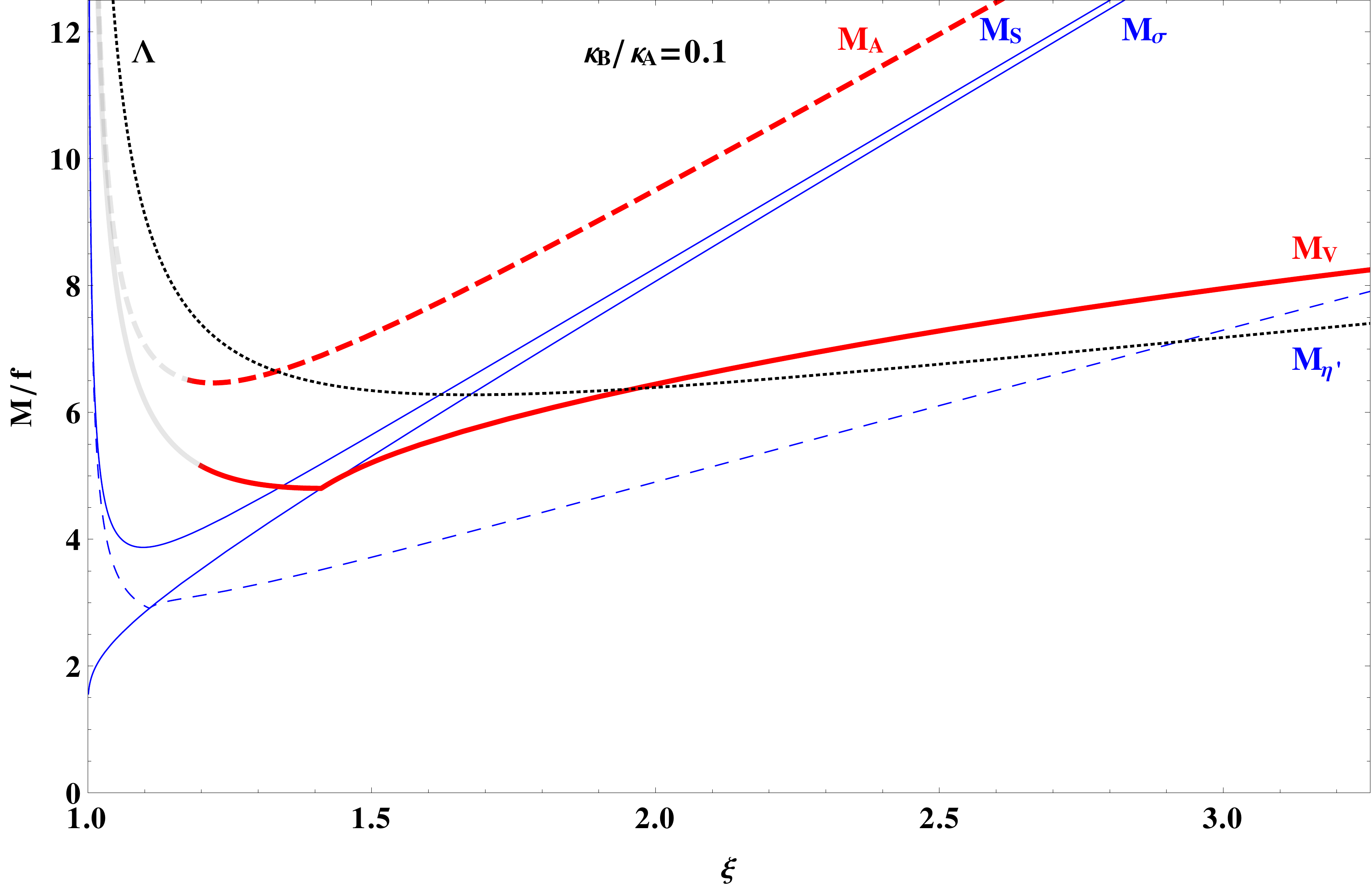}
\includegraphics[scale=0.25]{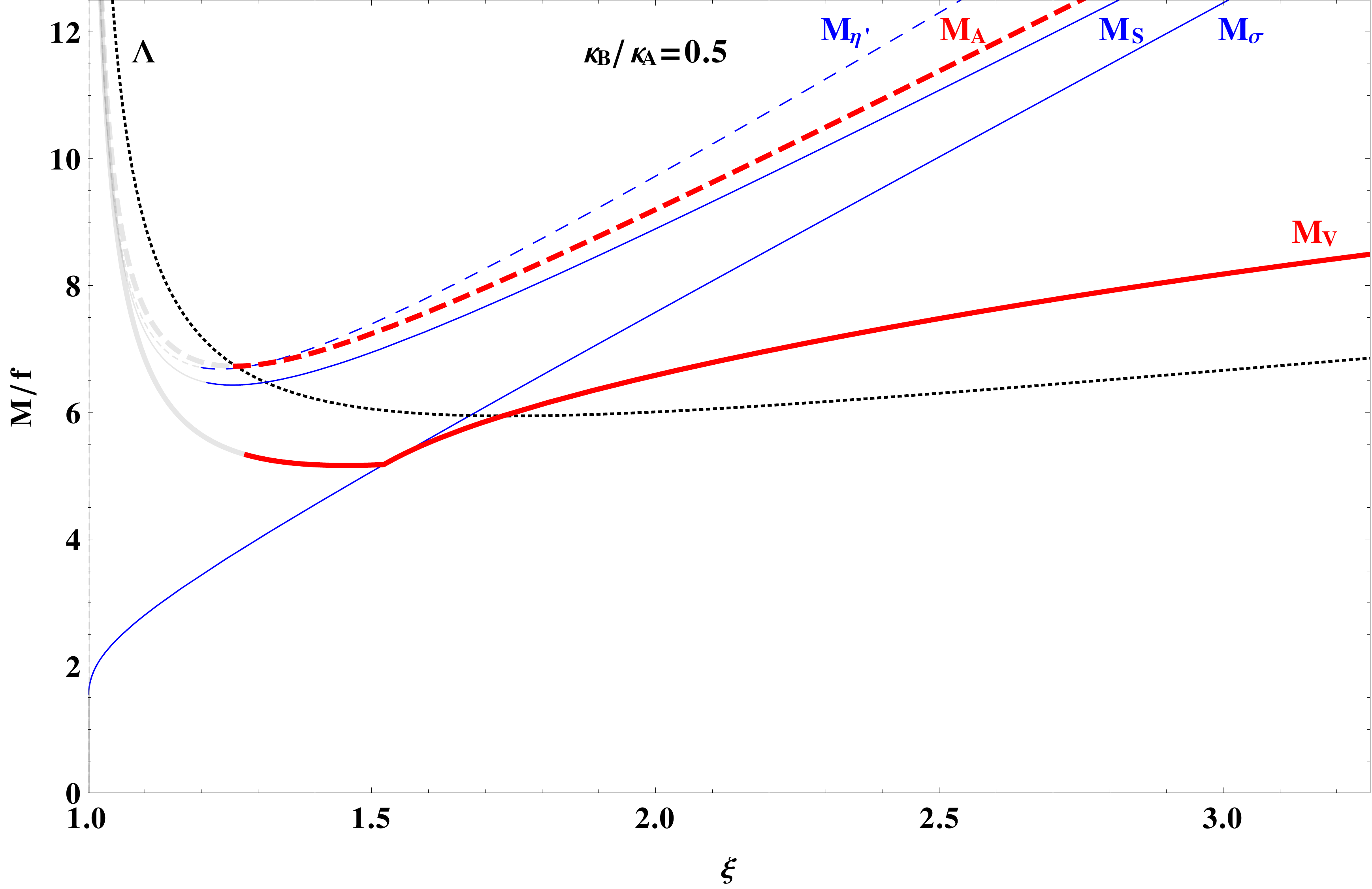}
\caption{The masses of the electroweak resonances in units of the Goldstone decay constant $f$, for $N=4$ (the masses
scale with $1/\sqrt{N}$), as a function of the coupling $\xi$, for $\kappa_B/\kappa_A=0.1$ (left-hand panel) and
$\kappa_B/\kappa_A=0.5$ (right-hand panel).
We displayed the full physical range for $\xi$, according to Fig.~\ref{FGLam}.
Each curve is shaded when the corresponding pole mass equation develops a large, unphysical imaginary part, 
$|{\rm Im}[g_\phi(M_\phi^2)]/{\rm Re}[g_\phi(M_\phi^2)] |>1$.
The dotted line is the cutoff of the constituent fermion loops.}
\label{M/FgA10}
\end{figure}

In Fig.~\ref{M/FgA10} we display the five independent resonance masses, $M_{\sigma,\eta^\prime,S,V,A}$, as a function of $\xi$,
for two representative values of $\kappa_B/\kappa_A$. While $M_\sigma = 2M_\psi$ grows over the entire range for $\xi$,
the other four masses follow a different pattern: they appear to be several times larger than $f$ when $\xi$ is very close to one (see the 
discussion in the next paragraph), 
then they steeply decrease to reach a minimum value$\sim (2-3)f$ for an intermediate value of $\xi$, and finally they grow roughly linearly 
for $\xi\gtrsim 1.5$.
We recall the two approximate mass relations, $M_S\simeq (M_\sigma^2 + M_{\eta^\prime}^2)^{1/2}$ and 
$M_A\simeq (M_V^2+3 M_\sigma^2/2)^{1/2}$, that hold neglecting pole mass differences in the loop form factor.
As a consequence, one has always $M_A>M_S>M_\sigma$, with a similar asymptotic value at large $\xi$. 
On the contrary, $M_V$ 
decreases until it becomes degenerate with $M_\sigma$, then it grows with a weaker slope.
Finally, $M_{\eta^\prime}$ may also become smaller than $M_{\sigma}$ at large values of $\xi$, 
but only for a sufficiently small value of $\kappa_B/\kappa_A$.
For example, taking $f=1$ TeV,  $N=4$ and $\kappa_B/\kappa_A=0.1$, the resonance masses for two representative values of $\xi$
are 
\bea
\xi=1.3~  &:~\quad&  M_A \simeq 6.6 ~{\rm TeV},~~  M_V\simeq 4.9~{\rm TeV},~~  M_S\simeq 4.6~{\rm TeV},~~M_\sigma \simeq 4.1~{\rm TeV},~~ M_{\eta^\prime} \simeq 3.3~{\rm TeV}~,
\nonumber\\
\xi=2.0~ &:~ \quad&  M_A \simeq 9.5~{\rm TeV},~~  M_V\simeq 6.4~{\rm TeV},~~  M_S\simeq 8.3~{\rm TeV},~~M_\sigma \simeq 8.1~{\rm TeV},~~ M_{\eta^\prime} \simeq 4.9~{\rm TeV}~.
\label{mass-spetrum-EW-benchmark}
\eea
In general, electroweak resonances lighter than  $\simeq 4 f \simeq 4$ TeV are possible in two cases: the scalar $\sigma$ becomes light 
when one approaches the critical coupling $\xi=1$,
where the mass gap vanishes; the pseudoscalar $\eta^\prime$ becomes light as $\kappa_B/\kappa_A$ tends to zero, where the anomalous
$U(1)_\psi$ symmetry is restored.
These two singlet states, together with the SM singlet Goldstone boson $G^{\hat 3}$, may be observed as the lightest scalar resonances at 
the LHC, beside the $125$ GeV Higgs boson.
In section \ref{mixing-singlets} we will discuss the mixing of $\sigma$ and $\eta^\prime$ with the analogous singlet states of the colour 
sector, a feature that will induce corrections to their masses.

A comment is in order on the region close to the critical coupling. In the limit $\xi\rightarrow 1$, 
one finds that $M_\sigma/f \sim [-\log(\xi-1)]^{-1/2}$ vanishes, 
while the other resonance masses diverge
relatively to $f$, $M_{V,A,S,\eta^\prime}/f \sim (\xi-1)^{-1/2}$.
The lightness of $\sigma$ may be interpreted as the signal that scale invariance is recovered below $\xi=1$, 
while all other resonances decouple in this limit.
However, we should remark that, for some of these heavy resonances, the NJL computation of their masses cannot be trusted close to the 
critical coupling, because the pole of the resummed propagator develops
a large, unphysical imaginary part. 
Recall, from the general discussion at the beginning of section \ref{Masses and couplings of scalar resonances}, that the curves in Fig.~
\ref{M/FgA10} are the solution 
of Eq.~(\ref{res_sol}),\footnote{ The function ${\rm Re\,} {\t B_0}(q^2, M_\psi^2)$ develops a cusp at $q^2 = 4 M_\psi^2$.
Through the definition of the masses $M_\phi$ adopted here, this cusp naturally shows up
in Fig.~\ref{M/FgA10} (and in Fig. \ref{fig_Sum_Rules} below) as soon as the value of a resonance mass 
goes through $2 M_\psi$. In practice, this only occurs for $M_V$ and $M_{\eta^\prime}$, at the cross-over from a bound state to a genuine 
resonance.}
where the imaginary part of $g_\phi(M_\phi^2)$ has been neglected.
The curves in Fig.~\ref{M/FgA10} are shaded when  $|{\rm Im}[g_\phi(M_\phi^2)]/{\rm Re}[g_\phi(M_\phi^2)] |>1$, where we consider that the 
corresponding result cannot be trusted anymore. 
This happens when $\xi\lesssim(1.2-1.3)$, for the vector and axial-vector resonances, with masses $M_{V/A}$ 
close to the cutoff of the NJL model.

Let us also comment on the complementary limit where $\xi$ is so large that  $M_\psi/\Lambda$ becomes of order one, 
as illustrated in Fig.~\ref{FGLam}.
In this case Fig.~\ref{M/FgA10} shows that the resonances become heavier than $\Lambda$ (except for $\eta^\prime$, if $\kappa_B/\kappa_A$ is small enough).
This is not necessarily problematic: while the mass $M_\psi$ of constituent fermions in the loops need to be smaller than
the loop cutoff $\Lambda$, external mesons heavier than $\Lambda$ do not harm the consistency of the NJL approximation.
Indeed, in QCD the NJL model predicts rather accurately resonance masses twice as large as the cutoff. 
Nonetheless, we notice that, for $M_\phi\sim \Lambda$,
the value of the two-point function $\t B_0(M_\phi^2,M_\psi^2)$ becomes sensitive to the regularisation chosen, 
defined in appendix \ref{loop-functions}, as the cutoff-dependent finite terms become sizeable.
As a consequence, we observe that the mass values in this region may vary up to a few $10\%$ in different regularisation schemes. This is an intrinsic theoretical uncertainty of the NJL approximation.
\\

The resonance masses in units of $f\equiv \sqrt{2} F_G$ may be compared with recent lattice 
studies of the same model~\cite{Arthur:2016dir,Arthur:2016ozw}, which provide scalar and vector masses 
in the same units.\footnote{Our normalisation of $f$, see footnote \ref{ftn7}, 
appears consistent with what is called $F_{PS}$ in the notations of Ref.~\cite{Arthur:2016dir} 
thus we compare our NJL predictions in units of $f$ directly with their numbers, assuming 
that the same normalisation has been used in those lattice calculations.}
Actually, the lattice simulations performed to date for this model are available only for an underlying 
$SU(2)$ gauge theory, thus equivalent to the special case $Sp(2)$ of our more general $Sp(2N)$ study. 
Let us recall that the meson masses scale as $M_\phi/f\sim 1/\sqrt{N}$, where the scaling originates 
solely from $f$ (this statement holds for a fixed value of the ratio $\kappa_B/\kappa_A$).
Therefore, the mass values illustrated for $N=4$ in Fig. \ref{M/FgA10}  get enhanced by a 
factor $2$ for $N=1$, and these rescaled values can be directly compared with the lattice
results.

The lattice prediction for the vector masses in the chiral limit is $M_V/f = 13.1\pm 2.2$, $M_A/f = 14.5\pm 3.6$ \cite{Arthur:2016dir}. 
The latter results, although affected with relatively large uncertainties, 
indicate a more moderate $V-A$ mass splitting than is generally expected from
the NJL model, see Eq.~(\ref{MVMA}), unless $M_\psi$ is rather small, which corresponds  
in the NJL framework to rather small values of $\xi$. More precisely, typically the previous 
central lattice values can be (approximately) matched for $\xi\simeq 1.1$, therefore not far 
above the critical NJL coupling value, where on the other hand the NJL calculation becomes less 
reliable, as already explained above, since
entering the $\xi$ range where the $V$ and $A$ width both become relatively large.
But accounting for the lattice uncertainties, the above values are also easily matched 
alternatively for rather large $\xi$ values, where the NJL prediction is also more reliable: for example 
for $N=1$ and $\xi=1.6$ [$\xi=1.9$], $M_V/f|_{NJL}\simeq 11~[\simeq 12.5]$, $M_A/f|_{NJL}\simeq 15.3~[\simeq 18]$.
[NB recall that the $V$ and $A$ masses are mildly dependent on $\kappa_B$, which enters
only indirectly through the mass gap. 
One should also keep in mind that the Fierz-induced relation (\ref{SVfierz4}) is 
assumed for the axial and vector coupling $\kappa_D$ in  Fig. \ref{M/FgA10}, and since    
the dominant contribution to the $V, A$ masses scales as $1/\kappa_D$, 
a somewhat smaller (larger) $\kappa_D$ would induce somewhat larger (smaller) $V, A$ masses, for a fixed
value of $\xi$]. 
At least one may tentatively conclude from this comparison that intermediate $\xi$ values, say $1.2\lsim \xi \lsim 1.6$ approximately, as 
well as very large $\xi >2$, appear more disfavoured. 

Concerning the lightest scalar masses, Ref.~\cite{Arthur:2016ozw} provides the very recent 
lattice estimates $M_\sigma/f =19.2(10.8)$,  $M_{\eta^\prime}/f = 12.8(4.7)$,  and $M_{S}/f =16.7(4.9)$, 
in the chiral limit (where the scalar non-singlet $S$ is called $a_0$ in Ref. \cite{Arthur:2016ozw}). 
Compared with Fig. \ref{M/FgA10} (rescaled for $N=1$)
and combined with the results for the $V$ and $A$ masses, $\xi$ values very close to $1$
appear disfavoured by the $\sigma$ mass, even when taking its lowest lattice value above, 
because in this region the NJL prediction for $M_\sigma$ is  
much smaller than $M_V$, as it is clear from $M_\sigma =2M_\psi$ (see also Fig.~\ref{M/FgA10}). 
The NJL (approximate) relation $M^2_S\simeq M^2_\sigma +M^2_{\eta'}$ (see Eq.~(\ref{Ma2})),
can be fulfilled within the large lattice uncertainties, although 
the rather high lattice central value of $M_{\sigma}$ is 
in tension with this relation.
So putting all together it may indicate
that relatively large values of $\xi\simeq 1.6-2$, well above the NJL critical coupling, are more favoured by lattice results. 
The $\eta^\prime$ pseudoscalar mass, in the NJL model, is very sensitive to the ratio $\kappa_B/\kappa_A$, see Eq.~(\ref{Metasimple}). 
Modulo the large lattice uncertainties, 
the comparison with lattice results appears to indicate 
intermediate values for this ratio, $\kappa_B/\kappa_A\simeq 0.2-0.4$, such that 
$M_{\eta^\prime}$ is comparable with $M_V$.

In conclusion the comparison of NJL and lattice results appears roughly consistent, 
at least the lattice results may be matched for some definite values of the NJL parameters $\xi$ and $\kappa_B/\kappa_A$,
with no strong tensions.  
But it appears still an essentially qualitative comparison at the present stage, 
given both the intrinsic NJL uncertainties amply discussed
previously, as well as the still relatively large lattice systematic uncertainties, specially for the 
scalar resonances: so unfortunately it cannot
be taken yet as giving tight constraints on the effective NJL model parameters. Note also that other
recent lattice simulations of composite Higgs model resonances are available in the 
literature (see e.g. \cite{DeGrand:2015lna,DeGrand:2016pur}), 
but are based on different gauge symmetries and/or global symmetry breaking 
pattern, thus not directly comparable with our results.

\subsection{Comparison with spectral sum rules}
\label{secWSR}

Several authors \cite{Bijnens:1993ap,Dmitrasinovic:1996ka,Klevansky:1997dk} have addressed the issue of spectral sum rules,
discussed in general terms in Section \ref{SR}, in the context of the NJL approximation applied to QCD. 
In this Section, we will study them in the context of the NJL approximation to the
underlying $Sp(2N)$ gauge dynamics of the present composite Higgs framework. The aim
will be to check whether these sum rules provide additional constraints on the parameters
of the model, namely $\xi$ and $\kappa_B/\kappa_A$.

It seems only natural to identify the spectral densities appearing in the sum rules displayed
in Eqs. (\ref{WSRVA}) and (\ref{scalSR}) with the discontinuities of the resummed NJL two-point 
correlators%
\footnote{At the level of one-loop two-point correlators, the spectral sum rule (\ref{WSRmexpl}) is trivially
satisfied, provided one identifies $m$ with $M_\psi$, due to the identity $\t \Pi_V (q^2) - \t \Pi_A (q^2) = - \t \Pi_A^L (q^2) $.
The identities
$$
\t \Pi_S (q^2) - \t \Pi_P (q^2) = \t \Pi_{S} (q^2) - \t \Pi_{\eta^\prime} (q^2) 
= \t \Pi_\sigma (q^2) - \t \Pi_G (q^2) =  2 (2N) M_\psi^2 \t B_0 (q^2 , M_\psi^2)
$$
allow only for the difference of the two last sum rules in Eq. (\ref{scalSR}),
involving $\t\Pi_{S{\mbox -}\eta^\prime} - \t\Pi_{\sigma{\mbox -}G}$, to be satisfied at one-loop.
The sum rule involving $\Pi_{S{\mbox -}P}$ is not expected to hold, since this
correlator does not constitute an order parameter for $SU(4)/Sp(4)$, see footnote \ref{fnte_SP}. \label{1loop_ids}
}
discussed in the preceding subsections, i.e.
\be
{\rm Im} \,\Pi_{V/A} (t) = 
\lim_{~\epsilon\to 0^+}
\frac{{\overline\Pi}_{V/A} (t+i\epsilon) - {\overline\Pi}_{V/A} (t-i\epsilon)}{2i}
,
\label{ImPI_VA}
\ee
or, in the singlet scalar and pseudoscalar channels,
\be
{\rm Im} \,\Pi_{S^0/P^0} (t) = 
\lim_{~\epsilon\to 0^+}
\frac{{\overline\Pi}_{\sigma/\eta^\prime} (t+i\epsilon) - {\overline\Pi}_{\sigma/\eta^\prime} (t-i\epsilon)}{2i}
,
\label{ImPI_SP}
\ee
and analogous relations between ${\rm Im} \,\Pi_{S/P} (t)$ and ${\overline\Pi}_{S/P} (t)$.
Before discussing the sum rules of Section \ref{SR} under these identifications,
let us recall that the sum rules themselves follow from the short-distance properties,
which reflect the properties of the underlying $Sp(2N)$ gauge dynamics, of
the two-point functions under consideration, and from general properties of quantum field theories,
here essentially invariance under the Poincar\'e group and the spectral property. The latter allow
to extend the definitions of the functions $\Pi_\phi(t)$ to functions in the complex $t$-plane,
with all singularities (poles and branch points) confined to the positive real axis. The former then allow to write down
unsubtracted dispersion relations for the appropriate combinations of two-point correlators,
from which the sum rules follow. The necessity to introduce a regularisation (here the cut-off $\Lambda$),
in order to render the one-loop correlators ${\t\Pi}_\phi(t)$ finite, and to perform the resummation shown in Fig. \ref{BS},
leads to functions ${\overline\Pi}_\phi (t)$ that will in general not respect all the required properties. 
For instance, with the choice of regularisation
adopted in the present study, ghost poles on the {\it negative} real $q^2$-axis will appear, as discussed at the beginning of 
Section~\ref{Masses and couplings of scalar resonances}. This situation is well known in the
context of the NJL approximation applied to QCD, where it has been examined quite extensively
by the authors of Ref. \cite{Klevansky:1997dk}, and we refer the reader to this article for additional details.

\begin{figure}[t]
\begin{center}
\includegraphics[scale=0.25]{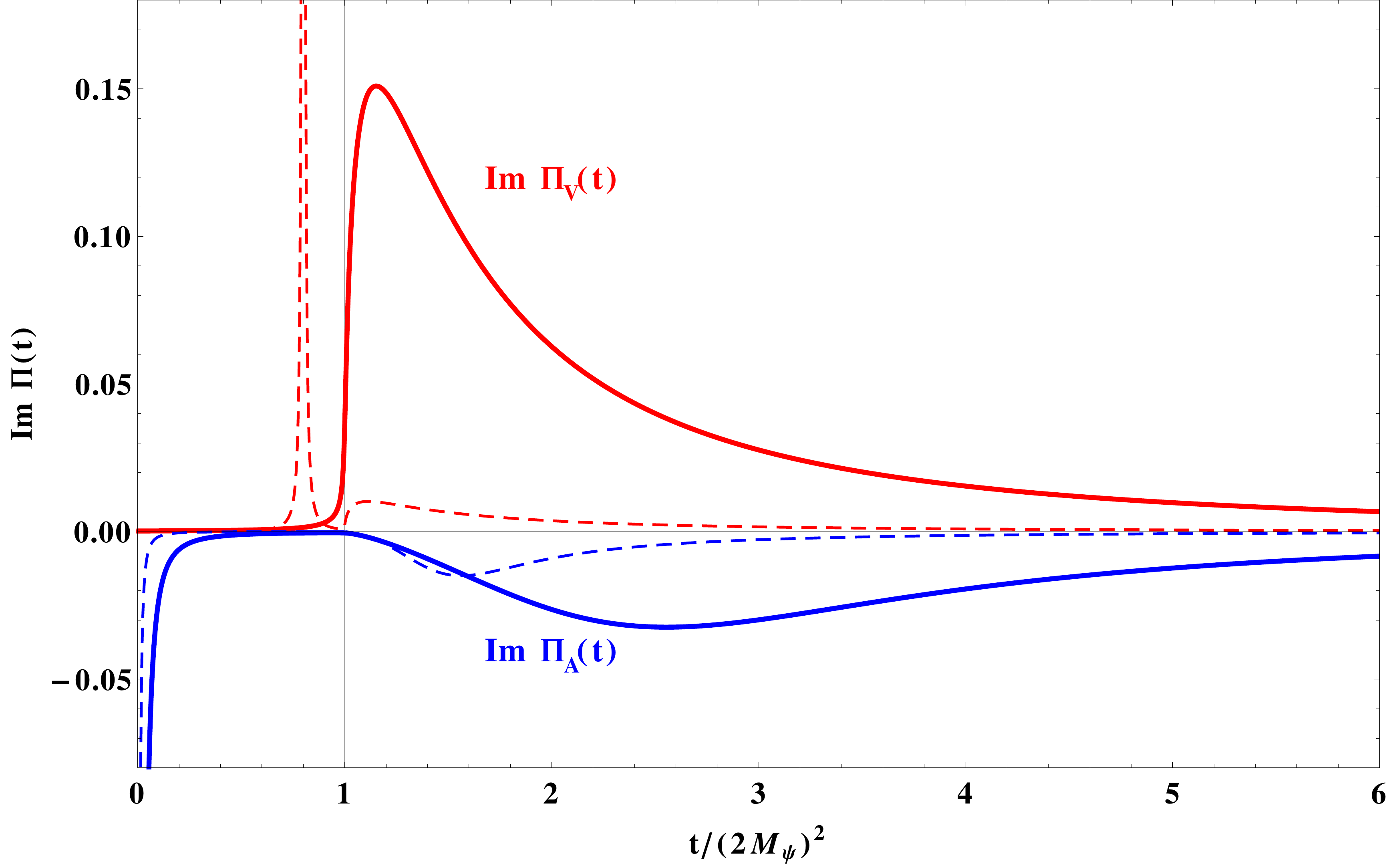}
\includegraphics[scale=0.25]{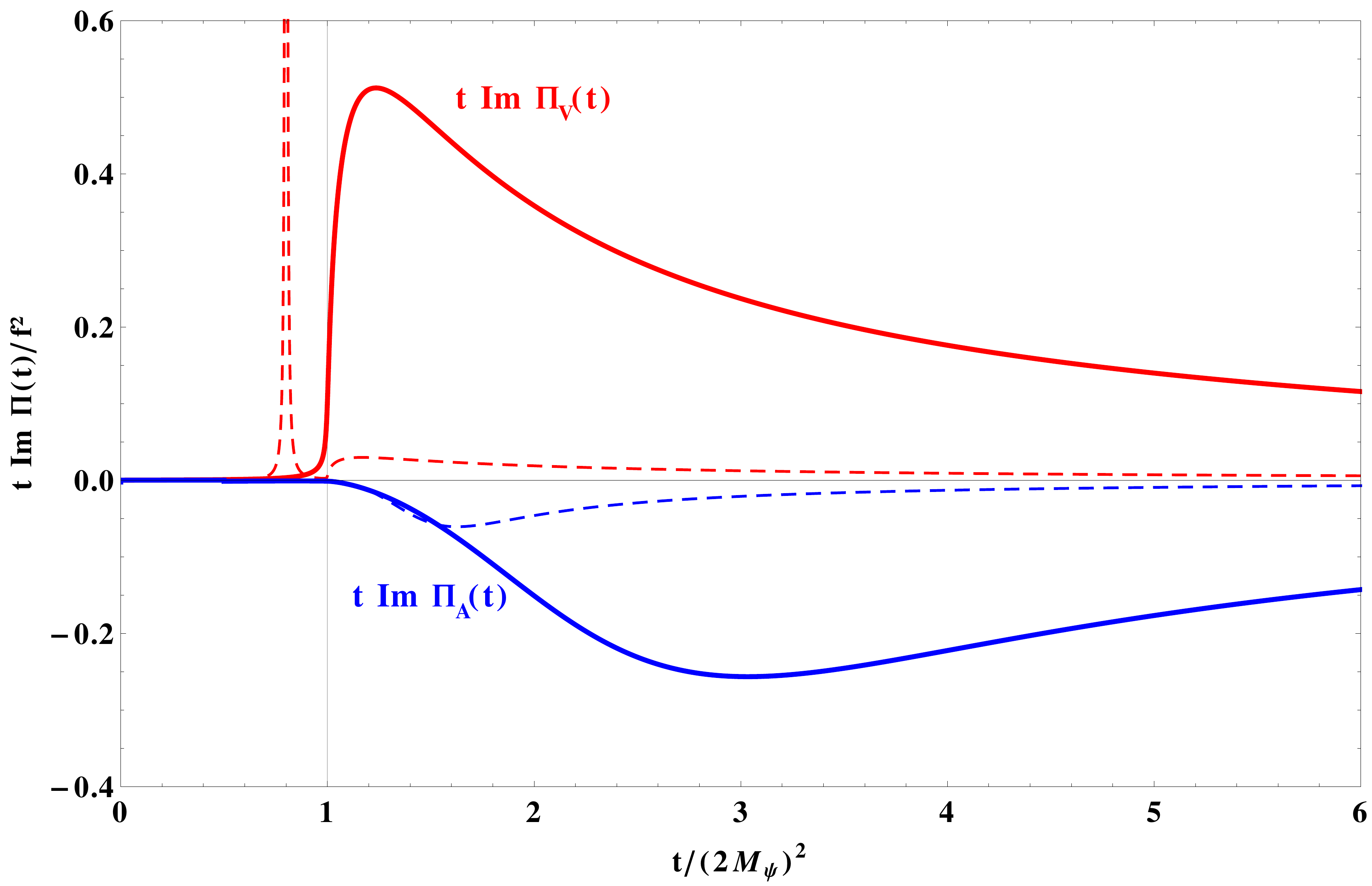} 
\end{center}
\caption{ The figure on the left shows the spectral functions ${\rm Im} \,\Pi_V (t)$ (upper curves, in red)
and $- {\rm Im} \,\Pi_{A} (t)$ (lower curves, in blue), as a function of $t/(2 M_\psi)^2$ .
The plotted quantities  are dimensionless and scale like $N$.
The solid and dashed lines correspond to $\xi = 1.3$ and $\xi = 2$,
respectively. The value of the parameter $\kappa_B/\kappa_A$ has been taken equal to 0.1 in all cases.
The narrow vector bound state below the
continuum starting at $t=(2 M_\psi)^2$ (materialised on the figures by the vertical line) 
is present in ${\rm Im} \,\Pi_{V} (t)$ when $\xi=2$,
but disappears for smaller values of $\xi$.
The pion pole appears clearly in ${\rm Im} \,\Pi_A (t)$, but the axial-vector resonance has a mass that is always
greater that $4M_\psi^2$, and therefore a narrow sub-threshold peak never occurs.
The figure on the right likewise shows the functions $t \,{\rm Im} \,\Pi_V (t)$ 
and $t \,{\rm Im} \,\Pi_A (t)$.
 The latter are in units of $f^2$ and consequently are $N$- independent.
}
\label{fig_spectral_VA}
\end{figure}

\begin{figure}[t]
\begin{center}
\includegraphics[scale=0.25]{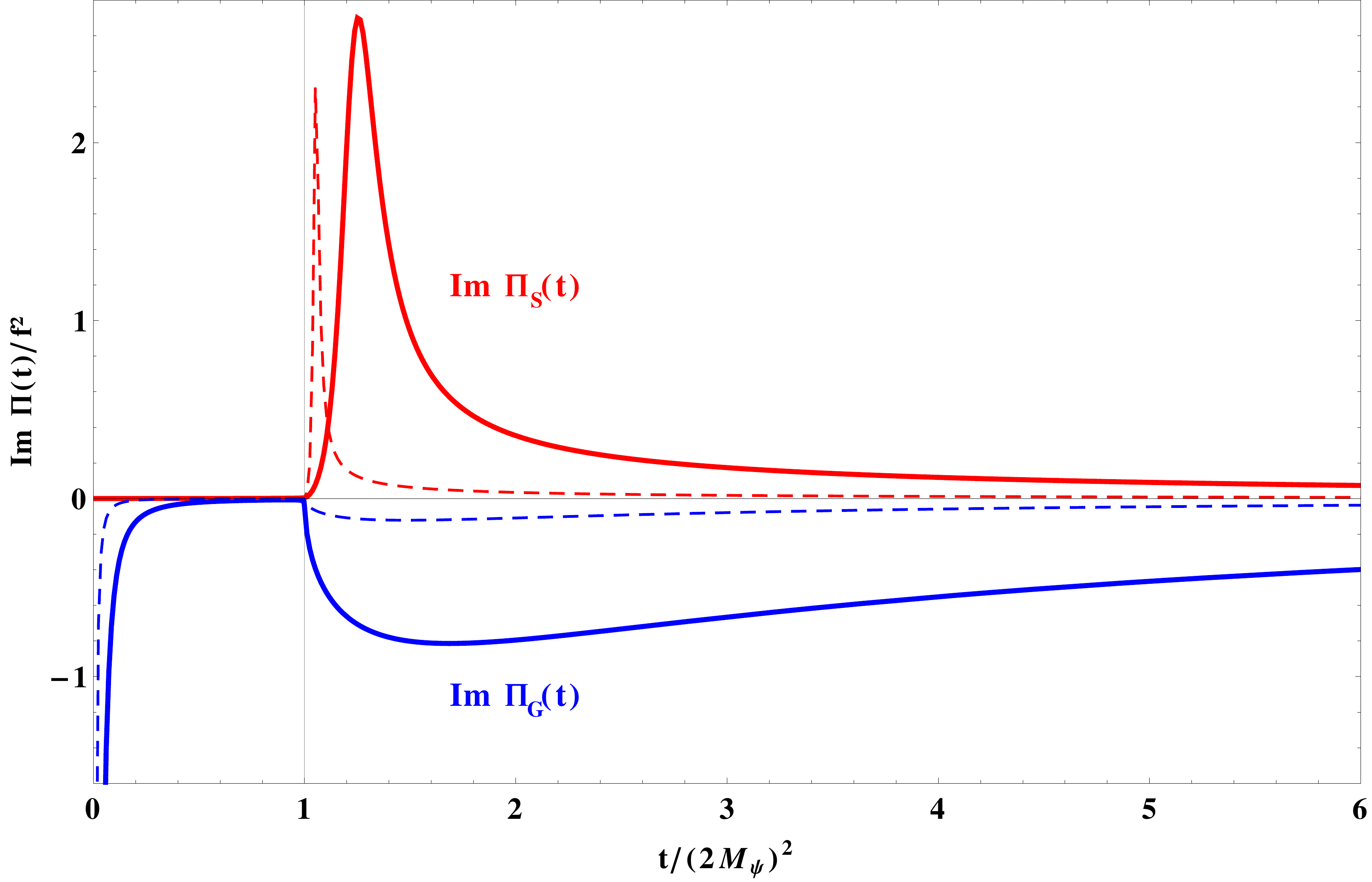}
\includegraphics[scale=0.25]{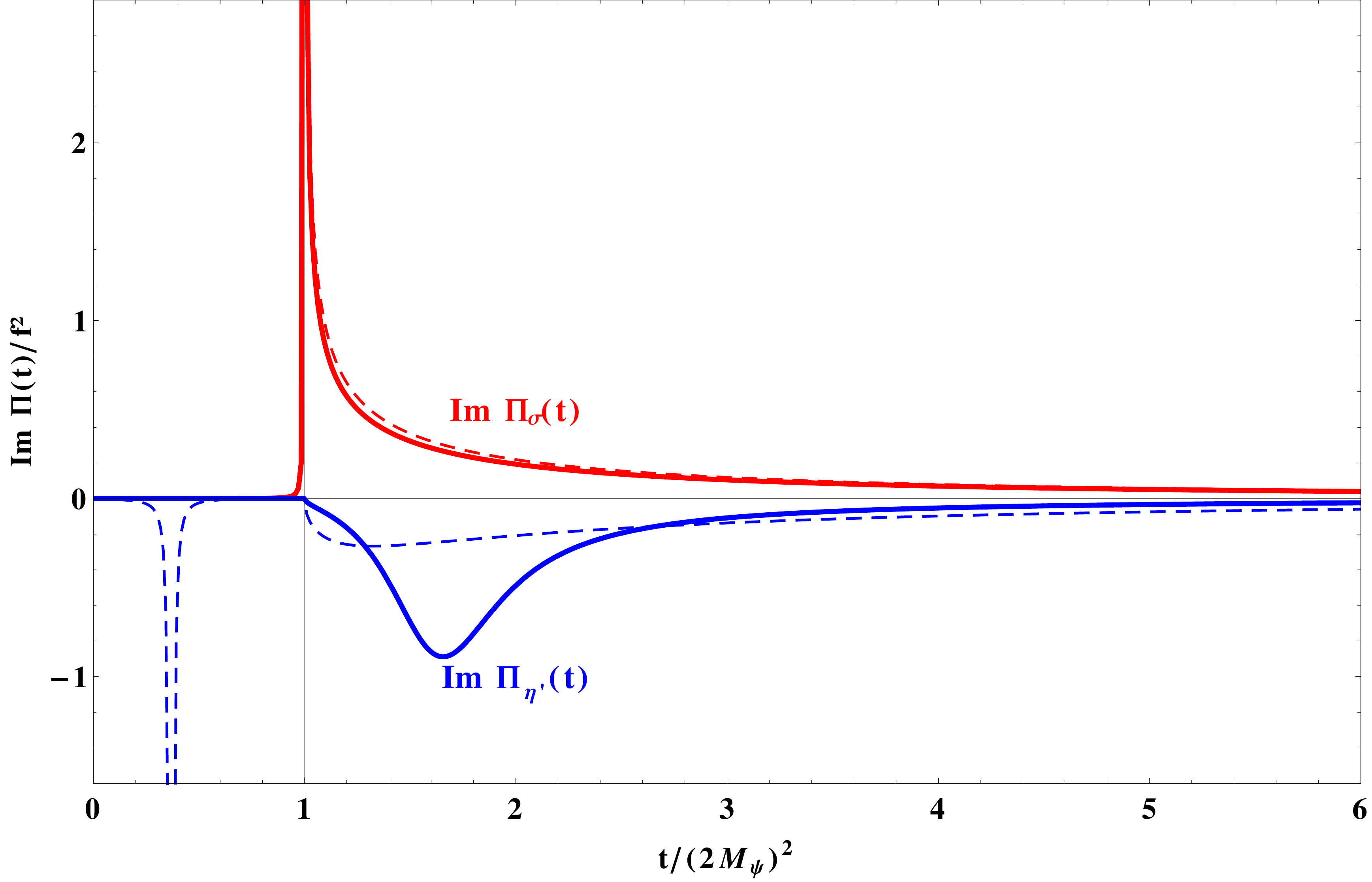} 
\end{center}
\caption{
The left-hand panel shows the non-singlet spectral functions ${\rm Im} \,\Pi_S (t)/10$ (upper curves, in red)
and $- {\rm Im} \,\Pi_{G} (t)$ (lower curves, in blue), as functions of $t/(2 M_\psi)^2$, for $\kappa_B/\kappa_A = 0.1$,
and for $\xi = 1.3$ (solid lines) and $\xi = 2$ (dashed lines). 
In the right-hand panel we fix  $\xi=2$ and show the 
singlet spectral functions ${\rm Im} \,\Pi_\sigma (t)$
(dashed  red) and $-{\rm Im} \,\Pi_{\eta^\prime} (t)$ (dashed  blue) for $\kappa_B/\kappa_A = 0.1$, as well as
${\rm Im} \,\Pi_\sigma (t)$
(solid red) and $-{\rm Im} \,\Pi_{\eta^\prime} (t)/20$ (solid blue) for  $\kappa_B/\kappa_A = 0.5$. 
The narrow $\eta^\prime$ bound state is present only for the smallest value of $\kappa_B/ \kappa_A$.
A narrow $\sigma$ pole 
appears in all cases right at the threshold $t = 4 M_\psi^2$.
 Note that the spectral functions are all expressed in units of $f^2$, such that they are dimensionless and have no $N$-dependence.
}
\label{fig_spectral_SP}
\end{figure}

The spectral densities resulting from the identifications in Eqs. (\ref{ImPI_VA})
and (\ref{ImPI_SP}) are shown in Figs. \ref{fig_spectral_VA} and \ref{fig_spectral_SP} 
(in order to make the figure more readable, we have kept $\epsilon$ in the definitions (\ref{ImPI_VA}) and (\ref{ImPI_SP})
very small, but finite).
It is most instructive to analyse them in conjunction with the spectrum of the mesonic
resonances, as given in Fig.~\ref{M/FgA10}, and with the general discussion at the beginning
of Section \ref{Masses and couplings of scalar resonances}. Figure \ref{fig_spectral_VA} shows the vector
and axial spectral functions for two different values of the parameter $\xi$. In the axial case,
one recognises the contribution from the pion pole at $t=0$, and no other narrow bound state. 
Only a rather broad resonance peak appears above the $t=4 M_\psi^2$ threshold, where
the continuum starts. This is in agreement with Fig.~\ref{M/FgA10},
which shows that $M_A$ is always greater than $M_\sigma = 2 M_\psi$. 
In the vector channel, a narrow bound state appears below the $2M_\psi$ threshold
for $\xi=2$, but is absent (it has moved to the real axis on the second Riemann sheet) for $\xi = 1.3$,
and is replaced by a resonance peak. Again, this agrees with Fig.~\ref{M/FgA10}, where one sees that
$M_V$ becomes greater than $2 M_\psi$ when $\xi$ takes values below $\sim 1.4$.

For the non-singlet scalar spectral density, shown on the left panel of Fig.~\ref{fig_spectral_SP}, there is no
narrow bound state lying below the threshold of the continuum, whatever the value of $\xi$. However,
the larger the value of $\xi$, the more the resonance peak moves closer to the threshold. 
The shape of ${\rm Im \,} \Pi_S(t)$ is also sensitive to $\kappa_B/\kappa_A$.
In the pseudoscalar non-singlet channel, only the massless pion pole shows up, and ${\rm Im \,} \Pi_P(t)$
is not sensitive to the value of $\kappa_B/\kappa_A$. 
The singlet scalar spectral density, shown on the right panel of Fig.~\ref{fig_spectral_SP}, presents a narrow
peak at the threshold, for any value of $\xi$ and $\kappa_B/\kappa_A$.
In the pseudoscalar singlet channel,
the features of the spectral function become also sensitive to this second parameter, 
as can already be inferred upon comparing the two panels of Fig. \ref{M/FgA10}.
In particular, a narrow sub-threshold bound state is only present for smaller values of 
$\kappa_B/\kappa_A$. 

\begin{figure}[tb]
\begin{center}
\includegraphics[scale=0.25]{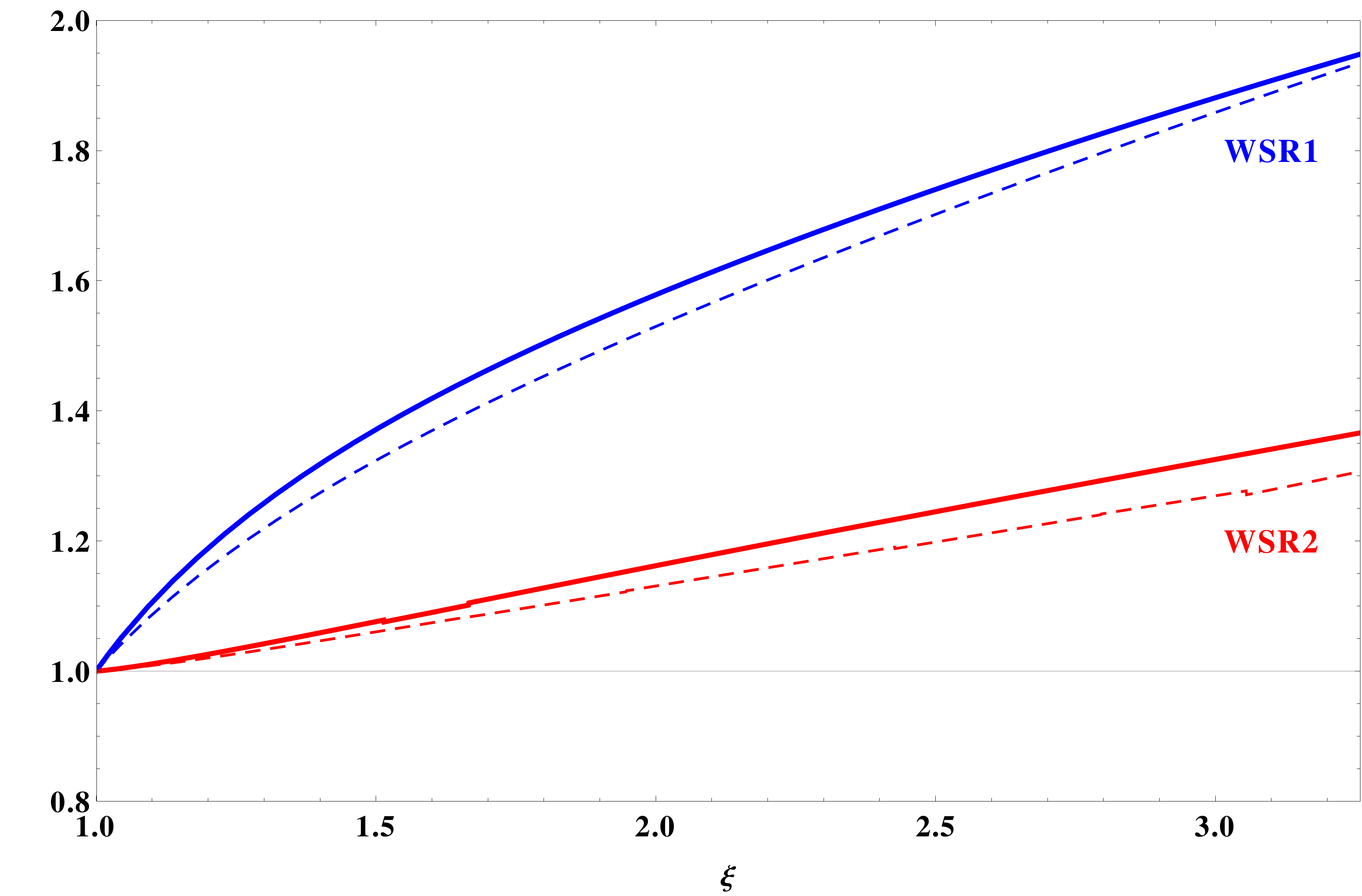}
\includegraphics[scale=0.25]{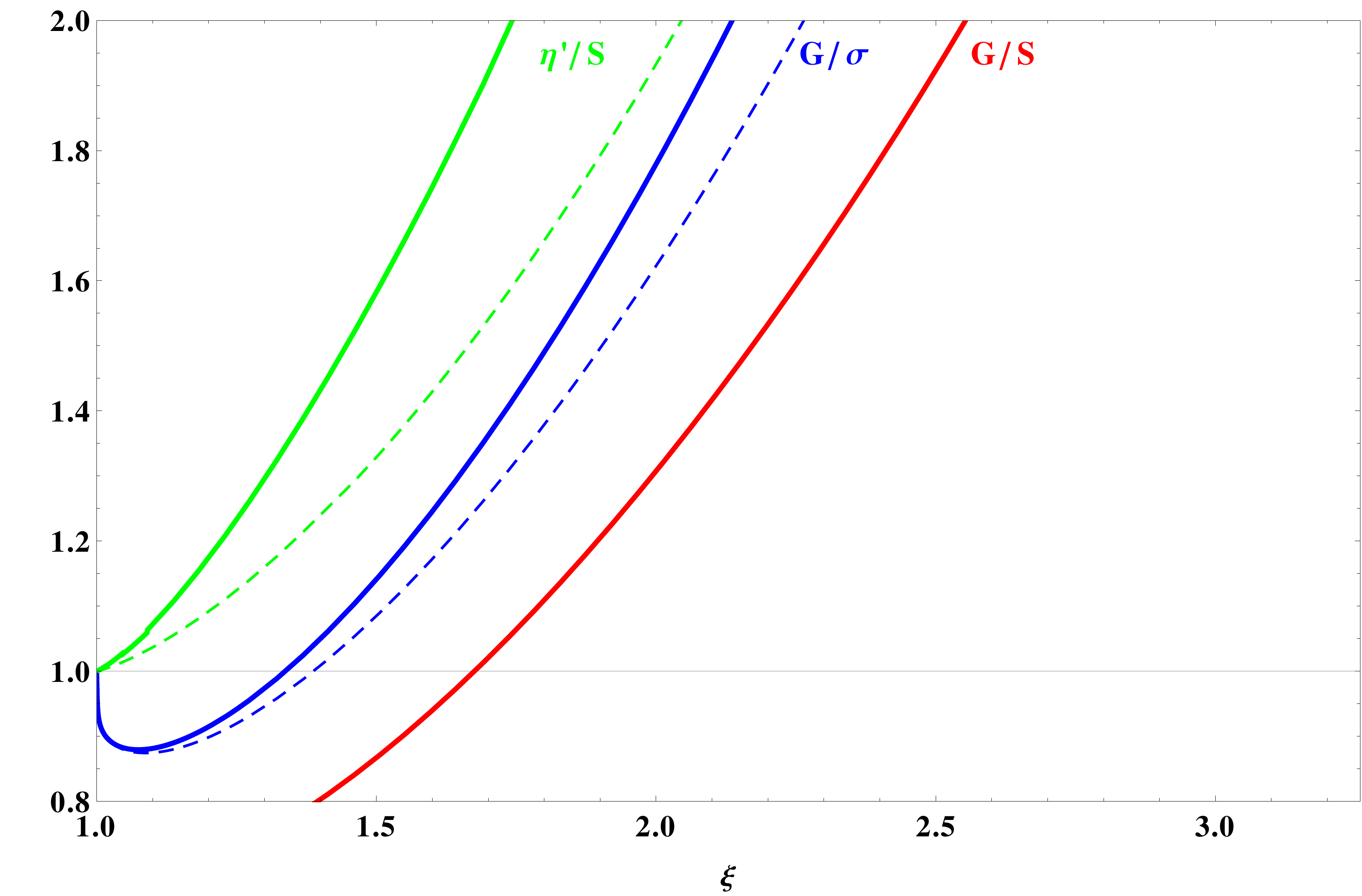}  
\end{center}
\caption{Left panel: the ratio of the integrals, taken over the whole positive
$t$-axis, 
 $\int dt~ {\rm Im\,}\Pi_V(t)/\int dt ~{\rm Im\,}\Pi_A(t)$ 
(blue, upper curves) and 
$\int dt ~t~{\rm Im\,}\Pi_V(t)/\int dt~ t~{\rm Im\,}\Pi_A(t)$ (red, lower curves), as a function of 
the parameter $\xi$, and for
$\kappa_B/\kappa_A = 0.1$ (solid lines) and $\kappa_B/\kappa_A = 0.5$ (dashed lines). 
Right panel: the ratio of the integrals, taken over the whole positive
$t$-axis, 
$\int dt ~{\rm Im\,}\Pi_{\eta^\prime}(t)/\int dt ~{\rm Im\,}\Pi_S(t)$ (green, upper curves),
$\int dt ~{\rm Im\,}\Pi_G(t)/\int dt ~{\rm Im\,}\Pi_\sigma(t)$ (blue, middle curves) and
$\int dt ~{\rm Im\,}\Pi_G(t)/\int dt ~{\rm Im\,}\Pi_S(t)$ (red, lower curve), as a function of the parameter $\xi$, for
$\kappa_B/\kappa_A = 0.1$ (solid lines) and $\kappa_B/\kappa_A = 0.5$ (dashed lines, not shown in the $G/S$ case).
 Note that the above ratios are independent from $N$.}
\label{fig_Sum_Rules}
\end{figure}

An illustration of the two Weinberg-type sum rules of Eq. (\ref{WSRVA}), 
as well as the sum rules of Eq. (\ref{scalSR}), is provided by Fig.~\ref{fig_Sum_Rules}. The integrals compared there, 
as functions of the coupling $\xi$ and for two values of $\kappa_B/\kappa_A$, 
run over the whole positive $t$-axis, which means that,
for the sake of illustration, the NJL description has been kept even beyond its expected range of validity.
Of course, it is certainly difficult to ascribe any physical
meaning to the spectral densities for values of, say, $t/\Lambda^2 \gsim 2$ [note that, for $\xi$ close to the critical coupling,
one has $2M_\psi\ll \Lambda$, therefore the NJL description holds up to a large value of $t/(2M_\psi)^2$]. 
Beyond this value of $t$, the NJL description ceases to be appropriate,
and we have to assume that the underlying $Sp(2N)$ gauge dynamics takes over. 
However, from the experience with QCD \cite{Peris:1998nj}, it is expected that the matching
between the two regimes is not very smooth. Keeping this proviso in mind, we show, on
the left-hand panel of Fig.~\ref{fig_Sum_Rules},  
the ratio of the integrals $\int dt ~{\rm Im} \,\Pi_V(t)$ and $\int dt ~{\rm Im} \,\Pi_A(t)$, 
as well as the ratio of the integrals $\int dt~ t~ {\rm Im} \,\Pi_V(t)$ and $\int dt~ t ~{\rm Im} \,\Pi_A(t)$. 
Similarly, the right-hand panel  shows the ratios of the integrals 
$\int dt ~{\rm Im} \,\Pi_{\eta^\prime} (t)$ and $\int dt ~{\rm Im} \,\Pi_S (t)$, and of the integrals
$\int dt ~{\rm Im} \,\Pi_G (t)$ and $\int dt ~{\rm Im} \,\Pi_\sigma (t)$.
If the sum rules were satisfied exactly for all values of $\xi$, all these curves would be a constant equal to one. 
This is obviously not the case. The general trend is that the departure from the sum rules
is more important for larger values of $\xi$. 
This is in line with Fig.~\ref{M/FgA10}, from which we infer that the continuum,
corresponding to $\sqrt{t} > 2 M_\psi$, starts close to the cut-off $\Lambda$ when $\xi\gtrsim1.5$,
therefore the NJL description becomes questionable
soon after the threshold. 
On the right-hand panel of Fig.~\ref{fig_Sum_Rules} we also show
the ratio of the integrals $\int dt ~{\rm Im} \,\Pi_G$ and $\int dt~ {\rm Im} \,\Pi_S$.
Since $\Pi_{S{\mbox -}P}$ is not an order parameter of the $SU(4)$ spontaneous breaking (see footnote \ref{fnte_SP}), 
there is no corresponding sum rule, and indeed this ratio deviates significantly from unity,
already for lower values of $\xi$. 
\\

In view of the difficulties to interpret the meaning of the sum rules, expressed
in  terms of the spectral densities provided by the NJL description through 
Eqs.~(\ref{ImPI_VA}) and (\ref{ImPI_SP}), one may consider an alternative
approach,
at least when ${\rm Im}\, \t\Pi_\phi (M_\phi^2)$ vanishes or is sufficiently
small so that it can be neglected. This happens, for instance, for the Goldstone state,
or for $\t\Pi_V (M_V^2)$ when there is a sub-threshold vector bound state.
In that case each correlator exhibits a single real pole, or narrow resonance
[except for ${\overline\Pi}_A(q^2)$, which exhibits both the Goldstone pole and the axial-meson resonance pole,  
the latter being not very narrow, though],
and one can saturate the sum rules with
these narrow states. Introducing, similarly to $F_G$ and to $G_G$ in Eqs. (\ref{FGdef}) 
and (\ref{GGdef}), respectively, decay constants defined as
\be
\la 0 \vert {\cal J}_\mu^{A}(0) \vert V^B(p;\lambda) \ra \equiv  f_V M_V  \: \epsilon^{(\lambda)}_\mu(p) \delta^{AB}~,
\qquad\qquad
\la 0 \vert {\cal J}_\mu^{\hat{A}}(0) \vert A^{\hat{B}}(p;\lambda) \ra \equiv f_A M_A  \: \epsilon^{(\lambda)}_\mu(p) \delta^{\hat{A} \hat{B}}~,
\label{fVfAdef}
\ee
where $\epsilon^{(\lambda)}_\mu(p)$  is the polarisation vector associated to $V$ or $A$,
with $\sum_\lambda \epsilon^{(\lambda)}_\mu(p) \epsilon^{(\lambda)*}_\nu (p) = - (\eta_{\mu\nu} - p_\mu p_\nu /M_{V,A}^2)$,
as well as
\be
\la 0 \vert {\cal S}^{\hat{A}} \vert S^{\hat{B}}(p) \ra= G_S \delta^{{\hat{A}} {\hat{B}}} ,
\qquad\qquad
\la 0 \vert {\cal S}^{0} \vert \sigma(p) \ra= G_\sigma ,
\qquad\qquad
\la 0 \vert {\cal P}^{0} \vert \eta^\prime(p) \ra= G_{\eta^\prime} ,
\label{defGS}
\ee
the sum rules become, in this narrow-width, single-resonance approximation,
\be
f^2_V M^2_V -  f^2_A M^2_A -F^2_G =0,
\qquad\qquad
f^2_V M^4_V - f^2_A M^4_A= 0,
\label{WSR}
\ee
and 
\be
G^2_\sigma -  G^2_G = 0,
\qquad\qquad
G^2_S - G^2_{\eta^\prime} = 0.
\label{SP_SR}
\ee

Now, taking the various expressions of the meson masses, decay constants,
as obtained from the NJL large-$N$ approximation above,
one can check to which extent these Weinberg-type and scalar sum rules
are actually saturated by the first resonance from each of the available spectra.
To proceed, one may first rewrite the {\em resummed} two-point correlators  
of Eq.~(\ref{PiVAsum})
in the pole-dominance form: 
from Eqs.~(\ref{PiVAsum}) and (\ref{fVfAdef}), the residues of the vector and axial-vector channels are defined by
\begin{equation}
f_{V/A}^2 M_{V/A}^2=\lim_{q^2\rightarrow M^2_{V,A}}  (q^2-M_{V/A}^2) ~ \overline{\Pi}_{V/A}(q^2)
=
\frac{-1}{(2K_{V/A})^2}\left[M^2_{V/A}\,\frac{{\rm d} \tilde{\Pi}_{V/A}(q^2)}{{\rm d} 
q^2}\biggr|_{q^2=M_{V/A}^2} \right]^{-1}~,
\label{fVAresidue}
\end{equation}
where in the second equality, we have expanded the denominator of $\overline{\Pi}_{V/A}(q^2)$ around the complex pole $M_{V/A}^2$ and used Eq.~(\ref{pole}).
Similarly to the definition of the resonance masses in Eq.~(\ref{res_sol}), one should however 
adopt a prescription to deal with the unphysical imaginary parts, NJL artefacts   
of the lack of confinement properties.
We adopt the following prescription: (i) the residues 
are evaluated at the real pole masses $M^2_{V,A}= {\rm Re} [g_{V,A}(M^2_{V,A})]$ 
defined by Eq.~(\ref{res_sol}),  
and (ii) we similarly define $f_{V,A}^2$ by the real parts of their right-hand-side expressions 
in Eq.~(\ref{fVAresidue}). Of course, in the range of parameter space where
the left-over imaginary contributions in Eqs.~(\ref{fVAresidue}) become large, it 
puts a definite limit on the reliability of the the NJL calculation, as will be specified below.  
According to this prescription, we obtain explicitly for the vector decay constant, 
\begin{equation}
f_V^2=-\frac{3 (2N)}{16 \kappa_D^2 M_{V}^4} Re \left[ \frac{1}{\tilde{B}_0(M_{V}^2,M_\psi^2)+(M_{V}^2+2 M_\psi^2)\tilde{B}^\prime_0(M_{V}^2,M_\psi^2)} \right]~.
\label{fVdef}
\end{equation}
The axial decay constant $f_A^2$ is obtained in a similar way by making 
the following replacements $M_V \rightarrow M_A$ and $(M_{V}^2+2 M_\psi^2) \rightarrow (M_{A}^2-4 M_\psi^2)$ 
in the previous equation.

Similarly, for the spin zero channels, the residues are defined by
\be
G_\phi^2\equiv - \lim_{q^2\rightarrow M^2_{\phi}} (q^2-M_\phi^2) \overline{\Pi}_{\phi}(q^2) 
= \frac{1}{(2 K_\phi)^2}
\left[\dfrac{{\rm d} \tilde{\Pi}_\phi(q^2)}{{\rm d}q^2}\biggr|_{q^2=M^2_{\phi}}\right]^{-1}~.
\label{limit}
\ee
From Eqs.~(\ref{PiSPsum}) and (\ref{defGS}), the scalar  decay constants are explicitly given by
\begin{equation}
G_{\sigma,S}^2=-\frac{1}{2 (2N)K_{\sigma,S}^2} 
~Re\left[\frac{1}{\tilde{B}_0(M_{\sigma,S}^2,M_\psi^2)+(M_{\sigma,S}^2 
-4 M_\psi^2) \tilde{B}^\prime_0(M_{\sigma,S}^2,M_\psi^2)} \right] ~,
\label{Gsigdef}
\end{equation}
while for the pseudoscalar decay constants we obtain
\begin{equation}
G_{G,\eta^\prime}^2
=-\frac{1}{2 (2N)K_{G,\eta^\prime}^2} ~Re \left[\frac{g_{A,a}^{-1}(M_{G,\eta^\prime}^2)}{\tilde{B}_0(M_{G,\eta^\prime}^2,M_\psi^2)+M_{G,\eta^\prime}^2 \tilde{B}^\prime_0(M_{G,\eta^\prime}^2,M_\psi^2)} \right] ~,
\label{Getadef}
\end{equation}
where the axial-vector pseudoscalar mixing 
(see section \ref{Resummed correlators and the Goldstone decay constant}) 
brings the factor $g_{A,a}^{-1}(M_{G,\eta^\prime}^2)$ for $G$ and $\eta^\prime$ respectively.

Generally, we cannot expect the sum rules in the narrow width approximation to be very well satisfied,
both because of the already discussed inherent approximations of the NJL framework, 
and also since the narrow width approximation itself is not justified in a substantial part of the
parameter range, as we will examine more precisely below. To be more specific, we will use the standard definition of the width,
\be
M_\phi \Gamma_\phi = \frac{{\rm Im}\,\t\Pi_{\phi}(M^2_\phi)}{{\rm Re}\,\t\Pi_{\phi}^\prime(M^2_\phi)}
~,
\label{res_width1}
\ee
with $\t\Pi_{\phi}^\prime(q^2)$ denoting the derivative of $\t\Pi_{\phi}(q^2)$ with respect to $q^2$.
By evaluating explicitly Eq.~(\ref{res_width1}) for the relevant resonances 
one may control the range of validity of the narrow width approximation.

Before a precise illustration of the deviations  
from the sum rules relations in Eqs.~(\ref{WSR}) and (\ref{SP_SR}) in the parameter space of the model, 
it is instructive to examine more closely the NJL expressions of the involved quantities, 
Eqs.~(\ref{fVdef}), (\ref{MV}) and (\ref{MA}). 
Namely, let us assume momentarily that we could      
crudely neglect the $q^2$ dependence of $\t B_0$, i.e. taking 
$\t B_0(M^2_V,M^2_\psi)\simeq \t B_0(M^2_A,M^2_\psi) \equiv\t B_0$ (therefore taking also
its derivative to vanish, $\t B^{\prime}_0(q^2)\simeq 0$). 
Within this approximation, the second sum rule in Eq.~(\ref{WSR}) is immediately satisfied, see Eq.~(\ref{fVdef}), while for the first sum rule,  one can write, after some simple algebra,
\be
f^2_V M^2_V -f^2_A M^2_A \simeq f^2_V (6 M^2_\psi)\left[ 1+{\cal O} \left(\frac{M^2_\psi}{M^2_V}\right)\right]
 \simeq -F^2_G \left[ 1+{\cal O} \left(\frac{M^2_\psi}{M^2_V} \right)\right]~, 
 \label{WSRapprox}
\ee
where in the first equality we used the fact that the relation in  Eq.~(\ref{MVMA}) becomes exact in this approximation,
and in the last equality we used Eqs.~(\ref{fVdef}) and (\ref{MV}) in the same approximation, 
and identified $F_G^2$ from its expression in Eq.~(\ref{FG4}).
This simple exercise shows explicitly and rather intuitively 
where the bulk of deviations from  the Weinberg sum rules (WSR) comes from: 
one infers that the sum rules in Eq.~(\ref{WSR}) will, in general, not be satisfied, since 
the quantities they involve are the pole masses,
$M^2_V= Re[M^2_V(M^2_V)]$ and  $M^2_A=Re[M^2_A(M^2_A)]$, the Goldstone decay constant $F^2_G=F^2_G(0)$,
and the vector decay constants $f_{V,A}^2$ in Eq.~(\ref{fVdef}), actually evaluated at the 
different $V,A$ pole masses 
and involving also the non-vanishing derivative $\t B_0^\prime(M^2_{V/A})$. 
Accordingly since the relevant expressions like Eq.~(\ref{fVdef}) are to be evaluated 
at {\it different} values of $q^2$, this implies not quite negligible differences in $\t B_0(q^2)$, 
and in its derivative. 
Only to the extent that they display a rather mild $q^2$-dependence will the narrow-width 
version (\ref{WSR}) of the sum rules approximatively hold~\footnote{We note that those finding and 
observations are qualitatively similar 
to the WSR results for the NJL model applied to low energy QCD in ref.~\cite{Peris:1998nj}, although
those authors used somewhat different approximations than ours.}. 
Moreover, the crudely neglected
terms ${\cal O}(M^2_\psi/M^2_V)$ in Eq.~(\ref{WSRapprox}) are actually not so negligible, 
the less when $\xi$ increases, just as $M^2_A/M^2_V$ also increases with $\xi$. 
Thus, we generally expect stronger deviations from Eq.~(\ref{WSR}) for larger $\xi$ values.
%

\begin{figure}[tb]
\begin{center}
\includegraphics[scale=0.25]{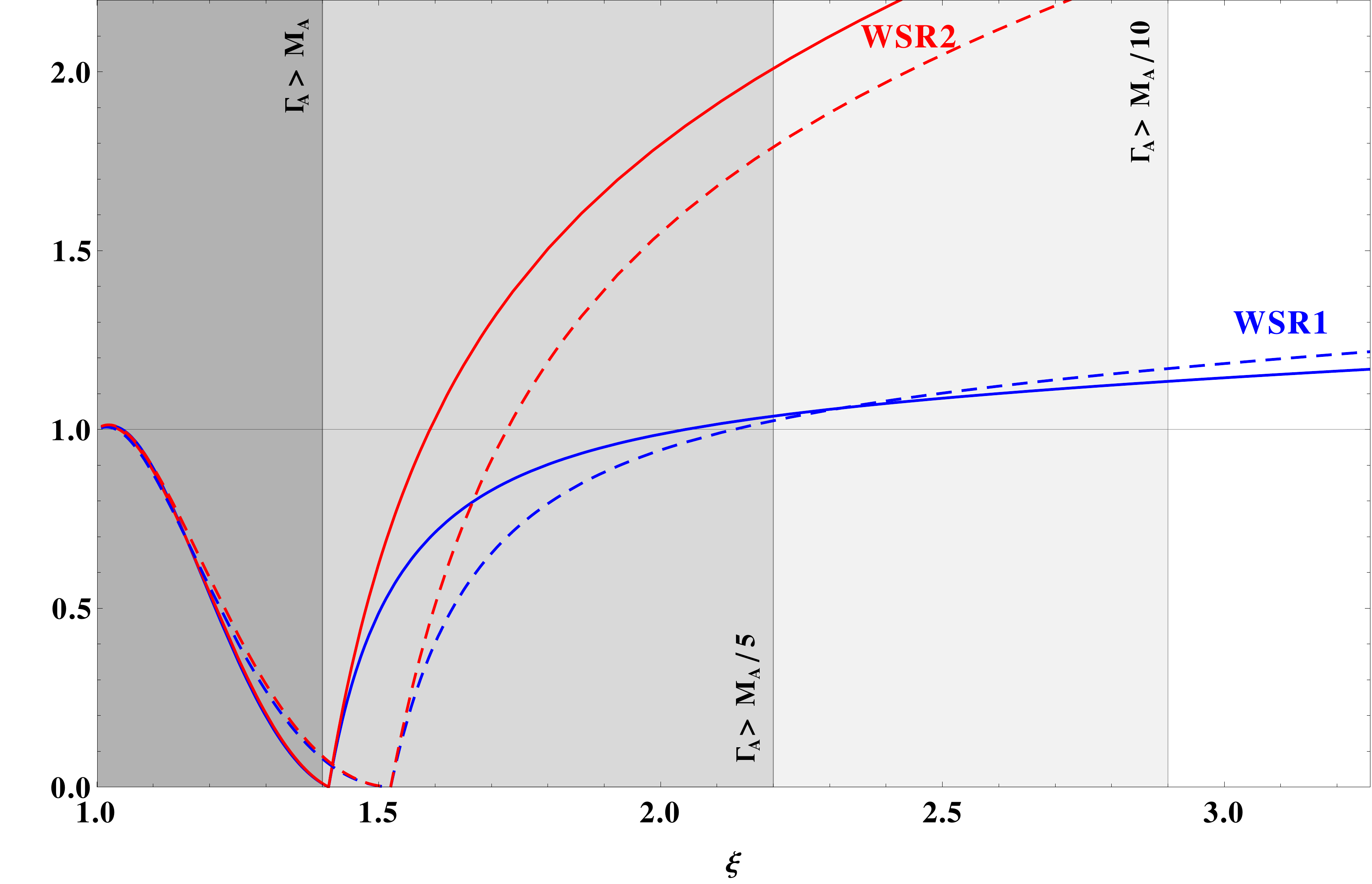}
\includegraphics[scale=0.25]{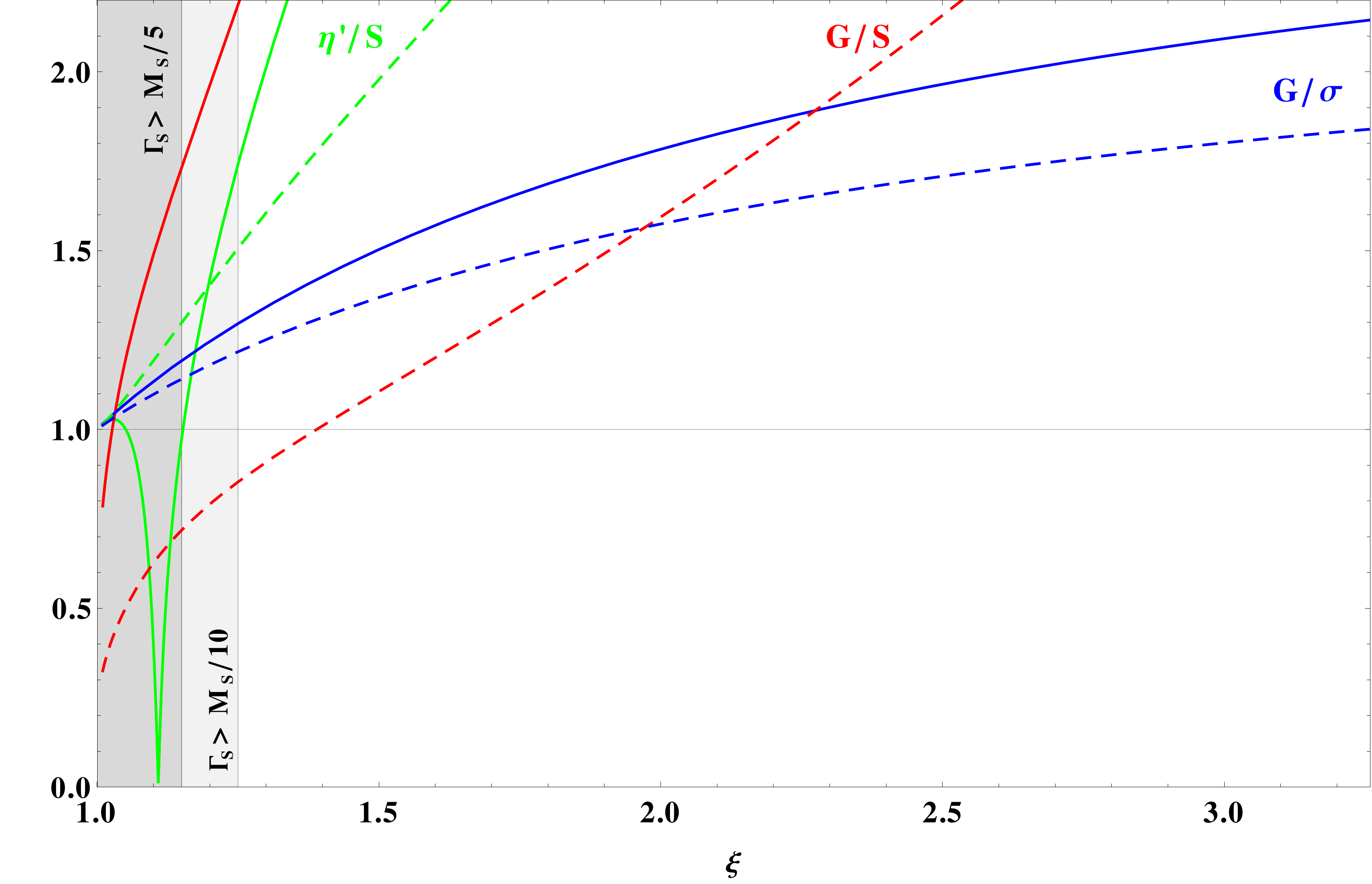}
\end{center}
\caption{ Left panel: the two ratios $(f^2_V M^2_V)/(F^2_G + f^2_A M^2_A)$ (WSR1, blue lines) 
and $(f^2_V M^4_V)/(f^2_A M^4_A)$ (WSR2, red lines)
as functions of the coupling $\xi$, for 
$\kappa_B/\kappa_A = 0.1$ (solid lines) and $\kappa_B/\kappa_A = 0.5$ (dashed lines). 
Right panel: the analog for scalar sum rules. Also indicated are the values of the most relevant 
resonance widths,
calculated from Eq.~(\ref{res_width1}) for $\kappa_B/\kappa_A = 0.1$. 
}
\label{wsrpole}
\end{figure}

In order to illustrate more precisely the deviations from the Weinberg-like sum rules of Eq.~(\ref{WSR}),
taking now the ``exact'' expressions of $f_{V/A}$, $M_{V/A}$ according to our NJL calculations
and prescriptions above, 
we consider the two ratios 
\be
\mbox{WSR}_1\equiv   
\frac{f^2_V M^2_V}{F^2_G + f^2_A M^2_A}~,
\qquad\qquad
\mbox{WSR}_2\equiv   
\frac{f^2_V M^4_V}{f^2_A M^4_A}
~,
\label{WSRNJL}
\ee
which would both equal unity if the sum rules were satisfied in their
narrow-width versions.
Similarly, for the scalar sum rules we consider the two  
ratios $G_G^2/G_\sigma^2$ and $G_{\eta^\prime}^2/G_S^2$.
The behaviour of these ratios with respect to $\xi$ and $\kappa_B/\kappa_A$ are illustrated in the left 
and right panels of Fig.~\ref{wsrpole} for the Weinberg and scalar sum rules respectively.
We also indicate some specific values of the relevant resonance widths, 
calculated from Eq.~(\ref{res_width1}) for the reference value $\kappa_B/\kappa_A = 0.1$. 
The corresponding shaded regions thus indicate approximately the
range where the narrow width approximation can be trusted or not. 
Note that the $V$ and $A$ widths are very weakly
sensitive to the values of $\kappa_B/\kappa_A$, so that the indicated ranges are also approximately valid
for $\kappa_B/\kappa_A=0.5$. In contrast the $\eta'$ and $S$ widths grow rapidly with $\kappa_B$, such
that the indicated limit $\Gamma_S/M_S =1/5$ ($\Gamma_S/M_S =1/10$)  is pushed, for $\kappa_B/\kappa_A=0.5$,
towards larger values of $\xi$, $\xi \simeq 1.7$ ($\xi \simeq 2$, respectively).

The two sum 
rules of Eq.~(\ref{WSR}) are actually reasonably satisfied in some specific ranges of $\xi$, respectively either for 
intermediate values $1.6\lsim \xi \lsim 2$, or for $\xi$ very close to $1$. Conversely the deviations
appear maximal in the range $\xi\simeq 1.2-1.6$ and again for very large $\xi$.
Most of these features can be understood more intuitively with the help of the above analysis. 
The intermediate range, where the deviations are the smallest,
corresponds to a range where, at the same time, the narrow width approximation is well justified,  
and the relevant pole-mass differences are still moderate such that the relevant $q^2$ arguments
of $\t B_0(q^2, M_\psi^2)$ are not very different. Then for very large values of $\xi$, while the $A$ width 
is becoming smaller, one enters the regime of increasingly large 
differences in the relevant $\t B_0(M^2_{A/V}, M_\psi^2)$ functions, thus increasing the deviations, although the first
WSR remains relatively well satisfied. 
The second WSR sum rule shows more rapidly increasing and important deviations for larger values of $\xi$, 
as intuitively expected since the fourth power of the masses 
enhances the increasing $M_A/M_V$ ratio. 
The WSR values are not very sensitive to the ratio 
$\kappa_B/\kappa_A$, but depend mostly on $\xi$:
a larger $\kappa_B$ value essentially shifts the values of the sum rules in Fig.~\ref{wsrpole}, 
as it implies larger values of $\kappa_A+\kappa_B$.
Conversely for decreasing values of $\xi$, the narrow width approximation becomes totally unreliable, 
say for $\xi \lsim 1.6$ in the case of $\Gamma_A$, where, correspondingly, the deviations are seen to be maximal.
Moreover, when approaching (from below) the threshold $M_V^2=4 M_\psi^2$, $\Gamma_V$ is vanishing, 
but $Re[\tilde{B}_0^\prime(M_V^2, M_\psi^2)]$ tends toward infinity, so that $f^2_V\to 0$, see Eq.~(\ref{fVdef}). 
This happens around $\xi \simeq 1.4~ (1.5)$ for $\kappa_B/\kappa_A=0.1~ (0.5)$.
This peculiar feature can be understood as follow. When moving towards the threshold from below, 
the residue of the vector resonance, $f_V^2 M^2_V$, tends to zero, because its contribution to the spectral function is progressively 
transferred from the sub-threshold to the continuum part of the spectral function.
Since in the pole dominance approximation one only considers the lightest resonances,
just below the threshold, the continuum contribution is not included within Eq.~(\ref{fVdef}), therefore 
the crossing of the threshold appears problematic in our NJL approximation.
Of course, this pathological behaviour is not present in Fig.~\ref{fig_Sum_Rules}, 
where we consider the complete two-point functions, which include also the continuum contributions.
Finally, very close to the critical coupling $\xi\simeq 1$,
although both $\Gamma_{V,A}$ are large, the mass gap in this region is 
relatively very small,  $M_\psi \ll \Lambda$, such that $M_A-M_V$ is minimal, and
$F_G\simeq M_\psi$ is also relatively small. Thus taking the real contributions prescriptions
according to Eq.~(\ref{fVdef}), one is again very close to the ideal approximation discussed above, 
leading to Eq.~(\ref{WSRapprox}). 

From these results, if considering that the best possible matching of the Weinberg-type sum rules, 
established on more general dynamical grounds, may be more important than the possible 
limitations of the NJL model approximation (somewhat   
in the spirit of Ref. \cite{Peris:1998nj}), one could be tempted to infer some preferred 
range of $\xi$ values, where both deviations are minimal 
(although as clear from the figure it is not possible to satisfy the two WSR 
exactly for the same value of $\xi$). 
However, given the limitations of the NJL dynamical approximation, partly responsible for
the non-perfectly matched Weinberg-type sum rules, we consider this 
only as an indicative trend rather than a genuine dynamical 
constraint on the couplings. 

Concerning next the scalar sum rules, note that the above relations in Eqs.~(\ref{Gsigdef}) and (\ref{Getadef}) 
do not lead to $G_G^2(q^2)-G_\sigma^2(q^2)=0$ and 
$G_{\eta^\prime}^2(q^2)-G_S^2(q^2)=0$, which would be valid only if all expressions were
evaluated at the same value of $q^2$.
This is due to the pseudoscalar axial mixing, i.e. a term proportional to $g_{A,a}(q^2)$ does not vanish in  the difference.
In addition, for $G_G^2(q^2)-G_S^2(q^2)$, there is a term proportional 
to $\kappa_B$ that indicates that this difference does not satisfy a convergent sum rule, consequently the discrepancy 
increases with $\kappa_B$. Indeed, as can be seen on Fig. \ref{wsrpole}, some of the scalar sum rules are approximately 
satisfied very close to $\xi=1$, but are rapidly and badly invalidated 
for larger values of $\xi$, even though the narrow width approximation is justified in this region. This is mainly
due to very large differences in the argument of the relevant functions $\t B_0(q^2, M_\psi^2)$, and also, 
as discussed above, due to the non-vanishing of $\kappa_B$. 
Note that, similarly to what is discussed above for the WSRs, the scalar sum rule associated to the $\eta^\prime$ 
may exhibit a pathological behaviour, when the lightest resonances do not incorporate the dominant contributions.
Indeed, the $\eta^\prime$ mass crosses the threshold for $\kappa_B/\kappa_A=0.1$ and the associated 
ratio $G_{\eta^\prime}^2/G_S^2$ tends to zero in this regime, which lies around $\xi=1.1$.

In summary, the mismatch between the NJL predictions and the spectral sum rules
resides in the gap between the contribution of the low-lying resonances and the full spectral functions.
Given these limitations in the comparison of our results with the spectral sum rules, 
and since our interest is mostly the phenomenology of the lightest composite states,
in the following we will keep studying the full range for the parameters $\xi$ and $\kappa_B/\kappa_A$.

\subsection{Evaluation of the oblique parameter $S$\footnote{We thank Alex Pomarol for encouraging us to estimate the ultraviolet correction to $S$ in the present model.}}
\label{S parameter}

In the absence of explicit symmetry breaking effects, like, for instance, the coupling
to the external electroweak gauge fields, the vacuum state $\vert {\rm vac}\rangle_0$ is
left invariant by the $Sp(4)$ subgroup of the $SU(4)$ flavour symmetry defined by the
generators $T^A$ satisfying Eq. (\ref{Tacom}), where $\Sigma_\epsilon$ stands for 
$\Sigma_0$ as given in Eq. (\ref{Sigma}). After electroweak symmetry breaking through misalignment, 
the vacuum state becomes $\vert {\rm vac}\rangle_v$. It is left invariant by a different $Sp(4)$
subgroup, whose generators $T^A_v = U_v T^A U_v^\dagger$ now satisfy\footnote{Similarly,
the generators of the coset space $SU(4)/Sp(4)$ corresponding to this new orientation
of the $Sp(4)$ subgroup are given by $T^{\hat A}_v = U_v T^{\hat A} U_v^\dagger$, and
satisfy $T^{\hat A}_v\Sigma_v - \Sigma_v \big( T^{\hat A}_v \big)^T=0$.}
\begin{equation}
T^A_v\Sigma_v + \Sigma_v \big( T^A_v \big)^T=0
,
\end{equation}
with $\Sigma_v$ and the $SU(4)$ transformation $U_v$ given by
\begin{equation}
\Sigma_v = U_v \Sigma_0 U_v^T 
,
\qquad
U_v = e^{i \sqrt{2} \langle h \rangle T^{\hat 1}/f} 
= \cos \left( \frac{\langle h \rangle}{2 f} \right) + 2 \sqrt{2} \, i \sin \left( \frac{\langle h \rangle}{2 f} \right) T^{\hat 1}
.
\end{equation}
The expression of the transformation $U_v$ conveys the information that the Higgs field $G^{\hat 1}$
takes a vev $\langle h \rangle$. The shift in the oblique parameter $S$ \cite{Peskin:1991sw}
induced by the composite electroweak sector is given by
\begin{equation}
\Delta S = 16 \pi \left. \frac{d \Pi^{(v)}_{3Y} (q^2)}{d q^2} \right\vert_{q^2 = 0}
,
\end{equation}
where the two-point correlator $\Pi^{(v)}_{3Y} (q^2)$ has the following expression
(cf. Appendix \ref{SU4-generators})
\begin{equation}
\Pi^{(v)}_{3Y} (q^2) \left( \eta_{\mu\nu} - \frac{q_\mu q_\nu}{q^2} \right) =
\frac{i}{2} \int d^4 x \, e^{i q \cdot x} {_v\langle} {\rm vac} \vert
T \{ \left( J^4_\mu (x) - J^3_\mu (x) \right) \left( J^4_\nu (0) + J^3_\nu (0) \right) \}
\vert {\rm vac} \rangle_v
.
\end{equation}
Expressing the generators $T^3$ and $T^4$ in terms of $T_v^A$ and $T_v^{\hat A}$,
$T^3 = \cos (\langle h \rangle / f) T^3_v - \sin (\langle h \rangle / f) T^{\hat 2}_v$,
$T^4 = T^4_v$, leads to\footnote{One can repeat the same exercice when in addition the singlet 
Goldstone boson $G^{\hat 3}$ takes a vev $\langle \eta \rangle$. This will leave the
expression for $\Delta S$ unchanged, the relation between $v$ and the two vev's
being given by 
$$\frac{v}{f} = \frac{\langle h \rangle}{\sqrt{\langle h \rangle^2 + \langle \eta \rangle^2}}
\sin \left( \frac{\sqrt{\langle h \rangle^2 + \langle \eta \rangle^2}}{f} \right).$$}
\begin{equation}
\Delta S = 8 \pi \frac{v^2}{f^2} \left. \frac{d }{d q^2} \left( q^2 \Pi_{V{\mbox -}A} (q^2) \right) \right\vert_{q^2 = 0}
, \quad
\frac{v}{f} = \sin \left( \frac{\langle h \rangle}{f} \right)
.
\end{equation}
Notice that the Goldstone pole at $q^2=0$ does not contribute to this expression.
The corresponding shift in the oblique parameter $T$ vanishes, due to custodial symmetry.

\begin{figure}[tb]
\begin{center}
\includegraphics[scale=0.25]{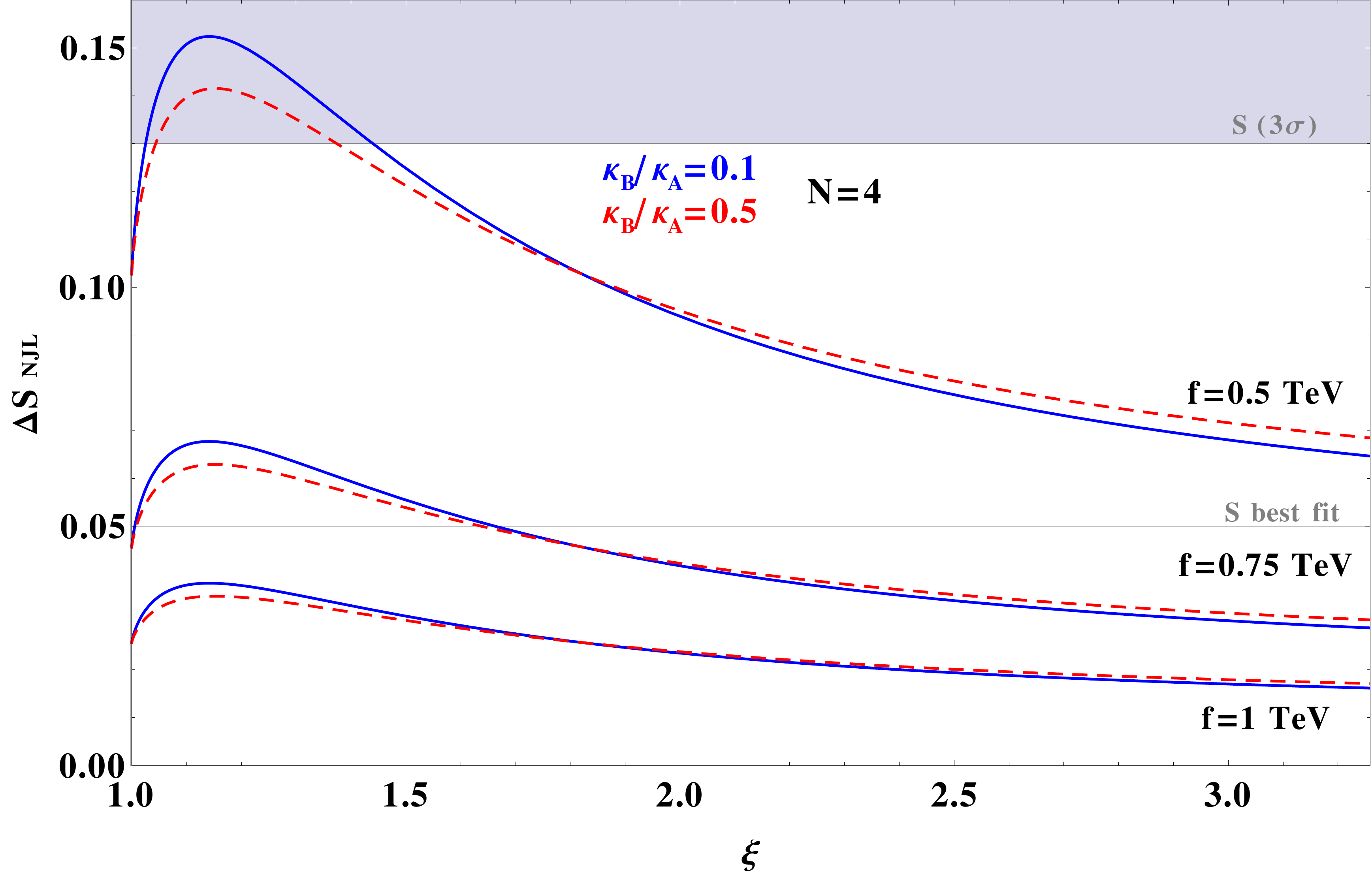}
\includegraphics[scale=0.23]{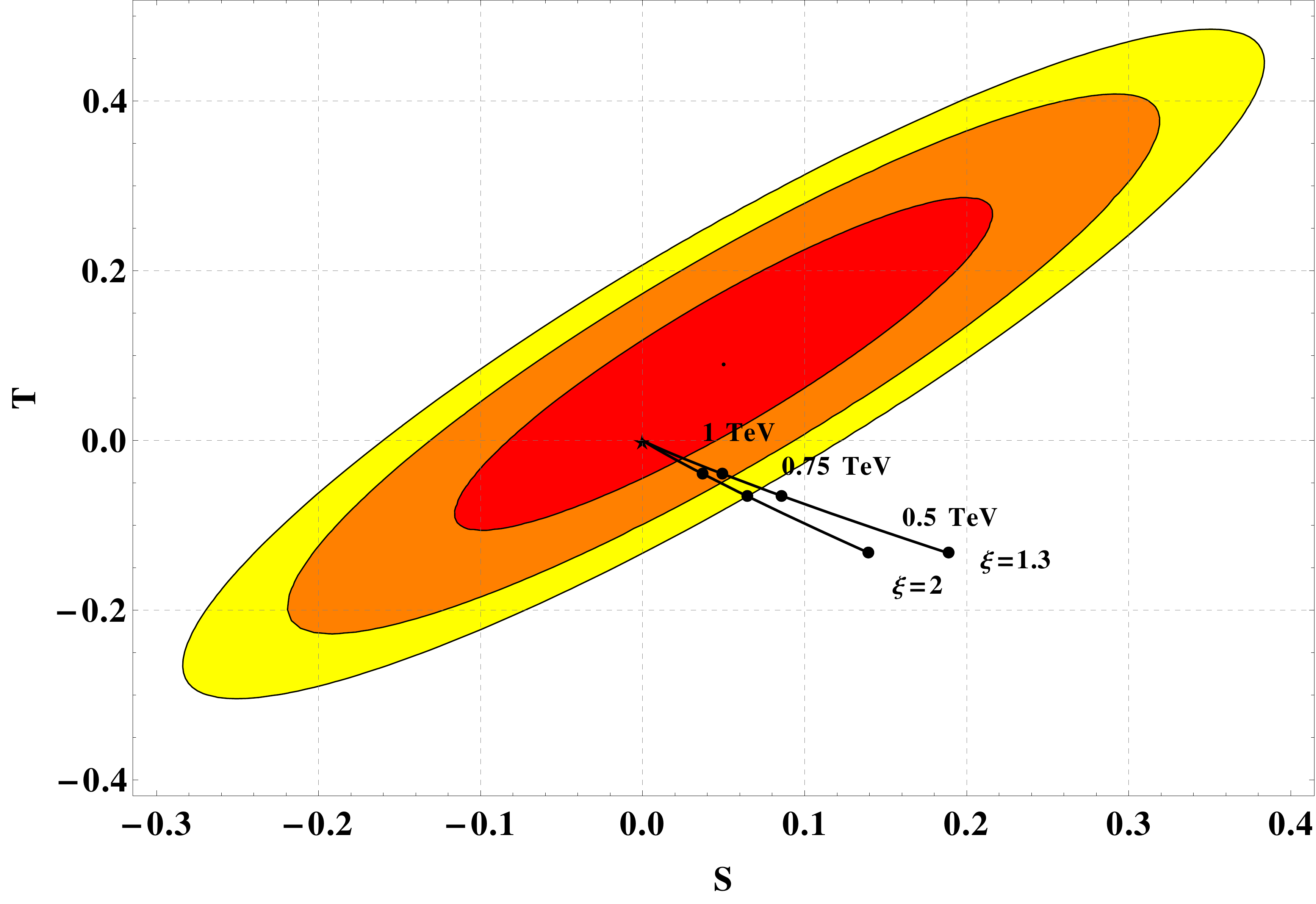}
\end{center}
\caption[long]{On the left, the contribution to the $S$ parameter from the composite electroweak sector in the NJL approximation [see Eq.~(\ref{S-NJL})] 
as a function of the dimensionless coupling $\xi$, and for three representative values of $f$, $f=(0.5, 0.75,1)$ TeV.
The value of the parameter $\kappa_B/\kappa_A$ has been taken equal to $0.1$ (solid blue curves) and to $0.5$ (dashed red curves), 
while the number of hypercolours is fixed to $N=4$ and the vector coupling is given by $\kappa_D=\kappa_A$. 
The best fit for $S$ is indicated by the horizontal line at 0.05 and the region above the $3 \sigma$ limit, assuming $T=0$, is shaded.
On the right, the preceding UV contribution, evaluated in the NJL approximation, as well as the IR contributions 
coming from the non-linear realisation of the EWSB (i.e. $\Delta S_{\rm NJL} + \Delta S_{IR}$ and $\Delta T_{IR}$),  as a function of $f$.
The black dots correspond to $f=0.5,0.75$ and $1$ TeV, and the curves stand for two representative values, 
$\xi=1.3$ and $\xi=2$, with $\kappa_B/\kappa_A=0.1$, $N=4$ and $\kappa_D=\kappa_A$.
The 68 \% (red), 95 \% (orange) and 99 \% (yellow) C.L. ellipses in the
$S-T$ plane are extracted from the fit of Ref.~\cite{Baak:2014ora}.
As stressed in the text, one expects in general additional contributions, which could significantly impinge on the values of $S$ and $T$.}
\label{S_plot}
\end{figure}

 In the NJL approximation the resummed correlator $\overline\Pi_{V-A}(q^2)$ is defined according to Eq.~(\ref{PiVAsum}), 
that implies 
\begin{equation}
\Delta S_{\rm NJL} = \frac{2N}{9\pi} \frac{v^2}{f^2} \left[ \dfrac{1}{2}+ g_A^2(0) - \dfrac 32 \left( \frac{1}{1 + x_\psi} - \ln \frac{1 + x_\psi}{x_\psi} \right)  ( 1 - g_A^2(0) )  \right]
= \frac{2N}{6\pi} \frac{v^2}{f^2} \left( 1+ {\cal O}(x_\psi)\right)~,
\label{S-NJL}
\end{equation}
where $x_\psi \equiv M_\psi^2/\Lambda^2$, 
the axial form factor $g_A(q^2)$ is defined in Eq. (\ref{gAdef}), and its value at $q^2=0$
reads 
\be
\dfrac{1}{g_A(0)} = 1 - \frac{\kappa_D/\kappa_A}{1+\kappa_B/\kappa_A} 2x_\psi \left(1-x_\psi  \ln\dfrac{1 + x_\psi}{x_\psi} \right)^{-1} \left( \frac{1}{1 + x_\psi} - \ln\dfrac{1 + x_\psi}{x_\psi} \right)
.
\ee
The left panel of Fig. \ref{S_plot} shows the variation of $\Delta S_{\rm NJL}$ as a function
of $\xi$, that is in one-to-one correspondence with $x_\psi$, according to Eq.~(\ref{gap2}).
As expected, $\Delta S_{\rm NJL}$ decreases when the strong sector decouples, i.e. with 
the increase of $f$. 
More precisely, for $\xi\to 1$ we have $x_\psi\rightarrow 0$ and $\Delta S_{\rm NJL} \simeq 2N/(6 \pi) (v^2/f^2)$. 
As $\xi$ increases, the factor $(1-g_A^2(0))$ becomes non-zero, and $\Delta S_{\rm NJL}$ first grows moderately, 
and then decreases as $x_\psi$ approaches one. 
In the range of parameter space where the narrow-width approximation applies, one may
saturate the above correlator with the first light resonances, see Eq.~(\ref{Q2PiVA}) with $q^2=-Q^2$, 
and in this case one obtains \cite{Peskin:1991sw,Knecht:1997ts}
$\Delta S_{\rm NJL} \simeq 8 \pi (v^2/f^2) (f_V^2-f_A^2)$.

The composite sector will also modify the couplings of the Higgs boson to the electroweak
gauge bosons by a factor $\sqrt{1-v^2/f^2}$. 
This modification  will upset the cancellation of logarithmic divergences
in the gauge-boson self-energies, and induce model independent shifts in both $S$ and $T$ \cite{Barbieri:2007bh}.
These contributions from low energies are given by \cite{Contino:2010rs,Panico:2015jxa}
\begin{equation}
\Delta S_{\rm IR} = \frac{1}{6 \pi} \frac{v^2}{f^2} \ln \left( \frac{\mu}{M_h} \right),
\qquad
\Delta T_{\rm IR} = - \frac{3}{8 \pi} \frac{1}{\cos^2\theta_W} \frac{v^2}{f^2} \ln \left( \frac{\mu}{M_h} \right)
= - \frac{9}{4} \frac{\Delta S_{\rm IR}}{\cos^2\theta_W} ,
\label{S-IR}
\end{equation}
One finds $\Delta S_{\rm IR} = (0.045 , 0.022, 0.014)$ and $\Delta T_{\rm IR} = (-0.17, -0.08, -0.05)$,
for $f = (0.5 , 0.75, 1)$~TeV, if the cut-off scale is taken equal to $4 \pi F_G = 2 \sqrt{2} \pi f$, 
leading to non-negligible contributions. Notice that Goldstone boson loops contribute to the low-$q^2$
end of the $\Pi_{V{\mbox -}A} (q^2)$ function, but only at sub-leading order in the $1/N$ expansion.
The NJL approximation only provides leading-order contributions, and thus cannot remove this
sub-leading (in the $1/N$ expansion) cut-off dependence in $\Delta S_{\rm IR}$ and $\Delta T_{\rm IR}$.

The right panel of Fig. \ref{S_plot} shows the combined contributions 
from Eqs.~(\ref{S-NJL}) and (\ref{S-IR}) to the $S$ and $T$ parameters as a function of $f$, 
for different values of $\xi$ and of $\kappa_B/\kappa_A$.
When  linear couplings between the top quark and the fermions of the strong sector are introduced, one expects in general additional 
contributions, which could significantly affect the $S$ and $T$ parameters.
These fermionic contributions, as well as other order $1/N$ corrections than $\Delta S_{\rm IR}$ and $\Delta T_{\rm IR}$, 
are beyond the scope of this paper. The right panel of 
Fig. \ref{S_plot} thus displays only a specific kind of contributions, and does by no means 
constitute a complete prediction of the model under discussion as far as $S$ and $T$ are concerned.

\section{Adding the coloured sector} 
\label{coloured-sector}

An appealing way to couple the SM fermions to the composite Higgs is to introduce a linear coupling between
each SM fermion and a composite fermion resonance with the same quantum numbers.
Such an approach, known as fermion partial compositeness \cite{Kaplan:1991dc,Contino:2004vy}, is especially
attractive in the case of the top quark: relatively light composite top partners allow to induce the required, large top Yukawa coupling.
In order for the composite sector to contain partners for the top (and possibly the other SM quarks), one needs to
introduce constituent fermions $X^f$ that are charged under the colour group $SU(3)_c$.
It is not possible to construct a `baryon' (a hypercolour invariant spin-1/2 bound state)
if $X^f$ transforms under the fundamental, pseudo-real representation of $Sp(2 N)$.
Following \cite{Barnard:2013zea}, we rather assume that $X^f$ transforms under the two-index, real representation of $Sp(2N)$
that is antisymmetric, $X^f_{ij}=-X^f_{ji}$, and traceless, $X^f_{ij}\Omega_{ji}=0$. This irreducible representation has
dimension $(2N+1)(N-1)$.
In order to embed a $SU(3)_c$ triplet-antitriplet pair, one has to introduce six such fermions, $f=1,\dots,6$.
Then, the theory acquires a flavour symmetry $SU(6)\supset SU(3)_c$, with $X^f \sim 6_{SU(6)} = (3+\bar 3)_{SU(3)_c}$.
The addition of such an $X$-sector modifies several results that we have derived for the $\psi$-sector in isolation,
because the underlying $Sp(2N)$ gauge dynamics connects the two sectors in a highly non-trivial way, as we now
describe.

\begin{table}[b]
\renewcommand{\arraystretch}{1.6}
\begin{center}
\begin{tabular}{|c|c|c|c|c|}
\hline
& Lorentz & $Sp(2 N)$ & $SU(6)$ & $SO(6)$  \\
\hline \hline
$X^f_{ij}$ & $(1/2,0)$ & ${\Yvcentermath1 \tiny \yng(1,1)}_{\, ij}$ & $6^f$ & 6 \\
\hline
 $\overline{X}_{fij} \equiv \Omega_{ik} X^\dagger_{fkl}\Omega_{lj}$ & $(0,1/2)$ & ${\Yvcentermath1 \tiny \yng(1,1)}_{\, ij}$ & $\overline{6}_f$ & $6$ \\
\hline\hline
$ M_c^{fg}\sim(X^f X^g)$ & $(0,0)$ & $1$ & $21^{fg}$ & $20'+1$ \\
\hline
$ {\overline M}_{c fg}\sim(\overline{X}_f\overline{X}_g)$ & $(0,0)$ & $1$ & $\overline{21}_{fg}$ & $20'+ 1$ \\
\hline \hline
$ a_X^\mu \sim (\overline{X}^f \overline{\sigma}^\mu X_f)$ & $(1/2,1/2)$ & $1$ & $1$ & $1$ \\
\hline
$(V^{\mu}_c,A^\mu_c)_f^g\sim(\overline{X}_f \overline{\sigma}^\mu X^g)$ & ~~$(1/2,1/2)$~~ & ~~$1$~~ & ~~$35^f_g$~~ & ~~$15+20'$~~ \\
\hline
\end{tabular}\end{center}
\caption{
The transformation properties of the elementary fermions, the spin-0 and spin-1 fermion bilinears, in the colour sector of the model.
Spinor indexes are understood, and brackets stand for a hypercolour-invariant contraction
of the $Sp(2N)$ indexes. }
\label{tabsu6}
\end{table}

Once both types of fermions $\psi^a$ and $X^f$ are in presence, the flavour symmetry group becomes $G = SU(4) \times SU(6) \times U(1)$,
where $U(1)$ is the non-anomalous linear combination of the two axial symmetries $U(1)_\psi$ and $U(1)_X$, which separately are both
anomalous with respect to $Sp(2N)$. The current corresponding to the $U(1)_\psi$ transformations and its divergence were already given in
Eqs. (\ref{U1_psi}) and (\ref{U1_psi_div}), respectively. In the case of the $U(1)_X$ transformations, the corresponding expressions read
[a sum over the flavour indices is understood, gauge and spinor indices are omitted]
\begin{equation}
\mathcal{J}^0_{X\mu} =  {\displaystyle\frac{1}{2}}
\left[ \left( \overline{X} \overline{\sigma}_\mu  X \right) - \left( X \sigma_\mu  \overline{X} \right) \right]~,
\label{U1_X}
\end{equation}
\begin{equation}
\partial^\mu \mathcal{J}^0_{X\mu} =
4 \sqrt{3} \, m_X {\mathcal P}^0_X + 2 (N-1) \,{\displaystyle\frac{N_f^X g_{HC}^2}{32\pi^2}}
\sum_{I=1}^{N(2N+1)} \epsilon_{\mu\nu\rho\sigma} G_{HC}^{I,\mu\nu} G_{HC}^{I,\rho\sigma}
~,
\label{U1_X_div}
\end{equation}
where the factor $N_f^X=3$ accounts for the number of flavours in the $X$-sector.
In the above, $\overline{X}$, as defined in Table \ref{tabsu6} below, transforms under the $Sp(2N)$
gauge group in the same way as $X$, and the gauge-invariant bilinear fermion contractions
between $X$ and $X$ are defined as 
\begin{equation}
(X^f X^g )\equiv X^f_{ij}\Omega_{jk} X^g_{kl} \Omega_{li} = {\rm tr} ( X^f \Omega X^g \Omega )~.
\label{X_inv}
\end{equation}
Contractions like $({\overline X}_f {\overline X}_g )$ and $({\overline X}_f X^g )$
are defined in the same way.
For later use we have also introduced a flavour independent mass term for the $X$ fermions,
\begin{equation}
{\mathcal L}^X_m = - 2 \sqrt{3}\, m_X {\mathcal S}^0_X~,
\label{X_mass}
\end{equation}
with
\begin{equation}
{\mathcal S}^0_X = \frac{1}{2} \,   \left[
({\overline X} T^0_X \Sigma_{0}^c  {\overline X} ) + (X \Sigma_{0}^c T^0_X X ) \right]~,
 \qquad\qquad
{\mathcal P}^0_X = \frac{1}{2i} \,  \left[
({\overline X} T^0_X \Sigma_{0}^c {\overline X} ) - (X \Sigma_{0}^c T^0_X X ) \right]~,
\label{Densities-SU6}
\end{equation}
in agreement with the general definitions given in Eq. (\ref{S_and_P}) and the normalisation
adopted there for the singlet scalar and pseudoscalar densities,  that is $T^0_X= 1\!\!1 /(2 \sqrt{3})$.
Note that the singlet contraction of two fermions in the (anti-)fundamental of $SU(6)$ is realised through the matrix 
\begin{equation}
\Sigma_{0}^c
=
\begin{pmatrix}
0 & 1\!\!1_3
\\
1\!\!1_3 & 0
\end{pmatrix}~,
\end{equation}
which determines the $SU(6)/SO(6)$ vacuum direction.
The two conditions in Eq.~(\ref{Tacom}) are satisfied with $\Sigma_\epsilon = \Sigma_{0}^c$ and the $SU(6)$ generators $T^F$ and $T^{\hat{F}}$ defined in appendix \ref{SU6-generators}.

Examining the respective $U(1)_{\psi}$ and $U(1)_{X}$ anomaly coefficients,
it is easily seen that the combination of the two axial singlet currents given by
\be
{\mathcal J}_\mu^0 = \ell({\Yvcentermath1 \tiny \yng(1)}) {\mathcal J}_{X\mu}^0 -
\frac{3}{2} \ell({\Yvcentermath1 \tiny \yng(1,1)}) {\mathcal J}_{\psi\mu}^0
=
\frac{3}{2} \ell({\Yvcentermath1 \tiny \yng(1,1)}) \left(\psi^a \sigma_\mu {\overline\psi}_a\right)
-  \ell({\Yvcentermath1 \tiny \yng(1)})
\left(X^f \sigma_\mu {\overline X}_f \right)~,
\label{axial_singlet}
\ee
is free from the gauge anomaly,
\be
\partial^\mu \mathcal{J}^0_{\mu} =
4 \sqrt{3} \, m_X {\mathcal P}^0_X~,
\label{U1_div}
\ee
where the Dynkin index $\ell (r)$ of the representation $r$ of the gauge group $Sp(2N)$
gives the normalisation of the $Sp(2N)$ generators $T^I (r)$ in this representation,
\be
{\rm tr} [T^I (r) T^J (r) ] = \frac{1}{2} \ell (r) \delta^{IJ}~,
\qquad\qquad
\ell({\Yvcentermath1 \tiny \yng(1)}) = 1~,
\qquad
\ell({\Yvcentermath1 \tiny \yng(1,1)}) = 2 (N-1)~.
\label{norm_gen}\ee
Consequently, the axial singlet transformation of {\it both}
the $\psi$ and $X$ fermions, with charges satisfying
\be
q_\psi =-3(N-1) q_X~,
\label{U1_charges}
\ee
is a true symmetry of the theory, even at the quantum level, in the limit where $m_X$ vanishes.

The introduction of fermions in the two-index antisymmetric representation of the $Sp(2N)$ gauge group
has another consequence. The first coefficient of the $\beta$-function of the gauge coupling $g_{HC}$
now reads
\be
b_0 = \frac{11}{3} C_2 ({\rm adj}) - \frac{4}{3} \sum_{i=\psi,X} N_f^i \ell(r_i)
= \frac{2}{3} (11 - 4 N_f^X) \left[ N + 1 - 2 \frac{4 N_f^X - N_f^\psi}{4 N_f^X - 11} \right].
\label{betaHC}
\ee
Therefore, as soon as $N_f^X\ge 3$, $b_0$ stays positive and asymptotic freedom is preserved
(at one loop) only if the number of colours $N$ is bounded from above,
\be
N < 2 \frac{4 N_f^X - N_f^\psi}{4 N_f^X - 11} - 1
\qquad [N_f^X \ge 3]
,
\label{Nlimit}
\ee
which, in the case at hand ($N_f^\psi=2$ and $N_f^X=3$), means $N\le 18$.
This upper bound prevents us from considering the limit $N \to \infty$
at the level of the fundamental hypercolour theory once the sector of $X$ fermions
has been introduced. Notice, however, that independently
from the existence of this upper bound on $N$, the anomalous contribution
on the left-hand side of Eq. (\ref{U1_X_div}) would not vanish in the 
't~Hooft limit $N \to \infty$, with $N g_{HC}^2$ staying constant.
Despite the absence of a well-defined large-$N$ limit at the level of
the fundamental theory, it remains useful to keep the naive counting in
powers of $1/N$ at the level of the NJL description of the dynamics,
since it allows, for instance, to identify contributions which will
be numerically suppressed even for already moderate values of $N$.
Therefore, when, in the sequel, we mention or use the $1/N$ expansion, it will thus always
be understood that it refers to the NJL context.

\subsection{The pattern of flavour symmetry breaking}\label{total-break}

Concerning the pattern of spontaneous symmetry breaking, there are now two possible fermion bilinears that may form a condensate.
A non-zero $\la\psi^a\psi^b\ra$ would break $SU(4)\times U(1)$ to $Sp(4)$, with NGBs transforming as $(5+1)_{Sp(4)}$.
A non-zero $\la X^f X^g \ra$  would break $SU(6)\times U(1)$ to $SO(6)$, with NGBs in the representation
$(20'+1)_{SO(6)} = (8+6+\bar 6 +1)_{SU(3)_c}$.
Light coloured scalars are phenomenologically problematic because of the strong bounds from collider searches.
An important contribution to their mass is induced by gluon loops, as discussed in section \ref{gauging}, 
in appendix \ref{SU6-generators} and in section \ref{Masses of coloured scalar resonances}. 
Another possibility to lift the coloured NGBs from the low energy spectrum is to introduce the mass term (\ref{X_mass}), 
which explicitly breaks $SU(6)\times U(1)$ to $SO(6)$.
Alternatively, if $SU(6)$ does not undergo spontaneous breaking, 
coloured NGBs would be absent. However, we will show below that the matching of anomalies would then require massless,
coloured fermions, that again call for a large radiative mass or for $m_X\ne 0$.

Since we have adopted the same fermion content as in Ref.~\cite{Barnard:2013zea}, let us stress some differences
with respect to the discussion of flavour symmetries in that paper.
First, the non-anomalous axial $U(1)$ symmetry was not discussed: we will show that it has several
phenomenological consequences. Second, the  colour triplet
and antitriplet components of $X^f$ were treated separately, and the global symmetry was identified with
$SU(3)\times SU(3)\times U(1)_V$, broken by a mass term to $SU(3)_c \times U(1)_V$.
However, these are just maximal subgroups of the complete global symmetry $SU(6)$, and of the complete unbroken
subgroup $SO(6)$, respectively.
The pattern is different from QCD, because there quarks and antiquarks transform under different representations of
the gauge group, while here the six copies of $X^f$ transform in the same way under $Sp(2N)$.
Note that $U(1)_V$ was introduced in Ref.~\cite{Barnard:2013zea} in order to provide top partners with the appropriate
SM hypercharge, but remarkably enough such a symmetry is automatically present, as one of the unbroken generators within $SO(6)$.

Once both the elementary fermions $\psi^a$ and $X^f$ are introduced, one can form several baryons.
As a consequence, the anomaly matching condition provides non-trivial constraints on the spontaneous
symmetry breaking, as discussed in section \ref{anomat}.
If one denotes by $V$ the conserved currents associated to the $H_m$ generators, and by $A$
the conserved currents associated to the generators of the coset $G/H_m$ (see section \ref{VW}),
one needs only consider the anomaly matching constraints that arise from the $\langle VVA \rangle$ correlators.
Then, to each fermion transforming in the representation $r$ of $G$ is associated an anomaly coefficient $A(r)$,
which is defined by
\be
2 {\rm tr}\left[T^{\hat A} (r) \{T^B (r),T^C (r)\}\right] = A(r) d^{{\hat A}BC}~,
\ee
where $T^A (r)$ and $T^{\hat A} (r)$ are the generators of $H_m$ and of $G/H_m$, respectively, in the representation $r$,
and $d^{{\hat A}BC}$ is an invariant tensor
that depends on $G$. 
The generators of  the fundamental representation $r_0$ are normalised  as in Eq.~(\ref{norm_gen}), and its anomaly coefficient is fixed to
$A(r_0)=1$.
The anomaly matching condition can be written as
\be
\sum_i n_i A(r_i) = \sum_i n'_i A(r_i) ~,
\ee
where the left-hand (right-hand) sum runs over the representations of the constituent (composite) fermions,
and $n_i$ ($n'_i$) are their multiplicities.
If this equality cannot be satisfied, then $G$ necessarily undergoes spontaneous symmetry breaking.

In the model under investigation, the possible trilinear baryons consist of
\be
\Psi^{abf}= (\psi^a \psi^b X^f)~,~~
\Psi^{ab}_f= (\psi^a \psi^b \overline X_f)~,~~
\Psi^{af}_b= (\psi^a \overline\psi_b X^f)~,~~
\Psi^{fgh}= (X^f  X^g X^h)~,~~
\Psi^{fg}_h= (X^f X^g \overline X_h)~,~~
\ee
plus their conjugates, where the brackets stand for a spin-1/2, hypercolour-singlet contraction
(multiple, independent contractions of this kind may be possible).
Each $\Psi$ decomposes in several irreducible representations $(r_4,r_6)$ of $SU(4)\times SU(6)$,
each corresponding to an independent baryon state:
for example $\Psi^{abf} \sim [(6,6)+(10,6)]$.
In addition, exotic baryons are also possible, formed by a larger, odd number of constituent fermions.

Let us begin with the $SU(4)^3$ anomaly. As $\psi$ lies in the fundamental representation of $SU(4)$,
its anomaly coefficient is $A_4(4)=1$. The $SU(4)$ representations contained in $\psi^a\psi^b$ or $\psi^a\overline\psi_b$ have coefficients
$A_4(1)=A_4(6)=A_4(15)=0$ and $A_4(10)=8$. Therefore, the anomaly matching between $\psi$ and the trilinear baryons $\Psi$ reads
\be
2N\cdot A_4(4)= 2N = \sum_{(r_4,r_6)} n_{(r_4,r_6)} A_4(r_4) \cdot \dim(r_6)  = n_{(10,6)} 6 \cdot 8 ~,
\label{AM4}
\ee
where the sum runs over the various massless baryon states, and $n_{(r_4,r_6)}$ are their multiplicities.
One can generalise the result to include exotic baryons: in full generality, hypercolour invariance requires
the total number of $\psi$ and $\overline\psi$ fermions to be even; then, in order to obtain a fermion, one needs
that the total number of $X$ and $\overline X$ is odd.
One can check \cite{Yamatsu:2015npn} that (i) the anomaly coefficient of any $SU(4)$ representation,
contained in $4\times \dots \times 4$ an even number of times, is a multiple of $8$, and (ii) the dimension of any $SU(6)$ representation,
contained in $6\times\dots\times 6$ an odd number of time, is a multiple of $2$.
As a consequence, the right-hand side of Eq.~(\ref{AM4}) generalises to a multiple of $2\cdot 8$,
and the matching is possible only for $N=8n$, with $n$ integer. An example with $N=8$
is provided by one exotic baryon $(\psi\psi XXX)\sim (10,20)$ plus three copies of $(\overline\psi\overline\psi X)\sim (\overline{10},6)$.
In summary, for $N\ne 8n$ $SU(4)$ necessarily spontaneously breaks to $Sp(4)$ and the
corresponding NGB decay constant $F_G$ is non-zero.
Strictly speaking, the other order parameters,
such as the condensate $\la\psi\psi\ra$, may still vanish, for instance if a discrete symmetry subgroup leaves the
vacuum invariant but not the $(\psi\psi)$ operator \cite{Dashen:1969eg}. This is, however, a rather unlikely situation to happen
\cite{Kogan:1998zc}, and we will assume that the spontaneous symmetry breaking of the $SU(4)$ flavour group
(towards its $Sp(4)$ subgroup) is due to the formation of a non-vanishing $\langle \psi\psi\rangle$
condensate. This corresponds actually to the dynamical situation described by the NJL framework, where
$SU(4)$ order parameters like the condensate are proportional to $F_G$.

Next, let us consider the $SU(6)^3$ anomaly. 
The crucial observation is that there are baryons, contained either in $(\psi\overline\psi X)$ or $(XX\overline X)$,
that transform under the representation $(1,6)$.
These states have evidently the same anomaly coefficient
$A_6(6)=1$ as the constituent fermion $X$, therefore the matching is trivially possible for any value of $N$:
\be
(2N+1)(N-1) \cdot A_6(6) = \sum_{(r_4,r_6)} n_{(r_4,r_6)} \dim(r_4) \cdot A_6(r_6) = n_{(1,6)} 1\cdot A_6(6) + \dots  ~,
\label{AM6}
\ee
where the ellipsis stands for the contribution from larger representations, which are not relevant in the present context.
As a consequence, from the point of view of the anomaly condition, the spontaneous breaking of $SU(6)$ is not a necessity, and in particular it allows the possibility that $\la XX\ra=0$. 
However, the mass inequalities mentioned in section \ref{inequalities} require, in the case where massless baryons are present in the bound state spectrum, 
massless spin-zero bound states, coupled to the currents associated with the generators of the $SU(6)/SO(6)$ coset, which is tantamount to the spontaneous breaking of $SU(6)$ towards $SO(6)$.

Note that the massless baryons required by anomaly matching carry colour and are phenomenologically excluded.
Once these baryons are made heavy by explicit symmetry breaking, there are no exact NGBs either, and again one
cannot tell whether the dynamics breaks spontaneously $SU(6)$ or not.
Indeed, in either case an explicit symmetry breaking mass term $m_X XX$ is required for specular reasons:
in the unbroken phase, one needs it to give a sufficiently large mass to the coloured baryons;
in the broken phase, the mass term is necessary to make the coloured NGBs sufficiently heavy.
Ref.~\cite{Cacciapaglia:2015vrx} argues that the mass of the top partners can be controlled by the parameter $m_X$,
if one assumes to be in the unbroken phase.

Finally, one should consider the anomalies involving the non-anomalous $U(1)$. The anomaly for $U(1)SU(6)^2$
is easily matched for any $N$, by the same set of baryons that matches the $SU(6)^3$ anomaly.
We also proved that the other anomalies involving $U(1)$, that is  $U(1)SU(4)^2$ and $U(1)^3$,
can be matched for any $N$ as well, but using a different set of baryons in each case.
It is highly non-trivial to match all $U(1)$ anomalies at the same time, and thus preserve this symmetry from spontaneous breaking.
As we have already argued though, it is quite unlikely that the spontaneous breaking of the $SU(4)$ flavour
symmetry happens without, at the same time, also triggering the spontaneous breaking of the $U(1)$ symmetry.

In the following sections, we will apply the NJL techniques to the complete model including the electroweak and the colour sector.
In particular, we will study the mass gap equations that determine $\la \psi\psi \ra$ and $\la XX\ra$ in terms of the coefficients
of the four-fermion operators. For $N\ne 8n$, only the phase $\la \psi\psi \ra\ne 0$ of the NJL model should be considered as a
good approximation of the full dynamics, while  $\la XX\ra$ is not constrained by the matching of anomalies. For $N=8n$, both
condensates may or may not vanish.

\subsection{Sum rules and pseudoscalar decay constants in the flavour-singlet sector}
\label{Sum rules and pseudoscalar decay constants in the flavour-singlet sector}

As a last point to be discussed in this section, let us recall that in section \ref{SR} we introduced the spectral sum rules for a
simple group $G$ that undergoes spontaneous breaking. That discussion applies to the $\psi$-sector alone, with coset $SU(4)/Sp(4)$,
as well as to the $X$-sector in isolation, with coset $SU(6)/SO(6)$. In the complete model, one can also construct
correlation functions involving simultaneously the two sectors and that are order parameters for the
whole symmetry group $SU(4) \times SU(6) \times U(1)$, i.e. involving also the non-anomalous axial singlet
transformations.
This leads to additional
sum rules that may constrain the resonance spectrum.
At the level of two-point functions, the relevant order parameters involving the two sectors are:
\bea
\Pi_{S^0}^{\psi X}  (q^2) & = &
i \int d^4 x \, e^{i q \cdot x} \langle {\rm vac} \vert T \{ {\mathcal S}^0_\psi (x) {\mathcal S}^0_X (0) \}  \vert {\rm vac} \rangle
~,
\nonumber\\
\Pi_{P^0}^{\psi X}  (q^2) & = &
i \int d^4 x \, e^{i q \cdot x} \langle {\rm vac} \vert T \{ {\mathcal P}^0_\psi (x) {\mathcal P}^0_X (0) \}  \vert {\rm vac} \rangle
~.
\eea
From them we derive two additional spectral sum rules, valid in the limit where $m_X$ vanishes:
\begin{equation}
\int_0^\infty dt ~{\rm Im} \Pi_{S_0}^{\psi X} (t) = 0 ~,
\qquad\qquad
\int_0^\infty dt ~{\rm Im} \Pi_{P_0}^{\psi X} (t) = 0 ~,
\end{equation}
which respectively constrain the spectrum of scalar and pseudoscalar singlets resonances.

One could examine the realization of these sum rules in the NJL framework, 
similarly to what we did for the electroweak sector in section \ref{secWSR}, 
for instance investigating whether
the first low-lying resonances in each channel saturate them. Here
we rather describe some of the expected features in general terms,
independently from the NJL approximation.
In the singlet pseudoscalar channel, we expect two states. The first one is the Goldstone boson $\eta_0$
produced by the spontaneous breaking of the non-anomalous axial $U(1)$ symmetry. The second one
is a massive pseudoscalar state $\eta^\prime$, which corresponds to the second
Goldstone boson that would be present in the absence of the gauge anomaly in the divergences of
the $U(1)_\psi$ and $U(1)_X$ currents. These states both couple to the (partially) conserved
$U(1)$ current, defined in Eq. (\ref{axial_singlet}) above,
\be
\langle {\rm vac} \vert {\mathcal J}^0_\mu (0) \vert \eta_0 (p) \rangle = i F_{\eta_0} p_\mu
~,
\qquad\qquad
\langle {\rm vac} \vert {\mathcal J}^0_\mu (0) \vert \eta^\prime (p) \rangle = i F_{\eta^\prime} p_\mu
~.
\label{defFeta0etap}
\ee
In the limit where $m_X$ vanishes, $F_{\eta_0}$ remains nonzero and $F_{\eta^\prime} \sim {\mathcal O} (m_X)$,
whereas for the masses $M_{\eta_0}^2 \sim {\mathcal O} (m_X)$ while $M_{\eta^\prime}^2$ does not vanish.
Of course, there are also couplings to the individual, non conserved, $U(1)_\psi$ and $U(1)_X$ currents,
defined in Eqs. (\ref{U1_psi}) and (\ref{U1_X}), respectively
\bea
&
\langle {\rm vac} \vert {\mathcal J}^0_{\psi\mu} (0) \vert \eta_0 (p) \rangle = i F_{\eta_0}^\psi p_\mu~,
&
\qquad\qquad
\langle {\rm vac} \vert {\mathcal J}^0_{\psi\mu} (0) \vert \eta^\prime (p) \rangle = i F_{\eta^\prime}^\psi p_\mu~,
\nonumber
\\
&
\langle {\rm vac} \vert {\mathcal J}^0_{X\mu} (0) \vert \eta_0 (p) \rangle = i F_{\eta_0}^X p_\mu~,
&
\qquad\qquad
\langle {\rm vac} \vert {\mathcal J}^0_{X\mu} (0) \vert \eta^\prime (p) \rangle = i F_{\eta^\prime}^X p_\mu~.
\label{defFeta}
\eea
According to the expressions given in Eqs. (\ref{U1_psi}), (\ref{U1_X}), and (\ref{axial_singlet}),
these four decay constants are related to the ones in the preceding equation through
$F_{\eta_0 , \eta^\prime} = F_{\eta_0 , \eta^\prime}^X - 3(N-1) F_{\eta_0 , \eta^\prime}^\psi$.
Both $\eta_0$ and $\eta^\prime$ states also couple to the singlet pseudoscalar densities,
\bea
&
\langle {\rm vac} \vert {\mathcal P}^0_{\psi} (0) \vert \eta_0 (p) \rangle =  G_{\eta_0}^\psi~,
&\qquad\qquad
\langle {\rm vac} \vert {\mathcal P}^0_{\psi} (0) \vert \eta^\prime (p) \rangle =  G_{\eta^\prime}^\psi~,
\nonumber
\\
&
\langle {\rm vac} \vert {\mathcal P}^0_{X} (0) \vert \eta_0 (p) \rangle =  G_{\eta_0}^X~,
&\qquad\qquad
\langle {\rm vac} \vert {\mathcal P}^0_{X} (0) \vert \eta^\prime (p) \rangle =  G_{\eta^\prime}^X~,
\label{defGeta}
\eea
and through Eq. (\ref{U1_div}) the two following relations hold:
\be
F_{\eta_0} M_{\eta_0}^2 = 4 \sqrt{3} \, m_X G_{\eta_0}^X~,
\qquad\qquad 
F_{\eta^\prime} M_{\eta^\prime}^2 = 4 \sqrt{3} \, m_X G_{\eta^\prime}^X~.
\label{FetaMrelation}
\ee

Although they do not lead to sum rules, it is both interesting and useful to
consider two-point correlators involving the axial singlet current and the
singlet pseudoscalar densities, defined in analogy to Eq. (\ref{PiAPdef})
for the non-singlet case,
\be
\Pi_{A^0P^0}^\psi (q^2) q_\mu =
\int d^4 x \, e^{i q \cdot x} \langle {\rm vac} \vert T \{ {\cal J}_\mu^{0} (x) {\cal P}^0_{\psi} (0) \}  \vert {\rm vac} \rangle
~,
\qquad
\Pi_{A^0P^0}^X (q^2) q_\mu =
\int d^4 x \, e^{i q \cdot x} \langle {\rm vac} \vert T \{ {\cal J}_\mu^{0} (x) {\cal P}^0_{X} (0) \}  \vert {\rm vac} \rangle
~.
\label{PiAPsingdef}
\ee
$\Pi_{A^0P^0}^\psi (q^2)$ and $\Pi_{A^0P^0}^X (q^2)$ are order parameters of $SU(4)\times U(1)$ and of $SU(6)\times U(1)$, respectively,
and in the limit where the current ${\cal J}_\mu^{0} (x)$ is conserved they are both saturated by the massless $\eta_0$
pole, as in Eq. (\ref{PiAP}). In the presence of the mass $m_X$, this is no longer true, and the Ward identities give
\be
q^2 \Pi_{A^0P^0}^\psi (q^2) = 4 \sqrt{3} m_X \Pi_{P^0}^{\psi X} (q^2) - 6 (N-1) \langle {\mathcal S}^0_\psi \rangle
~,\qquad\qquad
q^2 \Pi_{A^0P^0}^X (q^2) = 4 \sqrt{3} m_X \Pi_{P^0}^{X} (q^2) + 2 \langle {\mathcal S}^0_X \rangle
~.
\label{PiAPsing}
\ee
These lead, in particular, to the constraints
\be
4 \sqrt{3} m_X \Pi_{P^0}^{\psi X} (0) = 6 (N-1) \langle {\mathcal S}^0_\psi \rangle
~,
\qquad\qquad
4 \sqrt{3} m_X \Pi_{P^0}^{X} (0) =- 2 \langle {\mathcal S}^0_X \rangle
~,
\ee
as well as
\be
F_{\eta_0} G_{\eta_0}^\psi = 6 (N-1) \langle {\mathcal S}^0_\psi \rangle + {\mathcal O}(m_X)
~,
\qquad\qquad
F_{\eta_0} G_{\eta_0}^X = - 2 \langle {\mathcal S}^0_X \rangle + {\mathcal O}(m_X)
~,
\label{FGeta0}
\ee
which provide useful cross-checks for the NJL calculation.

\subsection{Effective couplings induced by the  hypercolour gauge anomaly}
\label{thooft}

In order to study, in the NJL framework, the
anomalous divergence of Eq.~(\ref{U1_X_div}), induced by the $Sp(2N)$ hypercolour gauge interaction,
let us first discuss the $X$-sector in isolation.
The sector of gauge configurations with unit winding number now induces $2 (N-1)$ fermionic
zero modes per flavour (in the present case, $N_f^X=3$) for the Dirac operator corresponding to the $X$ and ${\overline X}$ fermions
(the uninteresting case $N=1$ is, of course, discarded). Through the index theorem, these zero modes induce
a violation of the $U(1)_X$ charge by $12 (N-1)$ units, which, as already discussed in Section \ref{colourless-sector}
for the electroweak sector, has to be reproduced by the effective 't Hooft vertex.
In the case of an $Sp(4)$ gauge group ($N=2$), it is straightforward to construct an operator $\mathcal{O}_X$ that
induces this violation of the invariance under $U(1)_X$, while at the same time preserving the
invariance under the $SU(6)$ flavour group:
\be
\mathcal{O}_X 
= - \frac{1}{6!} \epsilon_{f_1\cdots f_6} \epsilon_{g_1\cdots g_6} (X^{f_1} X^{g_1})\cdots (X^{f_6} X^{g_6})
=  - {\rm det} (X^{f} X^{g}) ~,
\label{OX}
\ee
where the determinant is taken in the six-dimensional flavour space.
For $N>2$ and only $6$ Weyl fermions at our disposal, one obvious extension of the above
operator satisfying the required properties would consist in taking $\mathcal{O}_X^{N-1}$.
One should, however, be aware that, on the one hand, this simple procedure might not
comply with the properties of the 't Hooft vertex as arising from the Grassmann integration
over the fermionic collective coordinates\footnote{Useful introductions to instantons are
provided by Refs. \cite{Coleman:1978ae,Olive:1979ke,Vandoren:2008xg}}, and, on the other hand, that the 't Hooft vertex could also
involve derivatives of the fermion fields. An example where this second feature is known to
happen is provided by the case of an $SU(2)\simeq Sp(2)$ gauge group with fermions
in the adjoint representation \cite{Vainshtein:1982ic}. Delving more deeply into these
issues would, however, lead us too far astray. Moreover, dealing with a term involving
derivatives of the fermion fields is beyond the NJL framework as it is usually understood.
From the point of view of the latter, the term $\mathcal{O}_X^{N-1}$, possessing all
the required symmetry properties, is quite appropriate, and henceforth we will assume
that at the level of the NJL approach, it provides the required description of the
explicit breaking of the $U(1)_X$ symmetry by quantum effects.

The preceding discussion considered the $SU(6)$ sector in isolation and, apart
from some subtle aspects due to the representation of the gauge group under
which the $X$ fermions transform, has essentially paralleled the related discussion for
the $SU(4)$ sector in section \ref{colourless-sector}. We will now bring the two sectors together
and, as was already the case for the discussion of the anomaly matching conditions in 
section \ref{total-break}, we will find that when acting together the two sectors unravel
new features. Indeed, the structure of anomaly-driven effective terms is actually
different, as one should take into account that a combination of $U(1)_X$ and $U(1)_\psi$
transformations, as given in Eq.~(\ref{U1_charges}), remains non-anomalous.
This drastically changes the form of appropriate effective interactions generalising
the 't Hooft terms usually being given by a (flavour) determinant, since $\psi$ and $X$ are not in the same
representation.  Combining this information with the discussion
above and in Section \ref{colourless-sector}, the lowest dimensional operator
that breaks both $U(1)_\psi$ and $U(1)_X$ axial singlet symmetries, while preserving
the $U(1)$ symmetry generated by the combination (\ref{axial_singlet}), reads
\be
\mathcal{L}_{\psi X} =  A_{\psi X} \frac{\mathcal{O}_\psi}{(2N)^2} \left[\frac{\mathcal{O}_X}{[(2N+1)(N-1)]^6} \right]^{(N-1)} + h.c.~,
\label{Lano46}
\ee
with $\mathcal{O}_X$ defined in Eq.~(\ref{OX}) and $\mathcal{O}_\psi$ the antisymmetric four-fermion operator in
Eq.~(\ref{LSbasic}),
\be
\mathcal{O}_\psi =-\frac{1}{4} \epsilon_{abcd}(\psi^a \psi^b)(\psi^c \psi^d)~.
\ee
The constant $A_{\psi X}$ can be taken real  and positive by adjusting the phase of $\psi$.
Its normalisation in Eq. (\ref{Lano46}) has been conveniently chosen in order 
to compensate the different powers of 
$N$ contained in the condensates, see Eqs.~(\ref{psicond}) and (\ref{condensateXX}).
This normalisation, with an $N$-independent coefficient $A_{\psi X}$, 
would reproduce the correct behaviour of the $U(1)_{\psi,X}$ anomaly
in the large-$N$ limit, would the latter exist, see the discussion around Eqs.~(\ref{betaHC}) 
and (\ref{Nlimit}).
Indeed,  Eq.~(\ref{U1_X_div}) shows that the effect of the anomaly would not vanish in this limit, as
$(N-1) g^2_{HC} \sim (N-1)/N \sim 1$. As we will see in Section \ref{mixing-singlets}, a
trace of this feature appears in the mass of the $\eta^\prime$, which
is proportional to $A_{\psi X}$, $M^2_{\eta^\prime}\sim A_{\psi X}[1+ {\cal O}(1/N)]$.

After formation of the two condensates  $\la \psi \psi \ra$ and $\la X X \ra$, the interaction (\ref{Lano46}) will generate
effective four-fermion interactions for $\psi$ and $X$, as well as a mixed $\psi\psi X X$ term, upon replacing appropriate number 
of fermion bilinears  by their respective condensate (i.e. closing the loops).
To identify these four-fermion interactions, relevant for the computation of the meson spectrum, let us first consider for simplicity the $SU(6)\to SO(6)$ sector. The fermion bilinear can be decomposed as
\be
( X^f X^g ) \equiv 2 (T_X^0 \Sigma_{0}^c )^{gf} \left( X \Sigma_{0}^c T^0_X X\right) + 2 (T^{\hat F} \Sigma_{0}^c)^{gf} \left(X \Sigma_{0}^c T^{\hat F} X\right)~,
\ee
in terms of the $SO(6)$ singlet and the two-index symmetric traceless components.
Then, taking into account combinatorial factors,
the operator of Eq.~(\ref{OX}) can be decomposed as
\footnote{
The coefficient of $\left( X \Sigma_{0}^c T^0_X X\right)^6$ in ${\rm det} ( X^f X^g )$
is $2^6 \, {\rm det}(\Sigma_{0}^c T^0_X)=  - 1/27$,  and the coefficient of $\left( X \Sigma_{0}^c T^0_X X\right)^4
\left(X \Sigma_{0}^c T^{\hat F} X\right)\left(X \Sigma_{0}^c T^{\hat G} X\right)$ is
$ 2^6  \, {\rm det}(\Sigma_{0}^c T^0_X) (2 \sqrt{3})^2 \frac{1}{2}
\left[ {\rm tr} (T^{\hat F}) {\rm tr} (T^{\hat G}) - {\rm tr} (T^{\hat F} T^{\hat G}) \right]
=
 \frac{1}{9} \,\delta^{{\hat F}{\hat G}}$. }
\be
\mathcal{O}_X = \frac{1}{27} \left[ (X \Sigma_{0}^c T^0_X X)^6 -3 (X \Sigma_{0}^c T^0_X X)^4 \,(X \Sigma_{0}^c T^{\hat F} X)\,(X \Sigma_{0}^c T^{\hat F} X) +\cdots \right]~,
\label{Oxinv}
\ee
where a sum over the $SU(6)$ generators $T^{\hat F}$ belonging to the $SU(6)/SO(6)$ coset is understood.
For the $SU(4)\to Sp(4)$ sector, the similar appropriate decomposition into $Sp(4)$-invariant bilinears reads
\be
{\mathcal O}_\psi =    \left( \psi \Sigma_0 T^0_\psi  \psi\right)
\left(\psi \Sigma_0 T^0_\psi  \psi \right)  - \left(\psi \Sigma_0 T^{\hat A} \psi\right) \left(\psi \Sigma_0 T^{\hat A} \psi \right)~.
\label{Opsiinv}
\ee
Next we insert the results (\ref{Oxinv}) and (\ref{Opsiinv}) into the full effective Lagrangian Eq.~(\ref{Lano46}),
and obtain
\bea
\mathcal{L}_{\psi X} = 
 \frac{A_{\psi X}}{(27)^{N-1}} 
\! 
&& \,
\Bigg\{ \left(\frac{\psi \Sigma_0 T^0_\psi \psi}{2N}\right)^2 \left[\frac{X \Sigma_{0}^c T^0_X X}{(2N+1)(N-1)} \right]^{6(N-1)} 
- \left(\frac{\psi \Sigma_0 T^{\hat A}\psi}{2N} \right)^2   \left[\frac{X \Sigma_{0}^c T^0_X X}{(2N+1)(N-1)} \right]^{6(N-1)} 
\nonumber\\
&&
\, -3 (N-1) \left(\frac{\psi \Sigma_0 T^0_\psi \psi}{2N}\right)^2  
\left[ \frac{X \Sigma_{0}^c T^0_X X}{(2N+1)(N-1)} \right]^{6(N-1)-2} 
\left[\frac{X \Sigma_{0}^c T^{\hat F} X}{(2N+1)(N-1)} \right]^2  \Bigg\}  +\cdots~,
\label{L46}
\eea
where the ellipsis denotes other interaction terms, of no relevance for our purposes.
The overall constant $A_{\psi X}$  remains arbitrary,
but the ratio of the coefficients of the three effective $XXXX$, $\psi\psi\psi\psi$, and $\psi\psi XX$ terms are fixed.
All effective couplings in the singlet and non-singlet sectors are thus
related to the unique coupling $A_{\psi X}$ in Eq.~(\ref{Lano46}), times appropriate powers of
the two condensates and combinatorial factors 
(see section \ref{Mass gap equations and effective four-fermions couplings} below).

\section{Spectrum of meson resonances in presence of the coloured sector}
\label{The spectrum of mesonic resonances in the coloured sector}

In this section we will compute the condensates and the masses of mesons, once the coloured sector is added to the electroweak sector, 
including their mixing through Eq.~(\ref{L46}).
The two sectors share the same $Sp(2N)$ hypercolour gauge interaction, therefore one can, in principle, relate the sizes of 
the effective four-fermion operators in the two sectors. 
One may assume, in particular, that the effective interactions between hypercolour-singlet fermion bilinears originate from $Sp(2N)$ current-current operators 
(see appendix \ref{fierz}). In this approximation one can link, to some extent, 
the couplings of the coloured operators to the electroweak ones.
In this way the mass gap and the spectrum in the $SU(6)$ sector are connected to the ones in the $SU(4)$ sector.

{\subsection{The mass gap in a theory with two sectors}
\label{Mass gap equations and effective four-fermions couplings}

Let us begin with the coloured scalar operators, which are relevant for the mass gap and for the spin-zero mesons. 
Besides the anomalous operator (\ref{L46}), 
there is  one more independent four-fermion operator that describes the dynamics
in analogy with the electroweak sector Lagrangian in Eq.~(\ref{LSbasic}),
\begin{equation}
\mathcal{L}_{scal}^{X}=\frac{\kappa_{A6}}
{(2N+1)(N-1)}(X^f X^g)({\overline X}_f {\overline X}_g)
-\frac{1}{2} m_X \left[ (X \Sigma_{0}^c X)+({\overline X} \Sigma_{0}^c {\overline X}) \right]
~,
\label{L4F-scal-color}
\end{equation}
where the coupling constant $\kappa_{A6}$ is real and its normalisation 
by an inverse factor $(2N+1)(N-1)$ has been conveniently chosen to
compensate for the factors of $N$ induced by the trace over hypercolour in the 
$X$-fermion one-loop two-point functions (see appendix \ref{SDresum}).  
 In contrast with the electroweak sector, we also include in Eq.~(\ref{L4F-scal-color}) an 
explicit symmetry-breaking mass $m_X$, already introduced in Eq.~(\ref{X_mass}), which 
can be chosen real and positive by tuning the phase of $X$. Note that also $A_{\psi X}$ in Eq.~(\ref{L46}) 
can be chosen real and positive, by tuning the phase of $\psi$.
Such a mass term may be phenomenologically necessary 
to raise the masses of the coloured pNGBs,
in order to comply with direct collider detection limits \cite{Cacciapaglia:2015eqa}.
More generally, a non-zero $m_X$ leads to several qualitative effects that are worth to be explored.
As the contraction over $Sp(2N)$ indices in Eq.~(\ref{X_inv}) is symmetric in hypercolour space, 
the scalar bilinear $(X^f X^g)$ must be symmetric in flavour space, that is, it transforms as the $21_s$ representation of $SU(6)$, 
to be compared with $(\psi^a\psi^b)$, which transforms as the $6_a$ of $SU(4)$.
Since $21_{SU(6)}=(1+20')_{SO(6)}$, one can rewrite the Lagrangian (\ref{L4F-scal-color}) in the physical basis, as
\begin{equation}
{\cal L}_{scal}^X= 
\frac{2\kappa_{A6}}{(2N+1)(N-1)}\left[
(X \Sigma_{0}^c T^0_X X )(\overline{X} T_X^0 \Sigma_{0}^c \overline{X} )
+
(X \Sigma_{0}^c T^{\hat F} X ) (\overline{X} T^{\hat F} \Sigma_{0}^c\overline{X} )
\right]
-\frac{1}{2} m_X \left[ (X \Sigma_{0}^c X) + ({\overline
X} \Sigma_{0}^c {\overline X}) \right] 
,
\label{L-scalar-coloured}
\end{equation}
where $T^{\hat{F}}$ are the 20 broken generators spanning the $SU(6)/SO(6)$ coset.

Combining the effect of the operators in Eqs.~(\ref{LSphys}), (\ref{L46}) and (\ref{L-scalar-coloured}), one can derive a system of two coupled 
gap equations for the $SU(4)$ and $SU(6)$ sectors,
\begin{equation}
\left\{
  \begin{array}{lll}
1-4(\kappa_A+\kappa_B) \tilde A_0(M^2_\psi) = 0~, \\
1-4(\kappa_{A6}+\kappa_{B6}) \tilde A_0(M^2_X)-\dfrac{m_X}{M_X} = 0
,\\
  \end{array}
\right.
\label{gap-2sectors}
\end{equation}
which determine the dynamical masses $M_\psi$ and $M_X$ as functions of the four couplings 
$\kappa_{A,B,A6,B6}$ and of the mass $m_X$. 
More precisely, when $m_X \neq 0$ the scale $M_X$ is not entirely generated by the dynamics, see Fig.~\ref{gap-coloured}.
Just as in the electroweak sector, $M_\psi$ can be traded for $\langle \Psi\Psi \rangle$, see  Eq.~(\ref{psicond}), 
the NJL dynamical mass $M_X$ is also related to the condensate $\langle XX \rangle$ in the coloured sector,
\begin{equation}
\langle XX \rangle \equiv \dfrac{1}{\sqrt{N_f^X}}\langle S_0^X\rangle
=-2 (2N +1)(N-1) M_X \tilde A_0(M^2_X)
,
\label{condensateXX}
\end{equation}
where the factor $(2N+1)(N-1)$ 
comes from the trace over hypercolour.
The two mass gap equations are coupled because the first operator in Eq.~(\ref{L46}) induces both the $\kappa_B$
and $\kappa_{B6}$ terms in Eq.~(\ref{gap-2sectors}).
These contributions are obtained by closing all but one fermion bilinears into a tadpole loop, 
as illustrated in Fig.~\ref{gapL46} for the case of the $\psi$-sector.
This amounts to replacing each bilinear by the associated condensate,
and to add a combinatorial factor $2$ in $\kappa_B$, as one $\psi$-bilinear out of 2 is not closed,
and $6(N-1)$ in $\kappa_{B6}$, as one $X$-bilinear out of $6(N-1)$ is not closed. 
Therefore, the anomalous terms in the gap equations are related to the original anomaly coefficient $A_{\psi X}$ by
\begin{equation}
\kappa_B\equiv \frac{A_{\psi X}}{2\cdot27^{N-1}} \left[ \frac{4N^X_f \langle XX \rangle^2}{(2N+1)^2(N-1)^2}  \right]^{3(N-1)} 
\frac{2}{2N}
=[4 M_X \t A_0(M_X^2)]^{6(N-1)} ~\frac{A_{\psi X}}{2N} ~,
\label{kbkb6}
\end{equation}
\begin{equation}
\kappa_{B6}\equiv \frac{A_{\psi X}}{2\cdot 27^{N-1}}  
\left[  \frac{4N^\psi_f \langle \psi\psi \rangle^2}{(2N)^2}\right]  \left[ \frac{4N^X_f  \langle XX \rangle^2}{(2N+1)^2(N-1)^2} 
\right]^{3(N-1)-1} \frac{6(N-1)}{(2N+1)(N-1)}
=
\frac{4N}{2N+1} \frac{M_\psi^2}{M_X^2} \frac{\t A^2_0(M_\psi^2)}{\t A^2_0(M_X^2)} ~\kappa_B~.
\label{kbkb62}
\end{equation}
The combinatorial factors will be essential, among other things, in order to recover the 
singlet Goldstone boson, see section \ref{mixing-singlets}.
The effective couplings $\kappa_{B,B6}$  are normalised such as to contribute to the gap 
equations (\ref{gap-2sectors}) as for a single sector in isolation.
However, since they  are functions of both dynamical masses, $\kappa_{B,B6}=\kappa_{B,B6}(M_\psi^2,M_X^2)$,
the two gap equations are actually coupled in a non trivial way.

\begin{figure}[b]
\includegraphics[scale=0.8, trim= 0 140 0 60]{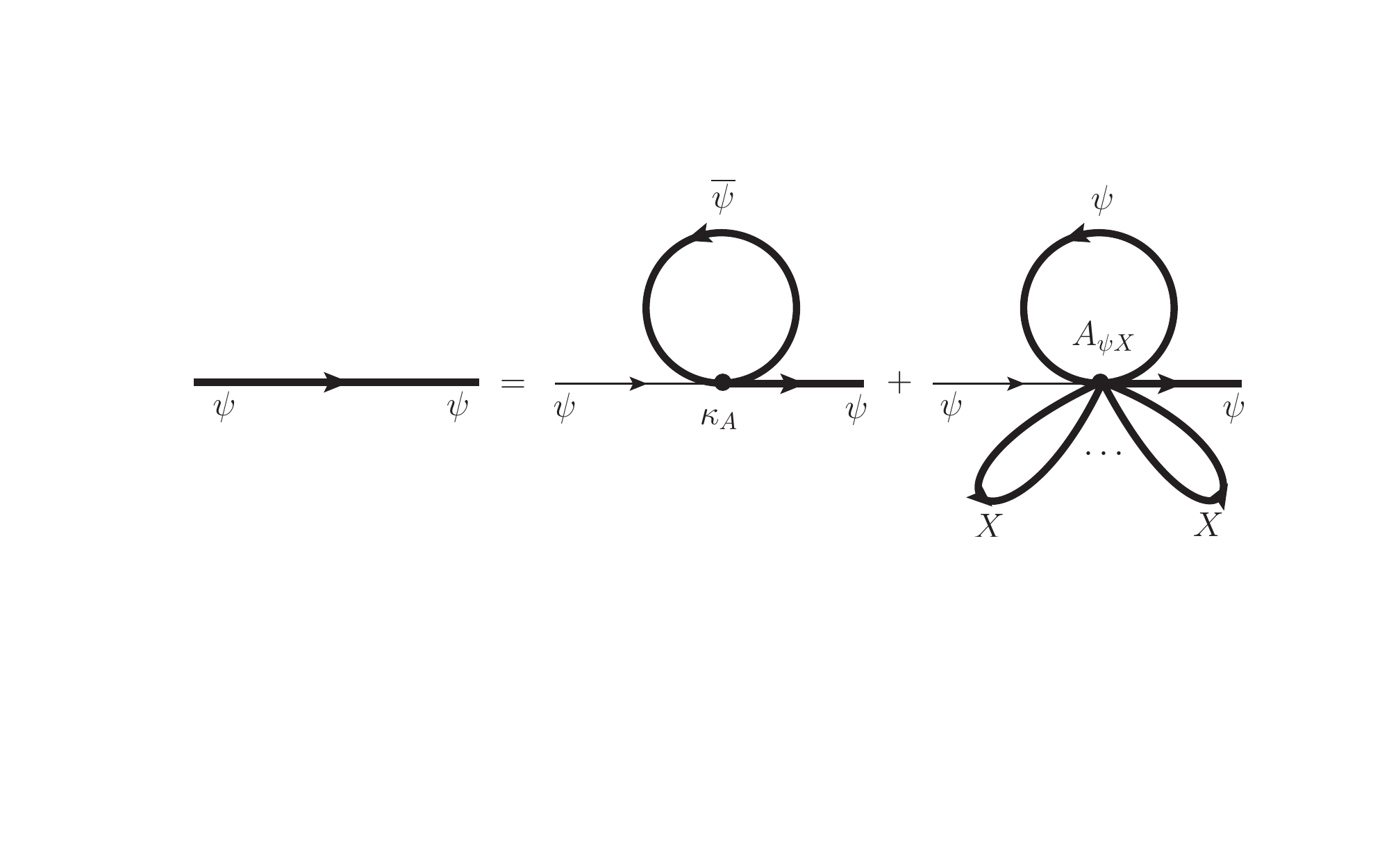}
\caption{Graphical illustration of the mass-gap equation in the $\psi$ sector.
The convention for the propagator lines are the same as in Fig.~\ref{figgap}.
The first term, proportional to $\kappa_A$, remains the same as in the electroweak sector 
in isolation. 
The second term, proportional to $A_{\psi X}$, is obtained by closing one loop of $\psi$-fermions and $6(N-1)$ loops of $X$-fermions in  Eq.~(\ref{L46}).
The mass-gap equation in the $X$-sector is obtained in an analogous way, with an additional term proportional to the explicit fermion mass $m_X$.}
\label{gapL46}
\end{figure}

Let us analyse in some detail the system (\ref{gap-2sectors}) of two coupled gap equations, because it is qualitatively different from
the canonical  NJL gap equation of QCD, and, to the best of our knowledge, it was not studied in the existing literature.
It is convenient to take the effective coupling $\kappa_B$ as the free parameter characterising the effect of the hypercolour anomaly,
that is, to express $\kappa_{B6}$ as a function of $\kappa_B$ according to Eq.~(\ref{kbkb62}).
This choice makes it easier to compare with the electroweak sector in isolation, 
and it also simplifies the algebraic form of the solutions of Eq.~(\ref{gap-2sectors}). 
As we have seen in section \ref{colourless-sector}, the $SU(4)$ sector forms a condensate
and a non-zero dynamical mass $M_\psi$ is generated 
when $\xi\equiv (\kappa_{A}+\kappa_{B})\Lambda^2/(4\pi^2)$ is above the critical value $\xi =1$. 
Similarly, in the chiral limit $m_X=0$, a non-zero dynamical mass $M_X$ is generated when 
$\xi_c\equiv (\kappa_{A6}+\kappa_{B6})\Lambda^2/(4\pi^2) > 1$.
Beyond that, the general resolution of the set of equations (\ref{gap-2sectors}) coupled through Eq. (\ref{kbkb62})
is very involved, especially for $m_X\ne 0$, and it can only be solved numerically.
Still, it is instructive to consider a few special cases.

\subsubsection{Case $m_X = 0$, $\kappa_B = 0$}

When  $\kappa_B = 0$, i.e. $A_{\psi X} = 0$, 
the two gap equations decouple. It is convenient to introduce dimensionless variables 
and functions in order to rewrite them in the form
\begin{equation}
\left\{
  \begin{array}{lll}
1-\xi_A \bar A(x_\psi) = 0~, \\
1-\xi_{A6} {\bar A} (x_X)= 0 ~,\\
  \end{array}
\right.
\label{gap-2sectors_decoupled}
\end{equation}
where $x_{\psi , X} \equiv M^2_{\psi , X} / \Lambda^2$, $\xi_{A,A6} \equiv (\Lambda^2/4\pi^2) \kappa_{A,A6}$, and ${\bar A} (x) \equiv 1 - x \ln ( 1 + 1/x )$.
The solutions of the two equations in (\ref{gap-2sectors_decoupled}) are simply related as
\begin{equation}
 x_\psi (\xi_A) = x_X \left(\xi_{A6}\right)~.
 \qquad\qquad
\end{equation}
The result is to restrict the range of the allowed values of $\xi|_{\kappa_B=0}= \xi_A$, as compared
to the case of one sector in isolation. Indeed, imposing that both conditions
$0 \le x_\psi (\xi_A) \le 1$ and $0 \le x_X (\xi_{A6}) \le 1$ be satisfied
simultaneously requires
\begin{equation}
 {\rm max} \left( 1 , \frac{\kappa_{A}}{\kappa_{A6}} \right) \le \xi \le 
 {\rm min} \left( 1 , \frac{\kappa_{A}}{\kappa_{A6}} \right)  \frac{1}{1 - \ln 2} \qquad\qquad (\kappa_B=0)
 ~.
\end{equation}
Hence, for $\kappa_A/\kappa_{A6} > 1$ the minimal value of $\xi$ is larger than unity,
whereas for $\kappa_A/\kappa_{A6} < 1$, the highest value allowed for $\xi$ is reduced,
see Fig.~\ref{gap-coloured}. These considerations do not depend explicitly on the value of $N$,
 although the actual values of $\kappa_A$ and of $\kappa_{A6}$,
being determined by the hypercolour dynamics, will depend on $N$.

Thus, although the two gap equations are decoupled, the presence of the second one
impinges on the possible values allowed for the coupling of the second one, and vice-versa.
This simply illustrates the fact that while the two gap equations may be decoupled, they
nevertheless share the same effective-theory cutoff $\Lambda$.

\subsubsection{Case $m_X=0$, $\kappa_B \neq 0$}

By treating $\kappa_B$ as an extra free parameter, 
the first equation in the system (\ref{gap-2sectors}) 
is formally identical to the gap equation for the electroweak sector in isolation,  Eq.~(\ref{gap2}),
with solution $x_\psi=x_\psi(\xi)$.
Then, rewriting $\kappa_{B6}$ as a function of $\kappa_B$ according to Eq.~(\ref{kbkb62}),
the second gap equation 
becomes a self-consistent relation for $x_X$, that depends on $N$, $\xi$, $\xi_{A6}$, and $\xi_B\equiv (\Lambda^2/4\pi^2) \kappa_{B}$:
\begin{equation}
\left\{
  \begin{array}{lll}
1 - \xi {\bar A} (x_\psi) = 0 ~, \\
{\cal G} (x_X , \xi_{A6}) \equiv x_X {\bar A} (x_X) \left[ 1 - \xi_{A6} {\bar A} (x_X) \right] = {\displaystyle{ \frac{4N}{2N+1} \xi_B  \frac{x_\psi (\xi)}{\xi^2} }}~.
  \end{array}
\right.
\label{gap-2sectors_gen}
\end{equation}
Note that the 
second equality assumes a consistent solution of the first equation, $x_\psi(\xi)$, which
requires $1 <\xi < 1/(1-\ln 2)$.
In practice 
we solve numerically the first equation for $x_\psi(\xi)$, then we use it as an input to solve numerically the second one for $x_X(\xi)$.

In Fig.~\ref{gap2fig} we plot ${\cal G} (x, \xi_{A6}) $ as a function of $x$, for a few representative values of $\xi_{A6}$, 
as well as the right-hand side of the second equation in (\ref{gap-2sectors_gen}), for two values of $N$ and $\xi_B$,
assuming for simplicity two equal mass gaps, $x_\psi=x_X=x$.
The intersection between the dashed and solid curves determines 
the solution $x_X=x_X(N,\xi,\xi_{A6}, \xi_B)$.
The function ${\cal G} (x, \xi_{A6})$ vanishes at $x=0$ and,
for any fixed value $0<x<1$, it decreases with $\xi_{A6}$.
For $\xi_{A6} \le 1$, ${\cal G} (x, \xi_{A6})$ increases in the whole interval  
$0 \le x \le 1$, while for $\xi_{A6} > 1$ it decreases to negative values for small $x$, 
then increases as $x$ moves towards unity, becoming positive before $x=1$, as long as
$\xi_{A6} < 1/(1-\ln 2)$.
On the other hand, the function $x_\psi (\xi)/\xi^2$
satisfies $0 \le x_\psi (\xi)/\xi^2 
\lesssim1/10$ for $0 \le x \le 1$. 
Since $\xi_B \ge 0$, there is therefore no solution to the second
equation in (\ref{gap-2sectors_gen}) in the interval $0 \le x_X \le 1$
when $\xi_{A6} \ge 1/(1-\ln 2)$. 
In contrast, for values $1 < \xi_{A6}<1/(1-\ln 2)$ there is always a non-trivial
solution with $x_X<1$, as long as the right-hand side of the second equation in (\ref{gap-2sectors_gen})
is sufficiently small.
Finally, for $0 < \xi_{A6}<1$ the occurrence of a solution
happens only for a sufficiently large $\xi_B$, also depending on $N$.
The latter properties actually reflect the 
critical value $\xi_{A6}+\xi_{B6} >1$, necessary in order to obtain
a non-trivial mass-gap, here somewhat disguised by the change of variables.
Note that for fixed values of $N$, $\xi$ and $\xi_B$, the value of $x_X$ 
increases with $\xi_{A6}$.

\begin{figure}[tb]
\includegraphics[scale=.3]{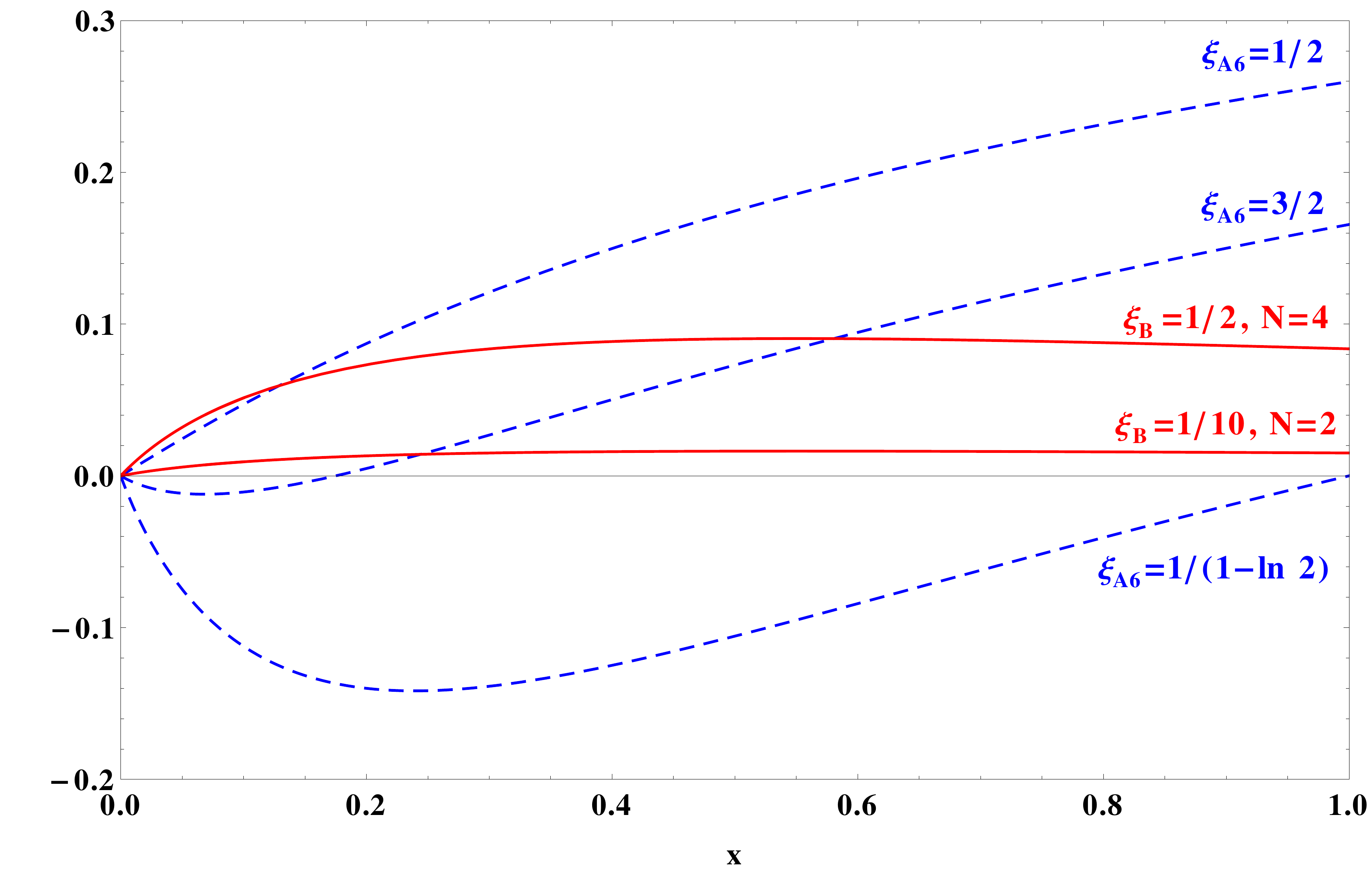}
\caption{Dotted curves: the function ${\cal G} (x , \xi_{A6}) $ for three representative values of $\xi_{A6}$ as indicated.
Thick curves: right-hand side of the second equation in (\ref{gap-2sectors_gen}) for two values of $N$ and $\xi_B$
as indicated, and taking $x_\psi=x$.}
\label{gap2fig}
\end{figure}

One can make one more step in the analytical study of the two coupled gap equations.
Moving the term proportional to $\xi_B$ in the first equation of (\ref{gap-2sectors_gen}) to its right-hand side, 
one may now eliminate $\xi_B$ between the two
equations, and obtain
\begin{equation}
 {\cal G} (x_\psi , \xi_{A})  = \left( \frac{1}{2} + \frac{1}{4 N} \right) {\cal G} (x_X , \xi_{A6}) ~.
\label{GxpsiGxX}
\end{equation}
A few simple remarks follow from this relation. First, if one of the masses, say $M_X$,
has been determined as a function of $\xi_A$, $\xi_{A6}$ and $\xi_B$, then the
relation of $M_\psi$ to $M_X$ involves only $\xi_A$, $\xi_{A6}$ and $N$, and not $\xi_B$. 
Second, this relation becomes rapidly independent of $N$ as $N$ increases.
Third, the relatively simple Eq.~(\ref{GxpsiGxX}) precisely gives the exact dependence of the ratio
of the two mass gaps, $M_X/M_\psi$,  as functions of the basic input parameters 
(although it is an implicit relation, due to the non-linearity in the masses $M_X, M_\psi$), as illustrated
for a few representative case in Fig.~\ref{gap-coloured}. More precisely, 
Eq.~(\ref{GxpsiGxX}) may be trivially expressed as 
\be
\frac{M_\psi^2}{M_X^2} = \left( \frac{1}{2} + \frac{1}{4 N} \right) 
\frac{ {\bar A}^2(x_X) [1-\xi_{A6} {\bar A}(x_X) ]} { {\bar A}^2(x_\psi) [1-\xi_A {\bar A}(x_\psi)]}\,.
\label{MpsiMX}
\ee
This indeed shows that, as long as $M_\psi^2, M_X^2 \ll \Lambda^2$ [which 
implies $\bar A(x_X)\simeq \bar A(x_\psi)$ since $ \bar A(x) \equiv 1 - x \ln ( 1 + 1/x )\simeq 1+M^2/\Lambda^2 \ln (\Lambda^2/M^2)$], 
one obtains $M_\psi < M_X$, at least for $\xi_A\simeq \xi_{A6}$.  
Indeed, the peculiar case of equal mass gaps, $x_\psi=x_X$, that is the one illustrated
in Fig.~\ref{gap2fig}, can only be obtained for significantly different values of  $\xi_A$ and $\xi_{A6}$ 
(for instance 
when  $N=4$, $\xi_{A6} =1/2$ and $\xi_B=1/2$, one has $x_\psi=x_X\simeq 0.13$, that corresponds to 
$\xi_A\simeq 0.9$).

In the above considerations we have kept $\kappa_A$ and $\kappa_{A6}$ (equivalently, $\xi_A$ and $\xi_{A6}$) 
arbitrary. Let us now examine more precisely a few typical situations concerning those parameters.  
When $\kappa_{A6}$ is larger than $\kappa_A$, the $SU(6)$ sector 
first forms a condensate for $\xi <1$ (see Fig.~\ref{gap-coloured}), and then $M_X>M_\psi$.
In the opposite case where $\kappa_{A6}$ is smaller than $\kappa_A$, the $SU(6)$ sector forms a condensate
for a value $\xi>1$, and $M_X<M_\psi$.
If $\xi_{A6} \gg \xi_A$, the mass gap grows rather fast, so that one eventually obtains
a very large $M_X\sim \Lambda$, and conversely a very large 
$M_\psi$ if $\xi_{A6} \ll \xi_A$.
Thus to obtain predictive calculations in both sectors from the NJL model, it requires that 
$\xi_A\sim \xi_{A6}$ are roughly of the same magnitude. 
In this way, there is a non-zero interval for the values of $\xi$ where the NJL predictions can be trusted 
($\xi, \xi_c>1$ and $M_{\psi,X}<\Lambda$) in both sectors.
Note that apart from these NJL consistency considerations, 
in principle no value of the  ratio $\xi_A/\xi_{A6}$ is theoretically excluded, 
but the case $M_\psi=0$ and $M_X \neq 0$ evidently 
does not describe a composite Higgs model since then the spectrum of resonances does 
not contain a pNGB Higgs doublet.
For $\xi_A=\xi_{A6}$, i.e. $\kappa_A = \kappa_{A6}$, and still for $m_X=0$, the ratio $M_X/M_\psi$ thus 
depends only of the value of $\kappa_B$ and $N$, as given precisely by the relation in 
Eq.~(\ref{MpsiMX}). 
When $\xi_B$ is close to zero, one gets $M_\psi \simeq M_X$, since 
the two gap equations are almost decoupled.
 Next, when $\xi_B$ increases, 
there is a complicated balance between the $N$, $M_\psi$ and $M_X$ dependence 
in Eq.~(\ref{kbkb62}), to determine 
$\kappa_{B6}/\kappa_B$, but the ratio $M_X/M_\psi$ is consistently determined from 
the relatively simple relation in Eq.~(\ref{MpsiMX}). This implies 
$\kappa_{B6}>\kappa_B$ and $M_X$ slightly above $M_\psi$, with a 
$M_X/M_\psi$ ratio that increases rather slowly with $\xi_B$, and 
is also a slowly increasing function of $N$. 
For instance  for $N=2$, $M_X/M_\psi\simeq 1.14-1.21$ for 
$\kappa_B/\kappa_A=0.01-0.5$.

Finally, let us briefly discuss the most general case $m_X \neq 0$.
The above considerations give of course only approximate relations, which however remains relatively good as long as $m_X$ remains moderate, 
$m_X \ll M_X$.
For $m_X\ne 0$ there is no critical coupling $\xi_c$ in the $SU(6)$ sector, as 
the minimal value of $M_X$ is obviously non-zero, being equal to $m_X$. 
A non-zero $m_X$ evidently leads to $M_X> M_\psi$ for equivalent coupling values in the two sectors,
see Fig.~\ref{gap-coloured}.

\begin{figure}[tb]
\includegraphics[scale=0.3]
{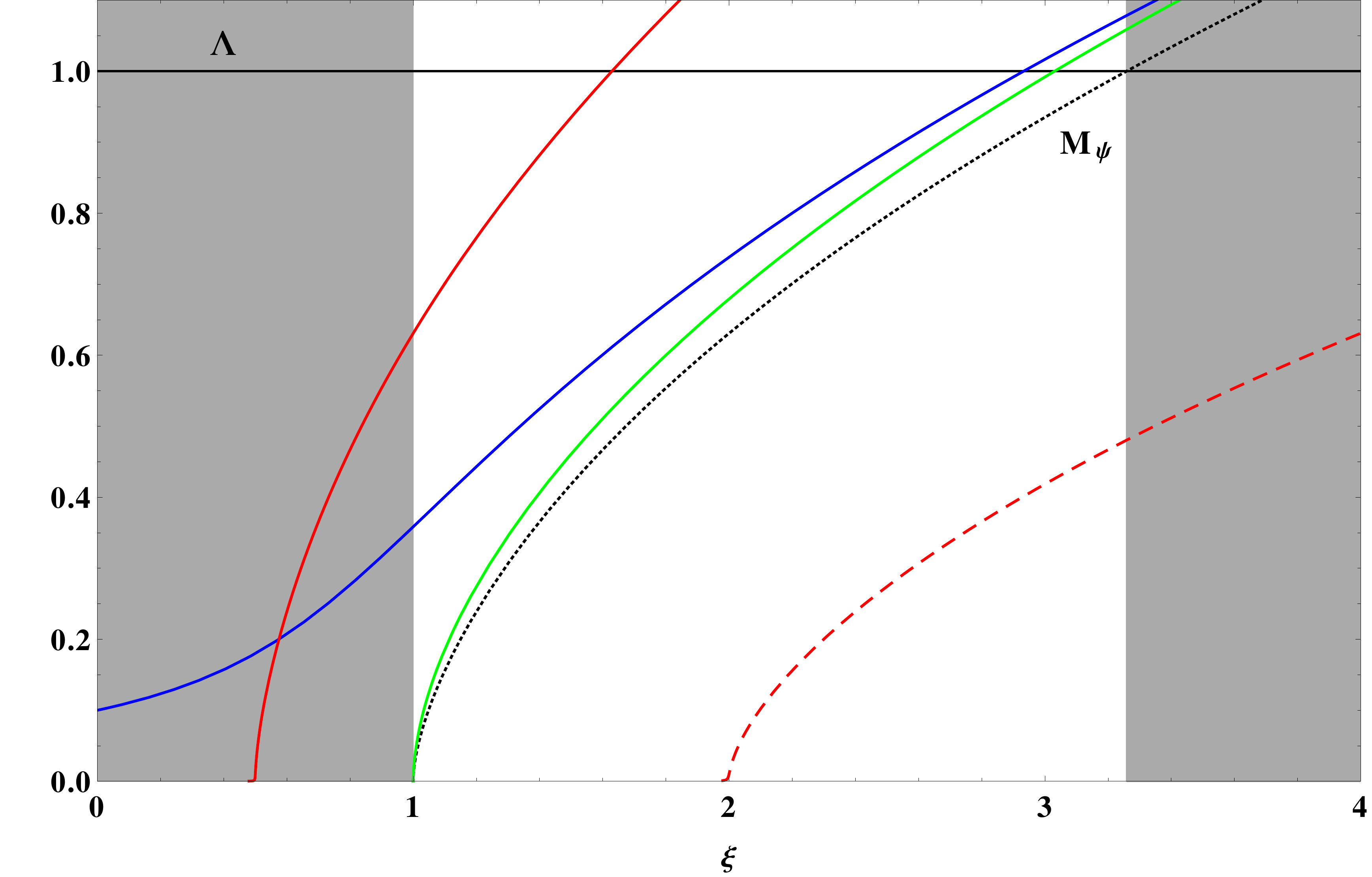}
\caption[long]{Comparison between the mass gap $M_\psi$ of the electroweak sector (black dotted line) and the mass gap $M_X$ of the coloured sector for few representatives cases.
When $\kappa_{A6}=\kappa_A$, $m_X=0$ and $\kappa_B/\kappa_A=0$, the two dynamical masses are equal, $M_\psi=M_X$.
To illustrate the behaviour of $M_X$ with respect to the free parameters of the theory ($\xi$, $\kappa_B/\kappa_A$, $\kappa_{A6}/\kappa_A$, 
$m_X$ and $N$) we illustrate small departures from this particular case.
The solid (dashed) red line corresponds to $\kappa_{A6}=2 (1/2)\kappa_A$ with $\kappa_B/\kappa_A=0$, $m_X=0$ and $N=4$.
In these cases, the critical coupling  of the coloured sector  is respectively smaller or larger 
than the one in the electroweak sector ($\xi=1$).
Next, the solid blue (green) line corresponds to $\kappa_{A6}=\kappa_A$, $N=4$ with $\kappa_B/\kappa_A=0$ ($\kappa_B/\kappa_A=0.1$) and  
$m_X=\Lambda/10$ ($m_X=0$).
In the case where there is an explicit symmetry-breaking mass $m_X$, there is no critical coupling in the 
coloured sector as the lowest value of $M_X$ is simply $m_X$.
Finally note that $M_X$ is almost independent of the number of hypercolour $N$. }
\label{gap-coloured}
\end{figure}

A couple of remarks are in order. In section \ref{mixing-singlets} we will see that the scalar singlet sector 
is consistent only for a very small value of 
$\kappa_B/\kappa_A$, see Eqs.~(\ref{kbcsig0}) and (\ref{kbcrit}).
This is due to the requirement of  vacuum stability, which
is not apparent in the mass-gap equations (\ref{gap-2sectors}). 
For example, this upper bound implies that a value $\xi_B=1/2$, as illustrated
in Fig. \ref{gap2fig}, is actually not possible. 
This in turns sets a lower bound on $\xi_{A6}$, 
in order to stay above the critical value, $\xi_{A6}+\xi_{B6} >1$, and to obtain a non-zero value of $M_X$. 
Let us now comment on the dynamical relation between $\kappa_B$ and 
the original anomalous parameter $A_{\psi X}$, given in Eq.~(\ref{kbkb6}),
and which involves $M_X$ and $N$. In the whole allowed range
$1< \xi < (1-\ln 2)^{-1}\simeq 3.25$, even when $M_X\simeq \Lambda$ for large $\xi$, 
the factor in square brackets in Eq.~(\ref{kbkb6}) is small in $\Lambda^3$ units,
essentially because of the loop-suppression, 
$4 M_X \t A_0(M_X^2) \simeq (4-8)\cdot 10^{-3} \Lambda^3$ 
(with moderate dependence on $\kappa_B/\kappa_A$ and $N$). 
This implies a strong suppression
of the effective coupling $\xi_B$ due to the large power $6(N-1)$ in Eq.~(\ref{kbkb6}), 
even for the minimal value $N=2$. Unfortunately, the original Lagrangian parameter
$A_{\psi X}$ originates from non-perturbative dynamics that is not under control at the present stage, 
so that its size is essentially arbitrary, see also the discussion in subsection \ref{thooft}
after Eq.~(\ref{OX}).
Therefore, we can just remark that, whatever the actual size of $A_{\psi X}$, the corresponding value of $\kappa_B$
is strongly suppressed by the dynamics. This may help to comply
with the upper bound from vacuum stability
on $\kappa_B/\kappa_A$, which behaves as $1/N$ for sufficiently large $N$, 
as we shall discuss in section \ref{mixing-singlets},
because the effective coupling $\kappa_B$ in Eq.~(\ref{kbkb6}) contains a power-$N$ suppression factor.

\subsection{Masses of coloured scalar resonances}
\label{Masses of coloured scalar resonances}

The scalar and pseudoscalar resonances associated to $X$-fermion bilinears transform under the flavour symmetry 
as $21_{SU(6)}=(1+20')_{SO(6)}$.
In analogy with the $\psi$-fermion sector, we can define a matrix $\overline{M}_c$ in flavour space,
\begin{equation}
\overline{M}_c= \frac{1}{2} M_X \Sigma_0^c +\left( \sigma_X +i \eta_X \right) \Sigma_0^c T^0_X 
+ \left(S_c^{\hat{F}}+ i G_c^{\hat{F}} \right) \Sigma_0^c T^{\hat{F}} ~,
\end{equation}
where the components $\sigma_X$ ($\eta_X$) and $S_c^{\hat{F}}$ ($G_c^{\hat{F}}$) are respectively the  
$SO(6)$-singlet and twenty-plet (pseudo)scalars.
The relevant operators for the computation of the spin-zero meson masses are those given in Eq.~(\ref{L-scalar-coloured}), plus
the effective four-fermions operators 
$\psi^4$, $X^4$ and $\psi^2 X^2$, which are induced by the anomalous Lagrangian of Eq.~(\ref{L46}), after spontaneous symmetry breaking,
\begin{eqnarray}
{\cal L}_{\psi X}^{eff} &=& \frac{\kappa_B}{2N} \left[\left(\psi \Sigma_0 T^0_\psi \psi\right) \left(\psi \Sigma_0 T^0_\psi \psi \right)
- \left(\psi\Sigma_0 T^{\hat A} \psi \right)\left(\psi\Sigma_0 T^{\hat A} \psi \right)+ h.c.\right]
\nonumber
\\
& + &
\frac{\kappa_{B6}}{(2N+1)(N-1)} \left[ \left( 6N-7 \right) \left(X \Sigma_{0}^c T^0_X X\right) \left(X \Sigma_{0}^c T^0_X X \right)
- \left(X \Sigma_{0}^c T^{\hat F} X \right)\left(X \Sigma_{0}^c T^{\hat F} X \right)+ h.c.\right]
\nonumber
\\
&+&
\frac{\kappa_{\psi X}}{2N} \left[\left(\psi \Sigma_0 T^0_\psi \psi \right)\: (X \Sigma_{0}^c T^0_X X)+ h.c. \right]~,
\label{effective-Lagrangian-2sectors}
\end{eqnarray}
where $\kappa_B$ and $\kappa_{B6}$, defined in Eq.~(\ref{kbkb6}) and (\ref{kbkb62}) respectively, 
are the same couplings that appear in the gap equations.
Note the factor $(6N-7)$ that multiples $\kappa_{B6}$, because here two $X$-fermion bilinears out of $6(N-1)$ 
are not closed into a loop, which implies a combinatorial factor $6(N-1)[6(N-1)-1]/2$.
The additional coupling $\kappa_{\psi X}$ is defined by
\begin{equation}
\kappa_{\psi X}\equiv \frac{A_{\psi X}}{27^{N-1}} 
\left[\frac{4 N^\psi_f \langle \psi\psi \rangle^2 }{(2N)^2}\right]^{\frac12}
\left[ \frac{4N^X_f \langle XX \rangle^2}{(2N+1)^2(N-1)^2} \right]^{3(N-1)-\frac 12} \frac{2\cdot 6(N-1)}{(2N+1)(N-1)}
=\frac{8\sqrt{6} N}{2N+1} \frac{M_\psi}{M_X} \frac{\t A_0(M_\psi^2)}{\t A_0(M_X^2)} ~\kappa_B~,
\label{kpsiX}
\end{equation}
and it controls the mixing between the $Sp(4)$ and $SO(6)$ (pseudo)scalar singlets 
$\sigma_\psi$ ($\eta_\psi$) and $\sigma_X$ ($\eta_X$), which will be treated separately 
in section \ref{mixing-singlets}.
Note that all three effective couplings vanish if $\langle XX \rangle=0$. When $\langle XX \rangle\ne0$ both $\kappa_{B6}$ and 
$\kappa_{\psi X}$ are fully determined as a function of $M_\psi$, $M_X$ and $\kappa_B$.
From  Eqs.~(\ref{L-scalar-coloured}) and (\ref{effective-Lagrangian-2sectors}) one can derive 
the four-fermion couplings for each physical channel, 
\be
K_{\sigma_X (\eta_X)}= 2 \frac{ [\kappa_{A6} \pm \left(6N-7 \right) \kappa_{B6}] }{(2N+1)(N-1)}~,
\qquad\qquad
K_{S_c (G_c)}= 2 \frac{ [\kappa_{A6} \mp\kappa_{B6}] }{(2N+1)(N-1)}~,
\label{KSG}
\ee
For convenience, all the relevant four-fermion couplings for the $X$-sector 
spin-zero and spin-one mesons  are collected in Table \ref{tab-coloured-functions},
together with the associated one-loop two-point functions.

\begin{table}[b]
\renewcommand{\arraystretch}{2.6}
\begin{center}
\begin{tabular}{|c|c|c|}
\hline
$\phi$   &   $K_\phi$   &   $\t\Pi^X_{\phi}(q^2)$
\\
\hline
\hline
 $G_c^{\hat{F}}$ & $\dfrac{2(\kappa_{A6} + \kappa_{B6})}{(2N+1)(N-1)}$ &
\multirow{2}{*}{$ \t\Pi_{P}^X(q^2) =(2N+1)(N-1) \big[\t A_0( M_X^2) -\frac{q^2}{2} \t B_0(q^2,M_X^2)\big]$}
\\
\cline{1-2}
$\eta_X$  & $\dfrac{2[\kappa_{A6} -(6N-7) \kappa_{B6}]}{(2N+1)(N-1)}$   & 
\\
\cline{1-3}
 $\eta_\psi-\eta_X$     &  $\dfrac{-\kappa_{\psi X}}{(2N)}$             &  \\
\hline
\hline
 $S^{\hat{F}}_c$  & $\dfrac{2(\kappa_{A6} - \kappa_{B6})}{(2N+1)(N-1)}$  &
\multirow{2}{*}{$ \t\Pi_{S}^X(q^2) =(2N+1)(N-1) \big[\t A_0( M_X^2) - \frac{1}{2} (q^2 - 4 M_X^2) \t B_0(q^2,M_X^2)\big]$}
\\
\cline{1-2}
 $\sigma_X$  & $\dfrac{2[\kappa_{A6} +(6N-7) \kappa_{B6}]}{(2N+1)(N-1)}$    & \\
 \cline{1-3}
   $\sigma_\psi-\sigma_X$       &  $\dfrac{\kappa_{\psi X}}{(2N)}$                      &    \\
\hline
\hline
 $V_c^{\mu F}$     & $\dfrac{-2 \kappa_{D6}}{(2N+1)(N-1)}$   &  $\t\Pi_{V}^X(q^2) =\frac{1}{3} (2N+1)(N-1) \big[- 2 M_X^2 \t B_0(0,M_X^2) + (q^2 + 2 M_X^2) \t B_0(q^2,M_X^2)\big]$ \\
\hline
\hline
 $A_c^{\mu {\hat{F}}}$     & $\dfrac{-2 \kappa_{D6}}{(2N+1)(N-1)}$               & $\t\Pi_{A}^X(q^2) = \frac{1}{3} (2N+1)(N-1) \big[- 2 M_X^2 \t B_0(0,M_X^2) + (q^2 - 4 M_X^2) \t B_0(q^2,M_X^2)\big]$ \\
 \cline{1-2}
  $a_X^\mu$       & $\dfrac{-2 \kappa_{C6}}{(2N+1)(N-1)}$                      & $\t\Pi_{A}^{X L}(q^2) = - 2 (2N+1)(N-1) M_X^2 \t B_0(q^2,M_X^2)$   \\
\hline
\hline
 $A_c^{\mu {\hat{F}}}-G_c^{\hat{F}}$     &               & \multirow{2}{*}{$\t\Pi_{AP}^X (q^2) = - (2N+1)(N-1) M_X \t B_0(q^2 , M_X^2)$} \\
 \cline{1-2}
  $a_X^\mu-\eta_X$       &                       &    \\
\hline
\end{tabular}
\end{center}
\caption{The four-fermion couplings $K_\phi$ in the $X$-sector, and the associated one-loop two-point functions $\tilde{\Pi}_\phi^X(q^2)$. 
The latter are related to the two-point functions of the $\psi$-sector as follows:  $\tilde{\Pi}_\phi^\psi(q^2)=\tilde{\Pi}_\phi(q^2, M_\psi^2,2N)$ and $\tilde{\Pi}_\phi^X(q^2)=\tilde{\Pi}_\phi[q^2, M_X^2,(2N+1)(N-1)]$,
 where $\tilde{\Pi}_\phi(q^2, M_\psi^2,2N)$ are defined in Table \ref{tab_phi}.
We also give the expression of the mixed (one-loop) pseudoscalar-longitudinal axial correlator, 
as well as those of the couplings mixing the singlet scalars of the two sectors, 
$\sigma_\psi$ and $\sigma_X$, and 
the singlet pseudoscalars $\eta_\psi$ and $\eta_X$.
The explicit calculation of the correlators $\t \Pi_\phi^X (q^2)$ is detailed in appendix \ref{SDresum}. }
\label{tab-coloured-functions}
\end{table}

%
We now calculate the masses of the scalar and pseudoscalar non-singlet resonances $S_c^{\hat{F}}$ and $G_c^{\hat{F}}$.
As already mentioned above, for the scalar and pseudoscalar singlet $\sigma_X$ and $\eta_X$, 
there is a mixing with the corresponding resonances $\sigma_\psi$ and $\eta_\psi$ of the electroweak sector,  
so that the whole singlet sector will be treated separately in section \ref{mixing-singlets}.

Concerning the  non-singlet pNGB $G_c$, we should also consider more generally 
a non-trivial pseudoscalar-axial vector mixing
for non-vanishing vectorial four-fermion couplings, as we anticipate will be introduced 
below in Section \ref{Masses of coloured vector resonances}, in analogy with
the electroweak sector discussed in Section \ref{Resummed correlators and the Goldstone decay constant}.
With the additional  explicit breaking mass term $m_X$ of Eq.~(\ref{L4F-scal-color}), 
the pseudoscalar axial-vector mixing formalism of Section \ref{Resummed correlators and the Goldstone decay constant} can easily be 
generalised with explicitly $m_X$-dependent resummed matrix correlator  
$\mathbf{\overline{\Pi}}_{G_c}(m_X)$, the analogue of Eqs. (\ref{Pisum22}) and (\ref{Pisum22fin}) for the coloured sector. 
Note that all of the one-loop two-point functions $\t \Pi(q^2,M_X^2)\equiv\t \Pi^X_\phi(q^2)$ of the $SU(6)$ sector 
can be obtained from those in table \ref{tab_phi} with the following replacements: $M_\psi\rightarrow M_X$ 
and $(2N)\rightarrow(2N+1)(N-1)$ (see appendix \ref{SDresum} for details).
Accordingly the pNGB obviously gets a nonzero mass, whose expression is obtained from the zero of the 
determinant, analogous to (\ref{DG}) for the $SU(4)$ sector,  as 
\be
D_{G_c} = \frac{m_X}{M_X} g_{A_c}^{-1} +2(\kappa_{A6}+\kappa_{B6}) \t B_0(p^2,M^2_X)\:p^2 \equiv 
2(\kappa_{A6}+\kappa_{B6}) \t B_0(p^2,M^2_X)(p^2-M^2_{G_c}). 
\label{DGc}
\ee

The calculation of the scalar $S_c^{\hat{F}}$ mass is simpler and follows the same derivation as 
for the scalar mass of the $SU(4)$ sector. Thus we obtain
\begin{equation}
M_{G_c}^2 = -(\frac{m_X}{M_X}) \frac{g_{A_c}^{-1}(M_{G_c}^2)}{2(\kappa_{A6}+\kappa_{B6}) \t B_0 \left( M_{G_c}^2,M_X^2 \right)} ~, 
\qquad\qquad
M_{S_c}^2= 4 M_X^2 -\frac{8 \kappa_{B6} \t A_0(M^2_X) +\frac{m_X}{M_X}  } 
{ 2(\kappa_{A6}-\kappa_{B6})  \t B_0 \left(M_{S_c}^2, M_X^2 \right)}~.
\label{Msignomix}
\end{equation}
where as before the pole masses are defined as $M_{G_c}^2=M_{G_c}^2(p^2=M_{G_c}^2)$.
 Accordingly, similarly to $M_\eta^2$ in Eq.~(\ref{Metabis}), when 
a non-vanishing coloured sector vector coupling $\kappa_{D6}$ is considered (see Section \ref{Masses of coloured vector resonances}), the pseudoscalar Goldstone mass $M_{G_c}^2$ is renormalised by the (inverse) axial form factor 
$g_{A_c}^{-1}(p^2\equiv M^2_{G_c}) \equiv 1 -2 K_{A_c} \t \Pi_A^{L X}(M^2_{G_c})$ where $K_{A_c}$ is defined in Table \ref{tab-coloured-functions}.

Note that there is another source of explicit symmetry breaking which may a priori lead to sizable contributions to the masses.
Indeed, when we switch on the SM gauge interactions, new contributions to the masses of the coloured states arise.
In the following, we will only consider the gauge corrections to the masses of the pNGB states, 
since the latter are the lightest resonances of the coloured sector. Therefore 
those corrections are more relevant than e.g. for the other scalar states.
The gauge contributions to the pNGB masses are given in general terms in section 
\ref{gauging} and in appendix \ref{SU6-generators} for the particular case of the $SU(6)$ sector.
The pNGB $G_c^{\hat{F}}$ decompose as an octet $O_c \sim 8_0$ and two sextet 
$(S_c +\overline{S}_c)\sim (6_{4/3} + \overline{6}_{-4/3})$ under $SU(3)_c \times U(1)_D$
[$U(1)_D$ is the hypercharge component in the $X$ sector, and is also defined in appendix \ref{SU6-generators}].
Consequently, there are two sources of gauge contributions which lead to a mass splitting between the octet 
and sextet components: one from the gauging of QCD and one from the gauging of the hypercharge.
However, from Eq.~(\ref{rad_qcd}) one can see that the QCD corrections are almost the same for the two components as $\Delta M^2_{O_c}/\Delta M^2_{S_c}|_{QCD}=9/10$.
For simplicity we will neglect this small difference.
In addition, the contribution coming from the gauging of $U(1)_Y$ is sub-dominant compared to the one from QCD, and we will safely neglect it.
This is due to the small value of the ratio $g^\prime/g_s$ at the energy scale of a few TeVs we are interested in.
Then the gauge contributions mainly originate from QCD and to evaluate the latter, 
we need to compute the integral in Eq.~(\ref{rad_qcd}) within the NJL framework.
To do that, we simply cut the integral at $Q^2= \Lambda^2$, 
where $\Lambda$ stands for the cutoff of the NJL model, and $F_{G_c}$ is given by the expression
\begin{equation}
F_{G_c}^2= -2 (2N+1)(N-1) M_X^2 \t B_0(M_{G_c}^2,M_X^2) g_{A_c} (M_{G_c}^2)~,
\end{equation}
which can easily be inferred  adapting Eqs.~(\ref{FG4}) and (\ref{FGsum}) to the $SU(6)$ sector.
Note that, for simplicity, the mass $M_{G_c}$ in the right-hand side 
is taken without gauge corrections.
The resulting radiative pNGB masses, obtained from Eq.~(\ref{rad_qcd}), are illustrated 
in the left panel of Fig.~\ref{Msu6gA10gB}, where by definition $M^2_{G_c}=\Delta M^2_{O_c}$, as $m_X=0$.
These numerical results will be discussed in more details in section \ref{spectrum-coloured}.
Let us just mention that this gauge-induced mass  could be 
sufficient by itself  to comply with the lower collider bounds \cite{Cacciapaglia:2015eqa}.

\subsection{Masses of coloured vector resonances}
\label{Masses of coloured vector resonances}

In order to calculate the masses of the vector and axial-vector resonances present in the coloured sector, we start from the following vector-vector four-fermion operators
\begin{equation}
{\cal L}_{vect}^X= \frac{\kappa_{C6}}{(2N+1)(N-1)} 
\left(\overline{X} \, T^0_X \, \overline{\sigma}^\mu X \right)^2
+\frac{\kappa_{D6}}{(2N+1)(N-1)} 
\left[\left(\overline{X} \, T^F \, \overline{\sigma}^\mu X \right)^2
+ 
\left(\overline{X} \, T^{\hat F} \, \overline{\sigma}^\mu X \right)^2 \right],
\label{kcd6}
\end{equation}
where as in the electroweak sector, due to the global $SU(6)$ symmetry, the four-fermions coupling $\kappa_{D6}$ of the vector channel is the same as the axial non-singlet channel.
From the above operators we obtain the vector and axial-vector four-fermions couplings $K_{V_c},K_{A_c}$ and $K_{a_X}$ (see table \ref{tab-coloured-functions})
and we derive the masses of the vector $V_c^F$ and axial $A_c^{\hat{F}},a_X$ resonances
\begin{equation}
M_{V_c}^2=-\frac{3}{4 \kappa_{D6} \t B_0(M_{V_c}^2,M_X^2)}+ 
2 M_X^2 \frac{ \t B_0(0,M_X^2)}{\t B_0(M_{V_c}^2,M_X^2)} -2 M_X^2 ,
\label{vector-colored-mass}
\end{equation}
\be
M_{A_c}^2=-\frac{3}{4 \kappa_{D6} \t B_0(M_{A_c}^2,M_X^2)}
+ 2 M_X^2 \frac{\t B_0(0,M_X^2)}{\t B_0(M_{A_c}^2,M_X^2)} 
+ 4 M_X^2~. 
\label{axial-vector-colored-mass}
\ee
Just like in the electroweak sector, if one neglects the $p^2$ dependence of the $\tilde{B}_0$ function, 
one obtains the usual NJL relation between the axial and vector masses, that is $M_{A_c}^2 \simeq M_{V_c}^2+6 M_X^2$.
The mass of the axial singlet $a_c^\mu$ is obtained by making the replacements 
$A_c^\mu \rightarrow a_X^\mu$ and $\kappa_{D6}\rightarrow \kappa_{C6}$ in Eq.~(\ref{axial-vector-colored-mass}).
Note that we have not considered the following operator
\begin{equation}
{\cal L}_{vect}^{\psi X}= \frac{\kappa_{\psi X}^V}{(2N)} \left(\overline{\psi}\, T^0_\psi \, \overline{\sigma}^\mu \psi \right)
\left(\overline{X} \, T^0_X \, \overline{\sigma}_\mu X \right)~,
\end{equation}
which induces a mixing between the axial singlets of the two sectors,  $a^\mu_\psi$ and $a_X^\mu$.
This mixing term respects all symmetries of the theory and should be present in general.
However, we neglected it as it is not generated by applying a Fierz transformation to 
the $Sp(2N)$ current-current operators in Eq.~(\ref{LUV0}). 

Note also that, in principle, the spin one masses receive SM gauge contributions as $V^\mu_c\sim 15_{SO(6)}=(1+8+3+\overline{3})_{SU(3)_c}$ and $A^\mu_c \sim 20'_{SO(6)} =(8+6+\overline{6})_{SU(3)_c}$.
However, following the discussion of section \ref{Masses of coloured scalar resonances} for the scalar masses, 
we will not consider such contributions here.

\subsection{The mass spectrum of the coloured resonances}
\label{spectrum-coloured}

In general the couplings of the four-fermion operators are free parameters.
However $\kappa_{A6}$ and $\kappa_{C6,D6}$ may be related 
if we assume that the dynamics is induced by $Sp(2N)$ current-current operators.
In this case, as in the $\psi$-sector, we find that the scalar and vector four-fermion couplings are equal, see 
appendix \ref{Fierz-transfoSP2N}. However, we also find that the size of these couplings relatively to the ones in the electroweak sector 
is not fixed by the current-current approximation. The reason is that, contrary to the  case of the $\psi$-sector, 
the $X$-sector current-current operator cannot be recast in terms of $Sp(2N)$ singlet-singlet operators only, see 
appendix \ref{Fierz-transfoSP2N}. 
Nonetheless in this section, for the sake of illustration, we will take equal couplings in the two sectors 
\begin{equation}
\kappa_{A6} =\kappa_{C6}=\kappa_{D6}= \kappa_A 
~.
\label{Fierz-SU(6)-sector}
\end{equation}
With this choice, as shown in Fig.~\ref{gap-coloured}, the range of validity of the NJL approximation 
is approximatively the same in the two sectors.

The resonance masses of the coloured sector are illustrated in Fig. \ref{Msu6gA10gB}.
To ease the comparison with the electroweak sector, the masses are in units of $f=\sqrt{2} F_G \gtrsim 1$ TeV, 
and are plotted as functions of the coupling  $\xi$ defined by Eq.~(\ref{gap2}).
 Note that in section \ref{The mass spectrum of the resonances}, for the $SU(4)$ sector in isolation,
the only constraint from vacuum stability was $\kappa_B/\kappa_A <1$: here we anticipate a similar but
stronger bound, see Eqs.~(\ref{kbcsig0}) and (\ref{kbcrit}) below. 
Consequently  
the value of $\kappa_B/\kappa_A$ is fixed to $0.01$ 
for illustration, which is safely below this upper bound in the case $N=4$.
Then, if one assumes that Eq.~(\ref{Fierz-SU(6)-sector}) holds, there is just one additional free parameter compared 
to the $SU(4)$ sector in isolation, namely the explicit symmetry-breaking mass term $m_X$. 
We illustrate two representative cases: 
one with no explicit breaking, $m_X=0$, and another one with explicit symmetry breaking, for which 
we take as a representative value $m_X=0.1 f$.

In the case with no explicit breaking (left panel of Fig.~\ref{Msu6gA10gB}), 
the behaviour of the masses is qualitatively similar to the $SU(4)$ sector, except for the pNGBs ${G_c}$.
This is due to the relations between the couplings of the four-fermion interactions: $\kappa_A=\kappa_{A6}$ and 
$\kappa_{B}\sim \kappa_{B6}\ll \kappa_{A}$.
The pNGB of the coloured sector receive 
a significant contribution to their masses from the gauging of the colour group, 
as discussed in section \ref{Masses of coloured scalar resonances}.
 As it can be seen from Fig.~\ref{Msu6gA10gB}, this contribution satisfies $\Delta M_{G_c}\gtrsim 1.3 f$, 
which is  enough to comply with the present collider bounds, 
as long as $f\gtrsim 1$ TeV.
Thus, we conclude that it is actually possible to introduce top quark partners  
without the need of an explicit mass term $m_X$ for the coloured fermions.
On the other hand, if we want to raise the mass of coloured pNGBs, 
while keeping a low mass scale of the theory, $f=1$ TeV, one needs to introduce a non-zero $m_X$, 
as illustrated in the right panel of Fig.~\ref{Msu6gA10gB} for $m_X=0.1 f$.
As all the coloured masses receive a contribution from $m_X$, 
for sufficiently large values of $m_X$ one could even decouple the coloured sector from the electroweak sector.
%

\begin{figure}[tb]
\includegraphics[scale=0.25]{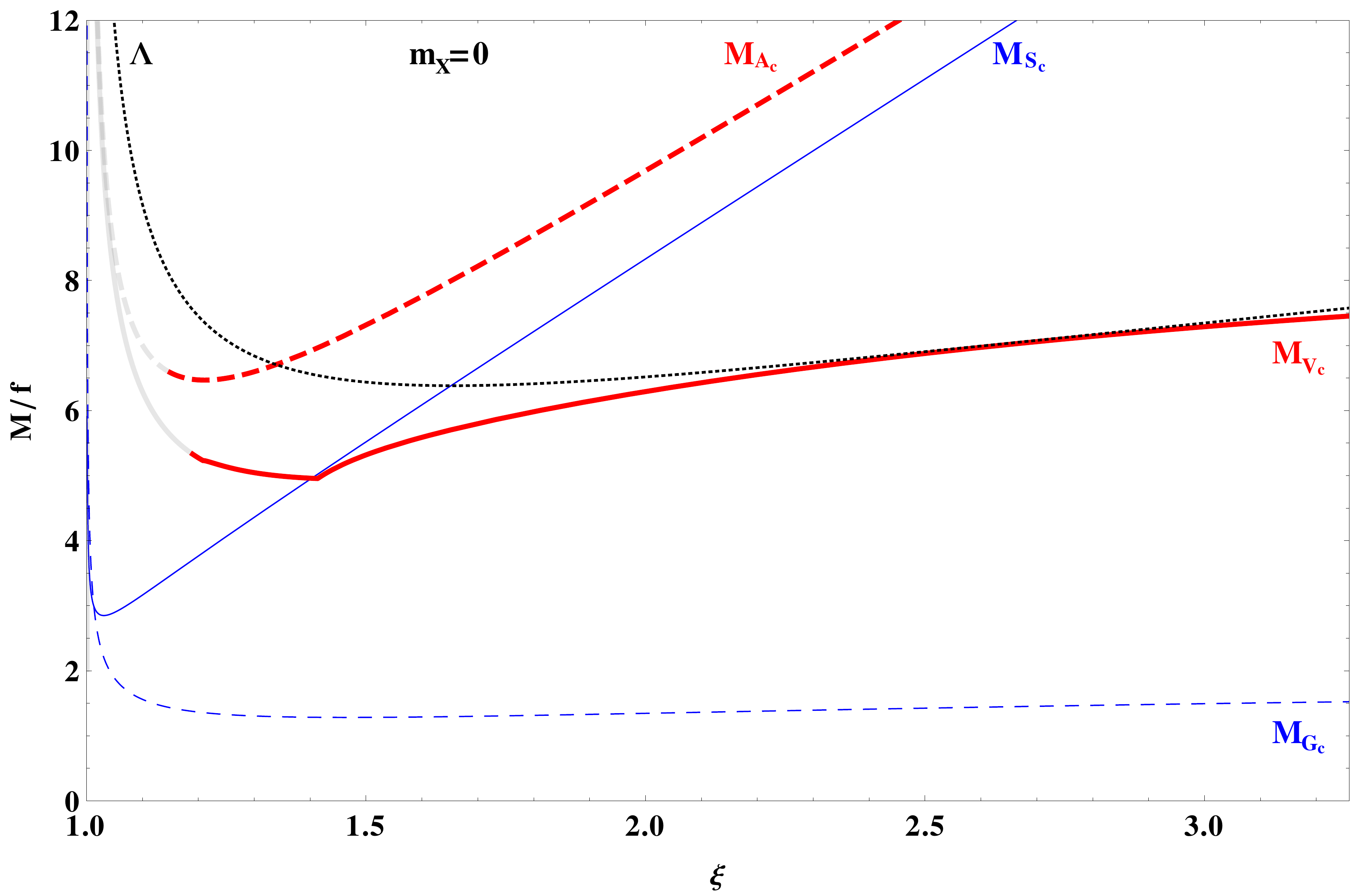}
\includegraphics[scale=0.25]{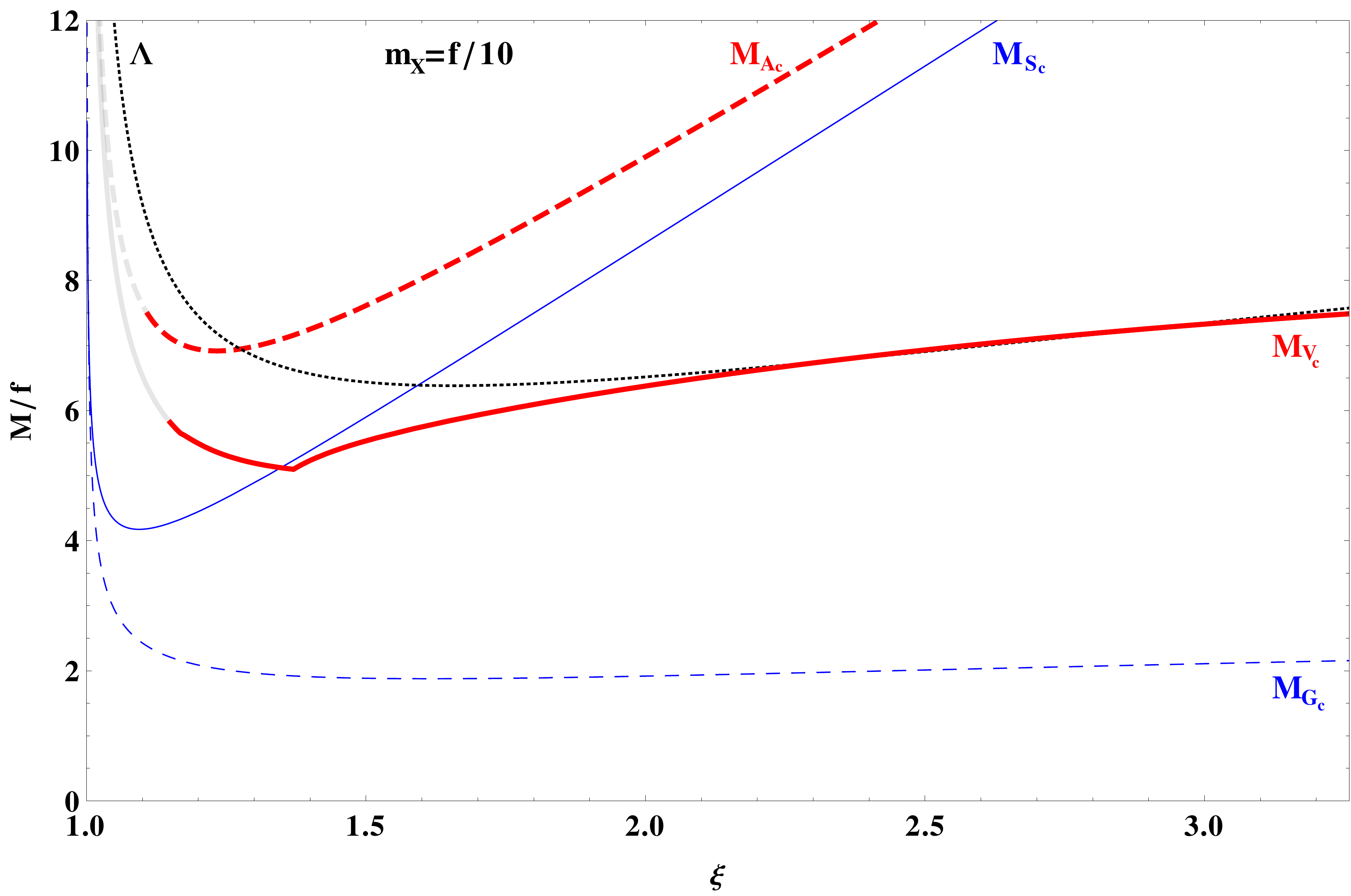}
\caption{The masses of the coloured resonances in units of the Goldstone decay constant 
$f\equiv \sqrt{2} f_G$, for $N=4$ (the masses scale as $1/\sqrt{N}$), 
as a function of the coupling $\xi$, for $\kappa_B/ \kappa_A=0.01$, $\kappa_{A6}=\kappa_A$, $m_X=0$ 
(left-hand panel) and $m_X=f/10$
(right-hand panel).
We displayed the full physical range for $\xi$, according to Fig.~\ref{FGLam}.
Each curve is shaded when the corresponding pole mass develops  a large, unphysical imaginary part, 
$|{\rm Im} g_\phi(M_\phi^2)/{\rm Re} g_\phi(M_\phi^2)|>1$, as defined from Eq.~(\ref{res_width}). 
The dotted line is the cutoff of the constituent fermion loops.
The Goldstone mass $M_{G_c}$ include the radiative corrections as discussed in section \ref{Masses of coloured scalar resonances}.}
\label{Msu6gA10gB}
\end{figure}

Finally, we display here the masses of the colour resonances for the same parameters as in Eq.~(\ref{mass-spetrum-EW-benchmark}),
$N=4$, $\xi=1.3$ and $\xi=2$, fixing $\kappa_B/\kappa_A=0.01$ and for the two representative values of $m_X$:
\bea
\xi=1.3,~m_X=0 \qquad~~~~ &:~& \qquad M_{A_c} \simeq 6.6 ~{\rm TeV},~~  M_{V_c}\simeq 5.1~{\rm TeV},~~  M_{S_c}\simeq 4.3~{\rm TeV},~~ M_{G_c} \simeq ~1.3{\rm TeV}~,
\nonumber
\\
\xi=1.3,~m_X =0.1 ~{\rm TeV}~ &:~& \qquad  M_{A_c} \simeq 7.0~{\rm TeV},~~  M_{V_c}\simeq 5.2~{\rm TeV},~~  M_{S_c}\simeq 4.9~{\rm TeV},~~ M_{G_c} \simeq 2.0 ~{\rm TeV}~.
\eea
\bea
\xi=2.0,~m_X=0 \qquad~~~~ &:~& \qquad  M_{A_c} \simeq 9.7 ~{\rm TeV},~~  M_{V_c}\simeq 6.3~{\rm TeV},~~  M_{S_c}\simeq 8.4~{\rm TeV},~~ M_{G_c} \simeq ~1.4{\rm TeV}~,
\nonumber
\\
\xi=2.0,~m_X =0.1 ~{\rm TeV}~ &:~&  \qquad  M_{A_c} \simeq 9.9~{\rm TeV},~~  M_{V_c}\simeq 6.4~{\rm TeV},~~  M_{S_c}\simeq 8.5~{\rm TeV},~~ M_{G_c} \simeq 1.8 ~{\rm TeV}~.
\eea

\subsection{Flavour-singlet sector}
\label{mixing-singlets}

The $\psi-X$ mixing in the (scalar and pseudoscalar) 
singlet sector, induced by the Lagrangian (\ref{L46}), 
is most conveniently treated in matrix formalism. Furthermore, since our model includes 
non-vanishing singlet axial-vector couplings both in the $SU(4)$  and $SU(6)$ sectors,  
we should take into account the additional pseudoscalar-axial mixing, similarly to the
case of the $SU(4)$ sector in isolation treated in section \ref{Resummed correlators and the Goldstone decay constant}. 
Accordingly, we shall consider $2 \times 2$ and $4 \times 4$ matrix equations for the correlators in the scalar
and pseudoscalar sectors, respectively. 

\subsubsection{ Scalar-singlet mixing}
\label{Scalar mixing and eigenstates}

Let us start with the scalar sector and consider the diagonal one-loop scalar-correlator matrix $\mathbf{\Pi}_{\sigma_\psi \sigma_X}$ 
and the matrix of scalar couplings $\mathbf{K}_{\sigma_\psi \sigma_X}$,
\be
\mathbf{\Pi}_{\sigma_\psi \sigma_X} =\begin{pmatrix}
\t \Pi_{S}^\psi & 0 \\ 0 & \t \Pi_{S}^X
\end{pmatrix}~,
\qquad\qquad
\mathbf{K}_{\sigma_\psi \sigma_X} = 
\begin{pmatrix}
K_{\sigma_\psi} &  K_{\psi X} 
\\
  K_{\psi X} & K_{\sigma_X}
\end{pmatrix}~,
\ee
where $K_{\sigma_\psi}$, $K_{\sigma_X }$ and $K_{\psi X}\equiv\kappa_{\psi X}/(2N)$ are collected in 
Tables \ref{tab_phi} and \ref{tab-coloured-functions}.
Note that when $K_{\psi X}=0$ (equivalently $A_{\psi X}=0$) there is no mixing between 
the singlets $\sigma_\psi$ and $\sigma_X$.
For simplicity, we have introduced the shorthand notations 
$\t \Pi^\psi_i \equiv \t \Pi_i(p^2,M_\psi^2)$ and $\t \Pi^X_i\equiv \t \Pi_i(p^2,M_X^2)$ 
for the one-loop correlators.
From the above matrices, one can now define the resummed matrix correlator $\mathbf{\overline{\Pi}}_{\sigma_\psi \sigma_X}$
\be
\mathbf{\overline{\Pi}}_{\sigma_\psi \sigma_X}= \mathbf{\Pi}_{\sigma_\psi \sigma_X} +  \mathbf{\Pi}_{\sigma_\psi \sigma_X}\, (2 \mathbf{K}_{\sigma_\psi \sigma_X}) 
\, \mathbf{\Pi}_{\sigma_\psi \sigma_X} +\cdots = 
(1\!\!1 - 2\mathbf{\Pi}_{\sigma_\psi \sigma_X} \,\mathbf{K}_{\sigma_\psi \sigma_X})^{-1}\: \mathbf{\Pi}_{\sigma_\psi \sigma_X},
\label{Pisigc}
\ee
and the resonance mass eigenvalues are obtained as the roots of the equation
$
\det (1\!\!1 - 2 \mathbf{\Pi}_{\sigma_\psi \sigma_X} \,\mathbf{K}_{\sigma_\psi \sigma_X}) = 0$, where
\begin{eqnarray}
\det  (1\!\!1 - 2\mathbf{\Pi}_{\sigma_\psi \sigma_X} \mathbf{K}_{\sigma_\psi \sigma_X}) 
&=& 1 - 2K_{\sigma_\psi} \t \Pi_{S}^\psi -2 K_{\sigma_X} \t \Pi_{S}^X +
4 \left(K_{\sigma_\psi} K_{\sigma_X}  -K_{\psi X}^2\right) \t \Pi_{S}^\psi \t\Pi_{S}^X 
\nonumber
\\
&=& c^S_0(p^2) + c^S_1(p^2) p^2 +c^S_2(p^2) 
(p^2)^2 ~.
\label{detsig}
\end{eqnarray}
The coefficients $c^S_i(p^2)$ are functions of the couplings $K_i$, 
and of the loop functions $\t A_0(M^2_\psi)$,   $\t A_0(M^2_X)$,
$\t B_0(p^2, M^2_\psi)$, and $\t B_0(p^2, M^2_X)$.
It is convenient to write the determinant as if it were a quadratic form in $p^2$, because the $p^2$-dependence of the coefficients 
$c^S_i(p^2)$, through the loop functions $\t B_0(p^2,M^2_{\psi,X})$, does not induce additional pole structure. 
Then, the scalar-singlet pole masses are obtained as the roots of this quadratic equation, 
evaluated at a self-consistent value of $p^2$,
\begin{equation}
M_{\sigma_0,\sigma'}^2 = {\rm Re} [g_{\sigma_0,\sigma'}(M_{\sigma_0,\sigma'}^2)]
~,\qquad\quad
g_{\sigma_0,\sigma^\prime}(p^2) \equiv 
\dfrac{-c_1^S(p^2) \pm \sqrt{[c_1^S(p^2)]^2- 4 c_2^S(p^2) c_0^S(p^2)}}{2 c_2^S(p^2)}~.
\label{sig0sigp}
\end{equation}  
The explicit expressions of the two scalar singlet masses $M^2_{\sigma_0}, M^2_{\sigma^\prime}$ 
are straightforwardly derived from the above equations, but are not very simple or telling, 
even in the chiral limit $m_X=0$, so that we refrain from giving them here. 
In the numerical illustrations below we use these exact expressions.

As we will examine quantitatively below, 
the lightest scalar mass $M_{\sigma_0}$ is  a {\em decreasing} function of $r\equiv \kappa_B/\kappa_A$, at least
as long as $M_{\psi,X} \ll \Lambda$, 
and it can even vanish at a critical value $r_c$, becoming formally tachyonic beyond. 
This critical value should therefore be considered as an intrinsic upper bound, since for $r\ge r_c$ 
the minimum of the effective scalar potential is destabilised, that is, the solution of the NJL mass-gap equations becomes unreliable.  
It is clear that $M_{\sigma_0}$ can only vanish if $c^S_0(0)=0$ 
in Eq.~(\ref{detsig}) (irrespectively of the additional  
$p^2$-dependence from the $\t B_0$ functions).
The latter condition determines $r_c$ as a function of the parameters $N$, $M_X$ and $M_\psi$, 
once one eliminates the coupling $\kappa_{A6}$ using Eq.~(\ref{gap-2sectors}), as well as $\kappa_{B6}$ and $\kappa_{\psi X}$ using Eqs.~(\ref{kbkb62}) and (\ref{kpsiX}). 
Then, in the chiral limit $m_X=0$, the condition $c^S_0(0)=0$ takes the form 
\be
1+ 2\left[1+\frac{f_6}{B_6(0)}\frac{A_4}{M^2_X} \frac{M^2_\psi}{M^2_X} \frac{2N(3N-4)}{2N+1}\right] r +
\left[1 -\frac{2f_6}{B_6(0)} \frac{A_4}{M^2_X} \frac{M^2_\psi}{M_X^2} \frac{2N(3N-2)}{2N+1} -\frac{6f_6}{B_6(0)B_4(0)} \frac{A_4^2}{M^4_X} \frac{2N(N-1)}{2N+1} \right] r^2 =0~,
\label{kbcsig0}
\ee
where $f_6\equiv 1+2B_6(0)M^2_X/A_6$}, and we are using the shorthand notations 
$A_4\equiv \t A_0(M^2_\psi)$, $A_6\equiv \t A_0(M^2_X)$, and similarly
for the functions $B_{4,6}(p^2)$. 
The mass of $\sigma_0$ vanishes as long as Eq.~(\ref{kbcsig0}), that is quadratic in $r$, has a real and positive root $r_c$, 
whose value depends on the dynamical masses $M_{\psi,X}$ and on $N$. 
For example, if one fixes $\kappa_{A6}=\kappa_A$, one finds that $\xi\lsim 1.4-1.5$ implies $\kappa_B/\kappa_A \le r_c \ll 1$
already for $N=2$, and the upper bound becomes more stringent proportionally to $\sim 1/N$.
For $m_X=0$ and $\xi=1.3$, one finds 
$r_c \simeq 0.103 $ for $N=2$, and
$r_c \simeq 0.024 $ for $N=4$.
However, for larger values of $\xi\gsim 1.7-1.8$, Eq.~(\ref{kbcsig0}) has no longer
a real positive root, instead $M_{\sigma_0}(\xi,r)$ has a positive minimum, at increasingly large values 
of $r$ as $\xi$ increases.
As we will see in the next subsection, there is another upper bound on $\kappa_B/\kappa_A$, 
Eq.~(\ref{kbcrit}), originating from the pseudoscalar-singlet mixing, also related to vacuum stability.
Assuming again $\kappa_{A6}=\kappa_A$, one finds that for $\xi\lsim 1.4$ the bound from  Eq.~(\ref{kbcrit}) has a numerical value very close to 
the solution $r_c$ of Eq.~(\ref{kbcsig0}), although its analytic form is different. 
For larger values of $\xi$, the bound from Eq.~(\ref{kbcrit}) is much more stringent and therefore
supersedes the condition $r<r_c$.
As we will examine in concrete illustrations below, these bounds put stringent restrictions  
on the singlet mass spectrum.
As further explained below for the pseudoscalar case, these constraints should be viewed
as an appropriate generalisation of the constraint $\kappa_B/\kappa_A < 1$, that applies to the
$SU(4)$ sector in isolation.

Concerning the scalar decay constants, defined as in Eq. (\ref{defGS}) with the obvious
replacement $S\to S_0^\psi, S_0^X$, they can be derived 
by generalising the  procedure  explained in section \ref{secWSR}.
They are defined by the residues of the diagonal elements of $\mathbf{\overline{\Pi}}_{\sigma_\psi \sigma_X}$
at the respective pole masses,
\be
(G^\psi_{\sigma_0})^2 \equiv -\lim_{p^2\to M^2_{\sigma_0}} 
 (p^2-M^2_{\sigma_0}) \mathbf{\overline{\Pi}}_{\sigma_\psi \sigma_X}^{11}(p^2)
~,\qquad\qquad
(G^X_{\sigma_0})^2 \equiv -\lim_{p^2\to M^2_{\sigma_0}} 
 (p^2-M^2_{\sigma_0}) \mathbf{\overline{\Pi}}_{\sigma_\psi \sigma_X}^{22}(p^2) ~,
\label{GSpsiX}
\ee 
and analogously for $\sigma_0\rightarrow \sigma^\prime$.
These decay constants enter in the scalar sum rules in combination with the other (pseudo)scalar decay constants.
We refrain here to give their explicit expressions, which are not simple.
The results obtained from Eq.~(\ref{GSpsiX}) can be crosschecked with the off-diagonal elements of $\mathbf{\overline{\Pi}}_{\sigma_\psi \sigma_X}$, as $G_{\sigma_0}^\psi G_{\sigma_0}^X= 
- \lim_{p^2\to M^2_{\sigma_0}}  (p^2-M^2_{\sigma_0}) \mathbf{\overline{\Pi}}_{\sigma_\psi \sigma_X}^{12}(p^2) $, and similarly for $\sigma'$.

\subsubsection{Pseudoscalar singlet mixing}
\label{Pseudoscalar singlet mixing and properties}

Considering now the more involved pseudoscalar sector, we start from the complete $4 \times 4$ matrix coupling
and correlator to account both for singlet mixing and pseudoscalar-axial singlet vectors $a^\mu_\psi, a^\mu_X$ mixing.
The latter mixing is treated similarly to the pseudoscalar axial-vector mixing for the Goldstone boson sector 
as considered in section \ref{Resummed correlators and the Goldstone decay constant}.
Accordingly we have
\be
\mathbf{K}_{\eta_\psi \eta_X}=
\begin{pmatrix}
K_{\eta_\psi}& -K_{\psi X} & 0 & 0
\\
-K_{\psi X} & K_{\eta_X} & 0 & 0
\\
0 & 0 & K_a & 0
\\
0 & 0 & 0 & K_{a_c}
\end{pmatrix}~,
\qquad
\mathbf{\Pi}_{\eta_\psi \eta_X}
=\begin{pmatrix}
\tilde{\Pi}_P^\psi & 0 & \sqrt{p^2} \tilde{\Pi}_{AP}^\psi & 0
\\
0 & \tilde{\Pi}_P^X & 0 & \sqrt{p^2}\tilde{\Pi}_{AP}^X
\\
\sqrt{p^2}\tilde{\Pi}_{AP}^\psi & 0 & \tilde{\Pi}_A^{L \psi} & 0
\\
0 & \sqrt{p^2}\tilde{\Pi}_{AP}^X & 0 & \tilde{\Pi}_A^{L X}
\end{pmatrix},
\label{KPieta}
\ee
where all the relevant pseudoscalar and axial-vector 
correlators and couplings for the $SU(4)$ and $SU(6)$ sectors are given respectively 
in Tables \ref{tab_phi} and \ref{tab-coloured-functions} (and we have used in Eq.~(\ref{KPieta}) 
the same short-hand notation as in section \ref{Scalar mixing and eigenstates}).
From the above matrices, we obtain the resummed two-point correlator defined as
\begin{equation}
\bold{\overline{\Pi}}_{\eta_\psi \eta_X}=  (\bold{1\!\!1}-  2\bold{\Pi}_{\eta_\psi \eta_X}\:\bold{K}_{\eta_\psi \eta_X})^{-1}\;
\bold{\Pi}_{\eta_\psi \eta_X}~.
\label{Pieta}
\end{equation}
According to the previous equation, the pseudoscalar mass eigenvalues are given 
by the zeros of the determinant of $\bold{1\!\!1} -2 \bold{K_{\eta_\psi \eta_X} \Pi_{\eta_\psi \eta_X}}$, which 
we give explicitly only in the chiral limit $m_X=0$ for simplicity. Note that the latter determinant 
keeps the form of a quadratic equation, apart from further $p^2$-dependence from the $\t B_0$ function
appearing in the coefficients. 
After using the relevant relations, Eqs.~(\ref{kbkb6}), (\ref{kbkb62}) and (\ref{kpsiX}),
and the mass gap equations (\ref{gap-2sectors}) in order to express all the effective four-fermion couplings $\kappa_i$
in terms of $\kappa_B$ alone, we obtain
\be
\det[ \bold{1\!\!1} -2\bold{K_{\eta_\psi \eta_X} \Pi_{\eta_\psi \eta_X}} (p^2)]
= p^2
\left[c^P_1(p^2)+ p^2 c^P_2(p^2) \right]
~,
\label{deteta}
\ee
where in notations similar to the scalar case, we define the relevant coefficients of the quadratic
equation as 
\begin{equation}
c^P_1(p^2)= 4\frac{\kappa_B A_4}{(2N+1)A_6 M_X^2}\,
\left[12 N(N-1) B_4(p^2) M_\psi^2 g_{a_c}^{-1}(p^2) +(2N+1) B_6(p^2) M_X^2 g_a^{-1}(p^2) \right]~,
\end{equation}
\begin{equation}
c^P_2(p^2) = -\frac{B_4(p^2) B_6(p^2)}{(2N+1)A_6^2 M_X^2}\,
\left[ 24 N(N-1) \kappa_B A_4 M_\psi^2 -(2N+1)(\kappa_A-\kappa_B) A_6 M_X^2 \right] .
\label{D2eq}
\end{equation}
The appearance of the axial singlet form factors $g_a$, $g_{a_c}$ is a result 
of the mixing between the singlet pseudoscalar axial-vector
\be
g_a^{-1}(p^2) = 1+ \frac{4\kappa_C}{2N} \t \Pi_A^{L \psi}(p^2)~,
\qquad\qquad
g_{a_c}^{-1}(p^2) = 1+ \frac{4\kappa_{C6}}{(2N+1)(N-1)} \t \Pi_A^{L X}(p^2)~.
\ee
The pseudoscalar analogue of the term $c_0^S(p^2)$ in the determinant 
of $\bold{1\!\!1} -2\bold{K_{\eta_\psi \eta_X} \Pi_{\eta_\psi \eta_X}}$
vanishes in the chiral limit $m_X=0$, as is explicit from Eq.~(\ref{deteta}),  
after non-trivial cancellations using the gap equations (\ref{gap-2sectors}), and 
Eqs.(\ref{kbkb6}) and (\ref{kbkb62}), thereby exhibiting the remaining singlet Goldstone boson associated with 
the non-anomalous combination of $U(1)_\psi$ and $U(1)_X$ transformations. 
Obviously, the other pseudoscalar singlet has a non-vanishing mass even for $m_X=0$, with
a relatively compact expression,
\begin{equation}
M_{\eta^\prime}^2= {\rm Re}[g_{\eta'}(M_{\eta'}^2)] +{\cal O}(m_X)~,
\qquad\qquad
g_{\eta'}(p^2) \equiv -\frac{c^P_1(p^2)}{c^P_2(p^2)}
~.
\label{Metap}
\end{equation}
Note that for sufficiently large $N$ (but keeping in mind $N \le 18$), 
$M^2_{\eta^\prime}$ is of order ${\cal O}(N^0)$, using that $\kappa_B\simeq 1/N$,
while the not-shown ${\cal O}(m_X)$  term is of order $1/N$.
This is naively compatible with the behaviour of the anomaly, 
which also goes like a constant for sufficiently
large values of $N$, see Eq.~(\ref{U1_X_div}) (considering that $g_{HC}^2\simeq 1/N$).

An important, interesting feature of the whole model emerges from the
examination of Eq.~(\ref{Metap}): for any $p^2$, the function $g_{\eta'}(p^2)$
has a {\em pole} at a particular value of $\kappa_B/\kappa_A$, as follows from Eq.~(\ref{D2eq}),
\be
\frac{\kappa_B/\kappa_A}{1-\kappa_B/\kappa_A} = 
\frac{1}{24} \frac{2N+1}{N(N-1)} \frac{A_6 M_X^2}{A_4 M_\psi^2} ~.
\label{kbcrit}
\ee
In other words, the $\eta^\prime$ mass grows rapidly and decouples 
when approaching from below the critical value of  $\kappa_B/\kappa_A$ defined by Eq.~(\ref{kbcrit}). 
This is not unexpected, as it is simply a generalisation of 
a property already observed in the $SU(4)$ sector in isolation. 
In the latter case, recall that the mass-gap equation (\ref{gap}) 
has solutions only for $\kappa^2_B <\kappa^2_A$, as discussed after Eq.~(\ref{gap2}):
as also explained in Ref.~\cite{Barnard:2013zea}, and apparent in Eqs.~(\ref{Lscal}) and (\ref{Sprop}),
for $\kappa_B > \kappa_A$ the effective potential is destabilised around the origin, already at tree level and, 
although one could expect a spontaneous symmetry breaking of some of the symmetries, one cannot 
perform a proper minimisation to determine the vacuum, within the NJL framework. 
This feature
is reflected also directly in the resonance mass spectrum, where the $\eta^\prime$ mass (for the $SU(4)$ sector in  isolation)
of Eq.~(\ref{Meta}) clearly has a pole for $\kappa_B=\kappa_A$ and becomes tachyonic for large $\kappa_B$.
Now the critical value in the full model, determined by Eq.~(\ref{kbcrit}), should be considered
accordingly as an absolute upper bound on $\kappa_B/\kappa_A$. It
takes a more involved dynamical form (depending also on the values of the mass gaps $M_\psi$ and $M_X$) precisely
because the mixing, as induced by the effective operators in 
Eq.~(\ref{L46}), couples the two sectors, mass gaps and couplings, in a non-trivial way
and involves $N$-dependent combinatorial factors.
Note that, upon using the relation (\ref{kbkb62}), the critical coupling in (\ref{kbcrit}) translates
into a simpler upper limit on $\kappa_{B6}$, approximately:
\be
\dfrac{\kappa_{B6}}{\kappa_{A}} < \frac{1}{6(N-1)} \dfrac{A_4}{A_6}~,
\label{kb6crit}
\ee
(upon neglecting higher order terms in $\kappa^2_{B6}$), in which the combinatoric
factor $6(N-1)$ can be understood upon
comparing with Eq. (\ref{effective-Lagrangian-2sectors}),
so that Eq. (\ref{kb6crit}) is a more transparent analogue of the limit $\kappa_B < \kappa_A$ in the 
$SU(4)$ sector in isolation (let aside  
the presence of the loop functions $A_4/A_6$, that reflects
the non-trivial dynamical connection between the two sectors).
The bottom line is that Eq.~(\ref{kbcrit}) gives a tight upper bound on $\kappa_B/\kappa_A$,  
due in particular to
the small coefficient $1/24$. To get an idea, consider the chiral limit $m_X=0$  and fix $\kappa_{A6}=\kappa_A$: as discussed
in section \ref{Mass gap equations and effective four-fermions couplings}, 
then $M_X$ lies slightly above $M_\psi$, with e.g. $M_X/M_\psi\simeq 1.15$ for $N=2$
and small $\kappa_B/\kappa_A$. Thus, neglecting for simplicity the relatively small
differences in the $\t A_0$ loop functions, Eq.~(\ref{kbcrit}) gives typically 
$\kappa_B/\kappa_A < 5/48 (M^2_X/M^2_\psi) \simeq 0.12$ for $N=2$, and the latter ratio 
decreases quite rapidly for larger $N$ due to the $\sim 1/N$ behaviour of  Eq.~(\ref{kbcrit}), for instance 
$\kappa_B/\kappa_A < 1/32 (M^2_X/M^2_\psi) \simeq 0.04$ for $N=4$.

More precisely, the physical upper bound on $\kappa_B/\kappa_A$ is even more stringent.
As the ``running'' mass $g_{\eta'}(p^2)$ grows rapidly 
when approaching from below the limiting value of $\kappa_B/\kappa_A$ defined by Eq.~(\ref{kbcrit}), 
the corresponding pole-mass self-consistent equation for $M^2_{\eta'}$, given in Eq.~(\ref{Metap}),
ceases to have a solution for a slightly smaller value of $\kappa_B/\kappa_A$. 
Moreover  
a large width develops much below this bound, which turns out to rapidly
exceed the pole mass. Accordingly, the NJL description of the $\eta'$ mass looses its validity
for even smaller values of $\kappa_B/\kappa_A$.
For a not too small $m_X \ne 0$, as discussed above,
$M_X$ can be substantially larger than $M_\psi$, therefore the bound in Eq.~(\ref{kbcrit}) is delayed
to larger $\kappa_B/\kappa_A$. Still, it remains quite constraining as long as $m_X$ remains moderate 
with respect to $\Lambda$. 
 A hierarchy among the mass gaps, $M_X\gg M_\psi$, can be also realised by taking  $\kappa_{A6} > \kappa_A$,
again relaxing the upper bound on $\kappa_B/\kappa_A$.
In summary, the detailed structure
of the mixing sets the maximal allowed value of $\kappa_B/\kappa_A$,  with important
consequences for the resonance mass spectrum, as we will illustrate below.

For $m_X \ne 0$, the exact expressions of the two pseudoscalar singlet masses 
$M_{\eta_0}, M_{\eta^\prime}$ (used in our numerical analysis) become rather involved: 
 Eq.~(\ref{deteta}) is modified to  
a ``quadratic'' polynomial equation in $p^2$ (i.e. upon formally neglecting
the additional $p^2$-dependence coming from the loop functions, entering the polynomial  coefficients). 
This is then more similar to the eigenvalue equation of the
scalar case above, see Eqs.~(\ref{detsig}) and (\ref{sig0sigp}), now with coefficients $c^{P}_i(p^2)$
which depends on $m_X$, 
where the coefficient of $(p^2)^0$ takes the form
\be
c^P_0 = 8 A_4 \kappa_B \frac{m_X}{M_X}\,g_a^{-1}\,g_{a_c}^{-1}\,.
\ee
Indeed, the pNGB $\eta_0$ mass is given to a very good approximation by the first
order expansion in $c^P_0$, namely
\be
M^2_{\eta_0} = -\frac{c^{P}_0(M^2_{\eta_0})}{c^{P}_1(M^2_{\eta_0})}~,
\label{Meta0}
\ee
which essentially captures its correct behaviour as long as $\kappa_B/\kappa_A$ is moderate and 
$m_X \ll \Lambda$.
For large values of $N$, $M^2_{\eta_0}$ is of order $1/N$.

Once having determined the $\eta_0$ and  $\eta^\prime$ masses, one can proceed to extract all relevant pseudoscalar decay constants from 
the pole mass residues of the matrix elements 
$\overline{\bold{\Pi}}_{\eta_\psi \eta_X}^{ij}(q^2)$ ($i,j=1,\cdots , 4$), where the resummed 
two-point correlator $\overline{\bold \Pi}_{\eta_\psi \eta_X}(q^2)$ is defined in Eq.~(\ref{Pieta}).
The procedure is similar to the one explained in section \ref{Resummed correlators and the Goldstone decay constant} for the simpler non-singlet case.
More precisely, from the definitions of the decay constants $F^{\psi (X)}_{\eta_0}$,  
$G^{\psi (X)}_{\eta_0}$ in Eqs.~(\ref{defFeta}) and (\ref{defGeta}),  
one obtains in general for $m_X\neq 0$
\bea
& \lim\limits_{q^2 \to M^2_{\eta_0}} (q^2-M^2_{\eta_0}) \overline{\bold \Pi}^{11(22)}_{\eta_\psi \eta_X}(q^2)  
\equiv -(G^{\psi(X)}_{\eta_0})^2~, 
\qquad\qquad
& \lim\limits_{q^2 \to M^2_{\eta_0}} (q^2-M^2_{\eta_0}) \overline{\bold \Pi}^{12,21}_{\eta_\psi \eta_X}(q^2)  
\equiv -G^{\psi}_{\eta_0} G^{\psi}_{\eta_0} ~, 
\label{PietaFG1}
\eea
\bea
& \lim\limits_{q^2 \to M^2_{\eta_0}} \dfrac{(q^2-M^2_{\eta_0})}{\sqrt{p^2}} 
\overline{\bold \Pi}^{13,31}_{\eta_\psi \eta_X}(q^2)  
\equiv  - \dfrac{G^\psi_{\eta_0} F^{\psi}_{\eta_0}}{2\sqrt{2}}~, 
\qquad\qquad
& 
\lim\limits_{q^2 \to M^2_{\eta_0}} \frac{(q^2-M^2_{\eta_0})}{\sqrt{p^2}} 
\overline{\bold \Pi}^{14,41}_{\eta_\psi \eta_X}(q^2)  
\equiv  - \dfrac{G^\psi_{\eta_0} F^{X}_{\eta_0}}{2\sqrt{3}}~,
\nonumber \\
& \lim\limits_{q^2 \to M^2_{\eta_0}} \dfrac{(q^2-M^2_{\eta_0})}{\sqrt{p^2}} 
\overline{\bold \Pi}^{23,32}_{\eta_\psi \eta_X}(q^2)  
\equiv  - \dfrac{G^X_{\eta_0} F^{\psi}_{\eta_0}}{2\sqrt{2}}~, 
\qquad\qquad
&
\lim\limits_{q^2 \to M^2_{\eta_0}} \frac{(q^2-M^2_{\eta_0})}{\sqrt{p^2}} 
\overline{\bold \Pi}^{24}_{\eta_\psi \eta_X}(q^2)  
\equiv  - \dfrac{G^X_{\eta_0} F^{X}_{\eta_0}}{2\sqrt{3}}~,
\label{PietaFG}
\eea
as well as 
\bea
&& \lim\limits_{q^2 \to M^2_{\eta_0}} \dfrac{(q^2-M^2_{\eta_0})}{q^2} 
\overline{\bold \Pi}^{33}_{\eta_\psi \eta_X}(q^2)  
\equiv  -\dfrac{(F^{\psi}_{\eta_0})^2}{8}~, 
\qquad\qquad
 \lim\limits_{q^2 \to M^2_{\eta_0}} \dfrac{(q^2-M^2_{\eta_0})}{q^2} 
\overline{\bold \Pi}^{44}_{\eta_\psi \eta_X}(q^2)  
\equiv  -\dfrac{(F^{X}_{\eta_0})^2}{12}~,  
\qquad\qquad
\nonumber \\
& &
\lim\limits_{q^2 \to M^2_{\eta_0}} \dfrac{(q^2-M^2_{\eta_0})}{q^2} 
\overline{\bold \Pi}^{34,43}_{\eta_\psi \eta_X}(q^2)  
\equiv  -\dfrac{F^{\psi}_{\eta_0} F^{X}_{\eta_0}}{4\sqrt{6}}~ ,
\label{PietaPiL}
\eea
where the factors $2\sqrt{2}$ and $2 \sqrt{3}$  take into account the normalisation of the $U(1)_\psi$ and $U(1)_X$ currents, respectively.
Similar expressions hold for the $\eta^\prime$ with the obvious replacement $\eta_0 \to \eta^\prime$.
Notice that the information on both diagonal and non-diagonal terms allow to extract unambiguously the signs of $G^{\psi (X)}_{\eta_0 
(\eta^\prime)}$ and $F^{\psi (X)}_{\eta_0 (\eta^\prime)}$. 
In the chiral limit, the pole of the $\eta_0$ migrates  from the longitudinal to the transverse axial correlator.
Consequently, in that case one can not extract the decay constants $F^{\psi (X)}_{\eta_0}$ 
from Eq.~(\ref{PietaPiL}), but only from Eq.~(\ref{PietaFG}).

In the following, for reasons of simplicity, we present analytical results only for the chiral limit $m_X = 0$. 
Let us consider the resummed axial longitudinal correlators, given by 
$q^2 \overline{\Pi}^L_{a_{\psi(X)}}(q^2)=8 (12) \overline{\bold \Pi}^{33 (44)}_{\eta_\psi \eta_X}(q^2)$ and 
$q^2 \overline{\Pi}^L_{a_{\psi} a_X}(q^2)=4 \sqrt{6} ~\overline{\bold \Pi}^{34,43}_{\eta_\psi \eta_X}(q^2)$, see Eq.~(\ref{PietaPiL}).
One can check that the linear combination corresponding to the conserved $U(1)$ current, 
vanishes for any $q^2$
\be
\overline{\Pi}_0^L(q^2) 
= 9(N-1)^2 \overline{\Pi}_{a_\psi}^{L} (q^2)
-6(N-1) \overline{\Pi}_{a_\psi a_X}^{L}(q^2) 
+ \overline{\Pi}_{a_X}^{L}(q^2) =0~,
\qquad\qquad
\overline{\Pi}_{a_\psi a_X}^{L} =\sqrt{\overline{\Pi}_{a_\psi}^{L} \overline{\Pi}_{a_X}^{L}}  ~.
\ee 
This is an important check, since the $U(1)$ current is conserved, despite the non-zero mass gap
spoiling the Ward identity at the naive one-loop level.
Then,  once fully resummed, there is no longitudinal part in the corresponding axial  two-points function, 
generalising, for the more involved singlet sector, the results obtained in section 
\ref{Resummed correlators and the Goldstone decay constant} for the simpler $SU(4)$ sector 
in isolation with (Goldstone) pseudoscalar-axial mixing. 
Coming now to the decay constants defined from Eqs.~(\ref{PietaFG1}) and (\ref{PietaFG}), 
using the gap equations (\ref{gap-2sectors}) and  the constraints among the effective couplings in 
Eqs.~(\ref{kbkb6}), (\ref{kbkb62}) and (\ref{kpsiX}), and after some algebra,  one obtains (in the chiral limit) 
\begin{equation}
(G_{\eta_0}^\psi)^2= \frac{-12(2N)^2 (N-1) A_4^2 M_\psi^2 g_a^{-1}(0) g_{a_c}^{-1}(0)}
{12 N(N-1) B_4(0) M_\psi^2 g_{a_c}^{-1}(0) +(2N+1) B_6(0) M_X^2 g_a^{-1}(0)} ~,
\qquad
(G_{\eta_0}^X)^2= \frac{(2N+1)^2 A_6^2 M_X^2}{6 (2N)^2 A_4^2 M_\psi^2} (G_{\eta_0}^\psi)^2~,
\label{Geta0psi}
\end{equation}
\begin{equation}
(F_{\eta_0}^\psi)^2
=\frac{-96 (2N)^2 (N-1)B_4^2(0) M_\psi^4 g_a(0)  g_{a_c}^{-1}(0)}
{12 N(N-1) B_4(0) M_\psi^2 g_{a_c}^{-1}(0) +(2N+1) B_6(0) M_X^2 g_a^{-1}(0)}
= \tilde{\Pi}_A^{L \psi}(0) g_a(0) \left[1- 4 \kappa_B\frac{A_4 B_6(0) g_a^{-1}(0)}{A_6\,c^P_1(0)} \right]~,
\label{Feta0psi}
\end{equation}
\begin{eqnarray}
&&(F_{\eta_0}^X)^2
= \frac{-24 (N-1)(2N+1)^2 B_6^2(0) M_X^4 g_{a}(0)^{-1} g_{a_c}(0)}
{12 N(N-1) B_4(0) M_\psi^2 g_{a_c}^{-1}(0) +(2N+1) B_6(0) M_X^2 g_a^{-1}(0)}
\nonumber
\\
&&\qquad\qquad\qquad\qquad\qquad\qquad\qquad
=\tilde{\Pi}_A^{L X}(0) g_{a_c}(0) \left[1- 24 \kappa_B \frac{(2N)(N-1) B_4(0) A_4 M_\psi^2 g_{a_c}^{-1}(0)}
{(2N+1) A_6 M^2_X\,c^P_1(0)} \right]~.
\label{Feta0X}
\end{eqnarray}
Notice from the second expressions of Eqs.~(\ref{Feta0psi}) and (\ref{Feta0X}) 
that the naive expressions of these decay constants, namely when the two sectors are in isolation, are respectively recovered for 
$M_X \to 0$ ($M_\psi\to 0$) as intuitively expected.
One can compute in a similar way the decay constants associated with the $\eta^\prime$.
We do not explicitly give them because the $\eta^\prime$  is not a pNGB and these expressions are 
rather involved.
The conserved $U(1)$ current ${\cal J}^\mu_0$ of Eq.~(\ref{axial_singlet}) implies
\be
F_{\eta_0 , \eta^\prime} = F_{\eta_0 , \eta^\prime}^X - 3(N-1) F_{\eta_0 , \eta^\prime}^\psi~.
\label{lincomb}
\ee
From Eqs.~(\ref{Feta0psi}) and (\ref{Feta0X}), we obtain the decay constant of the ${\eta_0}$ in the 
chiral limit
\be
F^2_{\eta_0} = -24 (N-1)\left[ 12 N (N-1) B_4 M_\psi^2 g_a(0) +(2N+1) B_6 M_X^2 g_{a_c}(0)\right] +{\cal O}(m_X)~,
\qquad\qquad
F^2_{\eta^\prime} = {\cal O}(m_X)~.
\label{Feta0}
\ee
As expected on general grounds (see section \ref{Sum rules and pseudoscalar decay constants in the flavour-singlet sector}), 
$F_{\eta_0}$ is non-zero in the chiral limit, while $F_{\eta^\prime}$ vanishes.
Furthermore, one can also check, after some algebra, that the generally expected
relations in Eq.~(\ref{FetaMrelation}) are indeed well satisfied (at least up to terms of higher orders 
in $m_X$) by our expressions above, which is a very non-trivial 
crosscheck of the NJL calculations. 
Likewise the general relations given in Eq.~(\ref{FGeta0}) are also well satisfied, 
providing an additional non-trivial crosscheck.

Actually, in the chiral limit the decay constants $F_{\eta_0}$ for the true Goldstone 
can be more directly calculated from the resummed transverse axial correlator $\overline{\Pi}_{a_\psi}(q^2)$ and $\overline{\Pi}_{a_X}(q^2)$
evaluated at $q^2=0$, in direct analogy with the non-singlet calculation of $F_G$.
From Eq.~(\ref{FGsum}), one obtains
\be
F_{\eta_0}^2  \equiv \lim\limits_{q^2\to 0}[-q^2 \overline{\Pi}_0(q^2)]= 
-\lim\limits_{q^2\to 0} q^2 [9(N-1)^2  \overline{\Pi}_{a_\psi}(q^2)+ \overline{\Pi}_{a_X}(q^2)]~,
\label{Feta0altern}
\ee
where the second equality comes from Eq.~(\ref{axial_singlet}), taking into account that  there is no mixing for the transverse contributions, i.e. $\overline{\Pi}_{a_\psi a_X}(q^2)=0$. 
The transverse resummed correlators are simply given by expressions similar to the one in Eq.~(\ref{FGsum}): 
$-q^2 \overline{\Pi}_{a_{\psi}}(q^2) = 8 \t \Pi^\psi_A(q^2) 
g_A(q^2)$ and $-q^2 \overline{\Pi}_{A_{X}}(q^2) = 12 \t \Pi^X_A(q^2) 
g_{A_c}(q^2)$.
Thus using the expression of the one-loop functions $\t \Pi^{\psi(X)}_A(0)$ from Table \ref{tab_phi}
and Table \ref{tab-coloured-functions}
directly gives
\be
F_{\eta_0}^2 = 9(N-1)^2 \left[-16(2N) M^2_\psi \t B_0(0,M^2_\psi) g_a(0)\right] +
\left[-24(2N+1)(N-1) M^2_X \t B_0(0,M^2_X) g_{a_c}(0)\right]~,
\ee
which is consistent with Eq.~(\ref{Feta0}).

\subsubsection{The mass spectrum of the singlet resonances}
\label{The mass spectrum of the singlet resonances}

We now study the mass spectrum of the scalar and pseudoscalar singlet resonances.
Before turning to the more involved case including the mixing between the resonances from the
electroweak and the coloured sectors, let us consider the instructive no-mixing case, where 
$A_{\psi X}=0$ and consequently $\kappa_B=\kappa_{B6}=\kappa_{\psi X}=0$.
From Eq.~(\ref{detsig}) we obtain for the scalar singlet masses
\begin{equation}
A_{\psi X}=0:
\qquad\qquad
M_{\sigma_0}^2=4 M_\psi^2 =M_{\sigma_\psi}^2~,
\qquad
M_{\sigma^\prime}^2=4 M_X^2 - \frac{m_X}{M_X} \frac{1}{2 \kappa_{A6} B_6(M_{\sigma^\prime}^2)} =M_{\sigma_X}^2~,
\end{equation}
which of course reproduce the masses in isolation.
As discussed above, in our benchmark case where $\kappa_{A6}=\kappa_A$ we have $M_\psi \leqslant M_X$, so that in the no-mixing case  
we have $M_{\sigma_0}^2\leqslant M_{\sigma^\prime}^2$ where the equality is valid for $m_X=0$.
In the same way, from Eq.~(\ref{deteta}) we obtain for the pseudoscalar masses
\begin{equation}
A_{\psi X}=0:
\qquad\qquad
M_{\eta_0}^2=0=M_{\eta_\psi}^2~,
\qquad
M_{\eta^\prime}^2=-\frac{m_X}{M_X} \frac{g_{a_c}^{-1}}{2\kappa_{A6}B_6}=
M_{\eta_X}^2~.
\end{equation}
Again, the latter expressions reproduce those in isolation,
and $M_{\eta_0}^2\leqslant M_{\eta^\prime}^2$, where the equality is valid for $m_X=0$.

\begin{figure}[tbp]
   \begin{minipage}[c]{.1\linewidth}
      \includegraphics[scale=0.24,trim= 0 0 0 0]{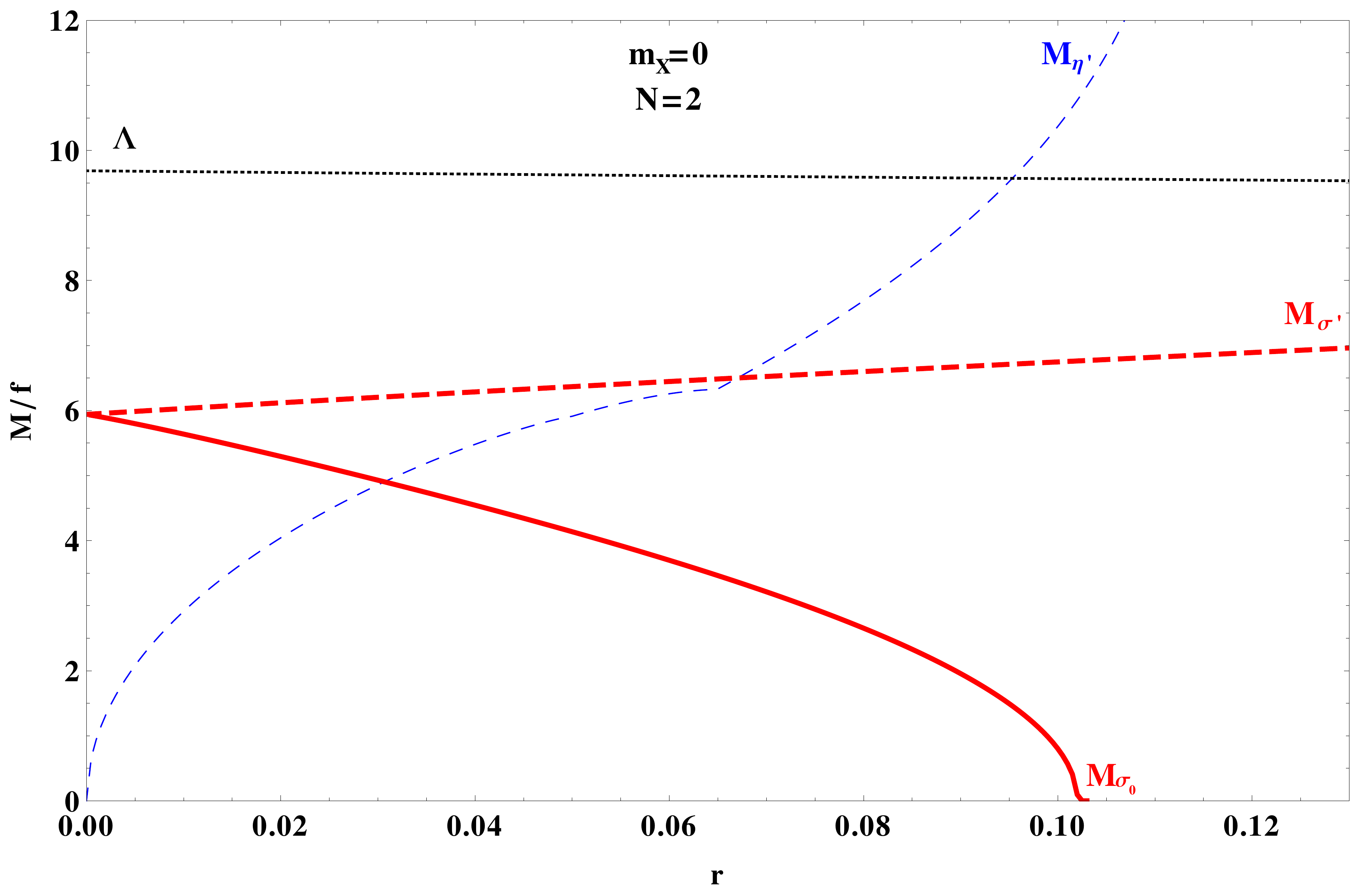} 
   \end{minipage} \hfill
   \begin{minipage}[c]{.5\linewidth}
      \includegraphics[scale=0.24,trim= 0 0 0 0]{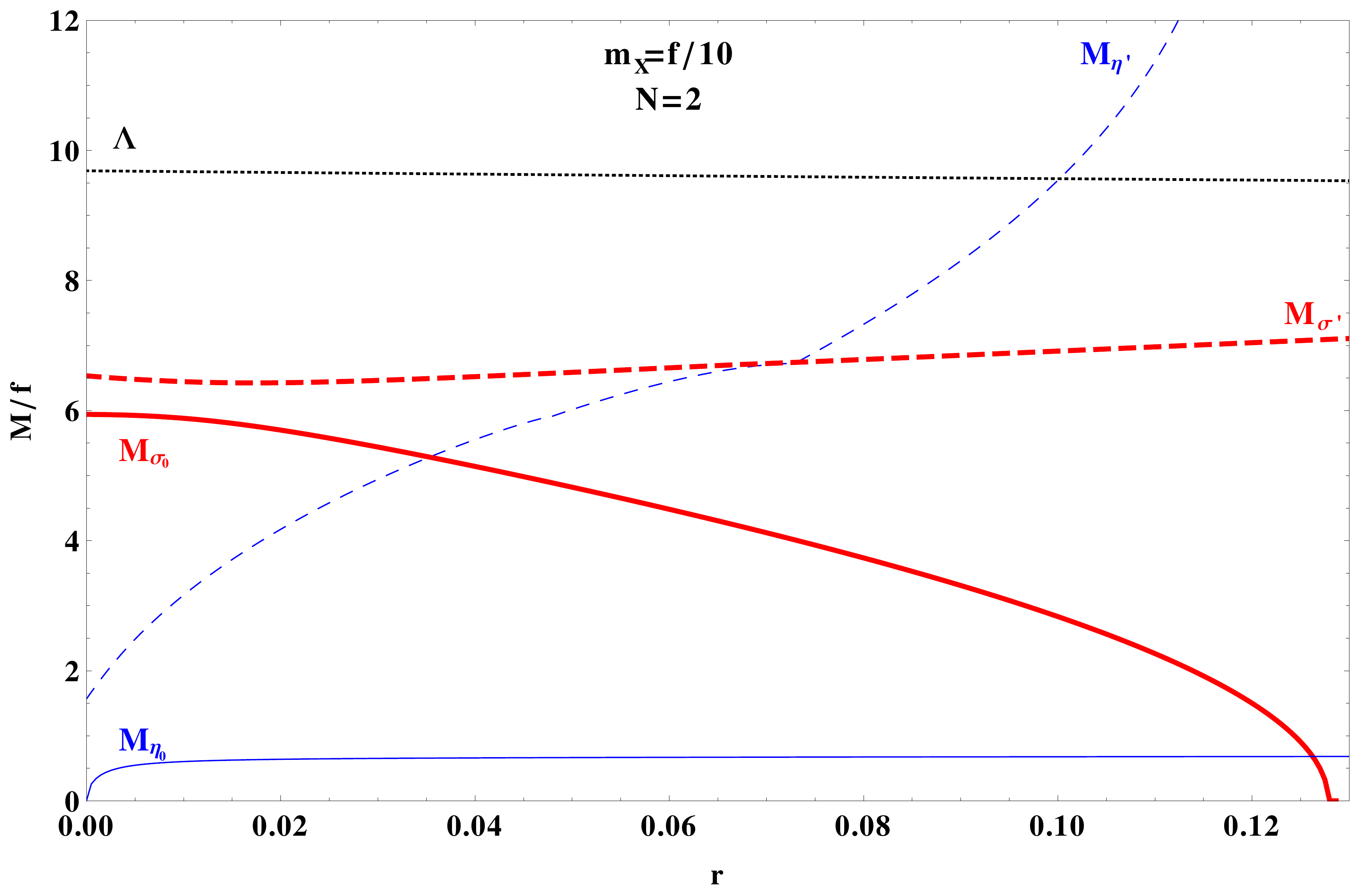}
   \end{minipage}\\
      \begin{minipage}[c]{.1\linewidth}
      \includegraphics[scale=0.24,trim= 0 0 0 0]{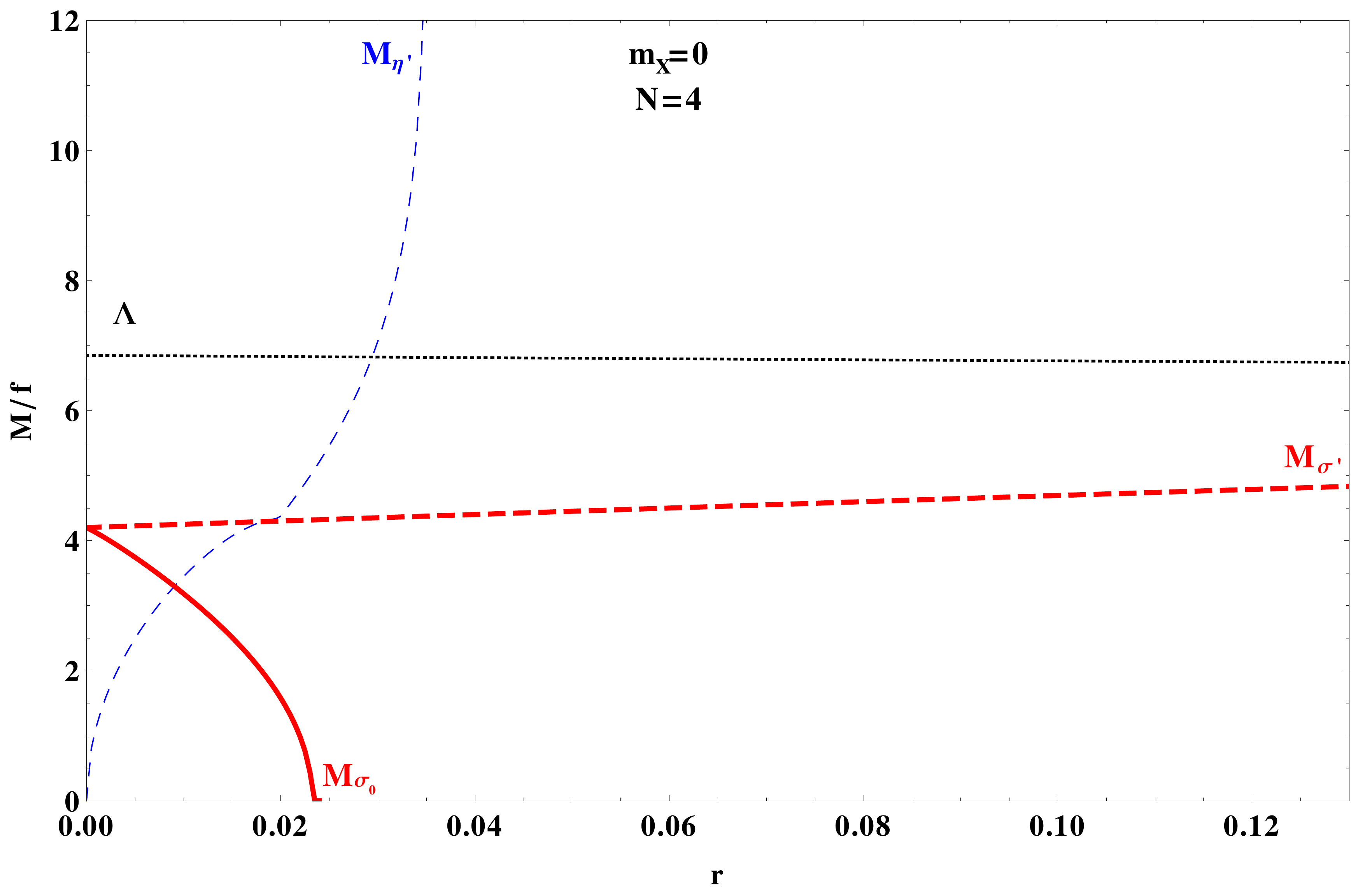}
   \end{minipage} \hfill
   \begin{minipage}[c]{.5\linewidth}
      \includegraphics[scale=0.24,trim= 0 0 0 0]{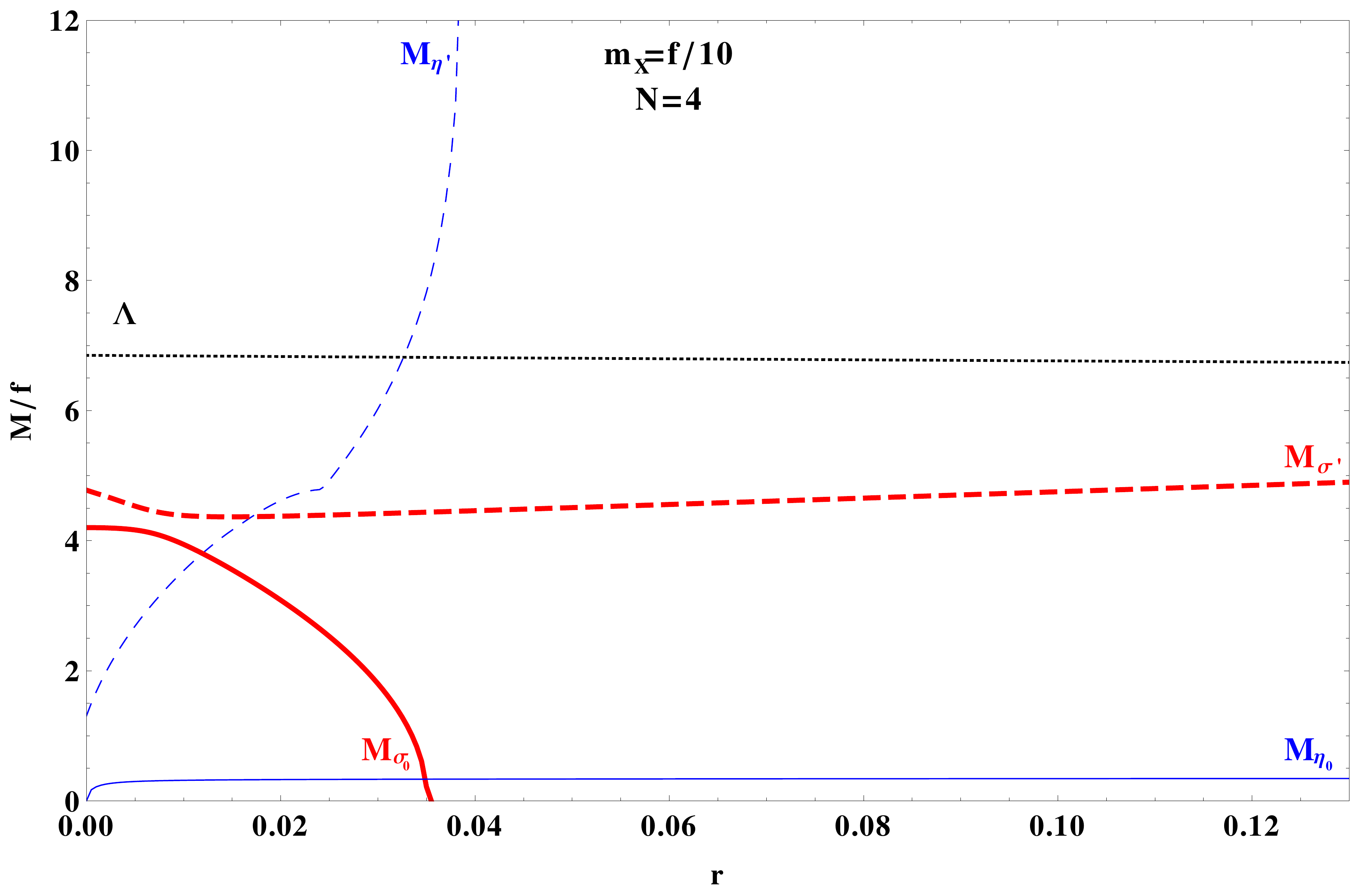}
   \end{minipage}
   \caption[long]{Singlet scalar and pseudoscalar meson masses in units of $f$, for a fixed value of the couplings $\xi=1.3 $ and  $\kappa_A=\kappa_{A6}$, as a function of $r\equiv \kappa_B/\kappa_A$,
for $N=2$ (top) and $N=4$ (bottom), and for $m_X=0$ (left) 
and $m_X =f/10$ (right). The Goldstone boson $\eta_0$ is massless in the chiral limit.}
\label{plot-kB-singlet}
\end{figure}

Once we switch on the mixing, important new features arise, as discussed above:
in particular, the upper bound on $\kappa_B/\kappa_A$ from Eq.~(\ref{kbcrit}), and the corresponding
rapid growth of $M_{\eta^\prime}$ when approaching from below the critical value of $\kappa_B/\kappa_A$. This is 
illustrated in Fig.~\ref{plot-kB-singlet}  for $N=2$ and $N=4$, as usual assuming $\kappa_{A6}=\kappa_A$. 
Consequently, the $\eta^\prime$ mass  may be of order $f$ for $\kappa_B/\kappa_A\ll 0.01$, 
but once $\kappa_B/\kappa_A$ grows to larger values, already well below
the bound of Eq.~(\ref{kbcrit}), $\eta'$ decouples rapidly.

Another interesting feature is implicit in the $\eta_0$ mass expression Eq.(\ref{Meta0}): namely,
$M_{\eta_0}$ rapidly reaches an asymptotic limit for moderate $\kappa_B/\kappa_A$ values, for fixed $N$, 
and this (approximate) maximum
decreases as $1/N$ for large $N$, as also illustrated in Fig. \ref{plot-kB-singlet}.
More precisely, in the approximation
of neglecting the differences in momenta of the loop functions, one obtains for large $N$ values
\be
M^2_{\eta_0}\simeq -\frac{A_6}{B_6} \frac{1}{3N} \frac{m_X}{M_X} \frac{M^2_X}{M^2_\psi}+{\cal O}(1/N^2)\,.
\label{approxMeta0}\ee
Of course $\eta_0$  being a pNGB, $M_{\eta_0}^2$ vanishes
linearly in $m_X$. This shows in addition that $M_{\eta_0}$ is approximately $\kappa_B/\kappa_A$-independent, once this ratio takes moderately large values,
as shown in Fig.~\ref{plot-kB-singlet}. Its mass can be well below $f$, for sufficiently large $N$ and/or small $m_X$. 

%

\begin{figure}[tb]
\includegraphics[scale=.25]{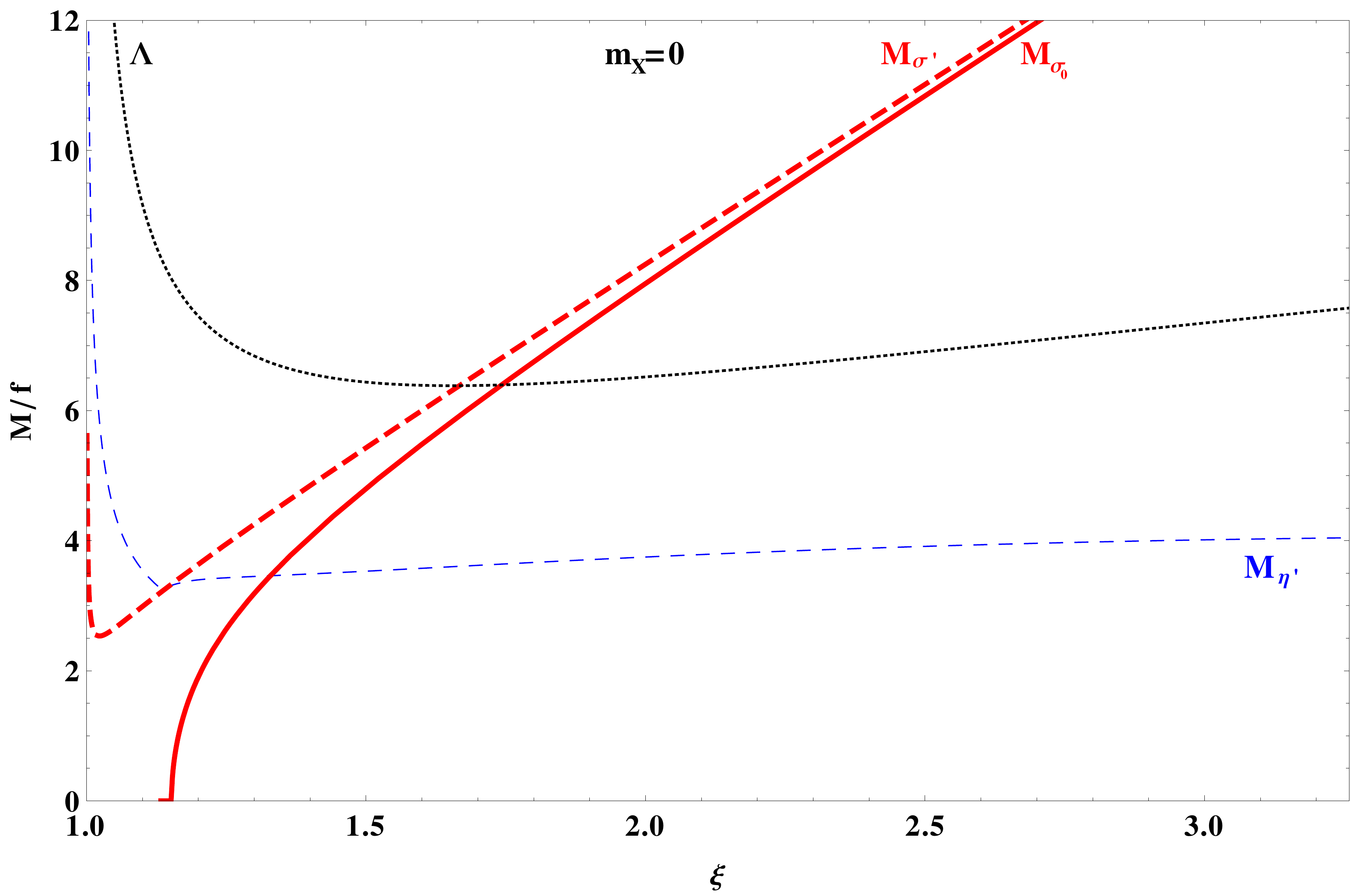}
\includegraphics[scale=.25]{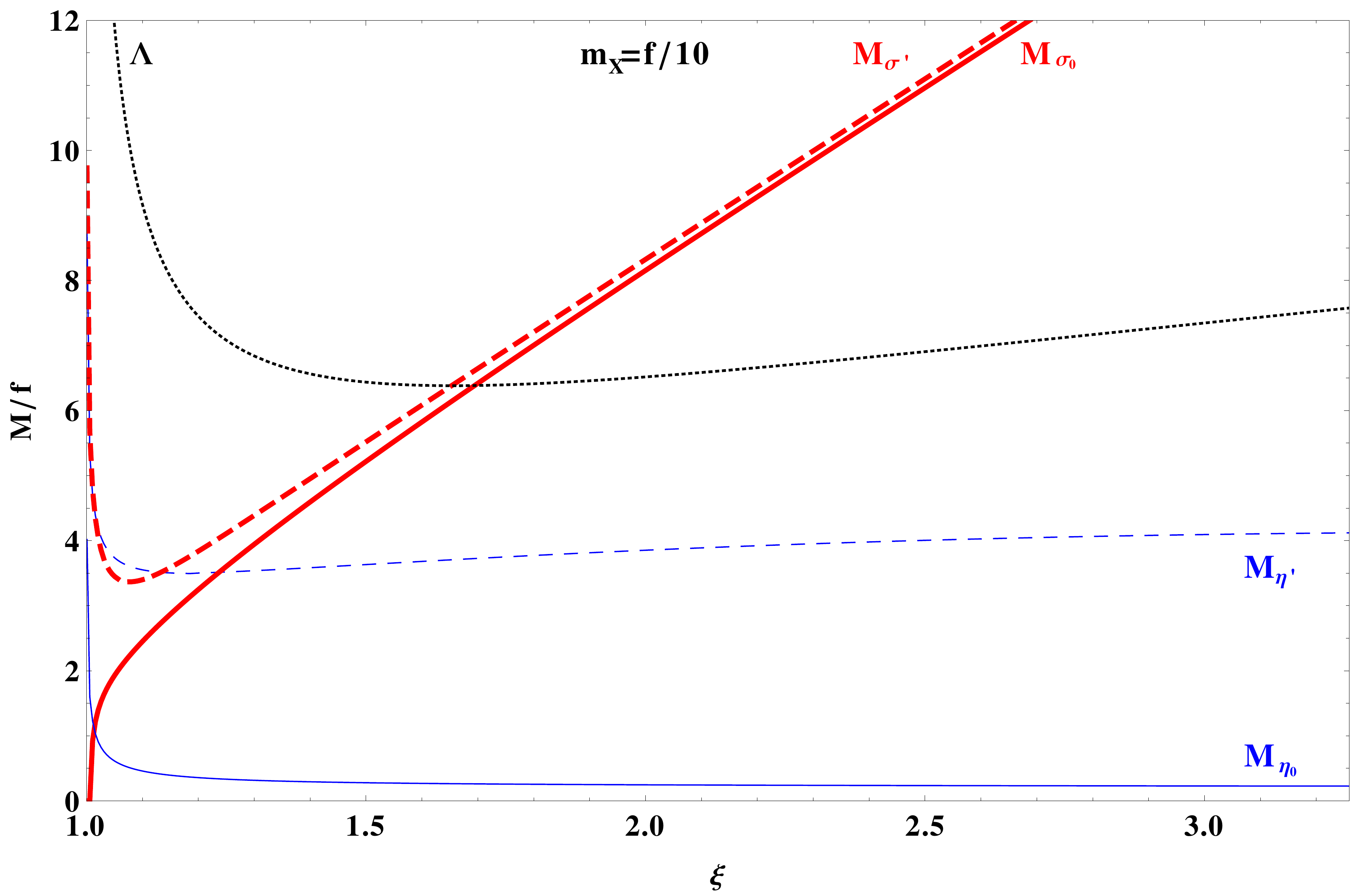}
\caption{Singlet scalar and pseudoscalar meson masses in units of $f$, as a function of $\xi$
for $N=4$, $\kappa_A=\kappa_{A6}$, $\kappa_B/\kappa_A=0.01$, $m_X=0$ (left panel) and $m_X=f/10$ (right panel).
The Goldstone boson $\eta_0$ is massless in the chiral limit.}
\label{plot-xi-singlet}
\end{figure}

The two {\em scalar} singlet masses are defined implicitly by Eq.~(\ref{sig0sigp}).
The heaviest state $\sigma^\prime$ always lies in the multi-TeV range, as illustrated in Figs.~\ref{plot-kB-singlet} 
and \ref{plot-xi-singlet}.
More interestingly,
as explained in section \ref{Scalar mixing and eigenstates}, for $\xi\lsim 1.7-1.8$
the lightest scalar mass $M_{\sigma_0}$ is a decreasing function of $\kappa_B/\kappa_A$ 
and vanishes at a critical  value given by the (positive) root of Eq.~(\ref{kbcsig0}). 
This critical value is different from the one defined by Eq.~(\ref{kbcrit}), but for $\xi\lsim 1.4$
it is numerically very close to the latter, more precisely 
it lies (slightly) below, for any $N\ge 2$. This is illustrated in Fig.~\ref{plot-kB-singlet} for $N=2$ and $N=4$.
Beyond the critical value of $\kappa_B/\kappa_A$, $\sigma_0$ becomes tachyonic
and the effective scalar potential is destabilised, therefore    
$M_{\sigma_0}$ can be very small just before reaching the critical value of $\kappa_B/\kappa_A$.
Recall, however,  that for $\xi\gsim 1.7$, 
the solution $M_{\sigma_0}=0$ at positive $\kappa_B/\kappa_A$ disappears,
being replaced by a minimum positive pass, that is reached for an increasing value of $\kappa_B/\kappa_A$ as $\xi$ increases. 
But, in this range for $\xi$, the bound
from  Eq.~(\ref{kbcrit}) is more stringent, restricting $\kappa_B/\kappa_A$ to be much smaller  
and therefore rendering non-physical  the behaviour of $M_{\sigma_0}(\kappa_B/\kappa_A)$ for larger values of  $\kappa_B/\kappa_A$.

Finally we also illustrate in Fig.~\ref{plot-xi-singlet} the $\xi$-dependence of the
scalar and pseudoscalar singlet masses, for representative values of $N$, and for $\kappa_B/\kappa_A$ fixed
safely below the upper bound in Eq.~(\ref{kbcrit}). Notice 
that $M_{\sigma_0}$ vanishes
for a sufficiently low value of $\xi$, where one saturates the condition of Eq.~(\ref{kbcsig0}), 
because the positive root of this equation decreases with $\xi$.
As a consequence, the whole meson mass spectrum should not be trusted for $\xi$ smaller than this critical value,
as the vacuum becomes unstable.

To conclude this section, let us briefly discuss the $\eta_0$ couplings to the SM gauge bosons.
 The collider phenomenology of this singlet  has already  been discussed in general in Ref.~\cite{Belyaev:2016ftv}.
As mentioned at the end of section \ref{gauging}, in the chiral limit
the anomalous coupling of a pseudo-Goldstone boson to a pair of gauge bosons is fully determined by the Wess-Zumino-Witten effective action.
While the $SU(4)/Sp(4)$ [$SU(6)/SO(6)$] pseudo-Goldstone bosons may couple only to the electroweak (colour) gauge bosons, the $\eta_0$ is
specially interesting as it couples to both, because it 
couples to both the $\psi$ and $X$-fermion number currents ${\cal J}^0_{\psi\mu}$ and ${\cal J}^0_{X\mu}$.
The two currents have a $U(1)_Y$ anomaly, and ${\cal J}^0_{\psi\mu}$ [${\cal J}^0_{X\mu}$] has a $SU(2)_L$ [$SU(3)_c$] 
anomaly as well.
Then, specialising Eq.~(\ref{WZW}) to our model, the $\eta_0$ couplings to the SM gauge bosons take the form
\be
\begin{array}{rcl}
{\cal L}_{eff,\eta_0}^{WZW} &=& - \dfrac{1}{ 16\pi^2} (2N) \left[-3(N-1)\right]\dfrac{\eta_0}{F_{\eta_0}}
\left(g^2 W_{i\mu\nu}\tilde{W}^{\mu\nu}_i +g'^2 B_{\mu\nu}\tilde{B}^{\mu\nu}\right) \\
&&- \dfrac{1}{ 16\pi^2} (2N+1)(N-1) \dfrac{\eta_0}{F_{\eta_0}}
\left(2 g_s^2 G_{a\mu\nu}\tilde{G}^{\mu\nu}_a + \dfrac{16}{3} g'^2 B_{\mu\nu}\tilde{B}^{\mu\nu}\right)  \\
&=& \eta_0\left[ k^0_{\gamma\gamma} 
e^2A_{\mu\nu}\tilde{A}^{\mu\nu} + k^0_{gg} 
g_s^2 G_{a\mu\nu}\tilde{G}^{\mu\nu}_a
+\dots\right] ~,
\end{array}
\label{WZWeta0}\ee
where the first (second) line is the contribution of the $\psi$ ($X$) fermion loops, and the dots 
stand for couplings involving the $Z$ or $W$ field strengths.
Here $\tilde{F}_{\mu\nu}\equiv \epsilon_{\mu\nu\rho\sigma} F^{\rho\sigma}/2$ and the coefficients $k^0_{\gamma\gamma,gg}$ 
are straightforwardly computed  using $B_{\mu\nu}\supset c_wA_{\mu\nu}$, $W_{3\mu\nu}\supset s_wA_{\mu\nu}$, and $e=gs_w=g' c_w$,
and similarly for couplings  involving the $Z$ or $W$ field strengths.
The decay widths into massless gauge bosons are
\be
\Gamma(\eta_0 \rightarrow \gamma\gamma)= 4\pi\alpha_{em}^2 M_{\eta_0}^3 (k^0_{\gamma\gamma})^2 ~,\qquad
\Gamma(\eta_0 \rightarrow gg)= 32\pi\alpha_s^2 M_{\eta_0}^3 (k^0_{gg})^2 ~.
\ee
Note that these rates are determined only by group theory factors, up to the decay constant $F_{\eta_0}$.
The latter can be computed in the NJL approximation, and the result is given in Eq.~(\ref{Feta0}).
Thus, the golden channel for the discovery of $\eta_0$ at the LHC is production via gluon-gluon fusion and decay 
into two gauge bosons: di-jet, di-photon, $\gamma Z$, $ZZ$ and  $WW$ final states.
We recall that the mass of $\eta_0$ is induced by the explicit breaking of the anomaly-free $U(1)$ symmetry:
this is due either to an explicit mass term for the constituent fermions, $m_X\ne 0$, 
or to the proto-Yukawa couplings of the SM fermions to the composite sector, that we do not specify in this paper.
Our NJL result for $M_{\eta_0}$ is given in  Eqs.~(\ref{Meta0}), (\ref{approxMeta0}).
The corrections to Eq.~(\ref{WZWeta0}), that strictly holds in the chiral limit, are expected to be subleading, as long as 
$\eta_0$ is significantly lighter than the non-Goldstone resonances.
Note that the ratio $\Gamma(\eta_0 \rightarrow gg)/\Gamma(\eta_0 \rightarrow \gamma\gamma) = 18(2N+1)^2/(N-4)^2 \cdot \alpha_s^2/\alpha_{em}^2 $
is independent from $F_{\eta_0}$ and $M_{\eta_0}$, and is larger than $2\cdot 10^4$ for any $N$. Thus a discovery appears more likely in the di-jet channel. Indeed, the alleged di-photon resonance at 750 GeV
could not be fitted by $\eta_0$, because the gluons-to-photons ratio is too large \cite{Belyaev:2015hgo}.

\subsection{Comments on spectral sum rules}
\label{sum-rules-full-model}

In this section, we comment on the spectral sum rules when both the electroweak and the coloured sectors are included.
We will not enter in the details here but rather focus on the main differences as compared to the electroweak sector in isolation.
The latter has been extensively discussed in section \ref{secWSR}.
A few modifications are worth noticing.
While in the electroweak sector the sum rule involving  $\Pi_{S-P}^\psi (q^2)$ is not expected to 
hold (see footnote \ref{fnte_SP}), in the coloured sector $\Pi_{S-P}^X (q^2)$  is an order parameter, therefore the first sum rule in Eq.~(\ref{scalSR}) is operative as well.
On the other hand, the presence of an explicit symmery-breaking mass term $m_X \neq 0$ 
spoils the convergence of the integrals in Eqs.~(\ref{WSRVA}) and (\ref{scalSR}),
so that one can only write the convergent sum rule of Eq.~(\ref{WSRmexpl}).
Therefore, the saturation of the coloured-sector sum rules is expected to worsen as $m_X$ increases.
Recall that the NJL approximation already implies large departures from the sum rules as shown, for the electroweak sector, in Figs.~\ref{fig_Sum_Rules} and \ref{wsrpole}.

Another qualitative difference is induced by the interplay between the two sectors.
Indeed, the mixings, defined by Eqs.~(\ref{Pisigc}) and  (\ref{Pieta}), between the (pseudo)scalar singlets of the 
two sectors modify the two-point (pseudo)scalar singlet correlators as compared to their expressions when considered in isolation.
As a consequence, the singlet two-point correlators develop two poles, corresponding to the $\sigma_0$ and $\sigma^\prime$ ($\eta_0$ and $\eta^\prime$) in the (pseudo)scalar case.
Let us assume that $m_X=0$ and take the example of the order parameters $\Pi_{S^0-P^0}^{\psi (X)}(q^2)$, 
which involves only the singlets densities $S^0_{\psi, X}$ and $P^0_{\psi, X}$.
The  corresponding sum rules are then given by 
\begin{eqnarray}
\int dt ~ {\rm Im} \overline{\Pi}_{S_0-P_0}^{\psi (X)}(t)  &\equiv &
\int dt \left[ {\rm Im} \overline{\bold\Pi}_{\sigma_\psi \sigma_X}^{11 (22)} (t)- {\rm Im} \overline{\bold\Pi}_{\eta_\psi \eta_X}^{11 (22)} (t) \right]=0
\nonumber
\\
&=& 
(G_{\sigma_0}^{\psi (X)})^2+ (G_{\sigma^\prime}^{\psi (X)})^2-(G_{\eta_0}^{\psi (X)})^2- (G_{\eta^\prime}^{\psi (X)})^2=0~,
\end{eqnarray} 
where the second line  has been obtained by assuming the saturation, in the narrow-width approximation, 
of the correlators by the first light resonances. The expressions of the scalar decay constants $G_i^{\psi (X)}$ can be obtained 
from sections \ref{Scalar mixing and eigenstates} and \ref{Pseudoscalar singlet mixing and properties}.

\begin{figure}[tb]
\includegraphics[scale=.25]{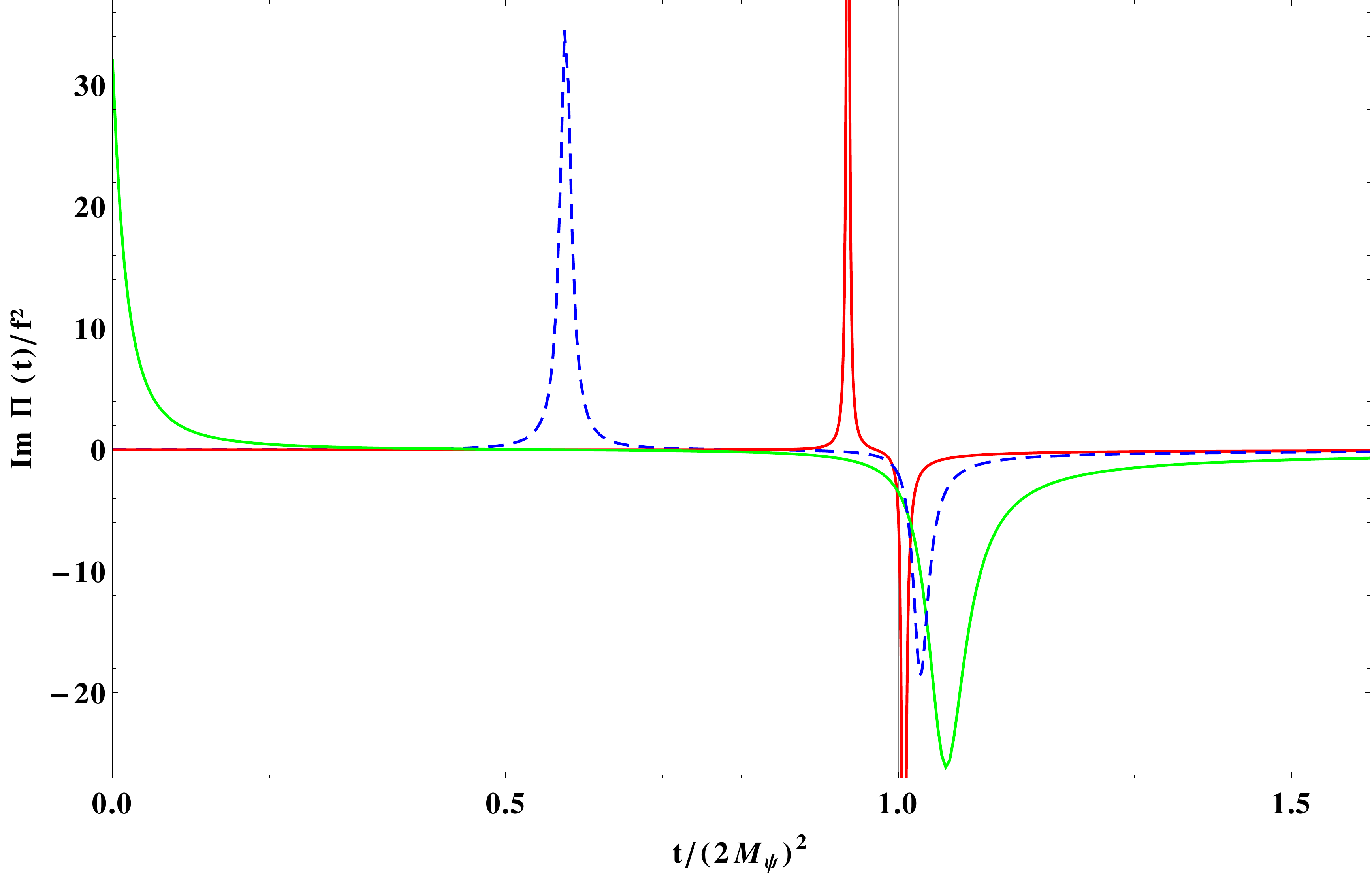}
\includegraphics[scale=.25]{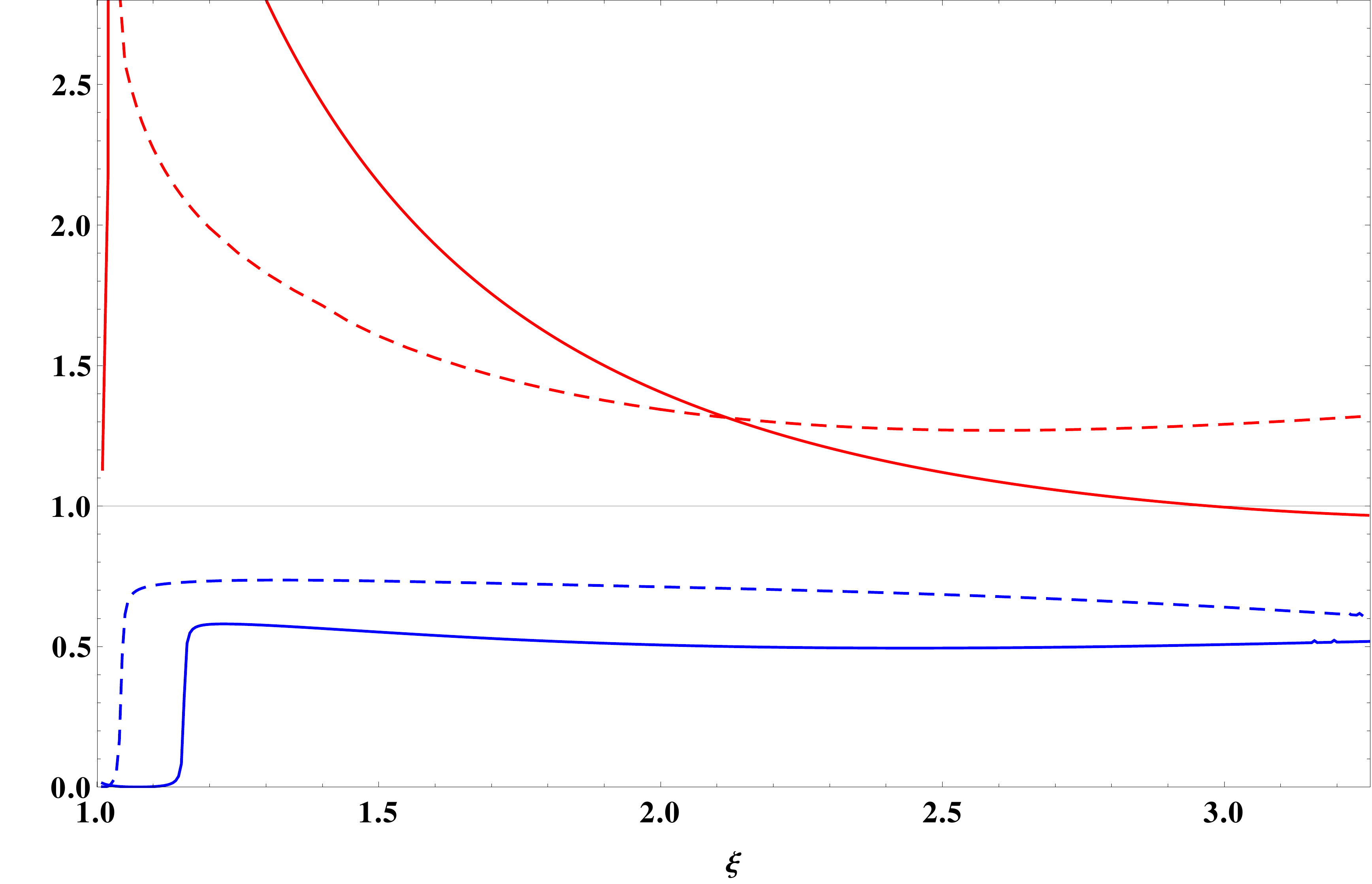}
\caption{Left panel: The spectral function ${\rm Im} \Pi_{S_0}^{\psi X}(t)$ as a function of $t/(2 M_\psi)^2$ for three values $\xi=1.15$ (solid green line), $\xi=1.3$ (dashed blue line) and $\xi=2$ (solid red line).
The other parameters are fixed to $N=4$, $\kappa_{A6}/\kappa_A=1$, $\kappa_B/\kappa_A=0.01$ and $m_X=0$.
One clearly sees the two poles, associated with the $\sigma_0$ and $\sigma^\prime$ scalar singlets, which become closer and closer as $\xi$ increases.
In the opposite limit where $\xi$ decreases, the $\sigma_0$ becomes lighter and lighter up to be massless for $\xi \simeq 1.15$ while the $\sigma^\prime$ always stays close to the threshold $4 M_\psi^2 \simeq 4 M_X^2$.
The residues of the poles have an opposite sign in agreement with the expectation from the associated sum rule.
 Right panel: The absolute value of the ratio of the integral $ \int^{t_0}_0 dt ~{\rm Im} \Pi_{S_0}^{\psi X}(t)/ \int_{t_0}^\infty dt ~{\rm Im} \Pi_{S_0}^{\psi X}(t)$ 
 (lower blue lines) as a function of $\xi$ for two values of the number of hypercolours $N=4$ (solid line) and $N=2$ (dashed line). 
As explained in the text, ${t_0}$ is the value above which the spectral density becomes negative.
Also shown is the absolute value of the ratio $(G_{\sigma_0}^{\psi} G_{\sigma_0}^{X})/(G_{\sigma^\prime}^{\psi} G_{\sigma^\prime}^{X})$ (upper red solid and dashed lines).
The other parameters are fixed to $\kappa_{A6}/\kappa_A=1$, $\kappa_B/\kappa_A=0.01$ and $m_X=0$.
 }
\label{spectral-density-mixing-singlet}
\end{figure}

When the two sectors are present, an additional $U(1)$ symmetry is also preserved, 
and leads to two additional sum rules (see section \ref{Sum rules and pseudoscalar decay constants in the flavour-singlet sector}).
For simplicity, in the sequel we focus only on the scalar sum rule, in order to avoid the complications coming from the pseudoscalar-axial mixing.
The corresponding sum rule takes the following form
\begin{equation}
\int dt  ~{\rm Im} \overline{\Pi}_{S_0}^{\psi X} \equiv
\int dt  ~{\rm Im} \overline{\bold\Pi}_{\sigma_\psi \sigma_X}^{12} (t)
= G_{\sigma_0}^{\psi} G_{\sigma_0}^{X}- G_{\sigma^\prime}^{\psi} G_{\sigma^\prime}^{X}=0~,
\label{sum-rule-scalar12}
\end{equation}
where in the last equality the saturation of the correlator by the first light resonances has been assumed.
Let us focus on this sum rule, as all the new features induced by the interplay between the two sectors are contained in the correlator $\overline{\Pi}_{S_0}^{\psi X}(q^2)$. 
First, one clearly sees the two poles associated to  $\sigma_0$ and $\sigma^\prime$  in the spectral density, which is displayed 
in Fig.~\ref{spectral-density-mixing-singlet} for different values of $\xi$ and $N=4$.
Increasing the value of $\xi$, the two poles become closer and closer
in agreement with Fig.~\ref{plot-xi-singlet}.
In principle there are two distinct thresholds above which the loops involving the fermions $\psi$ or $X$ develop an imaginary part.
However, as  the mixing parameter $\kappa_B/\kappa_A$ is small, these two thresholds are very close (see Fig.~\ref{gap-coloured}) and one can consider in a good approximation only one threshold located around $4M_\psi^2 \simeq 4 M_X^2$.
While in the spectral density the second pole associated to the $\sigma^\prime$ remains always close to this threshold,  one sees 
that the $\sigma_0$ pole moves continuously from $p^2\simeq 4M_\psi^2$ (for large values of $\xi$) down to $p^2=0$ (for $\xi \simeq 1.15$) when the $\sigma_0$ becomes massless (see Fig.~\ref{plot-xi-singlet}).
From Eq.~(\ref{sum-rule-scalar12}), one also sees that the residues of the two poles in the spectral density should have an opposite sign in order to respect the sum rule.
This is in agreement with the left panel of Fig.~\ref{spectral-density-mixing-singlet}.
As the scalar singlets are narrow and the continuum part of the spectral density is small, one expects the sum rule of Eq.~(\ref{sum-rule-scalar12}) 
to be well respected by the NJL approximation and the saturation by the first light resonances to be a good 
approximation.\footnote{Note that in the electroweak sector in isolation, the continuum of the scalar singlet density is also small and the pole is narrow.
However, there is  no sum rule involving only scalar singlets, so that the above argument does not apply.}

The saturation of the sum rule (\ref{sum-rule-scalar12}) is illustrated in the right panel of Fig.~\ref{spectral-density-mixing-singlet}.
 We plot the absolute value of the ratio of integrals $ \int^{t_0}_0 dt ~{\rm Im} \Pi_{S_0}^{\psi X}(t)/ \int_{t_0}^\infty dt ~{\rm Im} \Pi_{S_0}^{\psi X}(t)$,  
as a function of $\xi$ and for two different values of the number of hypercolours, $N=4$ and $N=2$.
In the true theory, this ratio is predicted to be one regardless of the value of the parameter  $t_0$.
In our NJL approximation of the strong dynamics, the result of the integration may depend on the value of $t_0$, that we conventionally choose as the value of $t$ where the spectral density vanishes.
In this way, one compares the positive and negative parts of the spectral densities,
in the same spirit as for the saturation of the sum rule with the two lightest resonances. 
To illustrate the latter, we plot
the absolute value of the ratio $(G_{\sigma_0}^{\psi} G_{\sigma_0}^{X})/(G_{\sigma^\prime}^{\psi} G_{\sigma^\prime}^{X})$,
that is obtained in the same way as in section \ref{secWSR}, but the explicit expression is more involved due to the mixing and we refrain from giving it here.
Below the critical value $\xi \simeq 1.15$ ($\xi \simeq 1.04$) for $N=4$ ($N=2$), this ratios becomes meaningless,  as the $\sigma_0$ pole disappears from the spectral density, such that a large departure from one is observed.
In summary, the right panel of Fig.~\ref{spectral-density-mixing-singlet} shows that the ratio of integrals (of decay constants) is smaller (larger) than one, but this departure from the sum-rule prediction
is reasonably small as long as $\xi$ is well above the instability region (see section \ref{secWSR} for a detailed discussion of the limitations of the NJL approximation with regard to the sum rules).

\section{Conclusion}
\label{conclusion}

The general idea of a composite, Nambu-Goldstone Higgs particle provides a very attractive framework for the EWSB.
We considered an asymptotically-free gauge theory confining at the multi-TeV scale and that has the potential to provide a self-consistent,
ultraviolet-complete framework to study the composite Higgs phenomenology.

The minimal model features four flavours of constituent fermions $\psi^a$, which condense as the hypercolour interaction becomes strong.
The first, remarkable result is that, unavoidably, the corresponding $SU(4)$ flavour symmetry breaks spontaneously to 
$Sp(4)$, as required in order to generate a NGB Higgs.
This follows from general results on vector-like gauge theories, reviewed in sections \ref{VW}-\ref{anomat}. Furthermore, such a dynamical symmetry breaking is successfully described by
a four-fermion operator, \`a la NJL: when the four-fermion coupling exceeds a critical value, a non-zero mass gap develops, as shown in section \ref{colourless-sector}.
The meson resonances are described by two-point correlators of fermion bilinears. The meson spins 
(zero or one) and their representations under the flavour group are determined by the
quantum numbers of the associated hypercolour-singlet fermion bilinears.
Following the standard NJL approach, 
we computed all the relevant two-point correlators, resummed at leading order in the number of hypercolours $N$: the meson mass is determined by the correlator pole,
while the residue at the pole fixes the meson decay constant. In section \ref{Resummed correlators and the Goldstone decay constant} 
we have shown that the NGB decay constant $f$ is almost ten times smaller
than the cutoff of the constituent fermion loops, therefore our effective theory is well under control up to meson masses of order $\sim10f$.
Recall that electroweak precision measurements require $f\gtrsim 1$ TeV and that fine-tuning in the composite Higgs potential is proportional to the ratio $v^2/f^2$.
In order to correlate the various meson masses, we made the hypothesis that the hypercolour dynamics is dominated by current-current interactions, see appendix \ref{$Sp(2N)$ current-current operators},
and we used Fierz transformations to relate the different four-fermion operators. In particular, in section \ref{Fierz-transfoSP2N} we derived some $Sp(2N)$ Fierz identities which, 
to the best of our knowledge, are not available elsewhere in the literature.

In section \ref{The mass spectrum of the resonances} 
we illustrated our results for the mass spectrum of electroweak mesons: for a reasonably small number of hypercolours, say $2N \lesssim 10$, 
the spin-one mesons are always heavier than $5f$, while the spin-zero mesons can be as light as $f$, and therefore accessible at the LHC,
in the following special cases. 
The singlet scalar mass $M_\sigma$
vanishes when the four-fermion coupling approaches its critical value, that is, when the condensate vanishes. 
The singlet pseudoscalar mass $M_{\eta^\prime}$
is induced by the axial anomaly: the anomalous contribution is expected to scale as $M_{\eta'}^2\sim 1/N$, 
but we did not attempt to quantify its absolute size.
Therefore, we cannot exclude a very light value for $M_{\eta^\prime}$. 
Note that these results for $\sigma$ and $\eta'$ hold for the electroweak sector in isolation:
the effects of the mixing with the singlets of the colour sector are summarised below.
The non-singlet scalar $S$ can also be light if both $\sigma$ and $\eta^\prime$ are,
as $M_S^2\simeq M_\sigma^2 + M_{\eta^\prime}^2$. 
In addition, one should keep in mind that the set of NGB is formed by the Higgs doublet plus a SM singlet $\eta$;
their masses arise only from SM loops, which we did not study here, and are expected to lie at or below the scale $f$.
In section \ref{secWSR}  we performed an important test of the accuracy of our methods, by comparing our results with spectral sum rules, 
that have to be satisfied by the exact two-point correlators. We thus identified the values of the four-fermion coupling that best
 reproduce the sum rules. Conversely, our results in the effective NJL approximation depart significantly from the sum rules, when the continuum part of the spectral function becomes sizable. 
We also compared our results with available lattice simulations for $N=1$, finding a fair agreement within the large error bars, with a preference for certain values of the four-fermion couplings;
however our methods are expected to be more accurate when $N$ is large.
 In section \ref{S parameter} we estimated the contribution of the composite sector to the oblique parameter $S$, demonstrating
that it is under control.
 
In order to provide composite partners for the top quark, one needs to introduce additional constituent fermions $X^f$, 
in a different hypercolour representation,
such that fermion-trilinear baryons can be formed, with the quantum numbers of the top quark. 
A gauge theory with fermions in two different representations
presents qualitatively new features, such as one non-anomalous $U(1)$ flavour symmetry, 
with an associated Nambu-Goldstone meson $\eta_0$. 
In section \ref{Sum rules and pseudoscalar decay constants in the flavour-singlet sector} we showed that this implies two additional sum rules, as well as a mixing between the singlet scalars and pseudoscalars of the two sectors. 
In addition, the axial anomaly should only generate operators that respect the non-anomalous $U(1)$ symmetry. As a consequence, we demonstrated
in section \ref{thooft} that the effect of the anomaly is described by an operator of very large dimension, 
involving $4+12(N-1)$ fermions.
Our analysis of this operator correctly takes into account all the symmetries of the model, and thus provides fully coherent results, 
and its large dimension may indicate that the effects of the anomaly are suppressed in such a scenario.
On the other hand, we cannot exclude that such suppression is an artefact of our approximation of the true dynamics, in terms of fermionic operators only.

The dynamics of spontaneous flavour symmetry breaking also complicates in the presence of two sectors. 
Our analysis of anomaly matching in  section \ref{total-break} shows that the condensate $\langle\psi\psi\rangle$ necessarily forms, with the possible exception
of the case when $N$ is a multiple of $8$. However the condensate $\langle XX\rangle$ may not form in the presence of light, coloured baryons.
Indeed, in section \ref{Mass gap equations and effective four-fermions couplings} 
we showed that the system of two coupled mass-gap equations is very sensitive to the relative size of four-fermion couplings in the two sectors.
As the NJL techniques can provide information on the spectrum of coloured mesons only in the case of a non-vanishing mass gap,
we focused on the region of parameters where a non-zero $\langle XX\rangle$ develops as well. 
Let us remark that the solution of the gap equations corresponds to a stable minimum of the effective potential only for some range of the four-fermion couplings,
and of course meson masses are under control only within this range.
In the present case, it turns out that the potential is stable (no tachyons) as long as the operators induced by the axial anomaly are suppressed with respect to the others, 
by a factor of ten to one hundred, as described in section \ref{mixing-singlets}.
Therefore, we concentrated on the mass spectrum in this region of parameters.

We computed the masses of coloured mesons with the same techniques described for the electroweak sector. 
The results are illustrated in section \ref{spectrum-coloured}. Once again, spin-one mesons are extremely heavy, above
$\sim 5 f$. The situation is much more interesting for the coloured NGBs $G_c$, 
organised a real QCD octet and a complex sextet, 
which are massless in the chiral limit. We computed the contribution to their masses from gluon loops, and we found 
$M_{G_c} \gtrsim 1.5 f$, as long as $2N \lesssim 10$.
This may be sufficiently large to comply with present collider searches.
Therefore, contrary to common lore, it is not strictly necessary to introduce an explicit mass term $m_X XX$. 
Nonetheless, we studied also the case $m_X\ne 0$, as some qualitative features of the mass gap and of the meson spectrum are very sensitive to this parameter.
In particular, the singlet pseudoscalar $\eta_0$  is an exact NGB in the chiral limit, therefore its mass is controlled by the size of $m_X$ (and by the size of the
couplings to external SM fermions), as discussed in section \ref{mixing-singlets}.
A prominent opportunity for the discovery of composite NGBs at the LHC is offered by their anomalous
couplings to two SM gauge bosons, determined by the Wess-Zumino-Witten term. We provided the general formula for these couplings, and we specifically discussed the phenomenological consequences for the $\eta_0$ state.
The mass of the other singlet pseudoscalar $\eta^\prime$ is extremely sensitive to the effective anomaly coefficient:
one may have $M_{\eta'} \lesssim f$ for $\kappa_B/\kappa_A\ll 0.01$, but as soon as  $\kappa_B/\kappa_A \sim 0.1$
this state decouples, $M_{\eta'} \gtrsim 10 f$.
Finally, the heaviest singlet scalar $\sigma^\prime$ always lies in the multi-TeV range, while 
the lightest singlet scalar $\sigma_0$ may be as light as $f$.
Indeed, we already remarked that the vacuum provided by the mass-gap equations is stable only within 
specific ranges of the effective four-fermion couplings. Whenever the latter are close to the boundary of the stability region, 
$M_{\sigma_0}$ vanishes.
 In section \ref{sum-rules-full-model} we commented on the spectral sum rules in the presence of two sectors, illustrating in particular
the interplay among the singlet spectral functions.

We presented the first thorough analysis of the spectrum of meson resonances, in a confining gauge theory
with fermions in two different representations of the gauge group. The main limitation of this study is the absence
of interactions with external fermion fields. The interest of such interactions is twofold: to generate Yukawa couplings
between the composite Higgs and the SM fermions, and to induce radiatively a Higgs potential that realizes EWSB.
As a matter of fact, the coloured sector of the model is engineered to contain fermion-trilinear bound states, 
 which may mix linearly with the SM fermions. 
The mass spectrum of these baryons and their couplings to the mesons can be computed
by generalising the techniques used in this paper. Indeed, in the QCD literature, several analytical predictions for the masses
and couplings of baryons are consistent with experiments and with lattice simulations. 
Thus, one may predict the properties of composite top quark partners that reside in definite representations of the flavour group, and then compute the Higgs effective potential
induced by the top sector loops. Such a theory has a lesser number of free parameters than a generic composite Higgs model with
no specific ultraviolet completion, therefore the challenge will be to reproduce the Higgs mass
with a minimal amount of fine tuning of the parameters. 
We aim to study the fermion bound states of the theory
in a separate publication \cite{baryons-paper}.

\section*{Acknowlegdments} 

This work has been carried out thanks to the support of the OCEVU Labex (ANR-11-LABX-0060) 
and the A*MIDEX project (ANR-11-IDEX-0001-02) funded by the "Investissements d'Avenir" French government program managed by the ANR.
MF acknowledges partial support from the European UnionÕs Horizon 2020 research and innovation programme, 
under the Marie Sklodowska-Curie grant agreements No 690575 and No 674896.

\appendix

\section{Generators of the flavour group and embedding of the SM group}
\label{generators}

In this appendix, we give explicit representations for the generators of the flavour groups
$SU(4)$ and $SU(6)$ and describe how the SM gauge fields are coupled to the
elementary fermion fields. There are general procedures to construct a basis of the Gell-Mann
type for any $SU(n)$ group, starting from the well-known representations of the generators
for the cases $n = 2$ and $n = 3$, see for instance \cite{Bouzas:2003ju}. 
The relations in Eq.~(\ref{Tacom}) allow to distinguish the generators $T^A$ for the unbroken subgroups, $Sp(4)$ and $SO(6)$, 
from the generators $T^{\hat A}$ in the corresponding coset spaces. 
For $n=2N_f$ flavours, choosing the $2 N_f \times 2 N_f$ matrix $\Sigma_\varepsilon$ in the
form
\be
\Sigma_\varepsilon = \begin{pmatrix}
0 & 1\!\!1 \\ \varepsilon 1\!\!1 & 0
\end{pmatrix}
,
\ee
the general solution of Eq. (\ref{Tacom}) can be
expressed as \cite{Peskin:1980gc}
\be
T^A = \begin{pmatrix}
{\cal A}^A & {\cal B}^A \\ {\cal B}^{A\dagger} & - ({\cal A}^A)^T
\end{pmatrix}
,
\qquad\qquad
T^{\hat A} = \begin{pmatrix}
{\cal C}^{\hat A} & {\cal D}^{\hat A} \\ {\cal D}^{{\hat A}\dagger} & + ({\cal C}^{\hat A})^T
\end{pmatrix}
,
\ee
where the $N_f \times N_f$ submatrices ${\cal A}^A$ and ${\cal C}^{\hat A}$ are hermitian,
with ${\cal C}^{\hat A}$ traceless,
whereas $({\cal B}^A)^T = - \varepsilon {\cal B}^A$ and $({\cal D}^{\hat A})^T = + \varepsilon {\cal D}^{\hat A}$.

\subsection{The $SU(4)$ sector}
\label{SU4-generators}

According to the preceding discussion, the 15 $SU(4)$ generators can be chosen
as follows. The 10 generators of the subgroup $Sp(4)$ read
\be
T^{1,2,3,4} = \frac{1}{2\sqrt{2}} \begin{pmatrix}
\sigma_{1,2,3,0} & 0 \\ 0 & - \sigma_{1,2,3,0}^T
\end{pmatrix},\quad
T^{5,6,7} = \frac{1}{2\sqrt{2}} \begin{pmatrix}
0 & \sigma_{1,3,0} \\  \sigma_{1,3,0} & 0
\end{pmatrix},\quad
T^{8,9,10} = \frac{1}{2\sqrt{2}} \begin{pmatrix}
0 & i \sigma_{1,3,0} \\  -i \sigma_{1,3,0} & 0
\end{pmatrix}~,
\ee
where $\sigma_i$, $i=1,2,3$ denote the Pauli matrices while $\sigma_0$ stands for the $2\times 2$ unit matrix.
The corresponding coset $SU(4)/Sp(4)$ is then generated by the 5 matrices
\be
T^{{\hat 1},{\hat 2},{\hat 3}} = \frac{1}{2\sqrt{2}} \begin{pmatrix}
\sigma_{1,2,3} & 0 \\ 0 & \sigma_{1,2,3}^T
\end{pmatrix},
\qquad
T^{\hat 4} = \frac{1}{2\sqrt{2}} \begin{pmatrix}
0 & \sigma_{2} \\ \sigma_{2} & 0
\end{pmatrix},
\qquad
T^{\hat 5} = \frac{1}{2\sqrt{2}} \begin{pmatrix}
0 & i \sigma_{2} \\ -i \sigma_{2} & 0
\end{pmatrix}
.
\ee
The set of generators
\be
T^{1,2,3}_{L,R} = 
\frac{T^{7} \mp T^6}{\sqrt{2}} , 
\quad -\frac{T^{10} \mp T^9}{\sqrt{2}} , 
\quad \frac{T^4 \mp T^3}{\sqrt{2}}
\label{EW-generators}
\ee
constitute a $SU(2)_L \times SU(2)_R$ subalgebra of $Sp(4)$, and provide
the generators for the electroweak interaction and the custodial symmetry.
With this convention, a multiplet $\psi^a$ in the fundamental of $SU(4)$ and of $Sp(4)$ decomposes
as $(\psi^1~\psi^3)^T \sim (1_L,2_R)$ and $(\psi^2~\psi^4)^T \sim (2_L,1_R)$.
The generator $T^{\hat 3}$ is associated with a NGB singlet under
$SU(2)_L \times SU(2)_R$, whereas the remaining four generators of the
$SU(4)/Sp(4)$ coset correspond to the Higgs bidoublet $H$, transforming as $(2_L,2_R)$. Under the diagonal
$SU(2)_V$ subgroup, generated by $T^a_L + T^a_R$, 
the generators $T^{\hat 2}$, $T^{\hat 4}$, $T^{\hat 5}$ transform as a triplet,
and $T^{\hat 1}$ as a singlet.

The external electroweak gauge fields $W_\mu^{1,2,3}$
and $B_\mu$ will then couple to the $\psi$ fermions through the combination
\be
-i{\cal V}_\mu\equiv -i g \left( W_\mu^1 T^1_L + W_\mu^2 T^2_L + W_\mu^3 T^3_L \right) - i g^\prime B_\mu T^3_R
~.
\label{Vsource}\ee
According to Eq.~(\ref{rad_masses}), the masses of the NGBs that are radiatively
induced by the gauging are given by
\be
\Delta M_H^2 =  \Delta M_{\hat 1,\hat 2,\hat 4,\hat 5}^2 
= - \frac{3}{4 \pi} \times \frac{1}{F_G^2}
\int_0^\infty d Q^2 \, Q^2 \, \Pi^\psi_{V{\mbox -}A} (-Q^2)
\times
\frac{1}{16 \pi} ( 3 g^2 + g^{\prime 2} )
~,
\qquad\qquad
\Delta M_{\hat 3}^2 = 0~.
\label{rad_ew}
\ee
Of course, this positive contribution to the Higgs doublet mass should be overcome by a negative one from the top quark couplings, in order to trigger EWSB.

One can estimate quantitatively $\Delta M_H^2$ from the explicit form of the correlator $\Pi_{V-A}^\psi (-Q^2)$ as computed in the NJL approximation.
If one assumes further that the lightest resonances saturate in good approximation the correlator (see section \ref{secWSR}), the integrand takes the simplified form
\be
-Q^2 \overline\Pi_{V-A}^\psi(-Q^2) \simeq F^2_{G} +f^2_{A} M^2_{A} \frac{Q^2}{Q^2+M^2_{A}}-f^2_{V} M^2_{V} 
\frac{Q^2}{Q^2+M^2_{V}} ~,
\label{Q2PiVA}
\ee
where the expressions of the resonance masses and decay constants are explicitly given sections \ref{Masses and couplings of vector resonances}, \ref{Resummed correlators and the Goldstone decay constant} and \ref{secWSR}.
Integrating Eq.~(\ref{Q2PiVA}) over $Q^2$ up to the NJL cutoff $\Lambda^2$, one obtains
\be
-\int_0^{\Lambda^2} dQ^2 Q^2 \overline\Pi_{V-A}^\psi(-Q^2) 
\simeq
\left(F^2_{G} +f^2_{A} M^2_{A} -f^2_{V} M^2_{V}\right)\Lambda^2 + 
f^2_{V} M^4_{V} \ln \frac{\Lambda^2+M^2_{V}}{M^2_{V}} -
f^2_{A} M^4_{A} \ln \frac{\Lambda^2+M^2_{A}}{M^2_{A}} ~.
\label{rad_qcdfin}
\ee
Assuming that the Weinberg sum rules (\ref{WSR}) hold, 
the first term proportional to $\Lambda^2$ vanishes while the remaining terms simplify and lead to
\begin{equation}
\Delta M_H^2 \simeq \frac{3}{64 \pi^2} \frac{1}{F_G^2} (3g^2+g^{\prime 2}) ~f_V^2 M_V^4 \ln \frac{M_A^2}{M_V^2}~.
\end{equation}
This estimation of $\Delta M_H^2$ is of course relevant only if the $V-A$ correlator is well saturated by the lightest resonances
and the Weinberg sum rules hold.
%

\subsection{The $SU(6)$ sector}
\label{SU6-generators}

We decompose the 35 $SU(6)$ generators according to the $SO(6)$ subgroup
and the coset $SU(6)/SO(6)$. 
We denote by $\lambda_a$, $a=1,2,\ldots 8$, the $SU(3)$ Gell-Mann matrices, and we also define
$\lambda_0 = \sqrt{2/3}\ diag(1,1,1)$.
A convenient basis for  the 15 unbroken generators is given by
\be
T^{1,\cdots ,8,9} = \frac{1}{2\sqrt{2}} \begin{pmatrix}
\lambda_{1,\cdots ,8,0} & 0 \\ 0 & - \lambda_{1,\cdots ,8,0}^T
\end{pmatrix},
\quad
T^{10,11,12} = \frac{1}{2\sqrt{2}} \begin{pmatrix}
0 & \lambda_{2,5,7} \\  \lambda_{2,5,7} & 0
\end{pmatrix},
\quad
T^{13,14,15} = \frac{1}{2\sqrt{2}} \begin{pmatrix}
0 & i \lambda_{2,5,7} \\  -i \lambda_{2,5,7} & 0
\end{pmatrix}.
\ee
The eight generators $T^{1,\cdots ,8}$ together with $T^9$ form a $SU(3)_C\times U(1)_D$ maximal subalgebra, 
that can accommodate the strong interaction gauge group,
as well as a part of the hypercharge gauge group $U(1)_Y$, with $Y=T^3_R + D$,
where $T^3_R$ is defined in Eq. (\ref{EW-generators}) and $D=(4/\sqrt{3})\cdot T_9$.
The 20 broken generators  read
\be\begin{array}{c}
T^{{\hat 1},\cdots ,{\hat 8}} = \dfrac{1}{2\sqrt{2}} \begin{pmatrix}
\lambda_{1,\cdots ,8} & 0 \\ 0 & \lambda_{1,\cdots ,8}^T
\end{pmatrix},~
\\
T^{{\hat 9},\cdots,{\hat{14}}}
= \dfrac{1}{2\sqrt{2}} \begin{pmatrix}
0 & \lambda_{1,3,4,6,8,0} \\ \lambda_{1,3,4,6,8,0} & 0
\end{pmatrix},~
\quad
T^{{\hat{15}},\cdots,{\hat{20}}} 
= \dfrac{1}{2\sqrt{2}} \begin{pmatrix}
0 & i \lambda_{1,3,4,6,8,0} \\ -i \lambda_{1,3,4,6,8,0} & 0
\end{pmatrix}.
\end{array}\ee
The generators $T^{{\hat 1},\cdots ,{\hat 8}}$ are associated to the NGBs multiplet $O_c\sim 8_0$ under $SU(3)_C\times U(1)_D$,
while $T^{{\hat 9},\cdots ,{\hat 20}}$ correspond to the NGBs $(S_c + \overline{S}_c) \sim (6_{4/3}+\overline{6}_{-4/3})$.

The constituent fermions $X$ transform as  $(3_{2/3}+\overline{3}_{-2/3})$ under $SU(3)_C\times U(1)_D$,
where the normalization of the $D$-charge is chosen to reproduce the correct hypercharge of top quark partners.
Therefore, the external colour gauge fields $G_\mu^{1,\cdots,8}$
and $B_\mu$ couple to the $X$ fermions through the combination
\be
-i g_c \sqrt{2} G_\mu^a T^a - i g^\prime \dfrac{4}{\sqrt{3}} B_\mu T^9 ~.
\ee
According to Eq.~(\ref{rad_masses}), the masses of the NGBs that are radiatively
induced by the gauging are given by
\be\begin{array}{l}
\Delta M_{O_c}^2 =  \Delta M_{\hat 1,\cdots, \hat 8}^2
= - \dfrac{3}{4 \pi} \times \dfrac{1}{F_{G_c}^2}
{\displaystyle\int}_0^\infty d Q^2 \, Q^2 \, \Pi^X_{V{\mbox -}A} (-Q^2)
\times
 \dfrac{3}{4 \pi} g_s^2 
~,\\
 \Delta M_{S_c}^2 =  \Delta M_{\hat 9,\cdots, \hat 20}^2
= - \dfrac{3}{4 \pi} \times \dfrac{1}{F_{G_c}^2}
{\displaystyle\int}_0^\infty d Q^2 \, Q^2 \, \Pi^X_{V{\mbox -}A} (-Q^2)
\times
\dfrac{1}{4\pi} \left( \dfrac{10}{3} g_s^2 + \dfrac{16}{9} g'^2 \right)~.
\end{array}
\label{rad_qcd}
\ee
The quantitative estimate of the integral of the $V-A$ two-point function is discussed in section \ref{Masses of coloured scalar resonances}.

\section{Loop functions}
\label{loop-functions}

The one-loop integrals relevant for our purposes are 
the one- and two-point functions,
\be
\t A_0(m^2)\equiv i\int\frac{d^4 k}{(2 \pi)^4}
\frac{1}{k^2 -m^2+i\epsilon}~,\qquad
\t B_0(p^2,m^2)\equiv i\int \frac{d^4 k}{(2 \pi)^4}
\frac{1}{\left(k^2 -m^2 \right)\left[\left(p+k\right)^2 -m^2 \right]}~.
\ee
[We adopted the notation $\t A_0$ and $\t B_0$ in order to avoid confusion
with the standard  one-loop functions $A_0$ and $B_0$ \cite{Passarino:1978jh}, which are defined in Euclidean metric and dimensional 
regularisation, and differ also from the above by an overall factor
$i(16\pi^2)$ in $D=4$ dimensions.]

In the context of the NJL model, the one-point function is regularised by introducing a cut-off $\Lambda$ on the Euclidean four-momentum,
\be
\t A_0(m^2) = \frac{\Lambda^2}{16 \pi^2}
\left[ 1- \frac{m^2}{\Lambda^2}\ln \frac{\Lambda^2+ m^2}{m^2}\right]~.
\label{A0def}
\ee
The zero-momentum two-point function is given by 
\be
\t B_0(0,m^2) =  \frac{d \t A_0(m^2)}{d m^2}
= \frac{1}{16 \pi^2}
\left[\frac{\Lambda^2}{\Lambda^2+m^2}  -\ln \frac{\Lambda^2+m^2}{m^2}\right]
=\frac{1}{16 \pi^2} \left[ 1 -\ln \frac{\Lambda^2}{m^2} +{\cal O}\left(\frac{m^2}{\Lambda^2}\right)\right]~.
\label{B00}\ee
For the finite, $p^2$-dependent part of the two-point function, we adopt the simple regularisation
\be
\t B_0(p^2,m^2) = \t B_0(0,m^2) + \frac{1}{32\pi^2} f \left(\frac{p^2}{4m^2}\right) ~,
\label{B0exp}
\ee
where
\be
f(r)  =  \left\{
\begin{array}{ll}
4 \left(\dfrac{1-r}{r} \right)^{1/2}\, \arctan \left(\dfrac{r}{1-r} \right)^{1/2} - 4 & {\rm~~~~(for~} 0 < r < 1) \\
\\
4 \left(\dfrac{r-1}{r} \right)^{1/2}\,\left[ \ln(\sqrt{r}+\sqrt{r-1}) -i\,\dfrac{\pi}{2}\right] - 4 & {\rm~~~~(for~} 1<r)\\
\\
4 \left(\dfrac{r-1}{r} \right)^{1/2}\,\left[ \ln(\sqrt{-r}+\sqrt{1-r}) \right] - 4 & {\rm~~~~(for~} r<0)~.
\end{array}
\right.
\ee
We remark that the finite terms are regularisation-dependent, therefore our expression may differ from analogous ones in the NJL literature at order $p^2/\Lambda^2$.

\section{Two-point correlators of fermion bilinears at one loop}
\label{SDresum}

In this appendix we present the detailed computation 
of the five one-loop two-point functions $\t \Pi_\phi (q^2,M_f^2)=\t \Pi_\phi^f(q^2)$ where $\phi=\{S, P, V, A, AP \}$ and $M_f$ is the dynamical mass of the hypercolour fermions $f=\psi, X$.
These two-point functions are crucial quantities in the NJL model as they are involved in the estimation of the masses and decay constants of the electroweak and coloured composite resonances (see sections \ref{The electroweak sector} and \ref{The spectrum of mesonic resonances in the coloured sector}).
For the two-component Weyl spinors, we follow the conventions of Ref.~\cite{Dreiner:2008tw} ($\psi$ and $\psi^\dagger$ propagate in the loops).
The Feynman rules appearing in the vertices can be extracted from the currents and densities given respectively in Eqs.~(\ref{Jdef}) and (\ref{S_and_P}).

Let us first focus on the electroweak sector.
In the scalar and pseudoscalar non-singlet channels we get
\begin{eqnarray}
i \t \Pi_{S(P)}^\psi(q^2) \delta^{\hat{A}\hat{B}}
&= & 
(-1)
\int^\Lambda \frac{d^4 k}{(2\pi)^4} 
Tr \left[i \Sigma_0 T^{\hat{A}} \Omega\Gamma_{S(P)}~ \frac{i \sigma \cdot k}{k^2-M^2_\psi} ~
i T^{\hat{B}}\Sigma_0 \Omega \Gamma_{S(P)}^\dagger ~\frac{i \overline{\sigma} \cdot (k+q)}{(k+q)^2-M^2_\psi} \right]
\nonumber
\\
&+&
(-1) 
\int^\Lambda \frac{d^4 k}{(2\pi)^4} 
Tr \left[i \Sigma_0 T^{\hat{A}} \Omega \Gamma_{S(P)} ~\frac{i M_\psi \Sigma_0 \Omega}{k^2-M^2_\psi} 
~i \Sigma_0 T^{\hat{B}} \Omega \Gamma_{S(P)} ~\frac{i M_\psi \Sigma_0 \Omega}{(k+q)^2-M^2_\psi} \right]~,
\label{Pi-general}
\end{eqnarray}
where the first (second) integral corresponds to the loop involving the kinetic (massive) part of the propagators.
The factors $\Gamma_{S(P)}= 1 ~(i)$, which distinguish the scalar and pseudoscalar channels, are a consequence of  Eq.~(\ref{S_and_P}).
These factors are the equivalent of the $\gamma_5$ matrix in Dirac notation and they give a relative sign between the two channels in the second term of Eq.~(\ref{Pi-general}), exactly like in QCD.
Similarly for the vector and axial-vector two points functions one obtains
\begin{eqnarray}
i \t \Pi^{\mu \nu, AB(\hat{A}\hat{B})}_{V(A)} (q^2,M_\psi^2)
&= & 
(-1)
\int^\Lambda \frac{d^4 k}{(2\pi)^4} 
Tr \left[i T^{A(\hat{A})} \overline{\sigma}^\mu~ \frac{i \sigma \cdot k}{k^2-M^2_\psi} ~
i  T^{B(\hat{B})}  \overline{\sigma}^\nu ~\frac{i  \sigma \cdot(k+q)}{(k+q)^2-M^2_\psi} \right]
\nonumber
\\
&+&
(-1) 
\int^\Lambda \frac{d^4 k}{(2\pi)^4} 
Tr \left[i  T^{A(\hat{A})}   \overline{\sigma}^\mu ~\frac{i M_\psi \Sigma_0 \Omega}{k^2-M^2_\psi} 
~ (-i T^{B(\hat{B})})^T  \sigma^\nu ~\frac{i M_\psi \Sigma_0 \Omega }{(k+q)^2-M^2_\psi} \right]~,
\label{Pi-general-vector}
\end{eqnarray}
where the functions $\t \Pi^{\mu \nu, AB(\hat{A}\hat{B})}_{V(A)} (q^2)$ are defined in Eq.~(\ref{PiV1loop}).
The vector and axial-vector channels only differ by the flavour trace [see Eqs.~(\ref{Tacom}) and (\ref{norm_T})] which again gives a relative sign between the two channels in the second integral.
Finally, for the axial pseudoscalar two-point function one has
\begin{eqnarray}
 i\t \Pi_{AP}^{\mu, \hat{A} \hat{B}}(q^2, M_\psi^2)\equiv i \t \Pi_{AP}^\psi(q^2)  p^\mu \delta^{\hat{A}\hat{B}}
&=&(-1)\int^\Lambda \frac{d^4 k}{(2\pi)^4} 
Tr \left[i T^{\hat{A}}   \overline{\sigma}^\mu ~\frac{i   \sigma \cdot k}{k^2-M^2_\psi} 
~i   T^{\hat{B}}\Sigma_0 \Omega \Gamma_{P} ~\frac{i M_\psi \Sigma_0 \Omega }{(k+q)^2-M^2_\psi} \right]
\nonumber
\\
&+& (-1)\int^\Lambda \frac{d^4 k}{(2\pi)^4} 
Tr \left[i T^{\hat{A}}   \cdot \overline{\sigma}^\mu ~\frac{i M_\psi \Sigma_0 \Omega}{k^2-M^2_\psi} 
~i  \Sigma_0 T^{\hat{B}}  \Omega \Gamma_{P}^\dagger  ~\frac{i   \sigma \cdot (k+q) }{(k+q)^2-M^2_\psi} \right]~,
\end{eqnarray}
where this time the integrals contain both the kinetic and the massive parts of the propagators.
Evaluating the Lorentz, flavour and hypercolour traces, one can check that the above equations are 
quite consistent with the ones given in table \ref{tab_phi}.
Note that the correlators in the singlet channels are obtained by replacing the generators $T^{\hat{A}}$ by the normalised identity matrix $T^0_\psi$ which only changes the flavour tensor structure of the loops, leading to the same result for the two-point functions $\t \Pi_\phi^f(q^2)$. 

Let us now turn to the correlators of the coloured $SU(6)$ sector.
The latter can be derived in complete analogy with the ones in the electroweak sector.
Besides the obvious replacements $M_\psi \rightarrow M_X$, $\Sigma_0\rightarrow \Sigma_0^c$ and $T^0_\psi \rightarrow T^0_X$, the major modification originates from the hypercolour traces.
Indeed, the  fermions $X$ are in the two-index antisymmetric and traceless representation of $Sp(2N)$.
Consequently, the hypercolour traces give a factor $(2N+1)(N-1)$ [instead of $(2N)$ 
\footnote{More precisely, due to the antisymmetry of the hypercolour singlet contractions, the corresponding traces of the electroweak sector contribute to the one-loop functions with a factor $\pm (2N)$ where the sign corresponds to a particular (massive or kinetic) loop in a given channel.
The minus sign is always compensate by the flavour trace which contains in that case $\Sigma_0^2=-1\!\!1$.
On the contrary, the hypercolour and flavour contractions in the coloured sectors are symmetric and always positive.}] which of course corresponds to the dimension of the hypercolour $X-$representation.
Note that this difference with respect to the electroweak sector can easily be inferred 
by considering the vector form $X^{\hat{I}}$ [$\hat{I}=1, \cdots ,(2N+1)(N-1)$] defined in Eq.~(\ref{XXrelation}).
Then, the one-loop two-point functions $\t \Pi^X_\phi(q^2)$, summarised in table \ref{tab-coloured-functions},  
are related to the ones in the electroweak sector as follow
\begin{equation}
\tilde{\Pi}_\phi^\psi(q^2)=\tilde{\Pi}_\phi(q^2, M_\psi^2,2N)~,
\qquad\qquad
 \tilde{\Pi}_\phi^X(q^2)=\tilde{\Pi}_\phi[q^2, M_X^2,(2N+1)(N-1)]~.
\end{equation} 

\begin{figure}[tb]
\includegraphics[scale=1.05, trim= 70 90 0 70]{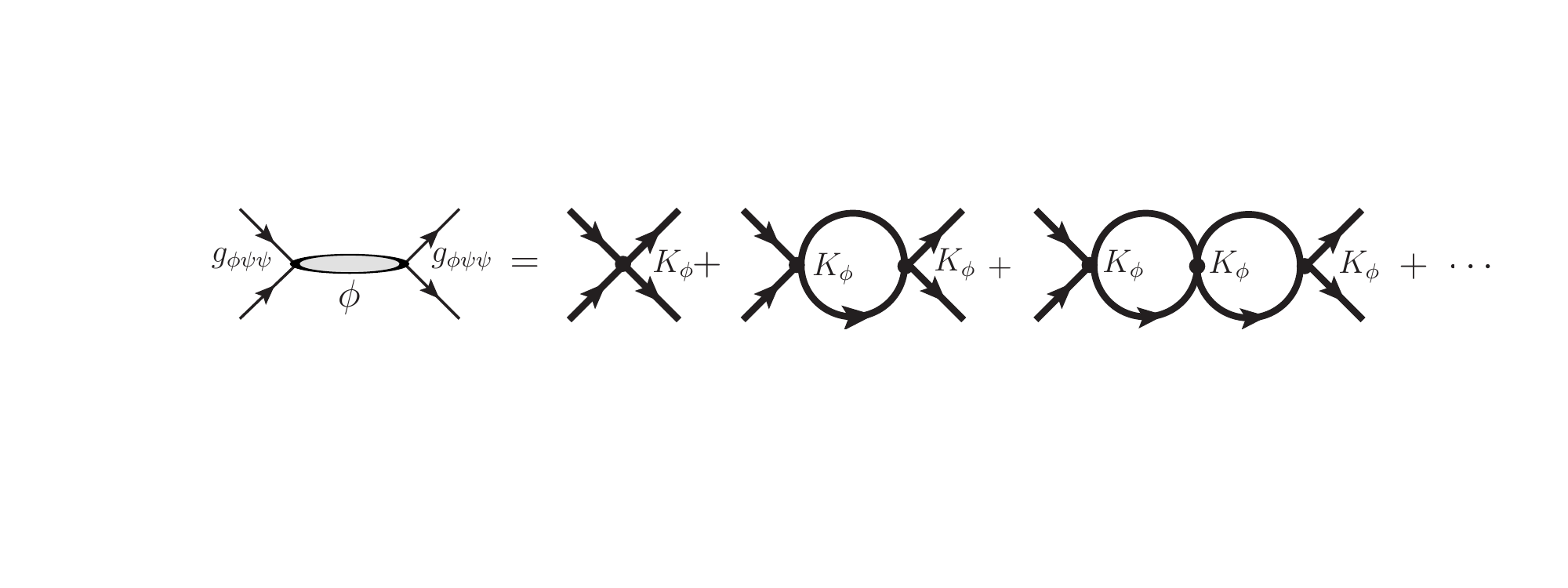}
\caption{Resummation of leading $1/N$ graphs for a mesonic T-matrix element, $\overline{T}_\phi$, corresponding to a composite meson exchange.}
\label{BS2}
\end{figure}

As explained in section \ref{Masses and couplings of scalar resonances}, the resummation of the above one-loop two-point functions,
at leading order in $1/N$,
gives the NJL resummed correlators, $\overline{\Pi}_\phi$, 
from which the masses and decay constants of the composite resonances are extracted.
Usually, in the NJL literature, one considers the $T$-matrix element $\overline{T}_\phi(q^2)$, rather than $\overline{\Pi}_\phi(q^2)$.
As illustrated in Fig.~\ref{BS2}, the geometrical series that defines $\overline{T}_\phi$ starts with the 
four-fermion interaction $K_\phi$, instead of the one-loop two-point function $\t \Pi^f_\phi(q^2)$, see Fig.~\ref{BS}.
Consequently the $T$-matrix element is given by
\be
\overline{T}_\phi(q^2) = 
\frac{ K_\phi}{1- \,2 K_\phi\,\t\Pi^f_{\phi}(q^2)}~.
\label{T-matrix}
\ee
The poles of $\overline{T}_\phi(q^2)$ and of $\overline{\Pi}_\phi(q^2)$ are of course identical and are given 
by $1= 2 K_\phi\,\t\Pi^f_{\phi}(M_\phi^2)$.
The only difference comparing Eqs.~(\ref{PiSPsum}) and (\ref{T-matrix}) comes from the numerators of the series, which 
lead different to residues.
The residues of $\overline{\Pi}^f_\phi$ have been extensively studied in sections \ref{The electroweak sector} and \ref{The spectrum of mesonic resonances in the coloured sector} while the residues of the $T$-matrix are the couplings $g_{\phi ff}$ of the physical resonance $\phi$ to the fundamental fermions $f$.
In analogy with Eq.~(\ref{limit}), these couplings are given by
\begin{equation}
g_{\phi ff}^{2}=- \lim\limits_{q^2\rightarrow M_\phi^2} (q^2-M_\phi^2) \overline{T}_\phi(q^2)
= \left[ 2 \left.\frac{d  \t\Pi_{\phi}^f(q^2)}{d q^2}\right|_{q^2=M_\phi^2} \right]^{-1}
~.
\end{equation}
They behave like $\simeq 1/\sqrt N$,  as expected from general large-$N$ considerations.

\section{Relating four-fermion operators by Fierz identities}
\label{fierz}

The couplings of the various four-fermions operators may be related under some assumption on the underlying dynamics (see Refs.~\cite{Klimt:1989pm,Buck:1992wz} for the case of QCD). 
In this way one can predict the relative strength of the various physical channels (spin-zero versus spin-one, electroweak sector versus colour sector, etc.).
We will start from $Sp(2N)$ current-current operators, that  encode the ultraviolet dynamics in the `ladder' approximation, 
that holds when $N$ is (moderately) large,
and we will use Fierz transformations to generate the various $Sp(2N)$ singlet-singlet operators.
We will also take this opportunity to summarise general results on Fierz transformations associated to  the $SU(N)$ and $Sp(2N)$ groups.

\subsection{Hypercolour current-current operators}
\label{$Sp(2N)$ current-current operators}

Let us derive the $Sp(2N)$ current-current operators from the covariant derivatives of the fermions $\psi$ and $X$.
They belong to the fundamental representation, $\psi \sim {\Yvcentermath1 \tiny \yng(1)}$, and to the two-index, 
traceless ($X_{ij} \Omega_{ji}=0$) and antisymmetric ($X_{ij}=-X_{ji}$) representation, $X\sim {\Yvcentermath1 \tiny \yng(1,1)}$. 
The covariant derivatives read
\begin{equation}
\left(D^\mu \psi \right)_i=\left[\partial^\mu \delta_{ij}-i g_{HC} (T^I)_{ij} {\cal G}^\mu_I \right] \psi_j~,
\label{Covariant1}
\end{equation}
\begin{equation}
\left(D^\mu X \right)_{ij}= 
\partial^\mu X_{ij}-i g_{HC} \left[ (T^I)_{ik} X_{kj}+ (T^I)_{jk} X_{ik}\right] {\cal G}^\mu_I
=
\left[ \partial^\mu \delta_{ik} \delta_{jl}-i g_{HC} 
(T^I_X)_{ijkl} {\cal G}^\mu_I \right] X_{kl}~,
\label{Covariant2}
\end{equation}
where ${\cal G}^\mu_I$ are the hypergluon fields, and $g_{HC}$ is the hypercolour gauge coupling. 
The hypercolour generators acting on $\psi_j$,
$(T^I)_{ij}$, and on $X_{kl}$,  $(T^I_X)_{ijkl} \equiv (T^I)_{ik} \delta_{jl} - \delta_{il}(T^I)_{jk} $, 
are normalised as 
\begin{equation}
{\rm Tr} (T^I T^J) \equiv \frac{1}{2}\ell\left( {\Yvcentermath1 \tiny \yng(1)} \right)\delta^{IJ} = \frac{1}{2}\delta^{IJ}  ~,
\qquad\qquad
{\rm Tr} ( T^I_X T^J_X )
\equiv
(T^I_X)_{ijkl} (T^J_X)_{klij} \equiv \frac{1}{2} \ell\left( {\Yvcentermath1 \tiny \yng(1,1)} \right)\delta^{IJ} =(N-1) \delta^{IJ}~.
\end{equation}
The non-derivative terms in Eqs.~(\ref{Covariant1}) and (\ref{Covariant2}) determine the coupling of the technigluons to the $Sp(2N)$-currents ${\cal J}^{\mu I}_\psi$ and ${\cal J}^{\mu I}_X$, which transform under 
the adjoint representation ${\Yvcentermath1 \tiny \yng(2)}$,
\begin{equation}
{\cal L}_{UV}=g_{HC}\left( {\cal J}^{\mu I}_\psi + {\cal J}^{\mu I}_X\right) {\cal G}_{\mu I} ~,
\label{LUVint}
\end{equation}
where 
\begin{equation}
{\cal J}^{\mu I}_\psi= 
\psi \left(\Omega T^I \right) \sigma^\mu \overline{\psi} ~, 
\qquad\qquad
{\cal J}^{\mu I}_X= 2\ {\rm Tr}\left[
X \left(\Omega T^I \right) \sigma^\mu \overline{X}\Omega \right]~.
\label{SP2N-currents}
\end{equation}
Here $\Omega_{ij}$ is the $Sp(2N)$ invariant tensor, the trace is taken over $Sp(2N)$ indexes, 
and the expression of ${\cal J}^{\mu I}_X$ has been simplified using  
${\rm Tr}\left[ X \Omega \sigma^\mu \overline{X} \left(\Omega T^I \right) \right] = - {\rm Tr}\left[X \left(\Omega T^I \right) \sigma^\mu \overline{X}\Omega\right]$.
It is understood that each fermion flavour $\psi^a$ ($X^f$) behaves equally with respect to the $Sp(2N)$ dynamics, 
that is, the $Sp(2N)$ currents are flavour singlets.
It will be useful to rearrange the fermion components $X_{ij}$ as a vector $X^{\hat{I}}$, with one index $\hat I$ of the representation $\Yvcentermath1 \tiny \yng(1,1)$,
\begin{equation}
X_{ij}= \sqrt{2}  (T^{\hat{I}}\Omega)_{ij} X^{\hat{I}} ~, 
\qquad\qquad
X^{\hat{I}}= -\sqrt{2} (\Omega T^{\hat{I}})_{ij} X_{ji} ~,
\label{XXrelation}
\end{equation}
so that the second current in Eq.~(\ref{SP2N-currents}) can be written in terms of the generators in the representation $\Yvcentermath1 \tiny \yng(1,1)$,
that are given by $SU(2N)$ structure constants,
\be
{\cal J}^{\mu I}_X = X^{\hat{I}}  (T^I_{\Yvcentermath1 \tiny \yng(1,1)} )^{\hat{I}\hat{J}}  \sigma^\mu  \overline{X}^{\hat{J}} ~,
\qquad\qquad
(T^I _{\Yvcentermath1 \tiny \yng(1,1)} )^{\hat{I}\hat{J}} \equiv i f^{\hat{I} I \hat{J}} = 2 {\rm Tr}\left( [T^{\hat{I}},T^I ] T^{\hat{J}} \right)~.
\ee

We assume that the confining strong dynamics can be described, in first approximation, by the exchange of one hypergluon which acquired a dynamical mass, 
which is the usual NJL assumption in QCD 
\cite{Klevansky:1992qe}. Then, the strong dynamics is supposed to generate, in the 'ladder' approximation,
 $Sp(2N)$ current-current operators only,
\begin{equation}
{\cal L}_{eff} = 
\frac{\kappa_{UV}}{2N} \biggl[
{\cal J}^{\mu I}_\psi {\cal J}_{\psi \mu}^I
+
{\cal J}^{\mu I}_X {\cal J}_{X \mu}^I 
+2
{\cal J}^{\mu I}_\psi {\cal J}_{X \mu}^I
 \biggr] ~,
\label{LUV0}
\end{equation}
where 
$\kappa_{UV}/(2N) \sim g_{HC}^2/\Lambda^2$ stands for the exchange of one `massive' hypergluon.
The large-$N$ scaling of the gauge coupling is $g_{HC}\sim 1/ \sqrt{2N}$, while $\kappa_{UV}$ and $\Lambda$ are $N$-independent.
The operators in Eq.~(\ref{LUV0}) are the product of fermion bilinears in the adjoint representation of $Sp(2N)$.
In order to study physical resonances, which correspond to $Sp(2N)$-singlet fermion bilinears, 
we need to rewrite these operators by using Fierz transformations in the Lorentz, flavour and  hypercolour spaces.
Note that the last operator in Eq.~(\ref{LUV0}) does not contribute to any meson resonance, because by a Fierz transformation one obtains only  `diquark-diquark' operators, such as $(\psi X) (\overline{\psi}\, \overline{X})$, 
which are not hypercolour singlets, and therefore are not relevant for our analysis.

The Fierz transformations of Weyl indices are determined by the well-known identities 
\begin{equation}
\left( \sigma^\mu \right)_{\alpha \dot{\alpha}} \left( \sigma_\mu \right)_{\beta \dot{\beta}} =
-\left( \sigma^\mu \right)_{\alpha \dot{\beta}} \left( \sigma_\mu \right)_{\beta \dot{\alpha}}=
2 ~\varepsilon_{\alpha \beta} \varepsilon_{\dot{\alpha} \dot{\beta}} 
~. 
\label{fierz-weyl}
\end{equation}
The $SU(N)$ and $Sp(2N)$ Fierz transformations, relevant for flavour and hypercolour indexes respectively, are presented in sections \ref{Fierz-transfos-SUN} and \ref{Fierz-transfoSP2N}
below.

\subsection{General properties of Fierz transformations}
\label{Fierz-general-section}

In this section we derive general properties of the coefficients in Fierz transformations. 
For a given irreducible representation $R$ of the symmetry group under consideration, let us construct the tensor products 
${\cal R} \otimes \overline{{\cal R}} = \sum_{\cal_A} {\cal R}_{\cal A}$ and ${\cal R} \otimes {\cal R} = \tilde\sum_{\cal A} {\cal R}_{\cal  A}$, 
where the index ${\cal A}$ runs over the irreducible representations contained in the product.
One can choose  \cite{Brauner} a set of matrices $\{\Gamma^{\cal A}_a \}$  ($\{\tilde\Gamma^{\cal A}_a \}$), with $a=1,\cdots, {\rm dim}{\cal R}_{\cal A}$, 
which form a basis of the vector space ${\cal R} \otimes \overline{{\cal R}}$ (${\cal R} \otimes {\cal R}$).
In the following, we will add a tilde wherever there is no conjugate in the tensor product.
Such matrices have size ${\rm dim}{\cal R}\times {\rm dim}{\cal R}$ and satisfy the orthogonality relations
\begin{equation}
{\rm Tr}(\Gamma_a^{\cal A} \Gamma_b^{\cal B})=\alpha ~\delta^{\cal A B} g^{\cal A}_{ab}~,
\qquad\qquad
{\rm Tr}(\tilde\Gamma_a^{\cal A} \tilde\Gamma_b^{{\cal B}\dagger})=\alpha ~\delta^{\cal A B} g^{\cal A}_{ab}~,
\end{equation}
where $\alpha$ 
is a normalisation constant and $g^{\cal A}_{ab}$ is a generic metric (in particular, $g^{\cal A}_{ab}g^{{\cal A}bc}=\delta_a^c$ and
$\Gamma^{a{\cal A}}\equiv g^{{\cal A}ab} \Gamma^{\cal A}_b$).
Any ${\rm dim}{\cal R}\times {\rm dim}{\cal R}$ matrix $M$ can be decomposed on the basis $\{\Gamma^{\cal A}_a \}$ as
\begin{equation}
M=\sum\limits_{\cal A} \sum\limits_a c^{a{\cal A}} \Gamma^{\cal A}_a
=\sum\limits_{\cal A}^\sim \sum\limits_a d^{a{\cal A}} \tilde\Gamma^{\cal A}_a~,
\qquad\qquad
c^{a{\cal A}}= \frac{1}{\alpha} {\rm Tr} (\Gamma^{a{\cal A}} M)~,
\qquad
d^{a{\cal A}}= \frac{1}{\alpha} {\rm Tr} (\tilde\Gamma^{a{\cal A}\dagger} M)~.
\end{equation}
Replacing the explicit form of $c^{a{\cal A}}$ and $d^{a{\cal A}}$ in $M$ we obtain  the completeness relations
\begin{equation}
\sum\limits_{\cal A} \sum\limits_a (\Gamma^{a{\cal A}})_{ij} (\Gamma_a^{\cal A})_{kl} =
\sum\limits_{\cal A}^\sim \sum\limits_a (\tilde\Gamma^{a{\cal A}})_{ij} (\tilde\Gamma_a^{{\cal A}\dagger})_{kl}
= \alpha~ \delta_{il} \delta_{kj}~.
\label{Fierz-completeness-relation}
\end{equation}
which are relevant to derive the Fierz coefficients.

Let us consider an interaction among four objects transforming as  
$({\cal R} \otimes \overline{{\cal R}})_{\cal A} ({\cal R} \otimes \overline{{\cal R}})_{\cal A}$, where the
subscripts indicate that each pair is contracted in the component ${\cal R}_{\cal A}$.  Then, the Fierz transformations can be written as
\begin{equation}
\sum\limits_a (\Gamma^{a{\cal A}})_{ij} (\Gamma_a^{\cal A})_{kl}
= \sum\limits_{\cal B} C_{{\cal A} {\cal B}} \sum\limits_b (\Gamma^{b{\cal B}})_{il} (\Gamma_b^{\cal B})_{kj}
= \sum\limits_{\cal B}^{\sim} D_{{\cal A} {\cal B}} \sum\limits_b (\tilde \Gamma^{b{\cal B}})_{ik} (\tilde \Gamma_b^{{\cal B}\dagger})_{jl}~,
\label{Fierz-general}
\end{equation}
where $C_{{\cal A} {\cal B}}$ and $D_{{\cal A} {\cal B}}$ are the Fierz coefficients for the channels $j\leftrightarrow l$ and $j\leftrightarrow k$, 
respectively.
In terms of `quarks' $\sim {\cal R}$ and `antiquarks' $\sim \overline{\cal R}$, one can dub them  
the `quark-antiquark' and  the `quark-quark' channels, respectively. Analogously, for the interaction $({\cal R} \otimes {\cal R})_{\cal A} (\overline{\cal R} \otimes \overline{\cal R})_{\overline{\cal A}}$,
the Fierz transformations read
\begin{equation}
\sum\limits_a (\tilde\Gamma^{a{\cal A}})_{ij} (\tilde\Gamma_a^{{\cal A}\dagger})_{kl}
= \sum\limits_{\cal B} \tilde C_{{\cal A} {\cal B}} \sum\limits_b (\Gamma^{b{\cal B}})_{il} (\Gamma_b^{{\cal B}T})_{kj}
= \sum\limits_{\cal B} \tilde D_{{\cal A} {\cal B}} \sum\limits_b (\Gamma^{b{\cal B}})_{ik} (\Gamma_b^{\cal B})_{jl}~,
\label{Fierz-general-bis}
\end{equation}

One can derive several, general constraints on the Fierz-coefficient matrices $C,D,\tilde C,\tilde D$. 
Applying twice a Fierz transformation on the same indexes 
the original contraction is recovered, therefore one obtains
\begin{equation}
\sum\limits_{\cal B} C_{{\cal A} {\cal B}} C_{{\cal B} {\cal C}}=\delta_{{\cal A}{\cal C}}~,
\qquad
\sum\limits_{\cal B}^\sim D_{{\cal A} {\cal B}} \tilde{D}_{{\cal B} {\cal C}}=\delta_{{\cal A}{\cal C}}~,
\qquad
\sum\limits_{\cal B} \tilde C_{{\cal A} {\cal B}} D_{{\cal B} {\cal C}}=  s_{\cal A} \delta_{{\cal A}{\cal C}}~,
\qquad
\sum\limits_{\cal B} \tilde D_{{\cal A} {\cal B}} D_{{\cal B} {\cal C}}=\delta_{{\cal A}{\cal C}}~,
\label{Fierz-sum}
\end{equation}
where $s_{\cal A} = +1$ ($ -1$) when the representation ${\cal R}_{\cal A}$ belongs to the (anti-)symmetric part of the tensor product ${\cal R} \otimes {\cal R}$,
and correspondingly the matrices $\tilde\Gamma^{\cal A}_a$  are (anti-)symmetric. Therefore, one has $C=C^{-1}$, while both $\tilde C$ and $\tilde D$ can be fully determined in terms of the matrix $D$.
The contraction associated to the singlet representation,  ${\cal R}_\bullet \subset {\cal R} \otimes \overline{{\cal R}}$,
can be chosen as $\Gamma^\bullet_{ij} = \delta_{ij}\sqrt{\alpha/{\rm dim}{\cal R}}$. 
Therefore,
Eq.~(\ref{Fierz-completeness-relation}) determines the first row of $C$ and $D$,
\be 
C_{\bullet {\cal A}}=\dfrac{1}{{\rm dim} {\cal R}}~,\quad \forall \ {\cal R}_{\cal A}\subset {\cal R} \otimes \overline{{\cal R}}~,
\qquad\qquad
D_{\bullet {\cal A}}=\dfrac{s_{\cal A}}{{\rm dim} {\cal R}}~,\quad \forall \ {\cal R}_{\cal A}\subset {\cal R} \otimes {\cal R}~.
\label{FierzSinglet}\ee
Indeed, from Eq.~(\ref{Fierz-general})  one can obtain explicit expressions of the Fierz coefficients,
\begin{equation}
C_{\cal AB}= \frac{1}{\alpha^2}\sum\limits_a {\rm Tr} [\Gamma^{a {\cal A}} \Gamma^{ {\cal B}}_b \Gamma^{{\cal A}}_a \Gamma^{b {\cal B}}]~,
\qquad\qquad
D_{\cal AB}= \frac{1}{\alpha^2}\sum\limits_a {\rm Tr} [\Gamma^{a {\cal A}} (\tilde\Gamma^{ {\cal B}}_b)^T (\Gamma^{{\cal A}}_a)^T \tilde\Gamma^{b {\cal B}\dagger}]~,
\label{Fierz-formal-expression}
\end{equation}
which are valid for every $b$. The direct computation of such expressions, however, may be very complicated in practice.
By summing over $b$ the two identities in Eq.~(\ref{Fierz-formal-expression}), one obtains quantities invariant under the exchanges ${\cal A} \leftrightarrow {\cal B}$ and $C\leftrightarrow C^{-1}$ ($D\leftrightarrow D^{-1}$), 
therefore one concludes that 
\begin{equation}
C_{{\cal A} {\cal B}} ~ {\rm dim}{\cal R}_{\cal B} = C_{{\cal B} {\cal A}} ~ {\rm dim} {\cal R}_{\cal A} ~,
\qquad\qquad
D_{{\cal A} {\cal B}} ~ {\rm dim}{\cal R}_{\cal B} = (D^{-1})_{{\cal B} {\cal A}} ~ {\rm dim} {\cal R}_{\cal A} ~.
\label{Fierz-dim}
\end{equation}
In particular, Eq.~(\ref{FierzSinglet}) implies $C_{{\cal A} \bullet}= C_{\bullet {\cal A}} ~ {\rm dim} {\cal R}_{\cal A}= {\rm dim} {\cal R}_{\cal A}/ {\rm dim} {\cal R}$.

In the special case of a (pseudo-)real representation ${\cal R}$, taking $\psi \sim {\cal R}$ and $\psi^\dagger \sim \overline{\cal R}$, one has $\overline{\psi}_i \equiv \psi^\dagger_j (\Omega_\epsilon)_{ji} \sim {\cal R}$,
where $\Omega_\epsilon$ is the invariant tensor establishing the equivalence of ${\cal R}$ and $\overline{\cal R}$,
which is symmetric ($\epsilon=+1$)  or antisymmetric ($\epsilon=-1$) in the case of real or pseudo-real representations, respectively. 
Therefore, the set of matrices $\{\Gamma^{\cal A}_a\}$  and $\{\tilde\Gamma^{\cal A}_a\}$ can be identified,
according to $\tilde\Gamma^{\cal A}_a = \Gamma^{\cal A}_a \Omega_\epsilon$.
In addition, the equality $\Omega_\epsilon \tilde\Gamma^{\cal A\dagger}_a = \epsilon \tilde\Gamma^{\cal A}_a \Omega_\epsilon$ holds, 
which implies in particular $(\psi \tilde\Gamma^{\cal A}_a \psi)^\dagger = \epsilon \overline\psi \tilde\Gamma^{\cal A}_a \overline\psi$.
Then, it is convenient to rewrite the Fierz transformations in Eq.~(\ref{Fierz-general}) [or, equivalently, Eq.~(\ref{Fierz-general-bis})]  
in terms of the interaction $({\cal R} \otimes {\cal R})_{\cal A} ({\cal R} \otimes {\cal R})_{\cal A}$,
\begin{equation}
\sum\limits_a (\tilde\Gamma^{a{\cal A}})_{ij} (\tilde\Gamma_a^{\cal A})_{kl}
= \sum\limits_{\cal B}^\sim C_{{\cal A} {\cal B}} \sum\limits_b (\tilde\Gamma^{b{\cal B}})_{il} (\tilde\Gamma_b^{\cal B})_{kj}
= \epsilon\sum\limits_{\cal B}^{\sim} D_{{\cal A} {\cal B}} \sum\limits_b (\tilde \Gamma^{b{\cal B}})_{ik} (\tilde \Gamma_b^{\cal B})_{jl}~.
\label{Fierz-general-ter}
\end{equation}
It follows immediately that the two sets of Fierz coefficients are related as 
\be
\epsilon D_{{\cal A} {\cal B}} = s_{\cal A}  C_{{\cal A} {\cal B}} s_{\cal B}~,
\label{CDsigns}
\ee
where $s_{\cal A,B} = \pm 1$ denotes, once again, the (anti-)symmetry of ${\cal R}_{\cal A,B}$ within ${\cal R} \otimes {\cal R}$.
In this (pseudo-)real case the singlet contraction corresponds to $\tilde\Gamma^\bullet_{ij} = (\Omega_\epsilon)_{ij}\sqrt{\alpha/{\rm dim}{\cal R}}$, therefore $s_\bullet = \epsilon$, and one recovers
Eq.~(\ref{FierzSinglet}).

\subsection{$SU(N)$ Fierz transformations}
\label{Fierz-transfos-SUN}

Let us derive the Fierz transformations associated to the fundamental representation of $SU(N)$ (see e.g. \cite{Buballa:2003qv}). 
In our model they are relevant for the flavour indexes, as the fermions $\psi^a$ and $X^f$ transform in the fundamental of $SU(4)$ and $SU(6)$,
respectively.

In the `quark-antiquark' channel, $(\overline{N}_a N^b)(\overline{N}_c N^d)\rightarrow (\overline{N}_a N^d)(\overline{N}_c N^b)$, one can employ
the completeness relation of Eq.~(\ref{Fierz-completeness-relation}) for $\overline{N}\otimes N$,
\begin{equation}
\sum_{I=1}^{N^2-1} (T^I)^{a}_{~b} (T^I)^{c}_{~d} + (T^0)^a_{~b} (T_0)^c_{~d}
= \frac{1}{2} \delta^{a}_{~d} \delta^{c}_{~b} ~,
\label{SUNcomp}
\end{equation}
where $T^I$ are the $(N^2-1)$ generators of $SU(N)$, $T^0\equiv  1\!\!1 / \sqrt{2N}$, and $\alpha=\ell (N)/2=\ell(\overline{N})/2= 1/2$
as we adopted the normalisation ${\rm Tr}(T^I T^J)=\delta^{IJ}/2$.
The first row of the Fierz-coefficient matrix $C_{\cal AB}$ is simply obtained by reshuffling the indexes in Eq.~(\ref{SUNcomp}),
\be
 (T^0)^{a}_{~b} (T^0)^{c}_{~d} =\frac{1}{N} (T^0)^{a}_{~d} (T^0)^{c}_{~b} +\frac{1}{N} \sum_I (T^I)^{a}_{~d}\:(T^I)^{c}_{~b}~,
\label{SUNfierz}\ee
The second row can be determined by imposing  $C^2\equiv 1\!\!1$, as follows from Eq.~(\ref{Fierz-sum}). Thus, one concludes that
\be
\begin{pmatrix}
\vspace*{0.1 cm} (T^0)^a_{~b} (T^0)^c_{~d} 
\\ 
\sum \limits_I (T^I)^a_{~b} \:(T^I)^c_{~d}
\end{pmatrix}
= C
\begin{pmatrix}
\vspace*{0.1 cm} (T^0)^a_{~d} (T^0)^c_{~b} 
\\ 
\sum \limits_I (T^I)^a_{~d} \:(T^I)^c_{~b}
\end{pmatrix}
= 
\begin{pmatrix}
\vspace*{0.2 cm}   \frac{1}{N} & \frac{1}{N} 
\\
\frac{N^2-1}{N} & -\frac{1}{N}
\end{pmatrix} 
\begin{pmatrix}
\vspace*{0.1 cm}
(T^0)^a_{~d} (T^0)^c_{~b} 
\\ 
\sum \limits_I (T^I)^a_{~d} \:(T^I)^c_{~b}
\end{pmatrix}~.
\ee

In the `quark-quark' channel,  $(\overline{N}_a N^b)(\overline{N}_c N^d)\rightarrow (\overline{N}_a \overline{N}_c)(N^b N^d)$, one needs also
the completeness relation for $N\otimes N$, that  involves $N(N+1)/2$ symmetric matrices $\Gamma_S^I$, and $N(N-1)/2$ antisymmetric matrices $\Gamma_A^I$,
\begin{equation}
 \sum_{I=1}^{N(N+1)/2} 
 (\Gamma^{I\dagger}_{S})^{ab} (\Gamma^I_S)_{cd}
 + \sum_{I=1}^{N(N-1)/2} 
 (\Gamma^{I\dagger}_{A})^{ab} (\Gamma^I_A)_{cd}
=\frac{1}{2} \delta^{a}_{~d} \delta_{~c}^{b}~.
\end{equation}
A convenient basis of (anti-)symmetric matrices is provided by $\Gamma^0 \equiv \Sigma_\epsilon T^0$, $\Gamma^I \equiv \Sigma_\epsilon T^I$, and $\Gamma^{\hat I} \equiv \Sigma_\epsilon T^{\hat I}$, 
where $(\Sigma_\epsilon)_{ab}$ is the invariant tensor of a maximal $SU(N)$ subgroup, 
which is $SO(N)$ in the case $\epsilon=+1$, and $Sp(N)$ in the case $\epsilon=-1$ (present only for $N$ even). Here the index $I$ runs over the subgroup generators only,
 and the index ${\hat I}$ spans the coset.
When $\epsilon=+1(-1)$, $\Sigma_\epsilon$ is a symmetric (antisymmetric) matrix and, according to Eq.~(\ref{Tacom}),  $\Gamma^0$ and $\Gamma^{\hat I}$ are symmetric (antisymmetric), while  $\Gamma^I$ are antisymmetric (symmetric).
Using this basis for the matrices $\Gamma^I_{S,A}$, one can construct explicitly the Fierz-coefficient matrix $D_{\cal AB}$,
\be
\begin{pmatrix}
\vspace*{0.1 cm}
(T^0)^a_{~b} (T^0)_{~d}^{c} 
\\ 
\sum \limits_I (T^I)^a_{~b} \:(T^I)_{~d}^{c}
\end{pmatrix}
= D \begin{pmatrix}
 \vspace*{0.1 cm} \sum \limits_I (\Gamma_S^{I\dagger})^{ac} (\Gamma_S^I)_{bd} 
 \\ 
 \sum \limits_I (\Gamma_A^{I\dagger})^{ac} (\Gamma_A^I)_{bd}
\end{pmatrix}
= 
\begin{pmatrix}
\vspace*{0.2 cm}  \frac{1}{N} & -\frac{1}{N} \\
\frac{N-1}{N} &  \frac{N+1}{N} 
\end{pmatrix} 
\begin{pmatrix}
 \vspace*{0.1 cm} \sum \limits_I (\Gamma_S^{I\dagger})^{ac} (\Gamma_S^I)_{bd} 
 \\ 
 \sum \limits_I (\Gamma_A^{I\dagger})^{ac} (\Gamma_A^I)_{bd}
\end{pmatrix}~.
\label{D-SU}
\ee
For example, the first row of  $D_{\cal AB}$ can be obtained from Eq.~(\ref{SUNfierz}) by contracting with $(\Sigma_\epsilon)^{dd'}(\Sigma_\epsilon)_{c'c}$,
and inverting appropriate pairs of (anti-)symmetrised indexes: the result agrees with Eq.~(\ref{FierzSinglet}).
The second row is determined e.g. by Eq.~(\ref{Fierz-dim}), up to an overall sign, that can be fixed once again by (anti-)symmetrising over appropriate indexes.

\subsection{$Sp(2N)$ Fierz transformations} 
\label{Fierz-transfoSP2N}

Let us derive the Fierz transformations associated to the hypercolour representations of the fermions $\psi_i$ and $X_{ij}$,
that is, ${\tiny \yng(1)}$ and ${\tiny \yng(1,1)}$ respectively. The group $Sp(2N)$ is a subgroup of $SU(2N)$, corresponding to the vacuum direction $\Sigma_-\equiv \Omega$,
defined in Eq.~(\ref{Omega}).
Taking advantage of Eq.~(\ref{Tacom}), one can decompose the $U(2N)$ completeness relation (\ref{SUNcomp}) into two parts, corresponding to the $Sp(2N)$ subalgebra and its coset,
\begin{eqnarray}
\sum \limits_{I=1}^{N(2N+1)} (T^I)_{ij} \:(T^I)_{kl} &=& \frac{1}{4} (\delta_{il}\delta_{kj} -\Omega_{ik}\Omega_{jl})~:\qquad Sp(2N)~,
\label{SPNcomp} 
\\
\sum \limits_{\hat{I}=1}^{(2N+1)(N-1)} (T^{\hat{I}})_{ij} \:(T^{\hat{I}})_{kl}  + (T^0)_{ij} \:(T^0)_{kl}
&=& \frac{1}{4} (\delta_{il}\delta_{kj} +\Omega_{ik}\Omega_{jl}) 
~:\qquad U(2N)/Sp(2N)~.
\label{SUN-SPNcomp}
\end{eqnarray}

The product of two fundamental representations of $Sp(2N)$ reads
\begin{equation}
\Yvcentermath1
{\small\yng(1)}\times {\small \yng(1)}= \left.\bullet\right._a ~+~ \left.{\small\yng(2)}\right._s ~+~ \left.{\small\yng(1,1)}\right._a ~,
\label{product-rep1}
\end{equation}
where the bullet stands for the singlet and the subscripts indicate whether the contraction is symmetric or antisymmetric under the exchange of the two factors.
These representations have dimensions
\begin{equation}
\Yvcentermath1
d \left({\tiny \yng(1)} \right)= 2N~,
\qquad
d \left(\bullet \right)= 1~,\qquad
d \left({\tiny \yng(2)} \right)= N(2N+1)~,
\qquad
d \left({\tiny \yng(1,1)} \right)=N(2N-1)-1=(2N+1)(N-1)~.
\label{dim1}
\end{equation}
Note that, for $N=1$, the two-index antisymmetric representation is absent.
The two indexes in ${\tiny \yng(1)}_i {\tiny \yng(1)}_j$ are contracted by an appropriate set of (anti-)symmetric matrices $\t\Gamma^a_{\cal A}$, that can be conveniently chosen as
\be
\t\Gamma_{\bullet} \equiv \Omega T^0 = \frac{\Omega}{\sqrt{4N}}~,
\qquad\qquad
  \t\Gamma^I_{\tiny\yng(2)}\equiv \Omega T^I~,
  \qquad\qquad  
  \t\Gamma^{\hat I}_{\tiny\yng(1,1)}\equiv \Omega T^{\hat I}~,
\ee
in one-to-one correspondence with 
the generators of $U(2N)$.
Multiplying (\ref{SPNcomp}) and (\ref{SUN-SPNcomp}) by $\Omega_{mi} \Omega_{nk}$ one obtains useful equalities to determine the Fierz transformations of $({\tiny \yng(1)}_i {\tiny \yng(1)}_j)({\tiny \yng(1)}_k {\tiny \yng(1)}_l)$.
Thus, the matrix of Fierz coefficients for the channel $(il)(kj)$, $C_{\cal AB}$, can be fully determined in agreement with the general results of section \ref{Fierz-general-section}:
\begin{equation}
\begin{pmatrix}
\vspace*{0.1 cm} (\Omega T^0)_{ij} (\Omega T^0)_{kl}
\\
\vspace*{0.1 cm} \sum \limits_I (\Omega T^I)_{ij} (\Omega T^I)_{kl}
\\
\sum \limits_{\hat{I}} (\Omega T^{\hat{I}})_{ij} (\Omega T^{\hat{I}})_{kl}
\end{pmatrix}
=
\begin{pmatrix}
\vspace*{0.2 cm} \frac{1}{2N} & \frac{1}{2N} & \frac{1}{2N}
\\
\vspace*{0.2 cm} \frac{2N+1}{2} & -\frac{1}{2} & \frac{1}{2}
\\
\frac{(2N+1)(N-1)}{2N} & \frac{N-1}{2N} & -\frac{N+1}{2N}
\end{pmatrix}
\begin{pmatrix}
\vspace*{0.1 cm} (\Omega T^0)_{il} (\Omega T^0)_{kj}
\\
\vspace*{0.1 cm} \sum \limits_I (\Omega T^I)_{il} (\Omega T^I)_{kj}
\\
\sum \limits_{\hat{I}} (\Omega T^{\hat{I}})_{il} (\Omega T^{\hat{I}})_{kj}
\end{pmatrix}~,
\label{fierzspqqbar}
\end{equation}
According to Eq.~(\ref{CDsigns}), the Fierz coefficients in the channel $(ik)(jl)$ are given by  $D_{\cal AB} = - C_{\cal AB}$ when both ${\cal A}$ and ${\cal B}$ are (anti-)symmetric contractions, 
and $D_{\cal AB} = C_{\cal AB}$ otherwise.

We can now determine the coefficients $\kappa_{A,C,D}$ of the four-fermion operators in the $\psi$-sector, which are defined by Eqs.~(\ref{LSphys}) and (\ref{L4Fv}), 
assuming that the dynamics is well approximated by the $\psi$-sector current-current operator of Eq.~(\ref{LUV0}), with coefficient $\kappa_{UV}$. 
Note that the 't Hooft operator with coefficient $\kappa_B$, defined by the second line of Eq.~(\ref{LSphys}), is not generated by the current-current interaction, as the latter preserves the anomalous $U(1)_\psi$ symmetry,
therefore the size of $\kappa_B$ is unrelated to $\kappa_{UV}$.
On the contrary, the sizes of $\kappa_{A,B,C}$ can be related to $\kappa_{UV}$ 
by performing the pertinent set of Fierz transformations over Lorentz, $SU(4)$ flavour, and $Sp(2N)$ hypercolour indexes.
Naively, with this procedure the current-current operator is recast into a sum over several operators: 
those with two hypercolour-singlet fermion bilinears, which correspond to physical meson states, 
plus those with two hypercolour-non-singlet fermion bilinears.
The former operators receive a coefficient
\begin{equation}
\kappa_A=\kappa_C= \kappa_D= \frac{2N+1}{4N} \kappa_{UV}~.
\label{link-couplings-electroweak}
\end{equation}
However, the latter operators could also contribute to these couplings, by further Fierz transformations. Therefore, the above equalities
cannot be firmly established on this basis.
Fortunately, there exists a unique way to express the current-current operator in terms of hypercolour-singlet fermion bilinears only, by using the identity
\begin{equation}
\sum\limits_I \left(\Omega T^I \right)_{ij}  \left(\Omega T^I \right)_{kl}=\frac{1}{4}
\left(\Omega_{il} \Omega_{kj}-\Omega_{ik}\Omega_{jl} \right)~,
\label{Fierz-identity-fundamental}
\end{equation}
which is obtained e.g. by considering the first row of Eq.~(\ref{fierzspqqbar}) and symmetrising over the indexes $(il)$, or equivalently by
multiplying the $Sp(2N)$ completeness relation (\ref{SPNcomp}) by $\Omega_{i'i}\Omega_{k'k}$.
Employing this relation we obtain
\begin{equation}
\kappa_A= \kappa_C= \kappa_D= \frac{1}{2} \kappa_{UV}~.
\label{link-couplings-electroweak-2}
\end{equation}
Therefore, in the current-current approximation, the scalar coupling $\kappa_{A}$ and the vector couplings $\kappa_{C,D}$ are equal and $N$-independent when $N$ becomes large, as $\kappa_{UV}$ is.
Notice that the naive relations in Eq.~(\ref{link-couplings-electroweak}) were correct al leading order in $1/N$.
The equality between vector and scalar couplings also holds in the standard NJL model for QCD \cite{Barnard:2013zea}.

Let us now analyse the product of two $Sp(2N)$ two-index traceless antisymmetric representations ${\tiny \yng(1,1)}$, that exist only for $N>1$,
and are relevant for the colour sector of our model.
The tensor product,
\begin{equation}
\Yvcentermath1
{\small\yng(1,1)}\times {\small\yng(1,1)}= \left.\bullet\right._s ~+~ \left.{\small\yng(2)}\right._a~+ ~ \left.{\small\yng(1,1)}\right._s~+ ~ \left.{\small\yng(2,2)}\right._s~+ ~ \left.{\small\yng(1,1,1,1)}\right._s~+ ~ \left.{\small\yng(2,1,1)}\right._a~,
\label{product-rep2}
\end{equation}
contains three four-index representations, of dimensions
\be
\Yvcentermath1
d \left({\tiny \yng(2,2)} \right)= \frac{N}{3} \left(4 N^3-7 N +3 \right),~
d \left({\tiny \yng(1,1,1,1)} \right)=\frac{N}{6} \left(4 N^3-12 N^2-N+3 \right),~
d \left({\tiny \yng(2,1,1)} \right)= \frac{1}{2} \left(4 N^4-4N^3-9N^2 +N + 2 \right).
\label{dim-X-fermion}
\ee
These numbers can be derived taking into account the symmetry properties of each representation in Eq.~(\ref{product-rep2}),
and subtracting the dimensions of the smaller representations, obtained by taking traces, as given in Eq.~(\ref{dim1}).
Note that, for $N=2$, the third, fifth and sixth representation on the right-hand side of Eq.~(\ref{product-rep2}) are absent: $5\times 5 = 1_s+10_a+14_s$. For $N=3$, the fifth representation only is absent:
$14 \times 14 = 1_s+21_a+14_s+90_s+70_a$. Finally, for $N>3$ all the components of the tensor product exist.

The indexes in ${\tiny \yng(1,1)}_{ij} {\tiny \yng(1,1)}_{kl}$ are contracted into the representation ${\cal R}$ by a set of tensors $(\t\Gamma^a_{\cal R})_{ijkl}$,
with $a=1,\dots,\rm{dim}{\cal R}$. 
Equivalently, one can use a single index running over the $(2N+1)(N-1)$ components of ${\tiny \yng(1,1)}$,
\begin{equation}
 X_{li} (\t\Gamma_{\cal R}^{a})_{ijkl} X_{jk} = X_{\hat{I}} (\t\Gamma^{a}_{\cal R})_{\hat{I} \hat{J}} X_{\hat{J}}~.
\end{equation}
where $X_{ij}$ and $X_{\hat I}$ are related by Eq.~(\ref{XXrelation}).
In this notation, the completeness relation reads 
\begin{equation}
\sum\limits_{\cal R}\sum_a (\t\Gamma^{a}_{\cal R})_{\hat{I} \hat{J}} (\t\Gamma^{a}_{\cal R})_{\hat{K} \hat{L}}
= \frac{1}{2} \ell({\Yvcentermath1 \tiny \yng(1,1)}) \delta_{\hat{I} \hat{L}}  \delta_{\hat{K} \hat{J}}
= (N-1)\delta_{\hat{I} \hat{L}}  \delta_{\hat{K} \hat{J}} ~,
\qquad
{\cal R}=\bullet, \Yvcentermath1 {\tiny  \yng(2), \yng(1,1),\yng(2,2), \yng(1,1,1,1),\yng(2,1,1)} ~.
\label{completeness-relation-X}
\end{equation}
In fact, the set of matrices $\{\t\Gamma^{a}_{\cal R} \}$ corresponds to the generators of the group $U\left[(2N+1)(N-1) \right]$,
normalised as ${\rm Tr}[\t\Gamma^{a}_{\cal R} \t\Gamma^{b}_{\cal R'}]= \frac{1}{2}\ell({\Yvcentermath1 \tiny \yng(1,1)}) 
\delta_{\cal R R'}\delta^{ab}$.
Let us provide the explicit form of these matrices for the smallest representations.
The singlet contraction is given by
\begin{equation}
(\t\Gamma_\bullet)_{ijkl}= \frac{1}{\sqrt{2N+1}} \Omega_{ij} \Omega_{kl}~,
\qquad\qquad
(\t\Gamma_\bullet)_{\hat{I} \hat{J}}= \frac{1}{\sqrt{2N+1}} \delta_{\hat{I} \hat{J}}~.
\label{Contraction1}
\end{equation}
The adjoint contraction, already employed in section \ref{$Sp(2N)$ current-current operators}, is given by
\begin{equation}
(\t\Gamma_{{\Yvcentermath1 \tiny \yng(2)}}^{K})_{ijkl}= 
(\Omega T^I)_{ij} \Omega_{kl} - \Omega_{ij}  (\Omega T^I)_{kl}~, 
\qquad\qquad
(\t\Gamma_{{\Yvcentermath1 \tiny \yng(2)}}^{K})_{\hat{I}\hat{J}} 
=-if^{\hat{I}\hat{J}K} = -2 {\rm Tr} ([T^{\hat I} , T^{\hat J} ] T^K)~.
\label{Contraction2}
\end{equation}
The two-index antisymmetric contraction has a similar structure, with the unbroken generators $T^I$ replaced by the broken ones $T^{\hat{I}}$, 
\begin{equation}
(\t\Gamma_{{\Yvcentermath1 \tiny \yng(1,1)}}^{\hat{K}})_{ijkl}= 
(\Omega T^{\hat{K}})_{ij} \Omega_{kl}
 +\Omega_{ij}  (\Omega T^{\hat{K}})_{kl}~, 
\qquad\qquad
(\t\Gamma_{{\Yvcentermath1 \tiny \yng(1,1)}}^{\hat{K}})_{\hat{I}\hat{J}}= d^{\hat{I} \hat{J} \hat{K}}
= 2 {\rm Tr} (\{T^{\hat{I}} , T^{\hat{J}} \} T^{\hat{K}})~.
\label{Contraction3}
\end{equation}
One can easily check that the symmetry properties of the contractions in Eqs.~(\ref{Contraction1}), (\ref{Contraction2}) and (\ref{Contraction3}) agree with those indicated in Eq.~(\ref{product-rep2}).

The singlet Fierz coefficients in the channel $(\hat{I} \hat{L}) (\hat{K} \hat{J})$, $C_{\bullet{\cal R}}$, are easily determined from the completeness relation (\ref{completeness-relation-X}), 
in agreement with Eq.~(\ref{FierzSinglet}).
The coefficients $C_{{\cal R}\bullet}$ are determined in turn by Eq.~(\ref{Fierz-dim}). Thus, we can write 
\begin{equation}
\begin{pmatrix}
\vspace*{0.1 cm}  (\t\Gamma_\bullet)_{\hat{I} \hat{J}} (\t\Gamma_\bullet)_{\hat{K} \hat{L}}
\\
\vspace*{0.1 cm} \sum \limits_{a} (\t\Gamma_{{\tiny \yng(2)}}^{a})_{\hat{I} \hat{J}} (\t\Gamma_{{\tiny \yng(2)}}^{a})_{\hat{K} \hat{L}}
\\
\vspace*{0.1 cm} \sum \limits_{a} (\t\Gamma_{{\tiny \yng(1,1)}}^{a})_{\hat{I} \hat{J}} (\t\Gamma_{{\tiny \yng(1,1)}}^{a})_{\hat{K} \hat{L}}
\\
\vspace*{0.1 cm} \sum \limits_{a} (\t\Gamma_{{\tiny \yng(2,2)}}^{a})_{\hat{I} \hat{J}} (\t\Gamma_{{\tiny \yng(2,2)}}^{a})_{\hat{K} \hat{L}}
\\
\vspace*{0.1 cm} \sum \limits_{a} (\t\Gamma_{{\tiny \yng(1,1,1,1)}}^{a})_{\hat{I} \hat{J}} (\t\Gamma_{{\tiny \yng(1,1,1,1)}}^{a})_{\hat{K} \hat{L}}
\\
\sum \limits_{a} (\t\Gamma_{{\tiny \yng(2,1,1)}}^{a})_{\hat{I} \hat{J}} (\t\Gamma_{{\tiny \yng(2,1,1)}}^{a})_{\hat{K} \hat{L}}
\end{pmatrix}
=
\begin{pmatrix}
\vspace*{0.1 cm}
\frac{1}{d \left({\Yvcentermath1 \tiny \yng(1,1)}\right)} & \frac{1}{d \left({\Yvcentermath1 \tiny \yng(1,1)}\right)} & \frac{1}{d \left({\Yvcentermath1 \tiny \yng(1,1)}\right)} & \frac{1}{d \left({\Yvcentermath1 \tiny \yng(1,1)}\right)} & \frac{1}{d \left({\Yvcentermath1 \tiny \yng(1,1)}\right)} & \frac{1}{d \left({\Yvcentermath1 \tiny \yng(1,1)}\right)}
\\
\vspace*{0.1 cm}
\frac{d \left({\Yvcentermath1 \tiny \yng(2)}\right)}{d \left({\Yvcentermath1 \tiny \yng(1,1)}\right)} & \cdots & \cdots & \cdots & \cdots & \cdots
\\
\vspace*{0.1 cm}
1 & \cdots & \cdots & \cdots & \cdots & \cdots
\\
\vspace*{0.1 cm}
\frac{d \left({\Yvcentermath1 \tiny \yng(2,2)}\right)}{d \left({\Yvcentermath1 \tiny \yng(1,1)}\right)} & \cdots & \cdots & \cdots & \cdots & \cdots
\\
\vspace*{0.1 cm}
\frac{d \left({\Yvcentermath1 \tiny \yng(1,1,1,1)}\right)}{d \left({\Yvcentermath1 \tiny \yng(1,1)}\right)} & \cdots & \cdots & \cdots & \cdots & \cdots
\\
\vspace*{0.1 cm}
\frac{d \left({\Yvcentermath1 \tiny \yng(2,1,1)}\right)}{d \left({\Yvcentermath1 \tiny \yng(1,1)}\right)} & \cdots & \cdots & \cdots & \cdots & \cdots
\end{pmatrix}
\begin{pmatrix}
\vspace*{0.1 cm}  (\t\Gamma_\bullet)_{\hat{I} \hat{L}} (\t\Gamma_\bullet)_{\hat{K} \hat{J}}
\\
\vspace*{0.1 cm} \sum \limits_{a} (\t\Gamma_{{\tiny \yng(2)}}^{a})_{\hat{I} \hat{L}} (\t\Gamma_{{\tiny \yng(2)}}^{a})_{\hat{K} \hat{J}}
\\
\vspace*{0.1 cm} \sum \limits_{a} (\t\Gamma_{{\tiny \yng(1,1)}}^{a})_{\hat{I} \hat{L}} (\t\Gamma_{{\tiny \yng(1,1)}}^{a})_{\hat{K} \hat{J}}
\\
\vspace*{0.1 cm} \sum \limits_{a} (\t\Gamma_{{\tiny \yng(2,2)}}^{a})_{\hat{I} \hat{L}} (\t\Gamma_{{\tiny \yng(2,2)}}^{a})_{\hat{K} \hat{J}}
\\
\vspace*{0.1 cm} \sum \limits_{a} (\t\Gamma_{{\tiny \yng(1,1,1,1)}}^{a} )_{\hat{I} \hat{L}} (\t\Gamma_{{\tiny \yng(1,1,1,1)}}^{a})_{\hat{K} \hat{J}}
\\
\sum \limits_{a} (\t\Gamma_{{\tiny \yng(2,1,1)}}^{a})_{\hat{I} \hat{L}} (\t\Gamma_{{\tiny \yng(2,1,1)}}^{a})_{\hat{K} \hat{J}}
\end{pmatrix}~.
\label{Fierz2}
\end{equation}
One needs further algebraic manipulations to determine the non-singlet Fierz coefficients $C_{\cal R R'}$, 
which anyhow will not be needed for our purposes.
For concreteness, let us display the explicit result in the case $N=2$, where there are only three representations in the tensor product
${\tiny \yng(1,1)}\times {\tiny \yng(1,1)}$. Using repeatedly the completeness relation and  the (anti-)symmetrisation over appropriate pairs of indexes,
we conclude that the matrix $C$ in the case $N=2$ takes the form
\begin{equation}
\begin{pmatrix}
\vspace*{0.1 cm}  (\t\Gamma_\bullet)_{\hat{I} \hat{J}} (\t\Gamma_\bullet)_{\hat{K} \hat{L}}
\\
\vspace*{0.1 cm} \sum \limits_{a} (\t\Gamma_{{\tiny \yng(2)}}^{a})_{\hat{I} \hat{J}} (\t\Gamma_{{\tiny \yng(2)}}^{a})_{\hat{K} \hat{L}}
\\
\vspace*{0.1 cm} \sum \limits_{a} (\t\Gamma_{{\tiny \yng(2,2)}}^{a})_{\hat{I} \hat{J}} (\t\Gamma_{{\tiny \yng(2,2)}}^{a})_{\hat{K} \hat{L}}
\end{pmatrix}
=
\begin{pmatrix}
\vspace*{0.1 cm} \frac{1}{5} & \frac{1}{5} & \frac{1}{5}
\\ 
\vspace*{0.1 cm} 2 & \frac 12 & -\frac 12
\\ 
\vspace*{0.1 cm} \frac{14}{5} & - \frac{7}{10} & \frac{3}{10}
\end{pmatrix}
\begin{pmatrix}
\vspace*{0.1 cm}  (\t\Gamma_\bullet)_{\hat{I} \hat{L}} (\t\Gamma_\bullet)_{\hat{K} \hat{J}}
\\
\vspace*{0.1 cm} \sum \limits_{a} (\t\Gamma_{{\tiny \yng(2)}}^{a})_{\hat{I} \hat{L}} (\t\Gamma_{{\tiny \yng(2)}}^{a})_{\hat{K} \hat{J}}
\\
\vspace*{0.1 cm} \sum \limits_{a} (\t\Gamma_{{\tiny \yng(2,2)}}^{a})_{\hat{I} \hat{L}} (\t\Gamma_{{\tiny \yng(2,2)}}^{a})_{\hat{K} \hat{J}}
\end{pmatrix}~.
\label{FierzN=2}
\end{equation}
The Fierz coefficients $D_{\cal R R'}$ in the channel $(\hat{I} \hat{K}) (\hat{J} \hat{L})$ are determined by Eq.~(\ref{CDsigns}), with $\epsilon=+1$
as $\tiny\yng(1,1)$ is a real representation.
Since we aim to rewrite the $X$-sector current-current operator of Eq.~(\ref{LUV0}) in terms of hypercolur-singlet fermion bilinears, 
the relevant Fierz coefficients are 
\begin{equation}
C_{{\Yvcentermath1 \tiny \yng(2)}~\bullet } =-D_{{\Yvcentermath1 \tiny \yng(2)}~\bullet } 
=\frac{N}{N-1}~.
\end{equation}

In analogy with the above procedure in the $\psi$-sector, one can try to determine the coefficients $\kappa_{A6,C6,D6}$ of the four-fermion operators in the $X$-sector, which are defined by Eqs.~(\ref{L4F-scal-color}) and (\ref{kcd6}).
If one applies a pertinent Fierz transformation, over Lorentz, $SU(6)$ and $Sp(2N)$ indexes, to the $X$-sector current-current operator in Eq.~(\ref{LUV0}), one obtains
\begin{equation}
\kappa_{A6}=\kappa_{C6}=  \kappa_{D6}= \kappa_{UV}~. 
\label{link-couplings-coloured}
\end{equation}
This indicates that the scalar and vector operators of the coloured sector receive the same coefficient, that is twice as large as for the corresponding operators of the electroweak sector,
see Eq.~(\ref{link-couplings-electroweak-2}).
However, at the same time $\kappa_{UV}$ also contributes to other operators, that involve hypercolour-non-singlet fermion bilinears,
therefore the above relations are ambiguous, as they rely on a specific recasting of the current-current operator, that is not unique. 
Another possible recasting is obtained by anti-symmetrising Eq.~(\ref{completeness-relation-X}), with respect to the pair of indexes $(\hat K\hat L)$,
to remove the symmetric components of Eq.~(\ref{product-rep2}),
\begin{equation}
\sum_a
(\t\Gamma_{{\Yvcentermath1 \tiny \yng(2)}}^{a})_{\hat{I}\hat{J}}
(\t\Gamma_{{\Yvcentermath1 \tiny \yng(2)}}^{a})_{\hat{K}\hat{L}}
+ \sum_a
(\t\Gamma_{{\Yvcentermath1 \tiny \yng(2,1,1)}}^{a})_{\hat{I}\hat{J}}
(\t\Gamma_{{\Yvcentermath1 \tiny \yng(2,1,1)}}^{a})_{\hat{K}\hat{L}}
=
\frac{(2N+1)(N-1)}{2} \left[(\t\Gamma_{ \bullet})_{\hat{I} \hat{L}} (\t\Gamma_{ \bullet})_{\hat{K} \hat{J}}
-
(\t\Gamma_{ \bullet})_{\hat{I} \hat{K}} (\t\Gamma_{ \bullet})_{\hat{J} \hat{L}}\right]
~.
\label{Fierz-2index-second-approach}
\end{equation}
This relation is the analog of Eq.~(\ref{Fierz-identity-fundamental}), associated to the tensor product ${\tiny \yng(1)}\times {\tiny \yng(1)}$. 
In general, this procedure does not allow to express the current-current contraction in terms of singlet-singlet contractions only, 
because the product ${\tiny \yng(1,1)}\times {\tiny \yng(1,1)}$ contains another antisymmetric representation, besides the adjoint.
The exception is the case $N=2$, where the second term on the left-hand side of  Eq.~(\ref{Fierz-2index-second-approach}) is absent.
If one neglects this second term even for  $N>2$,
the relation between the current-current operator and the singlet-singlet operators becomes
\begin{equation}
\kappa_{A6}=\kappa_{C6}=\kappa_{D6}=\frac{(2N+1)(N-1)^2}{2N}\kappa_{UV}~.
\label{relation-coupling-SU6-2}
\end{equation}
Note that these couplings can be much larger than those in Eq.(\ref{link-couplings-coloured}), when $N$ is large.
The problem is that the current-current operator contains terms leading in $1/N$, that cannot be written as singlet-singlet contractions only, except for $N=2$.
In the latter case, Eq.~(\ref{relation-coupling-SU6-2}) is exact and its right-hand side reads $5\kappa_{UV}/4$, to be compared with
Eq.~(\ref{link-couplings-electroweak-2}) in the electroweak sector.

We conclude that, for $N>2$, the strength of the coloured-sector couplings cannot
be fixed in terms of $\kappa_{UV}$, and we treat it as a free parameter. In particular, $\kappa_{A6}$ is independent from the strength of the electroweak-sector coupling $\kappa_A$:
in our phenomenological analysis we take $\kappa_{A6}\sim\kappa_A$, such that the domain of validity of the NJL calculations is similar in the two sectors, and the NJL predictions can be compared.
On the other hand, the equality between the scalar and vector couplings in each sector is a solid prediction of the current-current approximation, that holds independently from their absolute sizes.
Finally, we remind that all predictions discussed in this appendix depend on the validity of the effective Lagrangian of Eq.~(\ref{LUV0}), 
that relies on the `ladder' approximation for the hypercolour dynamics. 
Therefore significant departures from these predictions cannot be excluded.

\bibliographystyle{JHEP}
\bibliography{biblio}

\end{document}